\documentclass[11pt]{article}

\usepackage[skip=14pt plus2pt, indent=0pt]{parskip}
\usepackage{verbatim}
\usepackage{epsfig,graphics}
\usepackage {graphicx}
\usepackage {epsfig}
\usepackage {subfigure}
\usepackage {tabularx} 
\usepackage{rotate}	
\usepackage{slashed}
\usepackage{enumerate}

\usepackage{amsmath}

\usepackage{amsfonts}
\usepackage{amssymb}
\usepackage{graphicx}
\usepackage{xcolor}
\usepackage{cite}

\usepackage{fancyhdr}
\usepackage[hidelinks]{hyperref}

\usepackage{braket}
\usepackage{cancel}

\def\beq{\begin{equation}}
	\def\eeq{\end{equation}}
\def\beqa{\begin{eqnarray}}
	\def\eeqa{\end{eqnarray}}

\def\Mpl{M_{\text{pl}}}
\def\Mpd{M_{\text{pl,}\, d}}
\def\Mpf{M_{\text{pl,}\, 4}}
\def\Mpt{M_{\text{pl,}\,  10}}

\def\MKK{m_{\text{KK}}}
\def\LUV{\Lambda_{\text{UV}}}
\def\LQG{\Lambda_{\text{QG}}}
\def\Ns{N_s}
\def\Ms{M_s}
\def\Ntot{N_{\text{tot}}}
\def\MBH{M_{\text{BH}}}

\def\varphid{\varphi_{d}}
\def\gsd{g_{s,\,d}}
\def\Mt{m_{\text{tower}}}
\def\Mti{m_{\text{tower,}\, i}}
\def\peff{p_{\text {eff }}}
\def\Mteff{m_{\text {tower, eff }}}
\def\fdim{2^{\lfloor d/2 \rfloor}}

\catcode`\@=11   
\newdimen\@rotdimen
\newbox\@rotbox  

\def\@vspec#1{\special{ps:#1}}
\def\@rotstart#1{\@vspec{gsave currentpoint currentpoint translate
		#1 neg exch neg exch translate}}
\def\@rotfinish{\@vspec{currentpoint grestore moveto}}
\def\@rotr#1{\@rotdimen=\ht#1\advance\@rotdimen by\dp#1%
	\hbox to\@rotdimen{\hskip\ht#1\vbox to\wd#1{\@rotstart{90 rotate}%
			\box#1\vss}\hss}\@rotfinish}
\def\@rotl#1{\@rotdimen=\ht#1\advance\@rotdimen by\dp#1%
	\hbox to\@rotdimen{\vbox to\wd#1{\vskip\wd#1\@rotstart{270 rotate}%
			\box#1\vss}\hss}\@rotfinish}%
\def\@rotu#1{\@rotdimen=\ht#1\advance\@rotdimen by\dp#1%
	\hbox to\wd#1{\hskip\wd#1\vbox to\@rotdimen{\vskip\@rotdimen
			\@rotstart{-1 dup scale}\box#1\vss}\hss}\@rotfinish}%
\def\@rotf#1{\hbox to\wd#1{\hskip\wd#1\@rotstart{-1 1 scale}%
		\box#1\hss}\@rotfinish}%
\def\rotate{\@ifnextchar[{\@rotate}{\@rotate[l]}}
\def\@rotate[#1]#2{\setbox\@rotbox=\hbox{#2}\@nameuse{@rot#1}\@rotbox}

\catcode`\@=12

\topmargin
-1.5cm
\textwidth
15.5cm
\textheight
23.5cm
\oddsidemargin
0.7cm
\evensidemargin
0.7cm

\begin{document}
\makeatletter
\@addtoreset{equation}{section}
\makeatother
\renewcommand{\theequation}{\thesection.\arabic{equation}}
\pagestyle{empty}
\vspace{-0.2cm}
\rightline{IFT-UAM/CSIC-22-149}
\vspace{1.2cm}
\begin{center}
		
		\LARGE{ The Emergence Proposal  in Quantum Gravity \\  and the Species Scale \\
			[13mm]}
		
		\large{ A. Castellano$^\clubsuit$, A. Herr\'aez$^\diamondsuit$ and L.E. Ib\'a\~nez$^\clubsuit$
			\\[12mm]}
		\small{
			$^\clubsuit$ {Departamento de F\'{\i}sica Te\'orica
				and Instituto de F\'{\i}sica Te\'orica UAM/CSIC,\\
				Universidad Aut\'onoma de Madrid,
				Cantoblanco, 28049 Madrid, Spain}  \\[5pt]
			$^\diamondsuit$  {Institut de Physique Th\'eorique, Universit\'e Paris Saclay, CEA, CNRS\\
				Orme des Merisiers, 91191 Gif-sur-Yvette CEDEX, France} 
			\\[8mm]
		}
		\small{\bf Abstract} \\[6mm]
	\end{center}
	\begin{center}
		\begin{minipage}[h]{15.60cm}
				In the Emergence Proposal in Quantum Gravity it is conjectured that all light-particle kinetic terms are absent in the fundamental ultraviolet theory and are generated by quantum corrections in the infrared. It has been argued that this may provide for some microscopic understanding of the Weak Gravity and Distance conjectures. In the present paper we take the first steps towards a systematic study of Emergence in the context of string theory. We emphasize the crucial role of the species scale in any effective field theory coupled to gravity, and discuss its computation in string theory and general systems with light towers of states. We then introduce the notion of Emergence and  show how kinetic terms for moduli, gauge bosons and fermions may be generated. One-loop computations play an important role in Emergence, so we present detailed calculations in $d$ spacetime dimensions for the wave-function renormalization of scalars, vectors and fermions. We extend and check the Emergence Proposal in a number string vacua, including 4d $\mathcal{N}=2$ theories arising from type IIA on a CY$_3$, where the towers at strong coupling are comprised by D0 and (wrapped) D2-branes, and also elaborate on how instanton corrections would fit within the emergence picture. Higher dimensional examples are also discussed, including 6d and 7d models arising from F-/M-theory on an elliptic CY$_3$ or a $K3$ surface. We also consider 10d string theories and study in some detail the emergence mechanism in type IIA. We show as well how the flux potential in 4d may be obtained from the emergence prescription, by analyzing the corresponding decompactification limits to M-theory. We find that the required kinetic terms for the dual 3-form fields can arise upon integrating out towers of massive gravitini (and bosonic superpartners). Our analysis renders support to the Emergence Proposal, and to the idea that infinite distance singularities may arise in Quantum Gravity as an intrinsic infrared phenomenon.
		\end{minipage}
	\end{center}
	\newpage
\pagestyle{empty}
\renewcommand{\thefootnote}{\arabic{footnote}}
\setcounter{footnote}{0}


	
\tableofcontents
	
\pagestyle{empty}
\newpage
\setcounter{page}{1}
\pagestyle{plain}
\section{Introduction}
\label{s:intro}
	
The observed universe seems to be in a weak coupling regime, although it is not far away from strong coupling. For instance, the electro-weak (EW) coupling is $g_{\text{ew}}\simeq 0.7$ at the EW scale itself. One can then envisage a situation in which the underlying fundamental theory is actually strongly coupled in the ultraviolet (UV) and the observed perturbatively weak interactions in our Universe only arise in the infrared (IR) after following the renormalization group (RG) flow of the couplings below some fundamental UV scale. Thus, in some sense, one could say that those kinetic terms would be {\it emergent}. This may indeed happen if the theory is not asymptotically free below the fundamental scale due to e.g. the presence of a large enough number of very massive vector-like particles. The fact that the fundamental theory could be strongly coupled may be effectively described by postulating vanishing (or at least very small) kinetic `metrics' for all the light states in the theory, keeping the interaction terms fixed (and naturally being of $\mathcal{O}(1)$). Then one could hope to generate such kinetic terms deep in the IR by quantum loop corrections involving e.g. heavy particles. As appealing as this idea may seem, to actually make it work is however not as easy as it sounds since loop corrections to the field metrics are in general divergent and hence cut-off dependent (upon restricting the regime of validity of our theory below some energy scale, $\Lambda_{\text{UV}}$). Counterterms for these divergences at the cut-off scale would in principle be needed (if one tries to extend the theory up to the continuum limit), against the original assumption that no (significant) kinetic terms were present at the fundamental scale.
	
The possibility of an IR generation of kinetic terms has been recently reconsidered in the context of the {\it Swampland Program} \cite{Vafa:2005ui} (see \cite{Brennan:2017rbf,Palti:2019pca,vanBeest:2021lhn,Grana:2021zvf} for reviews). It has been argued that an emergence principle of kinetic terms for both moduli and gauge bosons in theories of Quantum Gravity (QG) may provide for a microscopic understanding of some of the most prominent Swampland conjectures, namely the Weak Gravity Conjecture (WGC) \cite{Arkani-Hamed:2006emk} and the Swampland Distance Conjecture (SDC) \cite{Ooguri:2006in,Ooguri:2018wrx,Grimm:2018ohb,Font:2019cxq}. Such a phenomenon in this specific context has been termed {\it Emergence} \cite{Palti:2019pca,Harlow:2015lma,Grimm:2018ohb,Corvilain:2018lgw,Heidenreich:2017sim,Heidenreich:2018kpg}. It has been shown in some cases that the possibility of setting to zero in the UV the kinetic term of say a $U(1)$ gauge EFT weakly coupled to Einstein gravity can be compatible with the generation of an IR kinetic term provided that there exists a tower of massive charged states, a prominent example being a Kaluza-Klein (KK) tower \cite{Corvilain:2018lgw}. One-loop effects associated to this tower of heavy modes then generate a low-energy kinetic gauge function of the form $1/g^2 \sim M_{\text{pl, 4}}^2/\Mt^2$ (in 4d), where $\Mt$ is the mass scale of the tower and $\Mpf$ denotes the four-dimensional Planck mass. Furthermore, this is qualitatively consistent with the magnetic WGC in 4d, which implies the existence of a cut-off for the effective description at a scale $\Lambda_{\text{wgc}} \lesssim gM_{\text{pl, 4}}$ \cite{Arkani-Hamed:2006emk}. Thus, the WGC would be, from this point of view, linked in a natural way to the emergence of gauge kinetic terms in the IR. On the other hand, it has also been shown that for massless moduli $\phi^i$ interacting with towers of fields with moduli-dependent masses (i.e. $\Mt=\Mt(\phi^i)$), a kinetic term may also be generated at the quantum level \cite{Hamada:2021yxy}. In fact, one obtains a one-loop structure for the non-linear sigma model metric (in Planck units) of the form ${g_{i j} \sim (\partial_i \Mt \partial_j \Mt/\Mt^2)}$, where ${\partial_i \Mt = \partial \Mt/\partial \phi^i}$. Integrating this expression along a geodesic in moduli space that reaches the singular point, one then obtains consistency with the behaviour predicted by the SDC, namely $\Mt \sim m_0\, \text{exp} (-\lambda \Delta_\phi)$, where $m_0$ is the mass of the tower in some reference point of moduli space, $\phi_0$, whilst $ \Delta_\phi$ denotes the geodesic distance measured with respect to $g_{i j}$.  These results are, to say the least, intriguing, and indeed in a strong formulation of the Emergence Proposal \cite{Harlow:2015lma,Grimm:2018ohb,Corvilain:2018lgw} it has been argued that all kinetic terms associated to massless fields should be generated in a similar way. In spite of the surprising connections between the Emergence Proposal and the already mentioned Swampland constraints, not much detailed work has been done yet regarding these ideas beyond some simple toy model examples. An exception to this is the 4d $\mathcal{N}=2$ vacua obtained from type IIB string theory on a Calabi--Yau three-fold, in the large complex structure limit studied in \cite{Grimm:2018ohb}. Hence, one would like to know, in particular, whether Emergence is a general phenomenon in Quantum Gravity (or just a lamppost effect), as well as to study how the kinetic terms in $d$-dimensional string theory vacua may `emerge', if at all. One would also like to address the question of whether other terms in the effective lagrangian, such as e.g. flux potentials in type II string theory, could also appear as an IR effect.
	
In this work we take the first steps towards a general systematic study of Emergence in the context of string theory, and try to address some of the open questions that had been posed up to now. To do so, we first discuss the crucial role that the {\it species scale} \cite{Dvali:2007hz,Dvali:2007wp}  plays in any Effective Field Theory (EFT) weakly coupled to Einstein gravity. The species scale may be defined as the scale, $\LQG$, at which strong quantum-gravitational effects cannot be neglected. Its definition seems to strongly depend both on the spectrum on the EFT under study, as well as the direction in moduli space that we sample. In the context of Emergence, it is crucial to use this scale, which is an intrinsically gravitational cut-off of our effective theories, as a \emph{physical} cut-off. This shows the intimate connection between QG and the notion of Emergence. 
	
In spite of being the first string theoretical set-up in which the Emergence Proposal was studied in some detail, the large complex structure limit of type IIB on a Calabi--Yau three-fold\cite{Grimm:2018ohb} (or its mirror large volume point in type IIA\cite{Corvilain:2018lgw}) also leaves some open questions about the generality of the proposal itself. A crucial point in this context is the study of the combined structure of towers of D3-brane bound states (or D0 and D2-D0 mirror duals) becoming light asymptotically, together with the role they play in the generation of the leading axion-dependent terms contributing both to the diagonal and off-diagonal entries of the gauge kinetic functions. We have found that these can be accounted for in the emergence framework as well, at least up to the same level of accuracy as the leading saxionic dependence. Additionally, it is a natural question to ask whether other corners of the vector multiplet moduli space of 4d $\mathcal{N}=2$ theories can be similarly analyzed from the prism of emergence, and after carefully examining them (see also \cite{Marchesano:2022axe}) we find a positive answer, being these limits of particular interest to recover certain dependences of the kinetic metrics which are subleading in the large volume point but appear to be dominant in such cases. Furthermore, the precise interplay between classical (infinite distance) limits obstructed by instanton corrections and the Emergence Proposal poses a challenge for the latter. We argue that these limits can also be accommodated within the emergence paradigm by studying a particular example in the hypermultiplet moduli space of the aforementioned 4d $\mathcal{N}=2$ theories, using key insights from \cite{Marchesano:2019ifh, Baume:2019sry,Alvarez-Garcia:2021pxo} (see also \cite{Hamada:2021yxy} for related discussions). 
	
To examine the generality of the Emergence Proposal in higher dimensions, we consider a 7d model from M-theory on a $K3$ manifold as well as a 6d $\mathcal{N}=(1,0)$ theory constructed from F-theory compactified on an elliptically-fibered CY$_3$ \cite{Vafa:1996xn,Morrison:1996na,Morrison:1996pp}, also studied in \cite{Lee:2018urn}.	In addition, we also examine the case of ten-dimensional string theories in which (part of) the moduli space is given just by the vacuum expectation value (VEV) of the dilaton field. Depending among other things on the sign of its VEV, one is driven either to strong or weak coupling, and thus different kinetic terms should be provided by distinct towers of states. We study in some detail the case of Emergence in 10d type IIA string theory, in which a thorough analysis can be performed. Indeed, whilst the dilaton metric can be obtained at weak (string) coupling by summing over the string oscillator modes, when approaching strong coupling instead one can reproduce it (along with the metrics for the Ramond-Ramond 1-form and 3-form) by integrating over massive gravitini, 3-form fields and spin-2 particles contained within the D0-brane tower. 
	
The last open issue that we treat in this work is the generation of flux potentials, making particular emphasis on the concrete set-up of 4d type IIA theories arising from Calabi--Yau compactifications. This is equivalent to the question of whether kinetic terms for the massless non-propagating 3-form fields in the theory can be emergent \cite{Font:2019cxq}. Since 3-forms couple naturally to membranes, one might have thought that in this case the membranes would be the ones playing the starring role within the emergence story. We argue, however, that at strong coupling no such exotic mechanism is actually needed, and  again  D0-brane towers seem to be the key to answer the original question. The crucial point is to realize that they contain also in 4d massive spin-$\frac{3}{2}$ particles (as well as other fields completing the appropriate supermultiplets), which couple to the massless 3-forms that we are interested in here, generating the desired asymptotic kinetic terms at one-loop. This is most clearly seen by decompactifying the theory first up to five dimensions in M-theory, and then decompactifying the whole Calabi--Yau to go to 11d supergravity.
	
Our analysis shows that the correct field dependence of all metrics in string theory vacua analyzed so far, which depend on the  infinite distance points in moduli space that we probe, may be generated (within the field theory realm) by integrating out the corresponding towers of massive states. Flux potentials, when written in the dual $(d-1)$-form formulation (in $d$-dimensions), seem to be also obtained via quantum corrections associated to their `kinetic terms'. The present work renders support to the emergence idea and the fact that infinite distance singularities may arise in QG as a purely IR phenomenon, in close analogy to the finite distance singularity of the conifold \cite{Strominger:1995cz}. In a separate paper we study possible phenomenological applications of the Emergence Proposal\cite{Castellano:2023qhp}.
	
\subsection*{Guide to read this paper}	
	
Given that this is a long paper (we apologize!), we provide here a hopefully useful guide describing its content. Thus, this work consists essentially in four main blocks distributed as follows: 
\begin{enumerate}[I.]
		
\item {\it Species Scale}. In section \ref{s:speciesscale} we introduce the species scale both for KK-like and stringy towers, and describe how it can be computed in the presence of multiple towers of states becoming light upon approaching some asymptotic limit in moduli space. The details of its precise computation in the weak coupling regime for different 10d superstrings and the bosonic string can be found in Appendix \ref{ap:speciesST}. We also discuss how in a weakly coupled string theory the number of species is related to the string coupling constant in $d$ spacetime dimensions.
		
\item {\it Emergence in Quantum Gravity}. In section \ref{s:EmergenceQG} we introduce the Emergence Proposal and present a series of general results and useful formulae for the rest of the paper. In particular:
		
     \begin{enumerate}[i.]
			
     \item In section \ref{ss:selfenergybosons} we compute the one-loop contribution from infinite towers of massive scalars and spin-$\frac{1}{2}$ fermions to the off-shell 2-point function (in $d$-dimensions) of moduli fields and $U(1)$ gauge bosons, and show how Emergence would apply for each case in turn. We also elaborate on how these results may provide for a microscopic understanding of the Swampland Distance Conjecture and the (magnetic) Weak Gravity Conjecture. Further details on the specific one-loop computations are relegated to Appendix \ref{ap:Loops}.
			
     \item In section \ref{ss:kineticfermions} we discuss how the emergence of fermionic kinetic terms (due to towers of Weyl fermions and charged scalars) may lead to large suppressions for Yukawa couplings in 4d, and how a \emph{magnetic} WGC for the Yukawa interactions can be formulated.
			
     \end{enumerate}
		
\item  {\it Emergence in String Theory}. In sections \ref{s:emergence4dN=2} and \ref{s:Emergenced>4} we present detailed studies of Emergence in concrete string theory vacua, chosen so as to illustrate different aspects of the emergence phenomenon, with towers of different nature becoming relevant. More precisely:
		
      \begin{enumerate}[i.]
			
      \item In section \ref{s:emergence4dN=2} we discus in detail the case of the emergence of kinetic terms in type IIA CY$_3$ compactifications at large volume, where the relevant towers come from D0-branes and D2-D0 bound states (some details regarding the D0 and D2-D0 field content are provided in Appendix \ref{ap:Dpbranecontent}, and details on the 5d M-theory limit are discussed in Appendix \ref{ap:5dMtheory}). We extend previous discussions in this set-up \cite{Grimm:2018ohb, Corvilain:2018lgw}, including also the dependence on axion fields and the connection with the Gopakumar-Vafa picture. We also discuss in section \ref{ss:hypermultiplet4d} how non-perturbatively obstructed infinite distance limits fit in with the Emergence logic by studying an example in which instanton corrections to the hypermultiplet metric become relevant,  using key results and insights from \cite{Marchesano:2019ifh,Baume:2019sry,Alvarez-Garcia:2021pxo}. We finally comment on some other infinite distance limits recently studied in \cite{Marchesano:2022axe}.
			
      \item In section \ref{s:Emergenced>4} we study higher dimensional examples. First, in section \ref{ss:emergence6d}, we revisit a 6d F-theory example analyzed in \cite{Lee:2018urn}, in which the relevant tower is associated to an emergent heterotic string coming from D3-branes wrapping holomorphic 2-cycles. We show how in order to match the singular behaviour, it is necessary to take into account the exponential degeneracy of the mass levels of the dual critical string. Then, in section \ref{ss:emergence7dN=2} we study a 7d theory obtained upon compactification of M-theory on (attractive) $K3$ surfaces, in which the tower now involves wrapped M2-branes which correspond to Kaluza-Klein modes in some dual frame. Finally, in \ref{ss:Emergence10dST} we discuss the Emergence mechanism in 10d string theories, focusing on type IIA, where the tower of D0-branes and the tower of string oscillator modes are the relevant ones at strong and weak coupling, respectively. Some of the details on the 11d M-theory limit at strong coupling are included both in Appendices \ref{ss:10dDO} and \ref{ap:MtheoryKKcompactS1}.
			
      \end{enumerate}
		
\item {\it Emergence of flux potentials}. In section \ref{s:Scalarpotential} we argue how flux potentials may arise as well via the generation of kinetic terms for $(d-1)$-forms ($d$ being the spacetime dimension) following the emergence prescription, and we study in detail the case of type IIA string theory on a CY$_3$. 
We show how the leading field dependence on the volume and the 4d dilaton can be generated from couplings of (non-dynamical) 3-forms to towers of massive gravitini. Some details regarding the gravitino replica couplings are given in Appendix \ref{ap:dimreductiongravitino}. Furthermore, we show how the emergent potentials are consistent with the AdS/dS Distance Conjecture and the (asymptotic) dS Swampland Conjecture.
		
\end{enumerate}
	
A minimal read of this paper would include, apart from the introduction and conclusions, blocks I, II.i, as well as the general discussion in block IV (i.e. its introduction and last subsection). With this, the reader can get a general picture of the Emergence Proposal (including the key role played by the species scale) and how it can actually work in toy model examples that roughly capture the relevant physics of more concrete set-ups. However, we very much encourage the reader to go into block III and the details of block IV, so as to see the precise realization of this surprising idea in some particular string theory settings, as they can provide key insights on the non-trivial interplay between the different ingredients that make everything work at the end of the day. Moreover, they also include some general discussions about possible generalizations of the principles observed in these concrete examples. The different set-ups that are presented  can somehow be read separately, so that even though some connections among them are drawn at different places, most of the content can be understood in a somewhat independent manner. Finally, block II.ii gives a complementary discussion that completes and extends that of block II.i, and also plays an important role for the upcoming paper \cite{Castellano:2023qhp}, about some phenomenological implications for the Standard Model of Particle Physics.
	
\section{The Species Scale in Quantum Gravity}
\label{s:speciesscale}	
In the framework of Effective Field Theory (EFT), when trying to compute any observable that could be measured in experiments or in order to explain some physical phenomenon, it is necessary to first establish the spectrum of the theory, namely the number of species, their quantum numbers and Lorentz type, as well as their interactions. Moreover, the regime of validity of such an EFT must also be specified, i.e. when this precise description breaks down and another one, possibly very different, takes over. This latter piece of information is typically captured by two seemingly unrelated quantities, the UV and IR cut-offs, $\LUV, \Lambda_{\text{IR}}$. The first one sets the maximum (centre-of-mass) energy density of any field configuration described by the EFT, whilst the second one imposes a maximum wavelength in the theory and hence has to do with the size of the region of spacetime upon which the theory must be restricted. In the following, we will be interested in the first of these cut-offs, in particular in the case where we couple an EFT to Einstein gravity.
	
Naively, since gravity is a non-renormalizable theory whose strength is measured by Newton's constant, $G_N$, one would expect the energy scale at which additional QG effects kick in and invalidate the effective description to be indeed the (reduced) Planck scale, $\Mpl=1/8 \pi G_N$. However, in recent years it has become clear that, specially in theories with a large number of species, there seems to be a much smaller scale around which new gravitational dynamics must appear, the so-called \emph{species scale}. In a $d$-dimensional EFT weakly coupled to Einstein gravity the species scale is given (up to numerical factors and logarithmic corrections) by \cite{Arkani-Hamed:2005zuc, Distler:2005hi, Dimopoulos:2005ac, Dvali:2007hz, Dvali:2007wp}
\begin{equation}
		\LQG\ \lesssim \ \frac {\Mpd}{N^{\frac{1}{(d-2)}}} \ , \label{species}
\end{equation}
where $N$ is the number of species with masses below such UV cut-off and $\Mpd$ denotes the $d$-dimensional Planck mass.\footnote{\label{fnote:hbar}In terms of the $d$-dimensional Newton's gravitational constant $G_N$, the Planck scale is defined by the relation $\Mpd^{d-2}=\hbar^{d-3}c^{5-d}/8\pi G_N$, with fundamental constants restored.} There are essentially two different classes of arguments, perturbative and non-perturbative, that motivate this scale as a fundamental QG scale. These were discussed in \cite{Dvali:2007hz, Dvali:2007wp} and we review them in the following.
	
Let us start with the perturbative considerations which imply the existence of a QG cut-off at a {\it species scale}. The idea hinges on studying the contribution of quantum corrections to the Einstein-Hilbert term of the action due to the interaction of gravity with $N$ particle species. Strictly speaking, GR being non-renormalizable, there is no sensible way to absorb the momentum dependence of loop corrections into a running coupling \cite{Anber:2011ut}. However, one can still estimate the scale at which amplitudes including loops become considerable and thus the perturbative series breaks down. Including such loop corrections from the $N$ species minimally coupled to the gravitational field \cite{Donoghue:1994dn,Aydemir:2012nz,Anber:2011ut,Calmet:2017omb,Han:2004wt} then results in a weakening of gravity by a factor of $1/N$. To see this, one can take the resummed one-loop propagator of the graviton in Lorentzian signature (for concreteness we consider a 4d flat background here although the computation may be easily extended to higher dimensions)
\beq \label{eq:gravitonpropagator}
	i\, \Pi^{\mu \nu \rho \sigma}= i \left(P^{\mu \rho} P^{\nu \sigma} + P^{\mu \sigma} P^{\nu \rho}-P^{\mu \nu} P^{\rho \sigma} \right)\, \pi(p^2)\ ,
\eeq
with $P^{\mu \nu}= \eta^{\mu \nu}-\frac{p^{\mu} p^{\nu}}{p^2}$ a projector operator onto polarization states transverse to $p^{\sigma}$ (i.e. it satisfies $P^{\rho}_{\sigma} P^{\sigma}_{\kappa}=P^{\rho}_{\kappa}$) and\footnote{Here $N$ is a weighted sum of light degrees of freedom. In 4d one has $N=N_s/3+N_f+4N_V$, with $N_s$ being the number of real scalars, $N_f$ the number of Weyl spinors and $N_V$ the number of vectors \cite{Han:2004wt,Aydemir:2012nz}.}
\beq \label{eq:selfenergy}
	\pi^{-1}(p^2)=2p^2 \left( 1-\frac{N p^2}{120 \pi \Mpf^2}\, \log (-p^2/\mu^2) \right )\, .
\eeq
Then, one can recover eq. \eqref{species} up to $\mathcal{O}(1)$ factors by finding the scale at which the perturbative expansion breaks down, which amounts to seek for the energy scale $\LQG$ at which the second term inside the parenthesis of the $\pi(p^2)$ factor becomes comparable to the first one, which represents the tree-level contribution. Notice that in the the one-loop propagator \eqref{eq:selfenergy} there is an additional energy scale, $\mu$, which is related to the renormalization of the quadratic terms in the curvature that appear typically at next order in the energy expansion of the EFT \cite{Aydemir:2012nz}. Ignoring such log factors, one indeed recovers eq. \eqref{species} from the above expression. However, they can be in principle included in the perturbative computation to calculate the corrections to the species scale. Thus, taking the renormalization scale $\mu$ to be fixed at or below $\Mpd$, one obtains the following implicit equation for $\LQG$
\beq 
	\Mpd^{d-2}\, \simeq\, N\, \LQG^{d-2}\, \log (\mu/\LQG)\, .
\eeq
This can be solved explicitly for $\LQG$ as follows
\beq 
	\LQG^{d-2} \simeq - (d-2) \frac{\Mpd^{d-2}}{N}\left\{ W_{-1} \left( -\frac{d-2}{N} \left[ \frac{\Mpd}{\mu} \right]^{d-2}\right)  \right\}^{-1}\, ,
\label{eq:speciesW-1}
\eeq
where $W_{-1}$ refers to the ($-1$)-branch of Lambert $W$ function.\footnote{\label{fn:Lambert} The Lambert $W$ function is defined as a solution to the equation
\beq
	\notag y\ e^y = x \iff y= W (x)\, .
\eeq
It has two real branches, namely  the principal branch, denoted  $W_0 (x)$ (defined for $x\geq 0$),  and the (-1)-branch, denoted $W_{-1}(x)$ (defined for $ -\frac{1}{e} \leq x < 0$). The asymptotic expansions that are relevant for this work take the form \cite{Lambert}
\beq
\label{eq:Wexpansion}
		\begin{split}
			&W_{0}(x\to \infty)\, =\, \log (x)-\log (\log(x))+\ldots\\
			&W_{-1}(x\to 0^-)\, = \, \log (-x)-\log (-\log(-x))+\ldots
		\end{split}
\eeq
} Given the asymptotic properties of $W_{-1}(x)$ at $x \to 0^{-}$ (c.f. eq. \eqref{eq:Wexpansion}), such solution for the species scale seems to approach zero in Planck units when $N \to \infty$, as it should be, and can be interpreted as correcting the $N$ dependence in eq. \eqref{species} by sending $N \to N\, \left|W_{-1} \left ( - \frac{d-2}{N}\right)\right|$, which is indeed a logarithmic correction roughly of the form $N\to N\, \log N$. This then justifies  the definition \eqref{species} up to logarithmic corrections. Such corrections both to the species scale and  to the number of species, $N$, may play a crucial role in some situations (e.g. when calculating the species scale for a tower of stringy states), in which case we will include them, whereas they can be safely ignored in some others, where they do not seem to change the results qualitatively. In the following, we will explicitly display them whenever relevant or necessary.
	
Let us now turn into the non-perturbative argument supporting the idea of the species scale \eqref{species} as the QG cut-off. It relies on the fact that Black-Holes (BHs) of size given by $\LQG^{-1}$ (thus much larger than the Planck length, $\ell_{\text{pl}}$) have a lifetime (due to Hawking radiation) roughly of order $\tau \sim \LQG^{-1}$, and hence they should already probe the microscopic theory of gravity \cite{Dvali:2007hz, Dvali:2007wp}. That is, the smallest size for a semi-classical BH is given by $\LQG^{-1}$ instead of $\ell_{\text{pl}}$. In order to see this, consider the decay rate of a $d$-dimensional, semi-classical BH of mass $\MBH$ into $N$ light species, which is given by \cite{Dvali:2007wp} 
\beq\label{eq:BHmassloss}
	\frac{d\MBH}{dt} \sim -N T_{\text{BH}}^2\, ,
\eeq
where $T_{\text{BH}}=\frac{1}{R_{\text{BH}}} \sim \left( \frac{\Mpd^{d-2}}{\MBH}\right)^{\frac{1}{d-3}}$ denotes the Hawking temperature of the BH. Using this latter relation, one can integrate the above equation for a BH of temperature $T_{\text{BH}} \sim \LQG$ as follows
\beq
	\tau \sim \frac{1}{N} \int_0^{\Mpd ^{d-2}/\LQG^{d-3}} \text{d}\MBH\ \left( \frac{\MBH}{\Mpd^{d-2}}\right)^{\frac{2}{d-3}} \sim  \LQG^{-1}\, ,
\eeq
where the upper integration limit corresponds to the mass of a BH with the aforementioned temperature and in the last step we have used \eqref{species}. This strongly supports the idea that semi-classical, neutral, non-rotating BHs must be larger than $\LQG^{-1}$, which should therefore be identified as the QG cut-off. Interestingly enough, as discussed in \cite{Dvali:2007hz}, one can also refine this non-perturbative computation to include the leading corrections to the species bound expression, which turn out to behave also as $\log N$ for large $N$, in agreement with our previous perturbative considerations. 
	
Having motivated the use of the species scale \eqref{species} as the natural UV cut-off for a theory of $N$ species weakly coupled to Einstein gravity, it is of particular interest in the context of the Swampland Distance Conjecture and the Emergence Proposal to compute it in the presence of towers of light particles. Indeed, according to the Emergent String Conjecture\cite{Lee:2019wij}, one expects essentially two possible scenarios for infinite distance limits: either a decompactification (possibly to a dual frame), or a weak coupling limit for an emergent critical string (not necessarily the original one). In these cases, the infinite tower of states becoming light asymptotically is comprised by KK modes or by the massive excitations of the critical string, respectively. Thus, we start by computing the species scale in the presence of each of these two types of towers.
	
\subsubsection*{Species scale with KK-like towers}
	
Consider a $d$-dimensional EFT describing the physics of $N_0$ light modes, weakly coupled to Einstein gravity, with Planck mass, $\Mpd$. In particular, let us think about this EFT as coming from the dimensional reduction of $(d+k)$-dimensional one. For simplicity, we consider this theory compactified on an isotropic, rectangular $k$-torus, $T^k$, of radius $R$ in $(d+k)$-dimensional Planck units,\footnote{That is, in our conventions here $R$ is dimensionless and the physical size of the $k$-torus in the $(d+k)$-dimensional Einstein frame is given by $\left( 2\pi R\, \ell_{d+k}\right)^k$, where $\ell_{d+k}$ denotes the $(d+k)$-dimensional Planck length.} but it should be noted that the argument can be easily generalized to more complicated cases. For this set-up, the Planck scales of the lower and higher dimensional theories are related as follows
\beq
	\Mpd^{d-2}=M_{\text{pl, d+k}}^{d-2}\, (2\pi R)^k\, .
	\label{eq:Planckscales}
\eeq
The mass scale of the $k$ corresponding  KK towers is\footnote{Each of these towers is indeed associated to internal momentum of the higher dimensional fields along the $k$ different directions in the compactification space. Notice that states with multiple KK charges do exist, and thus the degeneracy of states grows roughly as the product of the maximum excitation numbers (see section \ref{ss:MultipleTowers}).}
\beq
	\MKK\, \simeq\, \dfrac{M_{\text{pl, d+k}}}{R}\, .
\eeq
When approaching the (infinite distance) large volume point, namely in the limit $R\to \infty$, these towers  become light exponentially with the proper field distance, as predicted by the SDC. We can then compute the species scale associated to such a dense tower of light states through eq. \eqref{species}. Assuming that every light mode in the $d$-dimensional EFT has its own KK replica, one can estimate the total number of states $N$ below the UV cut-off as follows
\beq
	N\, \simeq\, N_0 \left( \frac{\LQG}{\MKK} \right)^{k}\, .
	\label{eq:numberKKmodes}
\eeq
Using eq. \eqref{eq:Planckscales} thus leads to the following species scale
\beq\label{eq:speciesscaleKK}
	\LQG^{d+k-2}\, \simeq\, \frac{\Mpd^{d-2} \MKK^k}{N_0}\, \simeq\, \frac{M_{\text{pl, d+k}}^{d+k-2}}{N_0}\, .
\eeq
Notice that we obtain precisely the species scale associated to the higher dimensional theory, including the $N_0$ massless modes present there. That is, the QG cut-off of the $d$-dimensional EFT, i.e. its associated species scale, gives precisely the right cut-off scale that one would expect in the full theory, namely the $(d+k)$-dimensional species scale. Let us emphasize the important role played here by the $N_0$ light species which are already present in the higher dimensional theory, and thus correct its QG cut-off from the naive $(d+k)$-dimensional Planck mass to the corresponding species scale. The intuitive result that the higher dimensional Planck mass actually provides for the QG cut-off is thus recovered when $N_0=\mathcal{O}(1)$, as expected.
	
Before turning into the computation of the species scale for stringy towers, let us remark a couple of key properties of general KK-like towers that make them qualitatively different from stringy ones, and play a crucial role in the computation of the species scale. First, there is the fact that there can be multiple KK-towers, with different charges and (possibly) different mass scales, which can lie below the species cut-off. We elaborate on this in section \ref{ss:MultipleTowers} but it is worth remarking it here as opposed to the case of stringy towers, for which only one critical string can become light at the same time. Second, the structure of KK towers is such that the degeneration of states in each step of the tower (i.e. for a given mass) grows at most polinomially, in contrast to the well-known exponential degeneration that appears for the high excitation levels of a string \cite{Green:2012oqa}.
	
\subsubsection*{Species scale from string towers}
	
We consider now the case in which the tower of states is provided by the oscillator modes of a fundamental critical string that becomes asymptotically tensionless. It is well-known that such a higher-spin tower presents a much denser spectrum than the typical KK-like towers. The reason for this is twofold: first, in Minkowski backgrounds, the stringy modes are characterized by a mass given roughly by the Regge excitation pattern 
\beq
	m_n^2\, \simeq\, (n-1) T_s\, ,
\label{eq:stringmasses}
\eeq
where $T_s=2 \pi \Ms^2$ denotes the string tension and $n \in \mathbb{N}$ refers to the excitation level of the string. Second, one has to take into account the high level density of states, $d_n$, associated to the very massive excited states. To leading order, for large $n$  this degeneration is an exponential of the form \cite{Green:2012oqa}
\beq
	d_n\, \sim\, e^{ \sqrt{n}}\, .
	\label{eq:leveldensity}
\eeq
To be precise, this expression should include a monomial prefactor, as well as some numerical constants in the exponential, which depend on the particular string theory under consideration as well as on the open or closed nature of the string\cite{Green:2012oqa}. However, since in the infinite distance limit one expects a diverging number of oscillator levels of the string falling below $\LQG$ (as we will confirm below), the use of this asymptotic expression for $d_n$ is justified. Nevertheles, we include these monomial (and model-dependent) prefactors in Appendix \ref{ap:speciesST}, where we perform the calculation of the species scale and number of species for type II, heterotic and bosonic string theories. As expected, their inclusion results in more refined expressions for these quantities, although the leading behavior relevant for our discussions is essentially captured by the approximation \eqref{eq:leveldensity}, such that we restrict to it for the remainder of this section.
	
Thus, in a $d$-dimensional EFT, we can compute the species scale associated to a tower of light modes coming from a fundamental string becoming asymptotically tensionless in Planck units (i.e. $\Ms / \Mpd \to 0$) as follows. First, we need to know the maximum excitation level $\Ns$, with mass at or slightly below the QG cut off, $\LQG$. Then, we need to count the accumulated number of string states up to such $\Ns$, that we denote now as $\Ntot$. The species scale thus fulfills
\beq
\label{eq:speciescale}
	\LQG^{d-2}\, \simeq\, \frac{\Mpd^{d-2}}{\Ntot}\, \simeq\, \Ns^{\frac{d-2}{2}}\, \Ms^{d-2}\, .
\eeq
Solving for $\Ns$ we arrive at
\beq
\label{eq:Ns}
	\Ns^{\frac{d-2}{2}} \sum_{n=1}^{\Ns} d_n \, \simeq\, \left ( \frac{\Mpd}{\Ms}\right )^{d-2}\, ,
\eeq
where we have substituted $\Ntot =\sum_{n=1}^{\Ns} d_n$ in terms of the sum of the density level of physical states for each step in the tower. Hence, one finds 
\beq
	\left ( \dfrac{ \Mpd}{\Ms}\right )^{d-2}\, \simeq\, \Ns^{\frac{d-2}{2}} \sum_{n=1}^{\Ns} \text{exp}(\sqrt{n}) \, \sim\, 2\, \Ns^{\frac{d-1}{2}} e^{\sqrt{\Ns}}\, ,
\label{eq:maxstringlevel}
\eeq
where the sum has been approximated by an integral, which is justified in the limit where $\Ns \to \infty$ (and in any case provides a lower bound for the series). We can then solve this expression explicitly for $\Ns$ to obtain
\beq
	\sqrt{\Ns}\, \sim\, (d-1)\, W_0 \left( \dfrac{1}{(d-1)\, 2^{\frac{1}{d-1}} } \left[ \dfrac{\Mpd}{\Ms} \right]^{\frac{d-2}{d-1}} \right)\, ,
\eeq
where $W_0$ refers to the principal branch of the Lambert $W$ function (see footnote \ref{fn:Lambert} above). The key result from the above expression is that it confirms that the maximum excitation level of the string that falls below the cut-off scale diverges when $\Ms/\Mpd \to 0$. Notably, by using the relevant expansion of the $W$ function from eq. \eqref{eq:Wexpansion}, such divergence can be seen to behave essentially in a logarithmic fashion, i.e.
\beq
	\sqrt{\Ns}\, \sim\, (d-2)\, \log \left(\dfrac{\Mpd}{\Ms}\right) + \mathcal{O}\left(\log \left( \log (\Mpd/\Ms) \right) \right)\, .
\eeq
This means that the species scale for the case of a critical emergent string in $d$ spacetime dimensions behaves as  
\beq
	\frac{\LQG}{\Mpd}\, \simeq\, \sqrt{\Ns}\, \frac{\Ms}{\Mpd}\, \sim\,  (d-2)\,  \dfrac{\Ms}{\Mpd}\  \log \left(\dfrac{\Mpd}{\Ms}\right)\, .
\eeq
Hence, the species cut-off associated to the stringy tower is the string scale itself, up to logarithm corrections. However, these corrections are crucial in order to make sense of the state counting between $\LQG$ and $\Ms$, given that they precisely encode the information about the (square root of the) maximum excitation level, $\Ns$. In other words, neglecting logarithmic corrections gives the right `intuitive' result for the species scale, $\LQG \sim \Ms$, but in order to count the number of species it is indeed necessary to include the logarithmic corrections coming from the exponential degeneration.

\subsubsection*{String theory as a theory of species and finiteness}
		
Recall that the $d$-dimensional dilaton is defined as 
\beq\label{eq:ddimdilaton}
	e^{-2 \varphid}\, =\, \frac{1}{\gsd^2}\, =\, \frac{1}{4\pi} \left( \frac{\Mpd}{ \Ms}\right)^{d-2}\, , 
\eeq
thus controlling the Planck-to-string scale ratio in $d$ dimensions. Hence, from the above analysis one finds the maximum oscillator level in the case of a weakly coupled stringy tower to be given by
\beq
	{\sqrt{\Ns}\,  \sim -2 \varphid \, =\, -2 \log (\gsd) }\, , 
\eeq
whilst for the total number of species below the cut-off one rather obtains
\begin{equation}
		\Ntot\, \simeq\, \dfrac{1}{(-2\varphid)^{d-2}} \ e^{-2\varphid}\, =\,  \dfrac{1}{\left\{ -2 \log (g_s)\right\}^{d-2}}\ \dfrac{1}{\gsd^2}\, .
\end{equation}
This is in agreement with the results from \cite{Dvali:2009ks,Dvali:2010vm}, which suggest that string theory can be seen as a theory of $N_{\text {eff}} \sim 1/g_{s,d}^2$ species coupled to gravity in the limit of small $g_{s, d}$. Notice that if the theory has $N_0$ massless modes to start with, this already gives a `universal' bound for the dilaton VEV in any spacetime dimension of the form
\beq
	N_0\, \lesssim \, \frac {1}{g_{s,d}^2}\, .
\eeq
Therefore, at finite (but weak) string coupling the massless spectrum has to be necessarily finite. On the other hand, only the singular limit $g_{s,d}\rightarrow 0$ allows for an infinite number of massless degrees of freedom associated to the critical string. Interestingly, this result is consistent with the conjecture of finiteness of the number of massless degrees of freedom put forward in \cite{Vafa:2005ui,Tarazi:2021duw,Hamada:2021yxy}.

\subsection{Multiple towers and the species scale}
\label{ss:MultipleTowers}
Let us now address the more general set-up in which several towers, with different charges and in principle different spectra are present. This case has been partially addressed already in \cite{Castellano:2021mmx} and we review and expand some of the results therein in the present section. As we will discuss later on, the following discussion can be specially relevant in the context of the Emergence Proposal since it is typically the case that not a single tower becomes light in asymptotic regimes, but several  of them. In particular, one could consider a situation where as we approach an infinite distance point in moduli space, multiple towers become light and lie below the species scale cut-off, such that all of them contribute to its computation, given that they all couple to gravity. However, it could also be the case that only a subset of these states interact with a particular gauge field or modulus scalar, so that if we want to apply the emergence mechanism we must take the effect of the other towers into account for the calculation of the cut-off scale associated to the EFT (i.e. the species scale) but not for the computation of the relevant kinetic term. Indeed, we will find examples of this later on.
	
\subsubsection*{Additive species}
In this scenario, one can think of two qualitatively different situations with regard to the light towers. First, consider a set-up in which there is no mixing between the multiple towers, e.g. when we have two (infinite) sets of states that couple to different fields and such that no states with both charges are present in the spectrum (of quasi-stable states). In this case, the key point is to realize that the total number of species below the species scale is given by $N_{\mathrm{tot}}= \sum_i N_i$, where $N_i$ is the number of states associated to the $i$-th tower. The species scale, given by eq. \eqref{species}, should then be computed by including all the light states from every one of the towers. Since the number of particles of any given set of states diverges in the asymptotic limit (in which we are interested here), $N_{\mathrm{tot}}$ will be in this case dominated essentially by one of the towers (unless all $N_i$ scale in the same way with respect to the relevant moduli, in which case the following calculation would just be modified by $\mathcal{O}(1)$ factors). In practice, we can calculate the species scale associated to each of the towers separately. So let us consider a tower comprised by states with masses
\begin{equation}
\label{eq:Mni}
		m_{n,\, i}\, =\, n^{1/p_i}\ \Mti\, .
\end{equation}
Thus e.g a standard KK tower has $p_i=1$.
We can compute the species scale as follows
\begin{equation}\label{eq:speciesscaleithtower}
		\LQG{}_{,i}\, \simeq\, N_i^{1/p_i}\Mti\, \simeq\, \dfrac{\Mpd}{N_i^{\frac{1}{d-2}}}\, .
\end{equation}
The physical value for the species scale is then $\LQG= \text{min}\{ \LQG{}_{,i} \}$, since it is dominated by the states associated to the tower with the lightest mass scale. For concreteness, let us label the leading tower (i.e. the one with the smallest corresponding species scale, c.f. eq. \eqref{eq:speciesscaleithtower}) by the index 1, characterized by the density parameter $p_1$ and mass gap $\frac{m_{\text{tower,}\, 1}}{\Mpd} \sim t^{-a_1}$, which goes to zero as a modulus $t$ goes to infinity. For all the other light towers, we consider $\frac{\Mti}{\Mpd} \sim t^{-a_i}$. Since the set of light states is dominated by the first tower, we have $N_{\text{tot}}=N_1+\ldots \simeq N_1$, and
\begin{equation}
		N_1\, \sim\, t^{\frac{a_1 p_1(d-2)}{ d-2+p_1}}\, , \qquad  \LQG\, \sim\, \Mpd\, t^{-\frac{ a_1 p_1}{d-2+ p_1}}\, .
\end{equation}
For consistency, we can recalculate the number of states associated to the other towers, namely the ones with $i\geq 2$, that lie below the physical species scale (i.e. the species scale calculated with the dominant tower, that is with $N_1$). These yield
\begin{equation}
		\tilde{N}_i\, \sim\, t^{\frac{(d-2) a_i p_i+ p_1 p_i (a_i - a_1)}{d-2+p_1}}\, ,
\end{equation}
and as expected $\tilde{N}_i \leq N_i$. Note that this expression contains negative contributions, so that it includes the case in which a tower is still too heavy to include any mode below the species scale, as well as the case where an asymptotically infinite number of them are present.
	
Let us remark also that the case of a stringy tower would correspond in this language to $p \to \infty$.\footnote{To be precise, one would need to include the exponential degeneration together with $p=2$, as we did in the previous section, but for the case at hand most of the calculations yield the correct result if we model such high degeneracy by just taking the limit $p \to \infty$. } Therefore, it is clear that such a dense tower would dominate any set-up in which it should be taken into account, and we would thus recover the result that $\LQG \sim \Ms$ (up to log corrections that are not included here), as expected.

\subsubsection*{Multiplicative species}
Second, we consider the case in which the different towers are such that states with mixed charges can be present. This is the case that is actually relevant in most set-ups. It is realized by e.g. several KK-like towers (or, as we discuss in some more detail in section \ref{sss:IIA-Ftheorylimit}, the dual version with towers of D0-D2 particles), since we can have states with non-vanishing momentum along several internal directions at the same time, or even in the presence of stringy spectra and KK towers, since the oscillator modes of the critical string also have their own KK replicas. In this case, the total number of states is not just additive as it was before, but instead it typically behaves like $N_{\text{tot}}\simeq \prod_i N_i$. This scenario is explained in detail in section 2 of \cite{Castellano:2021mmx}, but we review here the main results that are important for our current purposes. Thus, consider the case where we have $k \in \mathbb{N}$ generating towers of (asymptotically) light particles, with masses given again by eq. \eqref{eq:Mni}, such that states with non-vanishing occupation numbers under the different towers appear in the spectrum. Let us assume that their masses take the general form
\begin{equation}\label{eq:massmixedsprectra}
		m_{n_1\,  \ldots  n_k}^2\, =\, \sum_{i=1}^k n_i^{2 / p_i} \Mti^2\, .
\end{equation}
Then, we can calculate an effective mass and density parameter which read as follows \cite{Castellano:2021mmx}
\begin{equation}
		\label{eq:Meffpeff}
		\Mteff =\left(m_1^{p_1} m_2^{p_2} \ldots m_k^{p_k}\right)^{1 / \sum_i p_i}\, , \qquad \peff =\sum_i p_i\, ,
\end{equation}
with the species number and cut-off scale given in terms of these as
\begin{equation}
		\label{eq:NtotLQGeff}
		\Ntot\, \simeq\, \left( \dfrac{\Mpd}{ \Mteff} \right) ^{\frac{(d-2)\peff}{d-2+\peff}}\, , \qquad   \LQG\, \simeq\, \Mpd \, \left( \dfrac{\Mpd}{ \Mteff} \right)^{-\frac{\peff}{d-2+\peff}} \, .
\end{equation}
The maximum occupation number for each tower can also be computed by combining this last equation with $\LQG \simeq N_i^{1/p_i}\Mti$. Additionally, one can again parameterize the mass scale of the towers by $\frac{\Mti}{\Mpd} \sim t^{-a_i}$ and obtain the expression for the species scale and the number of species as a function of the modulus $t$ when it becomes large.
	
Let us emphasize that, contrary to the previous case, it is not true anymore that the leading tower determines always \emph{alone} the value of the species scale. In fact, even in the case of a parametrically smaller mass for one tower, other towers can still significantly contribute with a divergent number of states below the species scale, and therefore it is not enough to know which tower has the lightest mass scale, but rather all the towers that lie below the cut-off. In particular, notice that this is crucial to know not only the full spectrum of the low energy EFT that we will eventually want to integrate out in order to study the Emergence Proposal, but also for the calculation of the species scale itself, whose asymptotic expression can be completely changed by the presence of towers that lie below the cut-off but do not present the lightest mass scale. To study systematically which towers actually lie below the species scale and contribute to eqs. \eqref{eq:Meffpeff}-\eqref{eq:NtotLQGeff}, one can perform an iterative process by starting with the lightest one first, calculating its associated species scale, and checking whether it lies above the first step of the second lightest tower. If so, the second tower must be included, the species scale should be properly recalculated and then the third tower must be checked. The algorithm must be carried on until we find a tower that lies above the species scale and must therefore not be included (see \cite{Castellano:2021mmx} for details). If $i$ towers contribute to the species scale (and have associated effective mass scale and density parameter denoted by $m_{\text{tower,}\, (i)}$ and $p_{\text{eff,}\, (i)}$), the condition to check whether the $(i+1)$-th tower also lies below the such scale can be easily stated as follows \cite{Castellano:2021mmx}
\begin{equation}
		m_{\text {tower,}\, (i)}\, \leq\, m_{\text {tower, }\,  i+1}^{\frac{d-2+p_{\text{eff,}\, (i)}}{p_{\text{eff,}\, (i)}}}\, .
\end{equation}
If the inequality if fulfilled, the $(i+1)$-th tower lies above the species scale associated to the previous $i$ towers and can be safely ignored in the EFT, otherwise it must be included and the process continues until one tower satisfies the inequality. Once again, in the case in which the $i$-th tower is of stringy nature, the inequality is automatically fulfilled for any heavier one (since $p_{\text{eff,}\, (i)} \to \infty\, $), and the species scale is saturated by the stringy modes, as expected. However, notice that this still allows for the possibility of a limit in which the lightest towers are of KK-type to give rise to $\LQG \sim \Ms$ if the KK towers are not dense enough to saturate the species scale before a tensionless string kicks in.

\section{Emergence in Quantum Gravity}
\label{s:EmergenceQG}
	
The discussion about the species scale in the previous section may be understood as the generation of contributions to the kinetic terms for gravitation in the infrared, appearing through loops involving a large number of light species. The kinetic term of the graviton gets renormalized in such a way that  the Einstein-Hilbert term is subject to important quantum corrections. The proposal of Emergence in Quantum Gravity \cite{Palti:2019pca,Harlow:2015lma,Grimm:2018ohb,Corvilain:2018lgw,Heidenreich:2017sim,Heidenreich:2018kpg} is to take this is as a general phenomenon, such that the remaining massless particles get analogously their kinetic terms from one-loop corrections involving towers of states below the QG cut-off. Thus, to first approximation, the \emph{Emergence Proposal} may be stated as follows:
	
\textreferencemark\, \textbf{Emergence (Strong)}: \emph{In a theory of Quantum Gravity all light particles in a perturbative regime  have no kinetic terms in the UV. The required kinetic terms appear as an IR effect due to  loop corrections involving the sum over a tower of asymptotically massles states.}
	
\begin{figure}[tb]
		\begin{center}
			\includegraphics[scale=0.5]{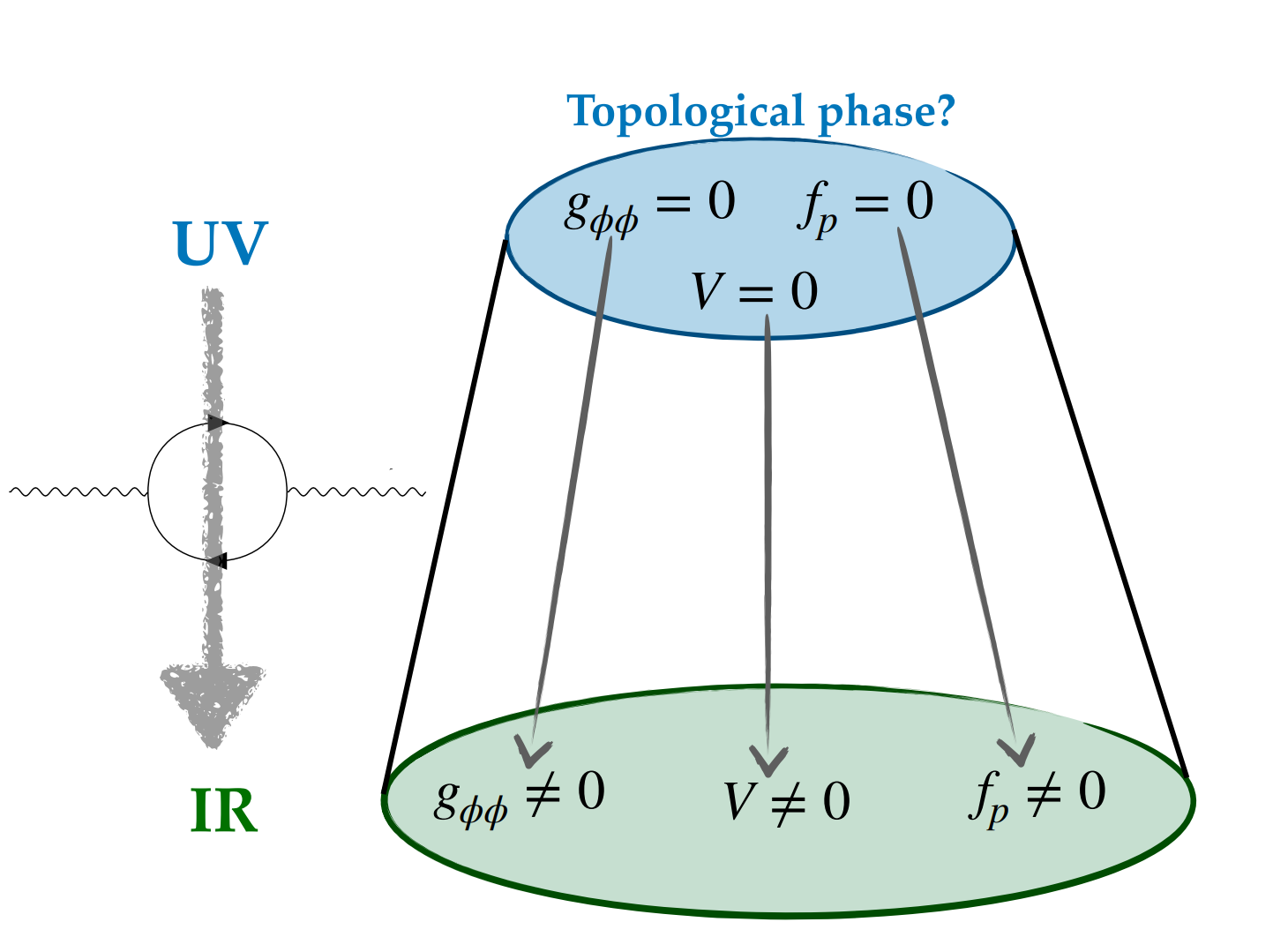}
			\caption{Scheme of the strong version of Emergence. The basic claim is that massless fields would not have any kinetic term in an underlying UV strongly coupled theory. The metrics would be generated by quantum loops involving the contributions due to (infinite) towers of states.}
			\label{fig:strongemergence}
		\end{center}
\end{figure}

It has been argued \cite{Palti:2019pca,Harlow:2015lma,Grimm:2018ohb} that the condition of vanishing kinetic terms  in the UV for the light degrees of freedom could perhaps suggest the existence of an underlying \emph{topological} fundamental theory, in which such particles do not propagate (see fig. \ref{fig:strongemergence}). Thus, there would be couplings for these non-propagating fields  but no geometric objects to start with, i.e. no kinetic terms (see \cite{Harlow:2015lma,Agrawal:2020xek}). On the other hand, one can be more conservative and formulate the following more general (but less ambitious) version of the Emergence Proposal:
	
\textreferencemark\, \textbf{Emergence (Weak)}: \emph{In a consistent theory of Quantum Gravity, for any singularity at infinite distance within the moduli space of the EFT, there is an associated infinite tower of states becoming massless which induce quantum corrections to the metrics matching the `tree level' singular behavior.}
	
Note that these statements are somewhat morally analogous to the resolution of the conifold singularity in the complex structure moduli space of type IIB string theory on a CY$_3$\cite{Strominger:1995cz}, in which the singularity in the (vector multiplet) field space metric can be understood as fully generated by integrating out the D3-brane BPS state that becomes massless precisely at the conifold point. Still, this is qualitatively different in the sense that in the emergence cases that we are discussing an infinite number of  states become asymptotically massless in the limit, yielding an infinite distance singularity, as opposed to the finite number of states for the conifold, which give rise to a finite distance one.
	
In the strong version of the proposal above one goes one step beyond by arguing that in fact the quantum induced metrics is all there is from an EFT point of view. In the present paper we  present compelling evidence in favour of the weak version  within string theory set-ups. Still, the results are also fully compatible with the strong one.

It is interesting to remark that the idea of emergent kinetic terms for gauge fields has a long history. Thus, in \cite{DAdda:1978vbw} it was pointed out how composite gauge bosons become dynamical in $\mathbb{CP}^{N}$ sigma-models. Inspired by this work and in the context of $\mathcal{N}=8$ supergravity, Cremmer and Julia \cite{Cremmer:1979up} suggested that composite gauge bosons  transforming in the adjoint of $SU(8)$ could become dynamical by acquiring kinetic terms and be used as a unification group. This idea was explored further in \cite{Ellis:1980cf,Ellis:1980tf}, in which they considered the possibility of embedding $SU(5)$ Grand Unified Theories (GUT) within this context. These considerations did not quite work in the end because of a number of reasons, most notably the endemic presence of anomalies. 

\begin{figure}[t]
		\begin{center}
			\subfigure[]{
				\includegraphics[height=4.2cm]{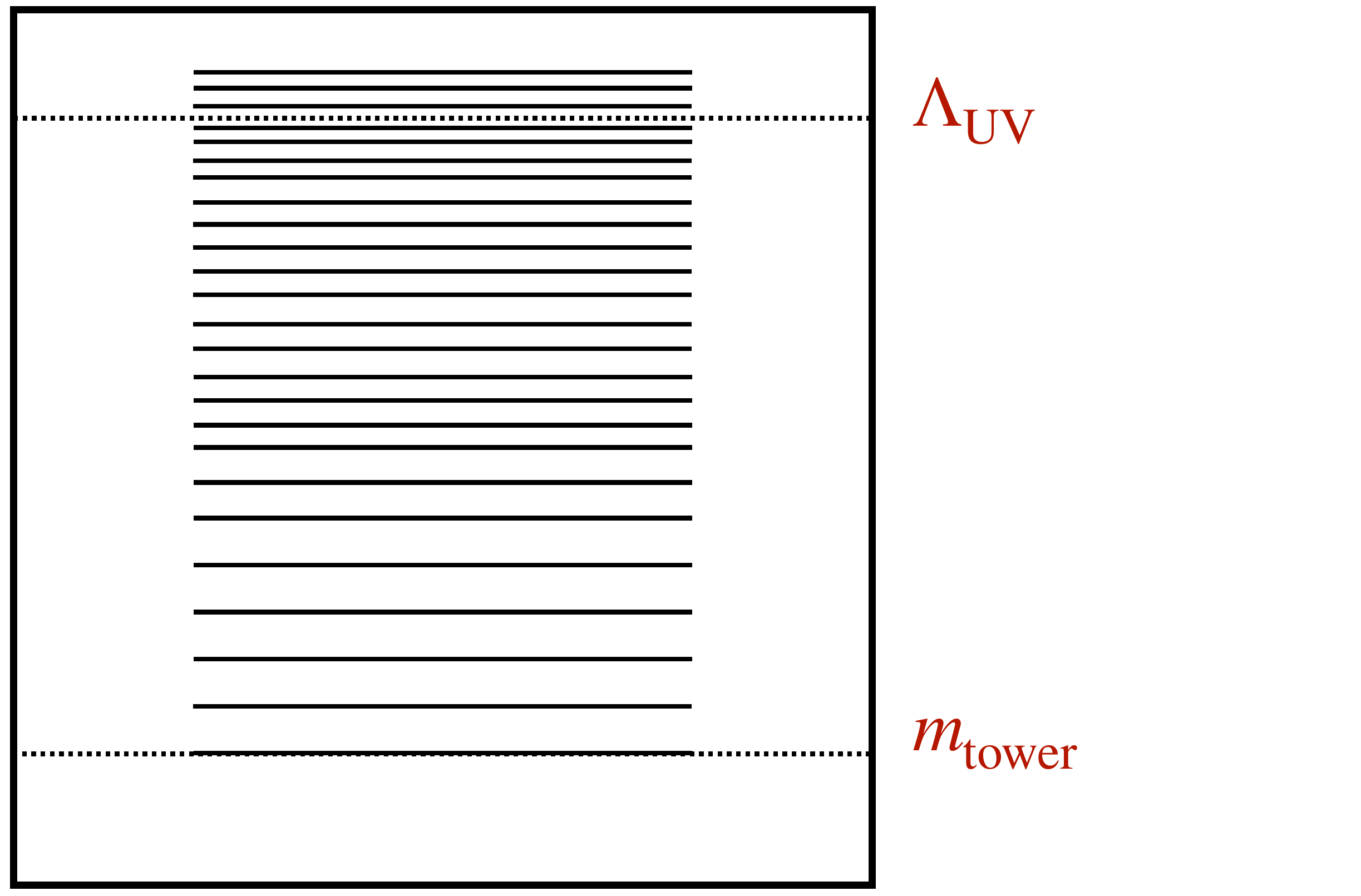} 
				\label{torresa}
			}
			\subfigure[]{
				\includegraphics[height=4.2cm]{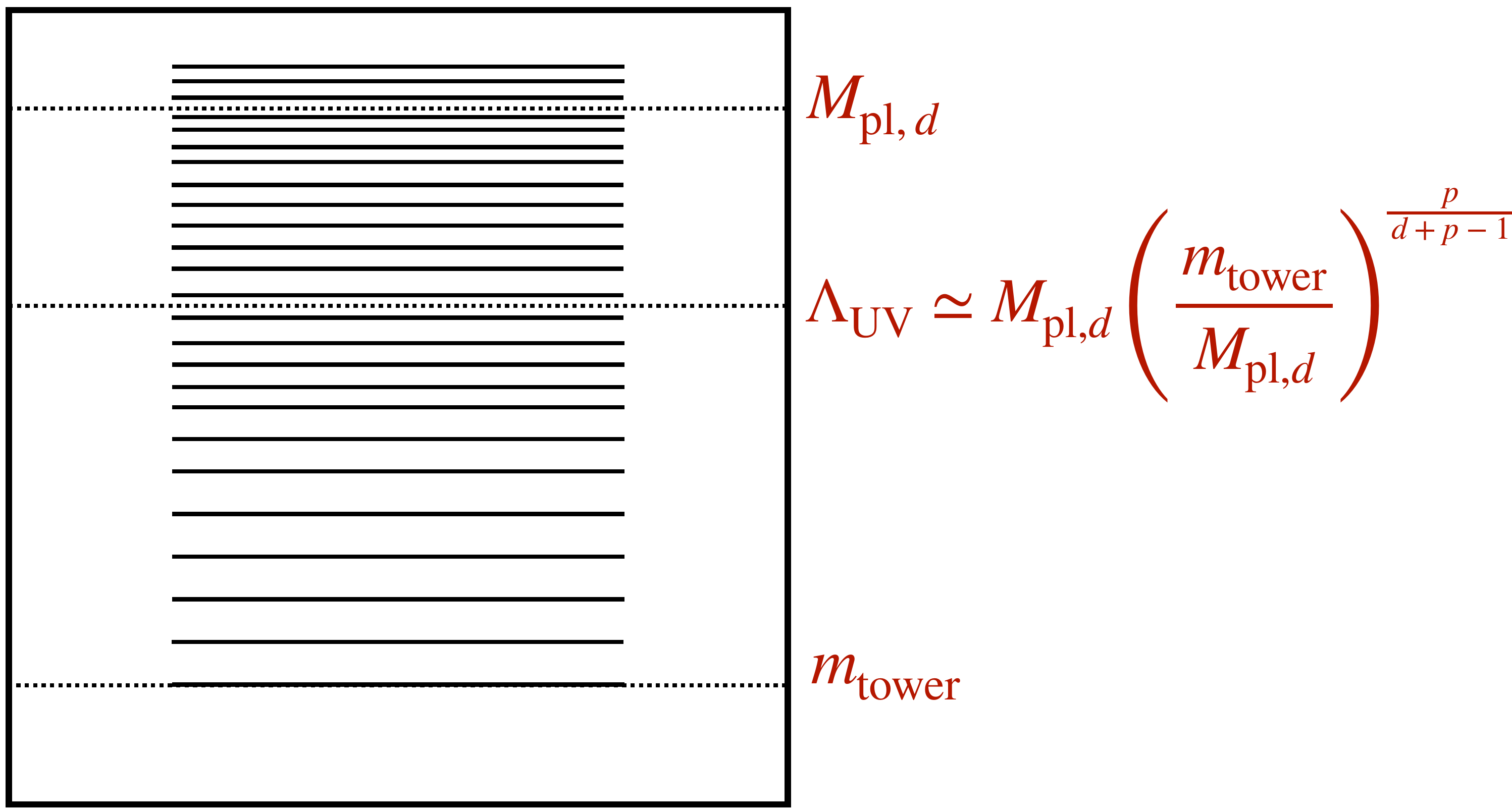}
				\label{torresb}
			}
			\caption{Emergence of metrics from a tower of states. (a) In the absence of gravity they are sensitive to the  UV cut-off, $\Lambda_{\text{UV}}$. (b) In the presence of gravity the UV cut-off must be identified with the species scale. The latter is related to the IR scale $\Mt$. Only  $\Mt$ and the Planck mass, $\Mpd$, appear in the EFT. }			
		\label{torres}
		\end{center}
\end{figure}

One attractive feature of the Emergence Proposal is that it gives us a microscopic rationale for the understanding of both the existence of the (magnetic) Weak Gravity Conjecture (WGC) and the Swampland Distance Conjecture (SDC) (see the reviews \cite{Brennan:2017rbf,Palti:2019pca,vanBeest:2021lhn,Grana:2021zvf} for their definition and current status). In order to give a flavour of why this would be so, let us consider here a simple toy model in $d > 4$ spacetime dimensions with a KK-like tower of charged particles labeled by both their $U(1)$ gauge charge $n \in \mathbb{Z}$ and mass $m_n=|n|\, \Mt$, with $|n| \leq N$ ($N$ being effectively very large). We also assume that these massive states couple to a single real modulus $\phi$ through its modulus-dependent mass, i.e. $m_n = m_n (\phi)$. Hence, let us first ignore gravity (see fig. \ref{torresa}) and just compute the one-loop contribution to the massless field metric $g_{\phi\phi}$. For concreteness, we take our infinite tower to be made from e.g. massive fermionic fields, $\Psi_n$, such that their (leading-order) coupling to the massless modulus is given by a Yukawa interaction of the type $(\partial_\phi \Mt)\, \phi \ \overline{\Psi^{(n)}} \Psi^{(n)}$. We thus find\footnote{We will ignore for the moment some subtleties associated to the relevant loop computations which will be explained with more detail in the next subsection. The results presented in here are essentially unchanged when performing the more careful analysis described below.}
\beq \label{eq:metricemergence}
	\delta g_{\phi\phi}\, \sim\, \sum_{n=1}^N n^2 \Lambda_{\text{UV}}^{d-4}  (\partial_\phi \Mt)^2\, \sim\, N \Lambda_{\text{UV}}^{d-2} \left(\frac {\partial_\phi \Mt}{\Mt}\right)^2\, ,
\eeq
where we have approximated the sum with an integral and we have introduced a UV cut-off $\Lambda_{\text{UV}} = N\, \Mt$ below which states belonging to the tower start to appear and need to be taken into account. Indeed, we see that a kinetic term is produced at the quantum level, but it is in principle divergent if one naively wants to take the continuum limit, i.e. $\Lambda_{\text{UV}} \to \infty$. However, the renormalization prescription will force us to have some (divergent) kinetic term already at the UV scale so as to be able to make finite  physical predictions at low energies. Thus, one cannot simply claim that kinematics is induced in the infrared. Something very similar happens as well with the $U(1)$ gauge kinetic term. Assuming that the heavy particles have some quantized charge $q_n=n$, one finds at the one-loop level the following inverse gauge coupling
\beq
	\delta \left(\frac {1}{g^2}\right)\, \sim\,  \sum_{n=1}^N n^2 \Lambda_{\text{UV}}^{d-4}\, \sim\, N \Lambda_{\text{UV}}^{d-2} \frac {1}{\Mt^2}\, ,
	\label{eq:gaugecouplingemergence}
\eeq
where again we have approximated the sum with an integral in the last step. As in the previous case, there is strictly speaking no `IR Emergence', since (divergent) kinetic terms must be already present in the UV regime.  

Let us reconsider the above computations but now in the context of Quantum Gravity (see fig. \ref{torresb}). In this case, it makes sense to identify the UV cut-off $\Lambda_{\text{UV}}$ with the species scale discussed in section \ref{s:speciesscale}, namely $\LQG \lesssim \Mpd/N^{1/d-2}$. By doing so, the explicit dependence on the momentum space cut-off disappears in the above two expressions and one is left with\footnote{In 4d the leading dependence on the cut-off appears through a logarithmic factor $\sim \log \left(\LQG^2/\Mt^2 \right)$, but after integration over the full tower the final result agrees with the naive expression \eqref{eq:emergencegeneralQG}, which presents no explicit logarithmic dependence on $\LQG$.}
\beq\label{eq:emergencegeneralQG}
	g_{\phi\phi}\, \lesssim\,  \Mpd^{d-2}\, \left(\frac {\partial_\phi \Mt}{\Mt}\right)^2\, , \qquad \frac {1}{g^2}\, \lesssim\, \Mpd^{d-2}\, \frac {1}{\Mt^2}\, .
\eeq
Consider for the moment the one-loop contribution to the gauge kinetic term given by the second expression above. One thus finds
\beq
	\Mt^2\, \sim\, (N\LQG^{d-2})\, g^2\, \lesssim\, g^2 \Mpd^{d-2}\, ,
\eeq
which has the qualitative structure of the magnetic Weak Gravity Conjecture for a $U(1)$ gauge field. In this sense, the emergence of the gauge kinetic term together with the species scale imply the (magnetic) WGC. Concerning the moduli, once we found the field space metric we can easily compute the distance in moduli space between any two given points, $\phi_a$ and $\phi_b$ (provided of course they lie ultimately at infinite distance from each other, which is where our computations have been done reliably). By doing so, we arrive at (in terms of say an affine parameterization, $\phi=\phi(\tau)$)
\beq
\label{eq:distancegeneral}
	\kappa_d\, \Delta\phi_{ab}\, =\,  \kappa_d\, \int_{\tau_a}^{\tau_b} \text{d}\tau \sqrt{g_{\phi \phi}\, \dot{\phi}^2}\, \sim \, \int_{\phi_a}^{\phi_b} \frac {\partial_\phi \Mt}{\Mt}\, d\phi\, \sim \, \log \left(\frac {\Mt (\phi_b)}{\Mt (\phi_a)}\right)\, ,
\eeq
where we denote $\dot{\phi}=\frac{d \phi}{d\tau}$ and we have substituted $\kappa_d^2\, g_{\phi \phi} \sim (\partial_\phi \Mt/\Mt)^2$ in the above expression and $(\kappa_d)^{-2}=\Mpd^{d-2}$. From here one obtains the sought-after exponential behaviour for the mass of the tower, namely $\Mt (\phi_a)/ \Mt(\phi_b) \sim e^{-\lambda (\kappa_d \Delta\phi_{ab})}$ with $\lambda$ some $\mathcal{O}(1)$ factor, which indeed yields the content of the Swampland Distance Conjecture. Note that the emergent kinetic terms above turn out to be essentially proportional to $N\LQG^{d-2}$, which itself is proportional to the kinetic term of the graviton. Therefore, getting full-fledged Emergence for the kinetic terms with $N\LQG^{d-2} \simeq \Mpd^{d-2}$ requires gravity to be fully emergent as well (or at least the one-loop emergent contribution to $\Mpd$ must be of the same order as the tree level one, in case the latter exists).
	
Even though the above analysis is framed within a very simple $d$-dimensional example (with $d>4$), we will see that similar results are obtained for more complicated tower structures and different spacetime dimensions, including also the case of stringy towers. The take-home message is that Emergence is closely connected with the WGC and SDC and that in this connection it is crucial that the cut-off in the loop computations is taken to be the species scale. However, the final expression for the metrics may be written in a way which is independent of the QG cut-off (or equivalently $N$) and only depends explicitly on IR data (i.e. the value of $\Mt$) as well as the Planck mass.

\subsubsection*{Classical metrics from quantum effects}
There are two basic questions that arise quite naturally from our discussion here. First, how comes that one can obtain a \emph{classical} metric by summing over \emph{quantum} contributions? And second, why is it that the obtained kinetic terms are independent of the number of species involved, $N$? To answer the first question it is convenient to keep track of the powers of $\hbar$ in the one-loop computations, both of the species scale and the emergent metrics (we still henceforth set $c=1$). In particular, notice that the integral over (internal) momentum in the loop calculation of the species scale carries a factor $\hbar^{d-3}$ in the denominator, such that what one actually gets is $G_N^{-1}\rightarrow \hbar^{d-3} G_N^{-1} = 8 \pi \Mpd^{d-2}$ in said computation (c.f. footnote \ref{fnote:hbar}).\footnote{Notice that instead of Newton's constant one often uses Einstein's gravitational coupling $\kappa_d^2=8\pi G_{N}/c^4$ to parameterize the strength of gravity in $d$ dimensions.} On the other hand, we also get the exact same factor in the denominator of the loop integrals associated to the emergence of kinetic terms for the other massless fields, such that we should indeed replace $\Mpd^{d-2}\rightarrow \Mpd^{d-2}/\hbar^{d-3}$ in every one of them. This is as it should be, since the pieces we want to reproduce are classical to start with, in the sense that they do not include explicit dependence on $\hbar$ (they do not depend on $\Mpd$ but rather on $\kappa_d$). Thus, in all the emergence computations that we perform in this paper one should actually understand that it is not $\Mpd^{d-2}$ which appears in the one-loop metrics but rather $\kappa_d^{-2}$ (although they obviously agree upon using natural units). The upshot is that, even though the Emergence Proposal is quantum in  nature, the species scale cut-off comes itself from a quantum computation and both quantum effects ultimately cancel each other, yielding a seemingly classical result.\footnote{The idea that loop corrections in gravity can lead to classical IR effects is a well-known fact in the literature, see e.g. the recent review \cite{Donoghue:2022eay} and references therein.} Something analogously happens with the number of species above, which does not appear explicitly in the final emergent metrics but rather has an effect captured by the Planck scale, according to eq. \eqref{species}.
	
\begin{figure}[tb]
		\begin{center}
			\includegraphics[scale=0.33]{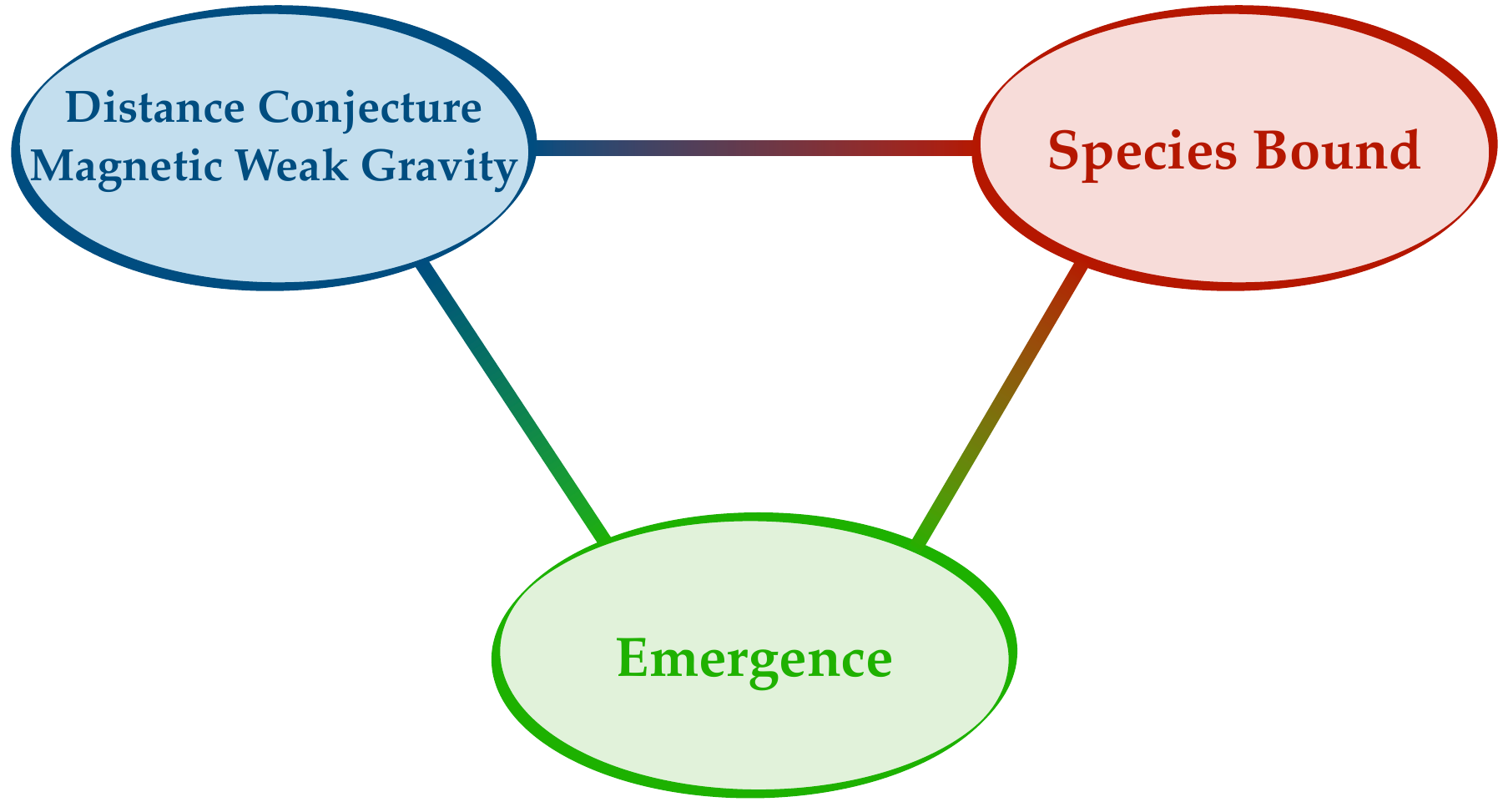}
			\caption{Connections between the Swampland Distance Conjecture, (magnetic) Weak Gravity Conjecture, the Species Bound and Emergence. Any two vertices in the diagram imply the third.}
			\label{fig:triangulo}
		\end{center}
\end{figure}

Above we have argued that combining Emergence and the Species Bound one can obtain the SDC and the magnetic WGC. However, the implications go in all three directions as schematically summarized in fig. \ref{fig:triangulo}. Any two vertices in the triangle imply the third one. Thus, we have already shown how Emergence plus the Species Scale imply the SDC and the magnetic WGC. On the other hand, from eqs. \eqref{eq:metricemergence} and \eqref{eq:gaugecouplingemergence} one concludes that Emergence together with the SDC and magnetic WGC imply the existence of the Species Bound (independently from its possible derivation in terms of graviton loops or rather non-perturbative BH physics, see section \ref{s:speciesscale}). Finally, one can also argue that it is possible to motivate Emergence in the above simplified setting starting from the SDC, the magnetic WGC and the Species Scale as follows. Thus, assuming the SDC in eq. \eqref{eq:metricemergence} one gets
\beq\label{eq:SDCWGC->Emergence}
	\delta g_{\phi\phi}\, \sim\, 
	N \LQG^{d-2} \left(\frac {\partial_\phi m_{\text{tower}}}{m_{\text{tower}}}\right)^2\, \sim\, N^3\LQG^{d-4}(\partial_\phi m_{\text{tower}})^2\, ,
\eeq
where we have substituted eq. \eqref{species} and we have used $\LQG \simeq N\, m_{\text{tower}}$. This may be rewritten as the sum over $N$ contributions
\beq
	\delta g_{\phi\phi}\, \sim\, \sum_{n=1}^N n^2 \LQG^{d-4}(\partial_\phi m_{\text{tower}})^2\, \sim\, \sum_{n=1}^N (\partial_\phi m_n)^2\LQG^{d-4}\, ,
\eeq
where $m_n=n\, m_{\text{tower}}$. But each contribution precisely matches the behaviour of a would-be one-loop graph involving fermions of mass $m_n$ and a Yukawa-like interaction between the tower and the scalar $\phi$ proportional to $\partial_\phi m_n$. Of course this is just a mathematical decomposition of eq. \eqref{eq:SDCWGC->Emergence}, which despite not being unique, is indeed consistent with the Emergence Proposal.

\subsubsection*{Emergence and String/M-theory}
	
In this paper we will use string/M-theory as a testing ground for the Emergence Proposal, so it is interesting to consider how Emergence fits in here in general terms. In sections \ref{s:emergence4dN=2} and \ref{s:Emergenced>4} we analyze several string theory examples of the emergence mechanism in four, six, seven and ten non-compact dimensions. Still, there are some general considerations that one can make before going into the details in each of them. To start with, the case of eleven-dimensional M-theory is rather special because it has no moduli space whatsoever, such that the approach for computing emergent kinetic terms close to infinite distance points that is typically taken cannot be directly applied here. Instead, and especially if the strong version of the Emergence Proposal turns out to be correct and applicable everywhere in Quantum Gravity, we would expect the kinetic terms of M-theory to be understood as well via Emergence but only with the full non-perturbative, UV-complete theory. Alternatively, one could think of 11d M-theory as arising from an asymptotic decompactification limit of 10d type IIA string theory, in which all kinetic terms may be emergent, as we will discuss below. Therefore, in a sense, M-theory would also fit within the emergence picture. 
	
Emergence is based on the quantum corrections to kinetic terms and one can wonder about what happens in theories with `too many' supercharges (i.e. more than 16), like supersymmetric string theory vacua in $d\geq7$, in which supersymmetry may imply strong cancellations within the loop computations. A first relevant perturbative computation is that of the species scale, which involves corrections to the 2-point function of the graviton due to all kind of light matter fields present in the theory and interacting gravitationally, as described in the previous section. However, due to the fact that in gravity there should be no `anti-screening', namely no negative contributions to the self-energy (at least for point-particles in $d>2$) those graphs could never cancel. This is expected on general physical grounds, but e.g. has been tested in four dimensions in \cite{Anber:2011ut,Donoghue:1994dn, Han:2004wt},\footnote{We acknowledge correspondence on this topic with J. Donoghue.} which suggests it may still be true in higher dimensions, given that one can always consider a trivial toroidal compactification down to 4d starting from any higher dimensional theory. This also agrees with the non-perturvative estimation of the species scale in terms of black holes described in section \ref{s:speciesscale}, which matches the perturbative one only if this cancellations never occur. In addition, upon assuming this behaviour one naturally concludes that all supersymmetric partners of the graviton, like graviphotons or gravitini, should also have their kinetic terms generated along similar lines. This would be the case e.g. for the RR $1$-form of 10d type IIA as well as for the graviphoton in 4d $\mathcal{N}=2$ CY$_3$ type II compactifications. These examples of Emergence, along with others, will be discussed in more detail in sections \ref{s:emergence4dN=2} and \ref{s:Emergenced>4}.  
	
Note that for the Emergence Proposal to be correct, it must always lead to `well-defined' kinetic terms. In particular, the \emph{overall} quantum corrections should have the appropriate sign. Therefore, if upon computing such kinetic terms and after having integrated out a tower of states one obtains the wrong sign, self-consistency of the emergence mechanism `predicts' that there should be an additional tower(s) of states that was overlooked and must be included in the computation. Thus, the proposal is far-reaching and should be testable in plenty of string theory vacua.
	
Let us finally comment on the close connection between Emergence and string dualities. Indeed, Emergence is associated to singular limits in the moduli space of the theories under study, those singularities being located at infinite distance. However, there is strong evidence (for Minkowski backgrounds at least) that in these limits one obtains either a decompactification or a string becoming asymptotically tensionless \cite{Lee:2019wij}. Hence e.g., the strings arising are in general different from the original fundamental string considered. Therefore, these singularities correspond in principle to different dualities which exist already in string theory, and the emergence process here discussed would provide (in its strongest formulation) for the kinetics terms of the emergent theory, which appear now typically at the {\it tree level} in the dual frame. A clear example of this process arises in weak coupling regimes of 2-form fields in 6d $\mathcal{N}=(1,0)$ theories, where a nearly tensionless string appears along the limit, as we discuss in section \ref{ss:emergence6d}.
	
Up to now we have only considered the generation of kinetic terms. What about the rest of the interactions in the IR? We are assuming all the way that interaction terms between massless fields and massive bosons/fermions already exist in the UV, since we are using them for the kinetic terms to arise. A particularly interesting issue is that of scalar potentials, which play a crucial role in determining the possible vacua of the theory. Could they be emergent as well? Indeed this may well be the case, as we will see later on in section \ref{s:Scalarpotential}, when studying the generation of the flux potential in type II CY$_3$ (orientifold) compactifications. It would also be interesting to investigate further whether e.g. non-perturbative superpotentials could be emergent as well. 

\subsection{Kinetic terms from one-loop corrections}
\label{ss:selfenergybosons}	
In this section we discuss the  computation of the wave-function renormalization that an infinite tower of bosonic and/or fermionic particles produce on a given modulus or $p$-form gauge field. For concreteness, we perform such computations for towers of spin-0 scalars and spin-$\frac{1}{2}$ Dirac fermions, but keeping in mind that we use them as a proxy to estimate the contribution of general towers of scalars and/or fermions to the quantum loops. We essentially outline the basic logic and main results here, whilst leaving the detailed calculations for the Appendix \ref{ap:Loops}. In particular, we perform all computations in general $d$ spacetime dimensions, and comment on qualitative differences for the contributions from the loops to the 2-point function depending precisely on $d$.
	
\subsubsection{Self-energy of a modulus}
\label{sss:selfenergymodulus}
\begin{figure}[t]
		\begin{center}
			\subfigure[]{
				\includegraphics[scale=1.3]{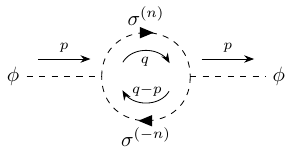} 
				\label{fig:scalarloopscalar}
			}\qquad \qquad
			\subfigure[]{
				\includegraphics[scale=1.3]{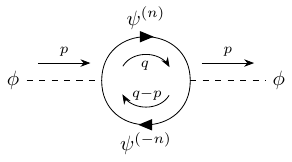}
				\label{fig:scalarloopfermion}
			}
			\caption{ Contributions to the wave-function renormalization for massless scalars from (a) scalar and (b) fermion loops.}			
		\label{fig:scalarpropagator}
		\end{center}
\end{figure} 

We begin by considering a real modulus field, $\phi$, coupled to a tower of massive scalars, $\sigma^{(n)}$, or fermions, $\psi^{(n)}$, through their mass terms. The precise form of the relevant piece of the action can be found in eqs. \eqref{eq:Skinphi}-\eqref{eq:Spsin}, and the precise trilinear couplings that enter into the Feynman diagrams contributing to the process, shown in fig. \ref{fig:scalarpropagator}, can be obtained by expanding the mass term of the states running in the loop up to linear order in the (fluctuation of the) modulus. Their strength is therefore given by
\begin{equation}
		\label{eq:scalarcouplings}
		\lambda_n=2m_n(\partial_\phi m_n)\,,  \qquad \mathrm{and} \qquad \mu_n=\partial_\phi m_n \, ,
\end{equation}
for the massive scalars and fermions, respectively.
	
In the context of Emergence, we are interested in computing the wave-function renormalization of $\phi$,  due to scalar and fermionic loops (see also \cite{Heidenreich:2018kpg, Hamada:2021yxy}). The idea is thus to extract the momentum-dependent part of the exact propagator of $\phi$ at one loop, which takes the following form (see Appendix \ref{ap:Loopsscalar} for details on this point)
\beq
	\label{eq:scalarpropagator}
	D(p^2)=\frac{1}{p^2-\Pi(p^2)}\, .
\eeq
Here, $\Pi(p^2)$ corresponds to the self-energy of the scalar field $\phi$. In the following, we will content ourselves with computing $\Pi(p^2)$ up to $\mathcal{O}(\hbar)$ in the effective action, such that we concentrate on the (amputated) one-loop Feynman diagram displayed in fig. \ref{fig:scalarpropagator}. (Recall that at tree-level $\Pi_0(p^2)\, = \, 0$.) Thus, the correction to the propagator, and hence to the modulus metric, is given by the term in $\Pi(p^2)$ proportional to $p^2$, namely
\begin{equation}\label{eq:metriccorrection}
		\delta g_{\phi \phi}= \frac {\partial \Pi(p^2)}{\partial p^2} \bigg\rvert_{p=0}\, .
\end{equation}

\subsubsection*{Scalar loop}
	
Let us begin by considering the contribution coming from a loop of massive real scalar fields, $\sigma^{(n)}$, corresponding to the Feynman diagram in fig. \ref{fig:scalarloopscalar}, which reads
\beq
	\Pi_n(p^2) \ = \frac{\lambda_n^2}{2} \int \frac {\text{d}^dq}{(2\pi)^d} \frac {1}{(q^2+m_n^2)} \frac {1}{((q-p)^2+m_n^2)}\, ,
	\label{eq:selfenergyscalar}
\eeq
with the coupling $\lambda_n$ defined in eq. \eqref{eq:scalarcouplings}. Since we are interested just in the correction to the exact modulus propagator, we need to extract the term linear in $p^2$ from the expression above, so that we take a derivative with respect to $p^2$ and evaluate the result at $p=0$. This leads to the following integral in momentum space
\beq
	\frac {\partial \Pi_n(p^2)}{\partial p^2} \bigg\rvert_{p=0}  = - \frac{\lambda_n^2}{2} \int \frac {\text{d}^dq}{(2\pi)^d} \frac {1}{(q^2+m_n^2)^3}\, .
	\label{eq:sigmaa}
\eeq
From this expression, one can already anticipate the different behaviour in terms of convergence of the momentum integral depending on whether $d$ is equal, lower than, or greater than 6. Even though for $d<6$ the loop integral is convergent for large $q$,  we introduce a momentum cut-off (that we will identify ultimately with the species scale), since this is required for later consistency once we fix the mass scale up to which we include the contribution from the states in the tower that we integrate out. The exact solution of \eqref{eq:sigmaa}, in terms of hypergeometric functions, takes the form displayed in eq. \eqref{eq:scalarloopscalarexact}, but we will focus here on the dependence of the leading term on the relevant energy scale (i.e. either the mass of the particle running in the loop, $m_n$, or the UV cut-off scale, $\Lambda$), since these are the ones from which we will extract at the end of the day the field dependence of the emergent kinetic terms (the rest are numerical factors with no dependence on any field at all). In particular, there are two relevant limits in which we are interested, namely the limit $\Lambda \gg m_n$, in which the mass of the particle running in the loop is negligible with respect to the UV cut-off, and $\Lambda\simeq m_n$, where the mass of the particle is roughly at the UV cut-off scale. The former is relevant for most states of KK-like towers in asymptotic limits, since as we have seen in the previous section the mass scale of the tower is typically asymptotically lighter than the species scale. The latter is relevant for the highest states in KK-like towers and for most states in stringy-like towers (up to logarithmic corrections), since the masses of the corresponding states are asymptotically of the same order as the species scale. Luckily for us, and as shown in Appendix \ref{ap:Loopsscalar}, both limits give rise to the same functional dependence on either $\Lambda$ or $m_n$, being essentially the numerical prefactors (which do not include any field dependence relevant for Emergence) the only difference between them.\footnote{Actually, the numerical prefactors as shown in Appendix \ref{ap:Loops} can be thought of as upper and lower limits for the contributions from each particle in the loop, since they include the two limiting cases for the mass of the relevant particles.} From the results summarized in tables \ref{tab:scalarloopscalarLambda>>m} and \ref{tab:scalarloopscalarLambda=m} we can extract the following asymptotic dependence for the dominant contribution to the 2-point function
\begin{equation}\label{eq:scalarloopscalarssummary}
		\frac {\partial \Pi_n(p^2)}{\partial p^2} \bigg\rvert_{p=0}\,   \sim 
		\left\{\begin{array}{lr}
			-  \dfrac{\lambda_n^2}{m_n^{6-d}} & \qquad\text{for } d< 6\, ,\\ \\ 
			-\lambda_n^2 \ \log \left( \dfrac{\Lambda^2}{m_n^2}\right) &\qquad \text{for } d= 6\, ,\\ \\ 
			-\lambda_n^2 \ \Lambda^{d-6}&\qquad \text{for } d>6\, ,
		\end{array}\right.
\end{equation}
where the precise meaning of $\sim$ is that we keep track of all the factors that are field dependent (and will eventually contribute to the field dependence of the emergent metric) and neglect only the numerical prefactors. Note that the computation for a scalar loop  was also done in \cite{Hamada:2021yxy} and their results are in agreement with ours.
	
Before proceeding with the fermionic loop, let us make some comments here about possible natural generalizations of the scalar case just discussed. First, as typically happens in supersymmetric theories, one could consider a tower of \emph{complex} massive scalars coupling to the modulus $\phi$ through its mass. This scenario reduces essentially to the one just described, since one can always write a complex field in terms of its real and imaginary parts, $\chi^{(n)}=\frac{\sigma^{(n)}_1+i\sigma^{(n)}_2}{\sqrt{2}}$, which both share the same (moduli-depenedent) mass and thus contribute to the modulus self-energy as summarized in \eqref{eq:scalarloopscalarssummary} above with an extra factor of 2. Second, one could even study the case in which the modulus itself is complexified, namely $\phi \to \Phi = \text{Re}\, \Phi + i\, \text{Im}\, \Phi$. The scalar charges \eqref{eq:scalarcouplings} would now be complex-valued, and similar considerations would lead to the one-loop generated metric $g_{\Phi \bar \Phi}$.\footnote{Notice that the fact that one obtains a hermitian metric at one loop, $\delta g_{\Phi \bar \Phi}$, is due to the pseudo-scalar nature of the imaginary part of the modulus, which prevents a term of the form $(\partial \text{Re}\, \Phi)(\partial \text{Im}\, \Phi)$ from appearing in the effective action.}

\subsubsection*{Fermionic loop}
	
We consider now the contribution to the propagator from a loop of massive fermions, $\psi^{(n)}$, with the trilinear couplings $\mu_n$ as defined in eq. \eqref{eq:scalarcouplings}. The relevant Feynman diagram, shown in fig. \ref{fig:scalarloopfermion}, reads as follows
\begin{equation}
		\label{eq:selfenergyfermion}
		\Pi_n(p^2) \, = \, -\mu_n^2  \int \frac {\text{d}^dq}{(2\pi)^d}\ \text{tr} \left (\frac {1}{i \slashed{q}+m_n}\ \frac {1}{i (\slashed{q}-\slashed{p})+m_n} \right) \, .
\end{equation}
After performing the relevant traces and rearranging terms (see around eq. \eqref{eq:scalarloopfermionstraces} for details), we get the following  expression for the one-loop contribution to the wave-function renormalization 
\begin{equation} \label{eq:wfrfermionloop}
		\frac{\partial \Pi_n(p^2)}{\partial p^2} \bigg\rvert_{p=0} \, = \,   -\mu_n^2\, \fdim  \int \frac {\text{d}^dq}{(2\pi)^d} \frac{1}{(q^2+m_n^2)^2} \ + \ 2 m_n^2 \, \mu_n^2\ \fdim \int \frac {\text{d}^dq}{(2\pi)^d} \frac{1}{(q^2+m_n^2)^3} \, .
\end{equation}
The first term is negative and it looks similar to the scalar contribution, with the difference that it naively diverges for $d\geq 4$ instead of $d\geq 6$. Performing a similar analysis as the one for the scalar loop, we obtain akin results (see Appendix \ref{ap:Loopsscalar}). Namely, for the two limits of interest, $\Lambda \gg m_n$ and $\Lambda \simeq m_n$, we get the same functional dependence on the coupling constants and the pertinent energy scales. The only difference between the two limits being the numerical prefactors, that once again play no role in the field-dependent part of the emergent metric. The detailed results are summarized in tables \ref{tab:scalarloopfermionLambda>>m} and \ref{tab:scalarloopfermionLambda=m}, which take the form (neglecting once again numerical prefactors)
\begin{equation}\label{eq:scalarloopfermionssummary}
		\frac {\partial \Pi_n(p^2)}{\partial p^2} \bigg\rvert_{p=0}\,   \sim 
		\left\{\begin{array}{lr}
			-  \dfrac{\mu_n^2}{m_n^{4-d}} & \qquad\text{for } d< 4\, ,\\ \\ 
			-\mu_n^2\ \log \left( \dfrac{\Lambda^2}{m_n^2}\right) &\qquad \text{for } d= 4\, ,\\ \\ 
			-\mu_n^2\ \Lambda^{d-4}&\qquad \text{for } d>4\, .
		\end{array}\right.
\end{equation}
The second term in eq. \eqref{eq:wfrfermionloop} is of the same form as the scalar contribution \eqref{eq:sigmaa}, with $\lambda_n= 2 m_n (\partial_\phi m_n)=2 m_n \mu_n$, including also a prefactor of $\fdim$ which takes into account the number of fermionic degrees of freedom in $d$ dimensions. Moreover, it has the opposite sign as the scalar contribution, so for supersymmetric theories both terms cancel and the leading contribution to the emergent metric comes from the first term in \eqref{eq:wfrfermionloop}. For example, in 4d, a single Dirac fermion seems to cancel the renormalization due to two complex or four real scalar fields, so that e.g. in 4d $\mathcal{N}=2$, a hypermultiplet only contributes to the modulus metric through the fermion loop.\footnote{A similar cancellation (this time exact) occurs for the would-be mass term generated for the modulus in case supersymmetry is present in our theory, which can be checked explicitly upon using our formulae, although one may need to consider some extra loop diagrams not contributing to the wave-function renormalization and thus not displayed in fig. \ref{fig:scalarpropagator}.} This suggests that following the leading order contribution coming from fermionic towers, as shown in eq. \eqref{eq:scalarloopfermionssummary}, is a good proxy for the leading terms in the emergent kinetic terms.

\subsubsection{Emergence of the moduli metrics}
\label{sss:emergencemodulimetric}
	
Armed with the above results, we study now in turn the emergence of the metric for real scalar moduli in the two relevant cases of Kaluza-Klein and string towers. We will see that, in spite of the apparently different dependence on the cut-off $\Lambda$ between scalar and fermion loop contributions, the final results for Emergence are essentially the same.
	
\subsubsection*{Moduli metric from  Kaluza-Klein towers}
	
Let us consider a KK-like tower of scalars with masses and associated species scale given by\cite{Castellano:2021mmx}
\beq
	m_n\, \simeq\, n^{1/p}\, \Mt \, , \qquad   \LQG\,  \simeq\, N^{1/p}\, \Mt \, ,
\eeq
so that e.g. a single KK tower would correspond to $p=1$. Note that an analogous analysis can be performed with fermionic towers and similar results are obtained. The  contribution of a single scalar to the wave-function renormalization of a modulus is given in \eqref{eq:scalarloopscalarssummary} and for $d>6$ it takes the form
\beq
	\delta g_{\phi\phi}^{(n)}\, \sim\,  \mathcal{A}_d\, \lambda_n^{2}\, \Lambda^{d-6}\, \sim\, 4\, \mathcal{A}_d\, m_n^2\, (\partial_\phi m_n)^2\, \Lambda^{d-6}\, ,
\eeq
with $\mathcal{A}_d$ a numerical prefactor depending only on $d$ that is not explicitly included in eq. \eqref{eq:scalarloopscalarssummary}. Its precise value for the two relevant limits is displayed in tables \ref{tab:scalarloopscalarLambda>>m} and \ref{tab:scalarloopscalarLambda=m}. After adding up the contribution from the states of the tower below the cut-off $\Lambda = \LQG $ (and by approximating the sum by an integral), we get
\beq 
	\delta g_{\phi\phi} \,=\, \sum_{n=1}^{N} \delta g_{\phi \phi}^{(n)}  \, \sim\, \frac {4 p \mathcal{A}_d}{p+4}\,  N^{\frac{4}{p}+1}\,  (\partial_\phi \Mt)^2\, \Mt^2\, \LQG^{d-6}\, \sim\, \frac {4 p \mathcal{A}_d}{p+4}\, \Mpd^{d-2} \left( \frac{\partial_\phi \Mt}{\Mt}\right)^2\, ,
\eeq
where we have used $N \simeq\Mpd^{d-2}/ \LQG^{d-2}$ as well as $\LQG\, \simeq\, N^{1/p}\, \Mt$. Thus, the non-trivial dependence on the characteristic mass of the tower, namely $\delta g_{\phi\phi} \sim 1/\Mt^2$, is recovered for any dimension, leading to the structure needed for the distance conjecture to hold a posteriori. Notice that the dependence on the tower density parameter, $p \in \mathbb{R}$, only enters through the numerical prefactor, so that it is to all effects `irrelevant' for the exponential behaviour (as long as we identify the UV cut-off with the species scale).

For the case with $d<6$ the contribution from one real massive scalar to the loop integral is of the form (c.f. eq. \eqref{eq:scalarloopscalarssummary})
\beq
	\delta 
	g_{\phi\phi} ^{(n)}\, \sim\, \mathcal{B}_d\, \lambda_n^2\,  m_n^{d-6}\, \sim\, 4\, \mathcal{B}_d\, n^{\frac{d-2}{p}}\, \Mt^{d-4}\, (\partial_\phi \Mt)^2\, .
\eeq
Again, $\mathcal{B}_d$ is the $d$-dependent numerical prefactor whose precise value is bounded by the ones in tables \ref{tab:scalarloopscalarLambda>>m} and  \ref{tab:scalarloopscalarLambda=m}. The total contribution from a tower up to the species scale is therefore
\beq 
	\delta g_{\phi\phi}\,=\, \sum_{n=1}^{N} \delta g_{\phi \phi}^{(n)}\, \sim\, \frac {4p\mathcal{B}_d}{d-2+p}\, N^{\frac{d-2+p}{p}}\, \Mt^{d-4}\, (\partial_\phi \Mt)^2\, \sim\, \frac {4p\mathcal{B}_d}{d-2+p}\, \Mpd^{d-2} \left( \frac{\partial_\phi \Mt}{\Mt}\right)^2\, ,
\eeq
so that essentially we arrive at the same expression as for $d>6$, with a different numerical coefficient. Similarly, the same behaviour is obtained for the marginal case $d=6$ up to a numerical prefactor.
	
\subsubsection*{Moduli metric from stringy towers}
As a second example, consider this time the coupling of a real modulus $\phi$ to the components of a (critical) string tower. For concreteness, we choose to do the computation with the fermionic modes of the tower only, although the same analysis may be repeated using the bosonic ones yielding similar results. Which specific type of fundamental string we consider is in practice irrelevant for the computation of the emergent metric, as we will see. Let us study first the case $d>4$. Consider the spectrum of a string with masses for fermion string excitations and corresponding degeneracies
\beq
	m_n^2 = 16 \pi^2 (n-1) M_s^2\, , \qquad  d(n)\, \sim\,  n^{-b}e^{a\sqrt{n}}\, ,
\eeq
where $a$ and $b$ are constants characteristic of each string. Thus e.g. for $d=10$ type II string one has $a=4\pi \sqrt{2}$ and $b=11/2$. For simplicity, we will take in this section $d(n) \sim e^{\sqrt{n}}$ since it is enough for our purposes and the results are already correct up to log corrections. We assume that the string scale $M_s$ depends on $\phi$ when measured in Planck units, as indeed happens in string theory (see eq. \eqref{eq:ddimdilaton}). Then the total contribution for the metric from fermion loops is 
\beq
	\delta g_{\phi \phi}\, \sim\, \sum_{n=1}^{N_s}\mu_n^2\, d(n)\, \LQG^{d-4} \sim \sum_{n=1}^{N_s} (\partial_\phi M_s)^2\, n\, d(n)\, \LQG^{d-4}\, .
\eeq
Recalling as well that $\LQG ^2 \simeq N_s\, M_s^2$, one finds
\beq
	\delta g_{\phi \phi}\, \sim\, M_s^{d-4}N_s^{\frac{d-4}{2}} (\partial_\phi M_s)^2\int^{N_s}_1 dn\, n\;  e^{\sqrt{n}} \sim  M_s^{d-4} N_s^{\frac{d-4}{2}} (\partial_\phi M_s)^2  N_s^{3/2}e^{\sqrt{N_s}}\, .
\eeq
Using now the expression \eqref{eq:maxstringlevel} we finally get
\beq\label{eq:modulusWFstringtow}
	\delta g_{\phi \phi}\, \sim\, \frac {(\partial_\phi M_s)^2}{M_s^2}\Mpd^{d-2}\, ,
\eeq
which is again the expected asymptotic behaviour. Note that the explicit dependence on $N_s$ (and hence on the species scale $\LQG$) drops out, and the result only depends on the mass of the lightest string excitation $M_s$, as it happened with the KK-like towers. It is easy to check that \eqref{eq:modulusWFstringtow} is also reproduced for $d\leq 4$. An analogous analysis may be performed using the bosonic loop instead of the fermionic one, yielding similar results.

\subsubsection{Self-energy of a gauge 1-form}
\label{sss:selfenergy1form}
\begin{figure}[t]
		\begin{center}
			\subfigure[]{
				\includegraphics[scale=1.3]{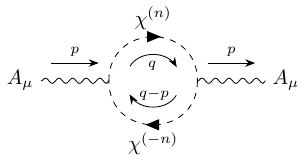} 
				\label{fig:1-formloopscalar}
			}\qquad \quad 
			\subfigure[]{
				\includegraphics[scale=1.3]{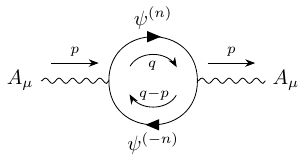}
				\label{fig:1-formloopfermion}
			}
			\caption{Wave-function renormalization for $U(1)$ gauge bosons due to (a) charged scalars and (b) fermion loops.}			
			\label{fig:1-formpropagator}
		\end{center}
\end{figure} 	
Let us now consider the contribution coming from loops of complex scalars, $\chi^{(n)}$, and Dirac fermions, $\psi^{(n)}$, to the propagator of a gauge 1-form, denoted $A_1$. In particular, we take the mass and charge of the $n$-th scalar or fermion to be given by $m_n$ and $q_n$, respectively, whereas the gauge coupling associated to the 1-form is denoted $g$. (c.f. eqs. \eqref{eq:SkinA1}-\eqref{eq:SpsinA1} for the details on the conventions and the precise action).
	
As in the modulus case of section \ref{sss:selfenergymodulus}, in order to compute the emergent kinetic terms for the 1-form we must study the relevant Feynman diagrams that contribute to the wave-function renormalization of $A_1$. Subsequently, we need to extract the momentum-dependent part of the exact propagator after taking into account the one-loop corrections coming from the scalar and fermion fields. We will use the  Lorenz gauge $\partial_\mu A^{\mu}\, =\, 0$, since it can be easily generalized to higher $p$-forms as $\partial_\mu A^{[\mu \nu_1 \ldots \nu_{p-1}]}\, = \, 0$. In this gauge, the propagator for the 1-form can be written as
\begin{equation}
		\label{eq:A1propagator}
		D^{\mu \nu} (p^2) \, = \, \left( \dfrac{p^2}{g^2} \delta ^{\mu \nu} - \Pi^{\mu \nu}(p^2) \right)^{-1}\, ,
\end{equation}
where $\Pi^{\mu \nu}(p^2)$ vanishes at tree level since it is the amputated Feynman diagram coming from the loops shown in fig. \ref{fig:1-formpropagator}. By imposing again our gauge choice, we can extract the tensorial dependence on
\begin{equation}
		\label{eq:A1loopamplitude}
		\Pi^{\mu \nu} (p^2) \, = \, \Pi(p^2) \delta ^{\mu \nu} \, ,
\end{equation}
so that we will be interested in the term linear in $p^2$ within $\Pi(p^2)$, as it gives precisely the correction to the propagator and thus to the gauge coupling, namely
\begin{equation}\label{eq:gaugecouplingcorrection}
		\delta\left(\dfrac{1}{g^2} \right) = \frac {\partial \Pi(p^2)}{\partial p^2} \bigg\rvert_{p=0}\, .
\end{equation}
We want to extract precisely the piece linear in $p^2$ from $\Pi(p^2)$ coming from the aforementioned loop corrections, both for the case in which the particle running in the loop is either bosonic or fermionic.
	
Before going into the systematics of the loop calculations, let us remark that for a general gauge $p$-form, we would have a propagator with Lorentz indices $D^{\mu_1 \ldots \mu_p}_{{\nu_1 \ldots \nu_p}}$, instead of $D^\mu_\nu$ in \eqref{eq:A1propagator}. Then, upon working in Lorenz gauge we would only need to replace ${\delta^{\mu}_\nu \, \to \, p! \, \delta^{[\mu_1}_{[\nu_1}\ldots \delta^{\mu_p]}_{\nu_p]}}$ in the previous equations in order to obtain the right Lorentz structure.
	
\subsubsection*{Scalar loop}
	
We start with the contribution due to the loop of complex charged scalars $\chi^{(n)}$, with mass $m_n$ and charge $q_n$, given by the Feynman diagram shown in fig. \ref{fig:1-formloopscalar}, which reads
\beq
	\Pi^{\mu \nu}_n(p^2) \, =\,   g^2 \, q_n^2 \int \frac {\text{d}^dq}{(2\pi)^d} \frac {(2q-p)^{\mu} (2q-p)^{\nu}}{(q^2+m_n^2)\left( (q-p)^2+m_n^2\right)} \, .
	\label{eq:A1scalar(ap)}
\eeq
Hence, the one-loop correction to the gauge field propagator is given by (see Appendix \ref{ap:Loops1-form} for details)
\begin{equation}
		\frac{\partial \Pi^{\mu \nu}_n(p^2)}{\partial p^2} \bigg\rvert_{p=0} \, = \, -g^2\,  q_n^2\,   \frac{4}{d} \, \delta^{\mu\nu} \, \int \dfrac{d^d q}{(2\pi)^d} \dfrac{q^2}{(q^2+m_n^2)^3} \, .
\end{equation}
Similarly to the discussion about the modulus case, it can be seen that this expression is expected to diverge for $d\geq 4$. However, we will still introduce a UV cut-off for any $d$, since the goal is to identify it with the species scale once we integrate the different states in the relevant towers up to that energy scale. The exact expression for the amplitude is computed in eq. \eqref{eq:1-formloopscalarexact}, but since we are mainly interested in the asymptotic dependence with the mass, the cut-off, and the charges, given that these are the ones that will generate the field dependence of the kinetic terms, we will only retain the leading order terms. The two relevant limits are again $\Lambda \gg m_n$ and $\Lambda \simeq m_n$, which roughly correspond to the KK and the stringy tower cases, respectively. In both cases, as discussed in detail in Appendix \ref{ap:Loops1-form} (see tables \ref{tab:1-formloopscalarLambda>>m} and \ref{tab:1-formloopscalarLambda=m}), the leading dependence in the aforementioned (eventually field-dependent) quantities is the same for any $d$, and only the numerical prefactors change. Hence, we can summarize the relevant part of the leading correction to the propagator as follows
\begin{equation}\label{eq:1-formloopscalarssummary}
		\frac {\partial \Pi_n(p^2)}{\partial p^2} \bigg\rvert_{p=0}\,   \sim 
		\left\{\begin{array}{lr}
			-  \dfrac{g^2\,  q_n^2}{m_n^{4-d}} & \qquad\text{for } d< 4\, ,\\ \\ 
			-g^2\,  q_n^2 \ \log \left( \dfrac{\Lambda^2}{m_n^2}\right) &\qquad \text{for } d= 4\, ,\\ \\ 
			-g^2\,  q_n^2\ \Lambda^{d-4}&\qquad \text{for } d>4\, .
		\end{array}\right.
\end{equation}

\subsubsection*{Fermionic loop}
	
Including now the Dirac fermions $\psi^{(n)}$, with mass $m_n$ and charge $q_n$, in the loop given by fig. \ref{fig:1-formloopfermion} we get
\begin{equation}
		\Pi^{\mu\nu}_n(p^2) \ = - (ig)^2 \, q_n^2  \int \frac {\text{d}^dq}{(2\pi)^d}\ \text{tr} \left (\frac {1}{i \slashed{q}+m_n}\ \gamma^\mu \ \frac {1}{i (\slashed{q}-\slashed{p})+m_n} \ \gamma^\nu  \right)\, .
\end{equation}
After performing the traces and selecting the terms that survive the angular integration (see discussion around eq. \eqref{eq:traces1-formloop}) we arrive at the following result for the piece that corrects the propagator
\begin{equation}
	\label{eq:1-formloopfermions2terms}
		\frac{\partial \Pi^{\mu \nu}_n(p^2)}{\partial p^2} \bigg\rvert_{p=0}  = \, -\fdim\, g^2\,  q_n^2 \, \delta^{\mu\nu}  \int \dfrac{d^d q}{(2\pi)^d} \dfrac{1}{(q^2+m_n^2)^2}\, + \,  \fdim\,  g^2\,  q_n^2\,   \frac{2}{d} \, \delta^{\mu\nu}  \int \dfrac{d^d q}{(2\pi)^d} \dfrac{q^2}{(q^2+m_n^2)^3}\, .
\end{equation}
The first term can be computed after introducing the UV cut-off $\Lambda$ and it gives the exact result presented in eq. \eqref{eq:1-formloopfermionexact}. In the two interesting limits, namely when $\Lambda \gg m_n$ or $\Lambda \simeq m_n$, it can be seen that the functional dependence with $g$, $q_n$, $m_n$ and $\Lambda$ is again the  same for a given $d$, where only the numerical coefficients in front are different (see tables \ref{tab:1-formloopfermionLambda>>m} and \ref{tab:1-formloopfermionLambda=m}). The pertinent leading expressions take the form
\begin{equation}\label{eq:1-formloopfermionssummary}
		\frac {\partial \Pi_n(p^2)}{\partial p^2} \bigg\rvert_{p=0}\,   \sim 
		\left\{\begin{array}{lr}
			-  \dfrac{g^2\, q_n^2}{m_n^{4-d}} & \qquad\text{for } d< 4\, ,\\ \\ 
			-g^2\, q_n^2\ \log\left( \dfrac{\Lambda^2}{m_n^2}\right) &\qquad \text{for } d= 4\, ,\\ \\ 
			-g^2\, q_n^2\ \Lambda^{d-4}&\qquad \text{for } d>4\, .
		\end{array}\right.
\end{equation}
The second term in eq. \eqref{eq:1-formloopfermions2terms} can be seen to be equal, but with opposite sign, to the scalar contribution, up to a relative prefactor that accounts for the difference in the number of degrees of freedom, given by $\fdim /2$. In parallel to the modulus case, this indicates that in the presence of unbroken supersymmetry the contribution from bosons would cancel against this second term coming from the fermionic loop, such that the first term in the fermion diagram seems to be again a good proxy for keeping track of the leading order correction to the kinetic terms of the massless fields.

\subsubsection{Emergence of $U(1)$ kinetic terms}
\label{sss:emergenceU(1)}
Having computed the general correction to the kinetic term of a gauge 1-form, we describe now how it can be employed to generate via Emergence the abelian gauge kinetic function in the two relevant cases of Kaluza-Klein and stringy towers. 
\subsubsection*{$U(1)$ kinetic terms from Kaluza-Klein towers}
Let us consider again a tower of the general form
\beq
	m_n\, \simeq\, n^{1/p}\, \Mt\, , \qquad  \LQG\, \simeq\, N^{1/p}\, \Mt\, ,
	\label{eq:abelianmasstower}
\eeq
where we assume that the tower involves particles with quantized charge $q_n$ under some $U(1)$. A crucial difference with the emergence for moduli scalars is the functional form of the charges with respect to the integer $n$ labelling the states in the tower, which can be somewhat model-dependent. For more generality, we will consider a useful parameterization given by\footnote{In this section we normalize the vector fields $A_{\mu}$ so that they have mass dimension one (c.f. eq. \eqref{eq:SkinA1}). Later on, in sections \ref{s:emergence4dN=2} and \ref{s:Emergenced>4}, we adopt the usual conventions in supergravity/string theory where all fields are dimensionless. Thus, upon doing so, eq. \eqref{eq:abelianquantizedcharges} catches some extra factors involving the appropriate Planck/string length.}
\beq
	q_n =  n^{1/r}\, ,
	\label{eq:abelianquantizedcharges}
\eeq
with $r=1,\infty$. For $r=1$ one recovers the charges of a standard KK-like tower under the KK photon, whereas $r \to \infty$ corresponds to the case in which all states in the tower present the same (constant) charge under the $U(1)$ field. String theory realizations of both kind towers can actually be found in concrete examples, as we will see e.g. for the D0 and D0-D2 towers analyzed in section \ref{s:emergence4dN=2}. Note that for $r=p$ one has consistency with a BPS tower whereas that is not the case if $r\not=p$. We have just seen that e.g. a single fermion of mass $m_n$ contributes to the gauge kinetic function of a $U(1)$ boson like (for $d>4$)
\beq
	\delta \left(\frac {1}{g^2}\right) \bigg\rvert_{n-\text{th}} \sim\,  q_n^2\,  \mathcal{C}_d\, \LQG^{d-4}\, ,
\eeq
with $\mathcal{C}_d$ a numerical prefactor which only depends on $d$. Its precise value for the two limits of interest is displayed in tables \ref{tab:1-formloopfermionLambda>>m} and \ref{tab:1-formloopfermionLambda=m}. Approximating the sum over the full tower by an integral leads to
\beq\label{eq:gaugeemergenceddimensions}
	\delta \left(\frac {1}{g^2}\right)\, \sim\, \sum_{n=1}^N n^{2/r} \mathcal{C}_d\, \LQG^{d-4}\, \sim\, \frac {\mathcal{C}_d\, r}{2+r}\, \left(\frac {\Mpd^{d-2}} {\Mt^2}\right)^{\alpha_{d,r}/\alpha_{d,p}}\, ,
\eeq
where we have used eqs. \eqref{eq:abelianmasstower} and \eqref{eq:abelianquantizedcharges} as well as the species bound, with
\beq
	\alpha_{d,p} = \frac {d-2+p}{2p(d-1)}\, , \qquad  \alpha_{d,r} = \frac {d-2+r}{2r(d-1)}\, .
	\label{eq:dosalphas}
\eeq
Note that for $r=p$ one finds $\alpha_{d,p} =\alpha_{d,r}$ and $1/g^2\sim 1/\Mt^2$, as expected for the particular case of a BPS tower. 

It is interesting to also consider the case in which only a sub-lattice of the full charge lattice is realized in the spectrum of a given tower. Therefore, let us consider an infinite set of particles labeled by $n\in \mathbb{Z}$ and charges given by $q_n=kn$, with $k$ being some fixed positive integer. Then the above analysis for $r=p=1$ still applies except for the replacement $1/g^2\rightarrow k^2/g^2$, so that one rather has
\beq
  m^2\, \lesssim\, k^2 g^2 \Mpd^{d-2}\, .
\eeq
Hence, we see that in this case the WGC is slightly weakened. Interestingly, this result is in agreement with the $\mathbb{Z}_k$-WGC as recently formulated in \cite{Buratti:2020kda}.

\subsubsection*{$U(1)$ kinetic terms from stringy towers}

Concerning the gauge kinetic function quantum-induced by the oscillation states of a stringy tower, the exact results of course depend on e.g. which type of string is involved in the limits we study as well as the precise origin of the $U(1)$ whose renormalization we care about. In practice, this means that it is difficult to provide completely general results. Still, we are going to discuss in here a particular structure which turns out to appear in large classes of e.g. heterotic compactifications down to $d<10$ spacetime dimensions. Consider, for concreteness, a tower of fermionic fields with some integer-valued charge $q$ at the $n$-th oscillator level of a given string theory compactification in $d>4$. We can try to estimate their contribution to the gauge kinetic function at the one-loop level as follows
\beq\label{eq:oneloopstringtowergauge}
	\delta \left(\frac {1}{g^2}\right)\bigg\rvert_{n-\text{th}} \sim\, \sum_q^{q_{\text{max}}} q^2\, d(q,n)\, \LQG^{d-4} .
\eeq
Here the function $d(q,n)$ parameterizes the degeneracy of each charge $q\in \mathbb{Z}$ present at the $n$-th oscillator level of the emergent string. However, since we know (c.f. section \ref{s:speciesscale}) that the \emph{total} degeneracy at each level $n$ behaves roughly as $d_n \sim e^{\sqrt{n}}$ (for large $n$), we will propose as an anzatz for the function $d(q,n)$ the following 
\beq
\label{eq:stringytowerdensity}
	d(q,n)\, \sim\, f(q)\, e^{\sqrt{n}}\, ,
\eeq
where $f(q)$ is some \emph{polynomial} function of the $U(1)$ charges, normalized as $\sum_q^{q_{\text{max}}} f(q) = 1$ so as to have total degeneracy equal to $d_n \sim e^{\sqrt{n}}$. Notice that we have assumed that there exists a maximum charge, $q_{\text{max}}$, at each oscillator level, which in principle could be some arbitrary function of $n$. However, in the following we will take $q_{\text{max}} \sim \sqrt{n}$, a choice which is motivated from our experience with the heterotic string.\footnote{This ultimately arises from the level matching condition on the left- and right-handed movers of the heterotic string, namely $\alpha'm^2/4=N_R =N_L+\textbf{Q}^2/2-1$ (see Appendix \ref{ss:heterotic10d} for more details on this issue).}
Hence, we obtain for each oscillator level the following one-loop contribution
\beq
	\delta \left(\frac {1}{g^2}\right)\bigg\rvert_{n-\text{th}} \sim\, n\, e^{\sqrt{n}}\, \LQG^{d-4}\, ,
\eeq
where we have approximated the sum over the gauge charges $q$ in eq. \eqref{eq:oneloopstringtowergauge} by an integral $\int_0^{q_{\text{max}}} \text{d}q\ f(q)\ q^2 \sim n$. Summing now over all oscillator levels of the tower up to $n_{\text{max}} = N_s$, one finds (again by approximating the sum by a definite integral over $n$)
\beq
	\delta \left(\frac {1}{g^2}\right)\, \sim\, \LQG^{d-4}  \int^{N_s}_1 \text{d}n\, n\, e^{\sqrt{n}}\, \sim\, \LQG^{d-4}\, N_s^{3/2}e^{\sqrt{N_s}}\, \sim M_s^{d-4}\, N_s^{\frac{d-1}{2}}\, e^{\sqrt{N_s}}\, ,
\eeq
where we have made use of the relation $\LQG \simeq \sqrt{N_s} M_s$. Finally, upon substituting eq. \eqref{eq:maxstringlevel}  in the expression above we get
\beq\label{eq:1-formemergencestringytowersgeneral}
	\delta \left(\frac {1}{g^2}\right)\, \sim\, \frac {\Mpd^{d-2}}{M_s^2}\, ,
\eeq 
for the resummed wave-function renormalization induced by the tower of string modes (up to the species scale). As a result, one finds once again that $1/g_{\text{IR}}^2 \sim 1/M_s^2$ (in Planck units), in the same way as happened for the KK-like tower before. One also recovers a magnetic WGC expression applied to this stringy set-up, with $M_s^2 \sim g^2\, \Mpd^{d-2}$. Notice, however, that the example analyzed here is to some extent just a string-theory-inspired toy model. We will discuss with more detail a particular realization of this scenario as arising from a 6d (emergent) heterotic string obtained from a singular limit within the (K\"ahler) moduli space of F-theory compactified on an elliptic CY$_3$ in section \ref{ss:emergence6d}. In that example, the structure of masses and charges discussed in the present section appears, and indeed an IR kinetic term for the relevant $U(1)$ gauge field, which depends inversely on the tension of the emergent heterotic string (in 6d Planck units), is obtained. 
	
\subsection{Kinetic terms for fermionic fields}
\label{ss:kineticfermions}
	
In section \ref{ss:selfenergybosons} above we restricted ourselves to study the emergence phenomenon associated to the kinetic terms of fields with \emph{bosonic} spin-statistics, namely scalars and 1-form gauge fields. We now turn to the generation of kinetic terms for light \emph{fermions} in our quantum gravity theories. In fact, this is an important case from the phenomenological point of view since most of the Standard Model (SM) particles are fermions.
	
\begin{figure}[t]
		\begin{center}
			\subfigure[]{
				\includegraphics[scale=1.3]{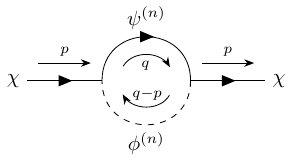} 
				\label{fig:fermionloopscalar}
			}\qquad \qquad 
			\subfigure[]{
				\includegraphics[scale=1.3]{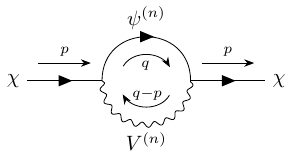}
				\label{fig:fermionloopvector}
			}
			\caption{One-loop diagrams contributing to the wave-function renormalization of light fermion fields in our theories.}
			\label{fig:kineticfermionsbas}
		\end{center}
\end{figure}

As first discussed in ref. \cite{Palti:2020tsy}, the fermionic case is different for essentially two reasons. To start with, the loops generically involve two different towers, one with fermions and the other with bosons, see fig. \ref{fig:kineticfermionsbas}. These two towers  are in general independent and may have different mass scales $m_{\mathrm{b}}$, $m_{\mathrm{f}}$  as well as different structure. Secondly, the coupling of the light fermions to the particles in the towers $Y_n$ appears to be somewhat model-dependent. Thus, recall that in the case of the kinetic terms for moduli scalars or gauge bosons, those couplings arose either from the moduli-dependence of the mass of the heavy fields or were rather determined by their gauge charge $q_n$, respectively. With light fermions, the analogous couplings to the massive towers are only fixed whenever the fermionic field corresponds to a gaugino in a supersymmetric theory, in which case the interaction is again controlled by the charge $q_n$. However, in a more general set-up one needs to specify the structure of the towers involved as well as their couplings, or perhaps make some simplifying assumptions about them in order to obtain specific results for the emergent kinetic term.
	
Before studying the emergence mechanism applied to fermionic fields, we need to first compute the relevant one-loop diagrams. We will always have in mind in what follows the 4d case, which has possible direct applications to the real world, although the computations will be performed in $d$ spacetime dimensions for completeness. We will moreover consider the case in which $\chi$ is a \emph{chiral} fermion (i.e. we restrict to even-dimensional spacetimes). This is to ensure that there is no mass term generated at the quantum level for our field, even without supersymmetry. The natural couplings would be thus of the general form
\begin{equation}\label{eq:fermion&bosonYukawas}
		Y_n\ \overline{\phi^{(n)}} \left(\psi^{(n)}\chi \right)\, , \qquad  {\tilde Y}_n \left(\psi^{(n)} \sigma^{\mu} \overline \chi \right) V_{\mu}^{(n)}\, ,   
\end{equation}
where the $\sigma$-matrices belong to the appropriate irreducible spinor representation of the Lorentz group in $d=2k$ dimensions acting on Weyl spinors, and $n \in \mathbb{Z} \setminus \lbrace 0 \rbrace$ labels the step in the tower. Such towers present in the bosonic sector either massive complex scalars or vector fields, $\phi^{(n)}, V_{\mu}^{(n)}$, respectively. The Weyl fermions $\psi^{(n)}$, on the other hand, pair up with their charge conjugate (say the one labeled by $-n$) so as to form massive Dirac fermions, i.e. $\Psi^{(n)}=\left(\psi^{(n)}, \overline{\psi^{(-n)}} \right)^{\text{T}}$. Moreover, both towers would be a priori `independent',\footnote{In principle one could consider two  independent  indices $m, n \in \mathbb{Z}$ to label the fields, $\lbrace \phi^{(m)}, \psi^{(n)} \rbrace$, as well as the interactions in eq. \eqref{eq:fermion&bosonYukawas}, $Y_{m, n}$. For simplicity, we choose to have only diagonal couplings, namely those between fields with $m=n$, as happens e.g. when there is some conserved charge or the towers involve supersymmetric partners.} meaning that their mass spectra are not in principle correlated, with the Yukawa-like couplings $\lbrace Y_n, \tilde Y_n \rbrace$ not having any further restriction.
	
It is interesting to point out \cite{Palti:2020tsy} that the  necessity of having both fermionic and bosonic towers at the same time in order to account for such an emergence mechanism in the present case is consistent with the existence of some sort of \emph{symmetry} relating both kinds of fields, like supersymmetry (or perhaps misaligned supersymmetry\cite{Palti:2020tsy}). Hence, the mere presence of light fermionic fields in our gravitational EFT points towards the existence of towers of massive states with different spin-statistics, given that it seems difficult to generate their associated kinetic terms with just one infinite set of e.g. fermions (or bosons) alone.
On the other hand, in a theory with a supersymmetric spectrum, the on-shell bosonic content equals that of the fermions, such that one would expect the towers to come along with their respective superpartners. Thus, if e.g. $\chi$ in fig. \ref{fig:kineticfermionsbas}(a) is a massless gaugino in a 4d $\mathcal{N}=1$ gauge theory, it would couple to both supersymmetric fields belonging to a charged chiral multiplet, which would present exactly the same mass. Moreover, this kind of Yukawa couplings associated to supersymmetrized gauge interactions involve the gauge charges precisely like their vector boson partners do\cite{Wess:1992cp}, namely the Yukawa couplings are of the form $Y_n \propto in$. Hence, let us turn first to the computation of the relevant loop diagrams.

\subsubsection{Self-energy of a Weyl fermion}
\label{sss:selfenergyfermion}	
Let us compute the wave-function renormalization induced on a chiral fermion $\chi$ by the tower of bosonic and fermionic particles interacting with the former through Yukawa-like couplings of the form discussed in \eqref{eq:fermion&bosonYukawas}. We will work all the time with Dirac fermions in what follows, such that in order to take into account that the massless spin-$\frac{1}{2}$ field $\chi$ is chiral, it is thus necessary to introduce the  chirality projector $P_{-}=\frac{1}{2}(1-\gamma^{d+1})$, where $\gamma^{d+1}$ is the proper generalization of the four-dimensional $\gamma^5$ to $2k$ dimensions
\beq
	\gamma^{d+1}=-i^{k-1} \prod_{\mu=0}^{2k} \gamma^{\mu}\, .
\eeq
In this way, one can directly work with a Dirac spinor $\mathcal{X}$, which in principle doubles the field content of our chiral fermion $\chi$, so that upon using $P_-$ one recovers the original chiral field, namely $\chi = P_{-}\ \mathcal{X}$. With this in mind, the interactions in eq. \eqref{eq:fermion&bosonYukawas} above can be rewritten in terms of $\Psi^{(n)}, \mathcal{X}$ as follows
\begin{equation}\label{eq:interactionsfermion2}
		Y_n\ \phi^{(n)} \left( \overline{\Psi^{(n)}} P_{-} \mathcal{X} \right) + \text{h.c.}\, , \qquad  {\tilde Y}_n\ \left(\overline{\Psi^{(n)}} \gamma^{\mu} P_{-} \mathcal{X} \right) V_{\mu}^{(n)} + \text{h.c.}\, .  
\end{equation}
The idea then is to extract the momentum-dependent part of the exact propagator of our massless fermion $\chi$ at one-loop after integrating out such heavy particles (see Appendix \ref{ap:LoopsWeylfermion} for details)
\beq\label{eq:Euclexactpropagator}
	S(\slashed{p})= \frac{1}{i\slashed{p}}\, P_{-} + \frac{1}{i\slashed{p}}\, P_{-}\, \left(i\Sigma(\slashed{p})\right)\, \frac{1}{i\slashed{p}}\, P_{-} + \ldots\ ,
\eeq
$i\Sigma(\slashed{p})$ being the (amputated) Feynman diagram in fig. \ref{fig:kineticfermionsbas}(a). Notice that this is nothing but the fermionic analogue of $\Pi(p^2)$ in eq. \eqref{eq:scalarpropagator}. Therefore, if one wants to isolate the quantum corrections to the kinetic term arising from the loops, one then needs to take a derivative of said quantity with respect to the external momentum $p^{\mu}$, and then set it equal to zero.
	
\subsubsection*{The loop computation}
In the following, we will particularize to the case in which the bosonic particle running in the loop is a spin-0 field, as in fig. \ref{fig:kineticfermionsbas}(a), although a similar analysis could be performed for the massive 1-form case. The computation thus involves a tower of Dirac fermions $\Psi^{(n)}$ of mass $m_n^{{\text{f}}}$ as well as another one made of complex bosonic scalars $\phi^{(n)}$ with mass given by $m_n^{{\text{b}}}$. Thus, the self-energy integral we need to examine for each step $n$ in the infinite tower is the following
\beq
	i\Sigma_n(\slashed{p}) \ = |Y_n|^2 \int \frac{\text{d}^dq}{(2\pi)^d} \frac{P_{-} \left( -i\slashed{q} + m_n^{\text{f}}\right) P_{+}}{q^2+(m_n^{\text{f}})^2} \frac {1}{(q-p)^2+(m_n^{{\text{b}}})^2}\ ,
	\label{eq:selfenergychi}
\eeq
whilst the relevant contribution to the fermionic propagator reads (see Appendix \ref{ap:LoopsWeylfermion})
\beq
	\begin{aligned}\label{eq:Euclwavefunctionfermion}
		\frac{\partial \Sigma_n(\slashed{p})}{\partial p^{\mu}} \bigg\rvert_{p=0} = \frac{-2 |Y_n|^2 \delta_{\mu \nu} \gamma^{\nu}\ P_{+}}{d} \int \frac{\text{d}^dq}{(2\pi)^d} \frac{q^2}{\left[ q^2 + (m_n^{{\text{f}}})^2 \right] \left[ q^2 + (m_n^{{\text{b}}})^2 \right]^2}\ .
	\end{aligned}
\eeq
Notice that by evaluating $i\Sigma_n(\slashed{p})$ at zero external momentum one can see that eq. \eqref{eq:selfenergychi} above vanishes, such that no mass term is generated quantum mechanically at one-loop, as it should be. Additionally, the self-energy includes the projector $P_{+}$, since it is associated to the \emph{chiral} massless fermion, $\chi$. One should also say that, similarly to what happened in the previous cases, the behaviour of the above momentum integral strongly depends on the dimension $d$ of our spacetime. In particular, it can be seen to diverge depending on whether $d\geq4$ or not. In any event, since we are interested in its consequences for the Emergence Proposal, we will impose again some UV cut-off which ultimately will be identified with the species scale, rendering the previous integral finite.
	
Now, in order to study with more detail the kind of corrections induced by the above diagrams and for future reference, we will first specialize to the easier case in which both towers present identical masses, namely we take $m_n^{{\text{b}}} = m_n^{{\text{f}}} = m_n$. For concreteness, let us show in here the explicit results for the case in which $\Lambda/m_n \gg 1$. The pertinent leading expressions take the form
\begin{equation}\label{eq:fermionloopsummary}
		\frac{\partial \Sigma_n(\slashed{p})}{\partial p^{\mu}} \bigg\rvert_{p=0}\,   \sim 
		\left\{\begin{array}{lr}
			-|Y_n|^2 \gamma_{\mu}\ P_{+}\ \frac{1}{m_n^{4-d}} & \qquad\text{for } d< 4\, ,\\ \\ 
			-|Y_n|^2 \gamma_{\mu}\ P_{+}\ \log\left( \dfrac{\Lambda^2}{m_n^2}\right) &\qquad \text{for } d= 4\, ,\\ \\ 
			-|Y_n|^2 \gamma_{\mu}\ P_{+}\ \Lambda^{d-4}&\qquad \text{for } d>4\, .
		\end{array}\right.
\end{equation}
These results are analogous to those found for the wave-function renormalization of gauge bosons in eqs. \eqref{eq:1-formloopscalarssummary} and \eqref{eq:1-formloopfermionssummary}. In any event, the more general expression for the one-loop contribution to the fermion self-energy (i.e. eq. \eqref{eq:Euclwavefunctionfermion}) in which the towers have different spins as well as masses can be found at the end of Appendix \ref{ap:LoopsWeylfermion}.
	
\subsubsection{Emergence of  fermion kinetic terms in 4d }
\label{sss:emergencefermion}

As stated above, we cannot give a general expression for the emergent kinetic term of a light fermion field without further specifying the structure of towers and couplings involved in the loop computations. Therefore, in this section we will restrict ourselves to the case of four spacetime dimensions with the aim of illustrating with a simple toy model how the generation of kinetic terms may arise for massless fermions. In order to be as general as possible, we consider here a model which does not impose identical mass gaps for both infinite towers of states from the beginning. In particular, we start with a couple of towers of fermions and bosons with a structure of the form
\beq
	m_n^{\mathrm{b}}\, =\, n\, m_{\mathrm{b}}\, , \qquad m_n^{\mathrm{f}}\, =\, n^{1/p}\, m_{\mathrm{f}}\, , \qquad  \LQG\, \simeq\, N\, m_{\mathrm{b}}\, \simeq\, N^{1/p}\, m_{\mathrm{f}}\, ,
\label{eq:dostorres}
\eeq
where we have allowed for different mass scales for the towers $m_{\mathrm{f}}$ and $m_{\mathrm{b}}$, in principle. For concreteness, we will assume $m_{\mathrm{f}}$ to be greater than $m_{\mathrm{b}}$, but notice that we could alternatively start with the reversed situation in which $m_{\mathrm{f}} \leq m_{\mathrm{b}}$, yielding identical results upon exchanging $m_{\mathrm{f}}$ and $m_{\mathrm{b}}$. The case with $p=1$ is indeed quite interesting, since depending on the precise value for $Y_n$ one can recover the wave-function renormalization of the fermionic components either in $\mathcal{N}=1$ chiral or vector multiplets. 
	
On the other hand, for $p\rightarrow \infty$ instead, one has a tower in which the mass gap between two adjacent levels is much smaller than the scale of the tower itself, namely $\Delta m^{\mathrm{f}}_n \ll m_{\mathrm{f}}$. Notice that with the above parameterization we can actually describe towers with different mass scales, for more generality. We moreover assume the massless fermion $\chi$ to couple to the states comprising both towers through the Yukawa-like couplings summarized in \eqref{eq:fermion&bosonYukawas}, thus leading to the one-loop diagrams displayed in fig. \ref{fig:kineticfermionsbas}. However, as already mentioned, the dependence of the Yukawa couplings  $Y_n$, ${\tilde Y}_n$ on the state labelling $n$ is, in general, not fixed by some gauge invariance of any sort. Hence, in order to accommodate different possibilities we consider here Yukawas of the general form\footnote{There is also the possibility of having $m_{{\text{b}}}=m_{{\text{f}}}$ and $Y_n = \mu_n = \partial_{\Phi} m_n$, which indeed captures the couplings of a supersymmetric tower to the fermionic component within some massless $\mathcal{N}=1$ chiral multiplet, leading to similar results as in section \ref{sss:emergencemodulimetric}.}
\beq
	Y_n\, \propto\, n^{1/r}\, .
\eeq
For $r=p=1$ one indeed recovers a situation in which both towers share the same mass (i.e. $m_{\mathrm{f}}=m_{\mathrm{b}}$ ) and the Yukawa charges grow like $Y_n \propto n$. This is precisely the case considered in \cite{Palti:2020tsy}, which mimics the behaviour involving some kind of gauge coupling. On the other hand, if $r\rightarrow \infty$, the Yukawa coupling does not depend on the state of the tower and it is just some fixed constant. Now, upon using the third relation in \eqref{eq:dostorres}, one sees that the species scale is related to the masses of the towers by
\beq
	\LQG^4\, \simeq\, N^{1/p}\, m_{\mathrm{b}}\, m_{\mathrm{f}}\, \Mpf^2\, ,
\eeq
where in four dimensions, the species scale is defined as $\LQG^2 \simeq \Mpf^2/N$. Thus, we obtain
\beq
	g_{\chi \chi}\, \sim\,  \sum^{N} n^{2/r}\, \sim\, N^{\frac{2}{r}+1}\, \sim\, \left(\frac {(m_{\mathrm{f}}\Mpf^2)^{1/3}}{m_{\mathrm{b}}}\right)^{\gamma_r}\, ,
\label{eq:metricafermiongeneral}
\eeq
where
\beq
	\gamma_{r,p}\, =\, \frac {3p(2+r)}{r(4p-1)}\, .
\eeq
 Note that considering different values for $r,p$ we find the $\gamma$-parameter to lie in the range $1\leq \gamma \leq 3$. For the particular case of $r=p=1$ one gets $m_{\mathrm{b}}=m_{\mathrm{f}}=\Mt$  and $\gamma_{1,1}=3$ yielding\cite{Palti:2020tsy}
\beq
	g_{\chi\chi}\, \sim\, \left( \frac {\Mpf}{\Mt}\right)^2\, ,
\eeq
which is analogous to the structure found for the gauge kinetic functions. (In particular, if we take $Y_n = -i \sqrt{2}\ q_n g$ \cite{Wess:1992cp}, as in the gaugino case, one obtains agreement with the 1-form wave-function renormalization discussed in section \ref{sss:emergenceU(1)}, as it should be.) For towers/charges with a different structure the more general result, i.e. eq. \eqref{eq:metricafermiongeneral}, applies. Still, the take-home message is that fermions may get large wave-function renormalization if at least one of the towers running in the loop is sufficiently light. 

\subsubsection*{Emergence of kinetic terms for charged scalars}
We already studied above the case of the emergence of kinetic terms  for a particular class of (massless) scalars, namely the moduli fields. There are however, additional scalars in string theory which are not moduli, like e.g. charged scalars in $\mathcal{N}=1$ chiral multiplets in 4d constructions. For moduli fields, the couplings to the towers are typically determined by the first derivative $\partial_\phi m_n(\phi)$ of the mass of each state in the tower with respect to the modulus $\phi$. However, as in the case of fermions discussed above, for generic scalars the coupling to the infinite towers is in general model-dependent.
	
The relevant loops in this case look like `SUSY rotations' of those shown in fig. \ref{fig:kineticfermionsbas}. They are therefore obtained by replacing $\chi \rightarrow {\tilde \chi}$ along with $\phi_n\rightarrow {\tilde \phi}_n$ (or $\Psi_n \rightarrow {\tilde \Psi}_n$) in fig. \ref{fig:kineticfermionsbas}(a) and similarly for fig. \ref{fig:kineticfermionsbas}(b). We refrain from presenting a complete calculation for this (related) case, since the relevant results may be obtained directly from those for fermions presented above.

\subsubsection{The  WGC for  Yukawa couplings in 4d}
\label{sss:WGCYukawa}
The stronger versions of the magnetic WGC propose that in the presence of a $U(1)$ gauge field along with charged particles, the limit $g\rightarrow 0$ is singular and should be accompanied by (infinite) towers of states becoming light. A similar question that could come to mind is what happens with Yukawa couplings, namely when some Yukawa coupling goes to zero, is there any associated tower of states becoming light? In the present discussion we would like to argue that, at least in the context of Emergence, the answer to this question is positive (see \cite{Palti:2020tsy} for similar ideas).
	
In the emergence picture one expects the non-vanishing Yukawa couplings present in the UV (if any) to be of order one, up to exponentially suppressed instanton corrections. In fact, this is what is generically found in specific string theory compactifications, see e.g\cite{Ibanez:2012zz}. Thus e.g. couplings involving three fields in four dimensions will have typically the form $Y^{\text{UV}}_{ijk}\simeq \mathcal{S}_{ijk}$, with all entries in $\mathcal{S}$ being essentially of $\mathcal{O}(1)$, such that no hierarchies of Yukawa interactions would be present at this level. Therefore, any hierarchical behaviour of the Yukawas (like the ones existing in the SM) would appear as an infrared effect. Indeed, the above considerations about the generation of kinetic terms for fermions may give us a hint regarding how this could come about. We will frame the present discussion for concreteness in an $\mathcal{N}=1$ supersymmetric setting, but most of the results should still be amenable to generalization to non-SUSY set-ups with appropriate changes. Thus, in such a supersymmetric scenario with a \emph{renormalizable} 4d superpotential, $W (\Phi)= \mathcal{S}_{ijk}\Phi^i\Phi^j\Phi^k$, the properly normalized Yukawa couplings would have the form
\beq
	Y_{ijk}\, =\, \mathcal{S}_{ijk}\,  (g_{i \bar i}g_{j \bar j} g_{k \bar k})^{-1/2}\, .
\eeq
where $g_{i \bar i}$ are the K\"ahler metrics of the corresponding (massless) chiral multiplets, which we have taken here to be diagonal for simplicity. We now assume that all these metrics for the massless fields are obtained via the emergence mechanism such that, according to the discussion above, we find
\beq
	Y_{ijk}\,  \gtrsim \,  \mathcal{S}_{ijk} \Mpf^{-\gamma}\left[ \left(\frac {m_{l,i}}{m_{h,i}^{1/3}}\right)^{\gamma_i/2}
	\left(\frac {m_{l,j}}{m_{h,j}^{1/3}}\right)^{\gamma_j/2}
	\left(\frac {m_{l,k}}{m_{h,k}^{1/3}}\right)^{\gamma_k/2} \right]\, ,
\eeq
where $\gamma =\sum_i\gamma_i/3$ and $\gamma_i=\frac{3p_i(2+r_i)}{r_i (4p_i-1)}$. Here the subindices $h,l$ stand for the heaviest and lightest masses in each of the loops, respectively. Now, all components in $\mathcal{S}_{ijk}$ being $\mathcal{O}(1)$, if we take some entry of $Y_{ijk}$ to approach zero, it will imply that some (or all) of the towers have to become nearly massless in Planck units. Hence, the above expression is in some sense the Yukawa analogue of the magnetic WGC. Notice  also that, depending on how light each of the $i,j,k$ towers (which need not be different) is, hierarchies in the Yukawa couplings could naturally arise, even though all components in the UV matrix (the $\mathcal{S}_{ijk}$) were originally of order one. Thus, one could argue that hierarchies in the Yukawa interactions may be a generic effect in \emph{emergent} EFTs, with essentially the same loops inducing the kinetic terms also providing for the hierarchical behaviour of the $Y_{i j k}$'s. In ref.\cite{Castellano:2023qhp} this fact will be exploited so as to address the Yukawa hierarchies observed in the Standard Model.
	
If the bosonic and fermionic towers present identical mass gaps and $g_{i \bar i}\sim 1/m_i^2$, as in the case in which they are SUSY partners of each other, one has $\gamma_i=\gamma=3$ and 
\beq
	\mathcal{S}_{ijk}\, m_im_jm_k\, \lesssim\, Y_{ijk} \, \Mpf^3\, ,
\eeq	
which is again reminiscent of a WGC-like expression for the Yukawa couplings. In particular, it was shown in \cite{Heidenreich:2017sim} that in a 4d $U(1)$ gauge theory coupled to gravity, the cut-off should be bounded as $\LQG \lesssim e^{1/3}\Mpf$, with $e$ being the gauge coupling. Hence, for any charged matter particle with mass $m$ below the cut-off one finds
\beq
	m^3\, \lesssim\, \LQG^3\, \lesssim\, e\, \Mpf^3\, ,
\eeq
which is a `gauge counterpart' of the above Yukawa constraint for the case of a single tower of states with mass scale $m$. This parallelism is in fact not a coincidence since, e.g., in a supersymmetric theory the gaugino presents Yukawa couplings (with respect to charged chiral multiplets) of strength controlled by $e$ as well. Since the gaugino kinetic term is equal to the gauge kinetic coupling, the same structure for this Yukawa coupling appears, as discussed above.	  
	
We end this section with an important comment concerning gauge coupling renormalization of light fermions. One could naively argue that by applying this idea to the interaction terms coupling the vector boson to some (conserved) current as in e.g. $({\bar \psi}\gamma^\mu A_\mu \psi)$, gauge couplings could also become hierarchical due to possible large corrections to the wave-function renormalization of the charged fermions. Of course, this cannot be the case since the Ward-Takahashi identities associated to the gauge interaction relate precisely the vertex and the wave-function renormalization factors, such that once we sit in a frame with canonically normalized fermions, the apparent hierarchies disappear. On the other hand, there are no a priori Ward-Takahashi identities for Yukawa couplings (unless they come from a gaugino interaction) and thus hierarchies can arise in general due to possible large anomalous dimensions.  
		
\section{Emergence in 4d $\mathcal{N}=2$}
\label{s:emergence4dN=2}
		
The most studied and probably best understood example of Emergence in Quantum Gravity so far is the one associated to the large complex structure (LCS) limit of type IIB compactifications on a CY three-fold $Y_3$ \cite{Grimm:2018ohb,Gendler:2020dfp}, or the dual large volume (LV) point in type IIA on the mirror $X_3$ \cite{Corvilain:2018lgw}. These set-ups seem to point towards an asymptotic decompactification to M-theory on $X_3$, described by an effective 5d $\mathcal{N}=2$ supergravity theory \cite{Cadavid:1995bk}, where the KK-like tower responsible for the breakdown of the original EFT was originally proposed to consist of BPS bound states of D3-branes, in the type IIB case, or bound states of D0-D2's in the dual type IIA setting. Let us remark that, from the higher dimensional perspective, both towers map to states obtained by wrapping M2-branes on supersymmetric 2-cycles of $X_3$, with non-vanishing momentum along the compact M-theory circle. In the following, we argue that the spectrum of towers relevant for this limit comprises not only such bound states, but also the KK replicas of \emph{every} five-dimensional field present in the asymptotic `resolved' 5d theory, which can be interpreted as an additional tower of just D0-branes from the 4d type IIA perspective.\footnote{\label{fn:LambdaminSDC}Incidentally, such D0-brane tower is precisely the one that verifies in this 4d $\mathcal{N}=2$ set-up the recently proposed lower bound\cite{Etheredge:2022opl} for the $\lambda$-parameter of the Swampland Distance Conjecture (see also the discussion around eq. \eqref{eq:4dSDC}).} 		
		
\subsection{Preliminary considerations}
\label{ss:preliminary}
		
Let us start by recalling here the bosonic part of the 4d $\mathcal{N}=2$ effective action that arises upon reducing the 10d type IIA supergravity theory on a Calabi--Yau three-fold $X_3$ (in the Einstein frame) \cite{Bodner:1990zm}
\begin{equation}\label{eq:IIAaction4d}
	\begin{aligned}
				\ S_{\text{IIA}}^{\text{4d}} = & \frac{1}{2\kappa^2_4} \int  \left( R \star 1 - \frac{1}{2} \text{Re}\, \mathcal{N}_{AB} F^A \wedge F^B-\frac{1}{2} \text{Im}\, \mathcal{N}_{AB} F^A \wedge \star F^B\right) \\
				& - \frac{1}{\kappa^2_4} \int \left( G_{a\bar b} dz^a\wedge \star d\bar z^b + h_{pq}d  q^p \wedge \star d  q^q\right)  \, ,
	\end{aligned}
\end{equation}
where $F^B=dA^B$, with $B=0, \ldots, h^{1,1}$, and we denote $\Mpf=(\kappa_4)^{-1}$ above. Here, the complex scalars $z^a=b^a+it^a$, $a=1,\ldots, h^{1,1}$ belong to the K\"ahler sector and comprise the vector multiplet moduli space, whereas the scalars in the various hypermultiplets (including e.g. the complex structure moduli) are denoted $q^p$. Notice that the gauge sector includes both the $U(1)$ gauge fields belonging to the vector multiplets as well as the graviphoton. The corresponding gauge kinetic matrix $\mathcal{N}_{AB}$  depends on the scalars $z^a$, and in the large volume regime its real part (which only depends on topological data and axions) takes the form \cite{Grimm:2005fa}
\begin{equation}\label{eq:gaugetopologicalterm}
			\text{Re}\, \mathcal{N}\, = \, \left(
			\begin{array}{cc}
				\frac{1}{3} \mathcal{K}_{abc}b^a b^b b^c & -\frac{1}{2} \mathcal{K}_{abc}b^b b^c  \\
				-\frac{1}{2}\mathcal{K}_{abc}b^b b^c & \mathcal{K}_{abc} b^c  \\
			\end{array}
			\right) \, ,
\end{equation}
whereas its imaginary part (which depends on the geometric moduli via the metric $G_{a\bar b}$) and its inverse read
\begin{equation}\label{eq:gaugekineticmatrix}
			\text{Im}\, \mathcal{N}\, = \frac{\mathcal{K}}{6} \, \left(
			\begin{array}{cc}
				1+4G_{a\bar b}b^a b^b & -4G_{a\bar b} b^b  \\
				-4G_{a\bar b} b^b & 4G_{a\bar b}  \\
			\end{array}
			\right) \, ,\qquad 
			(\text{Im}\, \mathcal{N})^{-1}\, = \frac{6}{\mathcal{K}} \, \left(
			\begin{array}{cc}
				1 &  b^a  \\
				b^a & \frac{1}{4} G^{a\bar b} + b^a b^b  \\
			\end{array}
			\right) \, .
\end{equation}
Here, $\frac{\mathcal{K}}{6}=\frac{1}{6} \mathcal{K}_{abc}t^a t^b t^c = \mathcal{V}(X_3)$ denotes the volume of the three-fold in string units, and the metric $G_{a\bar b}$ is special K\"ahler and hence determined by a K\"ahler potential $K_{\text{ks}}$ as follows
\beq\label{eq:kahlersectormetric}
		G_{a\bar b}=\partial_a \partial_{\overline{b}} K_{\text{ks}}=\partial_a \partial_{\overline{b}} \left(- \log \frac{4}{3} \mathcal{K} \right)\, .
\eeq
For future convenience, let us also state here that the form of the K\"ahler metric in the toroidal orbifold that we will use as a `toy model' simplifies considerably, since has a diagonal form
\beq\label{eq:toroidalkahlermetric}
		G_{a\bar b}= \frac{\delta_{ab}}{4} \frac{1}{(t^a)^2}\, .
\eeq
The limit in which we are interested in is the large volume point, where the overall volume of the CY$_3$ diverges in string units \cite{Corvilain:2018lgw}. Notice, however, that since we want to restrict to trajectories along the vector multiplet moduli space only (which is decoupled from the hypermultiplet sector, see discussion around eq. \eqref{eq:N=2modulispace}), such limit must be accompanied by a proper rescaling of the 10d dilaton, $\phi$, in order to keep the 4d dilaton, $\varphi_4$, fixed. (Recall that the 4d dilaton controls the 4d Planck mass, $\Mpf^2/M_s^2 = 4 \pi e^{-2\varphi_4}$.) 
		
As already mentioned, this particular limit corresponds to a decompactification to 5d in terms of M-theory on the same $X_3$, where the M-theory circle grows large. Therefore, it is important to properly identify which tower is responsible for the breakdown of the original EFT, and effectively implements the decompactification. By looking at how the spectrum behaves in this limit, we could have guessed that the tower is simply that of D0-branes, since they are the lightest objects (see e.g. \cite{Font:2019cxq}). However, as argued in \cite{Corvilain:2018lgw} by exploiting the properties of the monodromy action around the singularity, it seems that one also needs appropriate bound states involving a \emph{finite} number of D2-particles (wrapping minimal 2-cycles of the CY$_3$) and $n$ D0's, with $n \to \infty$ in the limit. From the point of view of Emergence, this latter fact can be motivated by looking at the gauge kinetic terms of the vectors in the 4d action (see eq. \eqref{eq:IIAaction4d}), where it is clear from eq. \eqref{eq:gaugekineticmatrix} that they seem to all become weakly coupled as we approach the large volume point. Thus, this would require an infinite tower of \emph{charged} states, which can only be fulfilled by turning on some non-zero D2 charge in the tower. Note that, as explained below, this also matches with the expectation coming from  the idea of including all states up to the species scale. 
		
However, one of the main points we want to convey in here is the necessity of including also the tower of particles made just from D0-branes, as opposed to considering only that of D2-D0 bound states. Notice that, whilst it is intuitively clear that bound states mixing $p$ D2-particles and $q$ D0's (with $(p,q)$ co-prime) exist and are stable, given that they minimize the central charge $Z_{\text{IIA}}=e^{K_{\text{ks}}/2} \mathcal{Z}_{\text{IIA}}\,$ controlling such D-particle masses, the same is not true for the tower of D0-branes alone, since they could only form bound states at threshold. Nevertheless, we will \emph{assume} these bound states to exist and show that the logical picture that arises from this assumption is fully coherent, in contrast to the situation that results from not including them.\footnote{In any event, proving that such bound states can be found is an interesting and difficult open problem by itself (see e.g. \cite{Sethi:1997pa} for early attempts in 10d) that we will not try to attack here.}
		
\subsubsection*{The species scale and the QG cut-off}	
We begin by considering the species scale in the presence of the aforementioned light towers of particles that become asymptotically light in 4d Planck units. First, notice that for a D-particle with $n_{2,a}\in \mathbb{Z}$ D2-brane charge (where the $a$ indicates the holomorphic 2-cycle wrapped by the brane) and $n_0\in \mathbb{Z}$ D0-brane charge, the moduli dependence of the mass is encapsulated in the normalized $\mathcal{N}=2$ central charge\footnote{We have set to zero the magnetic charges corresponding to of D4- and D6-particle states, which give rise to dyonic central charges \cite{Ceresole:1995ca}. The reason being that they do not become massless in the limit of interest (see e.g. \cite{Font:2019cxq}) and hence do not contribute to the one-loop computation that we perform in section \ref{ss:4doneloop}.}
\beq\label{eq:centralcharge}
		\frac{m_{n_{2p}}}{\Mpf} = \sqrt{8\pi } e^{K_{\text{ks}}/2} |\mathcal{Z}_{\text{IIA}}| = \sqrt{\frac{\pi}{ \mathcal{V}(X_3)}} |n_0+n_{2,a} z^a|\, .
\eeq
Recall that in a $d$-dimensional EFT weakly coupled to Einstein gravity the species scale in the presence of $N$ species is given by \eqref{species}. Now, since we claim that the asymptotic physics is governed by a 5d $\mathcal{N}=2$ theory, it is natural to expect that it should be around the Planck scale of the five-dimensional theory, $M_{\text{pl,}\, 5}$ which can be related to the 4d Planck mass by
\beq
		\label{eq:M5R5M4}
		M_{\text{pl,}\, 5}^2\ 2\pi R_5 = \Mpf^2\, ,
\eeq
where $R_5$ denotes the radius of the M-theory circle in 5d Planck units. The precise equivalence between M-theory and type IIA quantities can be seen explicitly from the dimensional reduction of eleven-dimensional supergravity on $X_3 \times S^1$\cite{Cadavid:1995bk}. Thus, e.g. the radius $R_5$ can be expressed in terms of the volume modulus of $X_3$ in string units as follows
\beq
		R_5^3 = \mathcal{V}(X_3)\, ,
\eeq
which via eq. \eqref{eq:M5R5M4} yields $ M_{\text{pl,}\, 5}^2\,  =\, \Mpf^2\,/(2\pi \mathcal{V}^{1/3})\, $. Furthermore, the volume scalar in M-theory, $\mathcal{V}_5$ (which belongs to the hypermultiplet moduli space), is related to the 4d dilaton as  
\beq\label{eq:4dilaton}
		\mathcal{V}_5(X_3)=e^{-2\varphi_4}=\frac{\mathcal{V}(X_3)}{g_s^2}\, .
\eeq
With these identifications at hand it is now easy to see how the masses of the D0- and D2-particles (see eq. \eqref{eq:centralcharge}) in 4d Planck units are translated naturally into 5d quantities
\begin{equation}\label{eq:massesD0D2}
	 \begin{aligned}
				\frac{m_{\text{D0}}}{\Mpf} & \sim g_s^{-1} e^{\varphi_4} \sim \frac{1}{\mathcal{V}^{1/2}} \sim \frac{1}{R_5^{3/2}} \sim \frac{m_{\text{KK},\, 5}}{\Mpf}\, ,\\
				\frac{m_{\text{D2}}}{\Mpf} & \sim e^{K_{\text{ks}}/2} |t^a| \sim \frac{M^a}{\mathcal{V}^{1/6}} \sim \frac{m_{\text{M2}}}{\Mpf}\, ,
	 \end{aligned}
\end{equation}
where in the last expression we have considered a single D2-brane wrapping just one 2-cycle and we have set the corresponding axion VEV $b^a$ to zero. Notice that in order to identify the mass of the D2-particles with that of the M2-branes wrapping the same cycles in the Calabi--Yau, we have split the K\"ahler coordinates into rescaled ones $M^a=t^a/\mathcal{V}^{1/3}$, which are subject to the so-called very special geometry constraint \cite{Corvilain:2018lgw}
\beq
		\mathcal{F}=\frac{1}{3!} \mathcal{K}_{a b c} M^a M^b M^c \stackrel{!}{=} 1\, ,
\eeq
along with the volume of the three-fold, $\mathcal{V}$ (see section \ref{ap:5dMtheory} for more details on the 5d vector multiplet moduli space). 
		
Now, given that we are dealing with a decompactification limit on a circle, and that the KK scale is identified with the masses of the D0-branes (see eq. \eqref{eq:massesD0D2} above), one can readily see that the species scale associated to that tower \emph{alone} gives the expected 5d Planck scale, i.e.
\beq \label{eq:D0tower}
		N_{\text{tot}} \simeq \frac{\Lambda_{\text{QG}}}{m_{\text{D0}}} \Rightarrow \frac{\Lambda_{\text{QG}}}{\Mpf} = \left( \frac{m_{\text{D0}}}{\Mpf} \right)^{1/3} \sim \mathcal{V}^{-1/6} \sim \frac{M_{\text{pl,}\, 5}}{\Mpf}\, ,
\eeq
with $N_{\text{tot}} \sim \mathcal{V}^{1/3}$ divergent, as expected. 
\begin{figure}[tb]
		\begin{center}
			\includegraphics[scale=0.17]{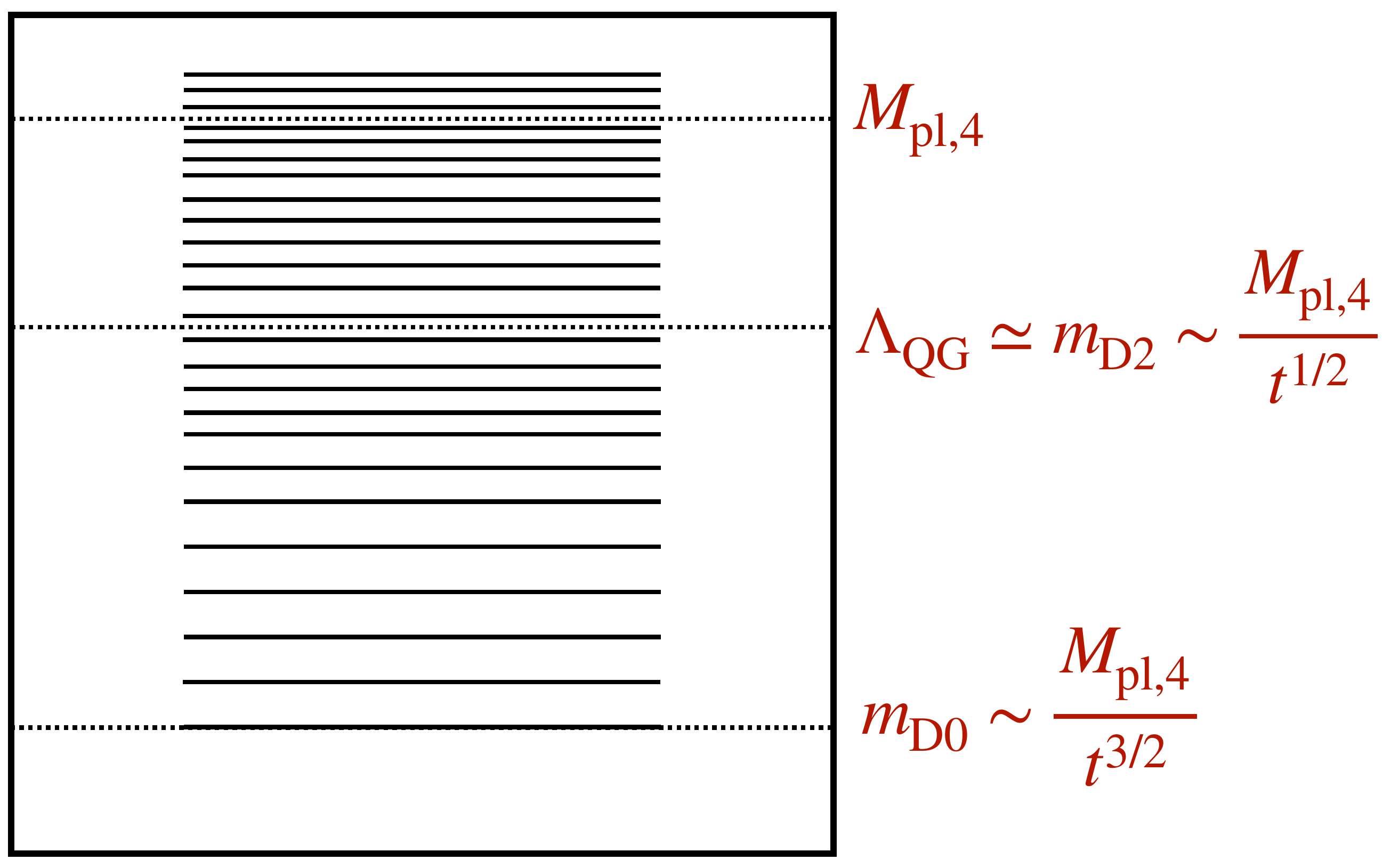}
			\caption{Mass scales associated to the D0- and D2-particles, together with the species scale, in the limit $\mathcal{V}\, =\, t^3 \, \to\,  \infty$.}
			\label{fig:D2D0towers}
		\end{center}
\end{figure}
However, following the procedure described in section \ref{ss:MultipleTowers} to compute the species scale in the presence of multiple towers, once we have calculated the scale associated to the lightest tower, we should compare it with the mass scale of the subsequent lightest tower. In the large volume limit, the next tower would be the one given by the D2-D0 particles. It can be seen from eq. \eqref{eq:massesD0D2} that the mass of a wrapped D2-brane is already of the order of the five-dimensional Planck scale, and thus of the order of the species scale that we just computed (at least when we consider a simple homogeneous scaling of the 4d K\"ahler coordinates that brings us directly to the large volume point without probing other singular limits first). In fact, if we look at the normalized central charge of bound states of $n$  D0-branes (with $n \lesssim N_{\text{tot}} \sim \mathcal{V}^{1/3}$) and fixed D2 charge (say $n_{2,a}=1$), and embed into it the tower of D0 charges that would give us eq. \eqref{eq:D0tower}, we see that the heaviest state of this mixed spectrum essentially has a D0 mass which scales as that of the single D2-particle (see fig. \ref{fig:D2D0towers} for a summary of the relevant scales). Thus, taking this tower into account along with the one of D0-branes alone essentially provides us with a species scale of the same order of the previously calculated one (i.e. the 5d Planck mass), since the species are \emph{additive} in this case. To sum up, both the tower of only D0's and the tower of D0-D2's with fixed (order one) D2 charge and running D0 charge contribute to the species scale and do not lie above it, so that they must both be integrated out according to the emergence prescription. On the contrary, we must not include a tower with running D2 charge since that would be asymptotically above the species scale.
		
\subsubsection*{The D0 and D0-D2-brane field content}
		
Having discussed the objects that must be included in the EFT defined up to the species scale in the large volume limit, let us now turn our attention to their corresponding field-theoretic content. One of the main points we want to stress here is that a \emph{stable} tower of BPS bound states of $n$ D0-branes must exist (for any $n\in \mathbb{Z}$ sufficiently light such that gravity can `resolve' them\footnote{Far from the infinite-distance singularity, the D0-branes are well-described by semi-classical solutions with large horizons. However, when we go sufficiently close to the large volume point, their Compton wavelength exceeds its Schwarzchild radius and thus the semi-classical description breaks down.}), and their particle content must correspond to massive KK replicas of the massless fields appearing in the 5d $\mathcal{N}=2$ EFT arising from compactifying M-theory on the same Calabi--Yau three-fold. In addition to that, we will review how the field content of the other relevant tower, namely the bound states of a single D2-brane and $n$ D0's, is given in the simplest cases by a massive charged $\mathcal{N}=2$ hypermultiplet for each step of the tower.
		
There are several arguments that support these claims, which stem either from the duality between type IIA string theory and M-theory, or rather from a more concrete super-quantum mechanical analysis. In order to not complicate unnecessarily the main discussion in this section, we summarize here the main intuition behind the field content of the BPS towers, and leave the details for Appendices \ref{ss:4dDO} and \ref{ss:4dD0D2}.
		
A quick way to motivate our claim is to remark that the large volume limit sends us to the 5d decompactified theory in which the M-theory circle grows large. Hence, the spectrum should include the KK-replicas of all `light' fields in the higher dimensional theory that lie below the 5d species scale. This means, in particular, that we should find the massive KK modes associated to the 5d massless fields, namely a spin-2 multiplet, ($h^{1,1}-1$) vector multiplets, and ($h^{2,1}+1$) hypermultiplets \cite{Cadavid:1995bk}, where $h^{p,q}$ refers to the Hodge numbers of the three-fold $X_3$ (see Appendix \ref{ap:5dMtheory} for details). From the four-dimensional perspective, such KK tower is nothing but the tower of bound states of D0-branes, whose mass is indeed identical to that of the 5d KK-tower (see first line of eq. \eqref{eq:massesD0D2}).
		
On the other hand, regarding the spectrum associated to the D0-D2 tower, it is clear from the above picture (and also from the second equation in \eqref{eq:massesD0D2}) that it should correspond to the KK replicas of the massive supermultiplet arising from an M2-brane wrapping the corresponding non-contractible 2-cycle in 5d M-theory. There is, however, an extra difficulty associated to this tower due to the fact that its field-theoretic content strongly depends on the moduli space of the cycle wrapped by the supersymmetric 2-brane along with its possible flat connections. In any event, for the simplest case in which such moduli space is trivial (i.e. just a point), one can see that each step in the D0-D2 tower is associated to a 4d massive hypermultiplet (see discussion around eq. \eqref{eq:D2qnumbers}).

\subsection{Emergence of the gauge kinetic function}
\label{ss:4doneloop}
Having identified the physics taking over in the asymptotic limit and which specific towers are responsible for such phase transition, our aim in this section is to show how the emergence mechanism works in detail from the 4d perspective. One could use this analysis to reproduce the metric in \eqref{eq:toroidalkahlermetric} via Emergence using the general results from section \ref{sss:emergencemodulimetric}. We do not go into the details of this computation here and focus instead on the more interesting case of the emergence of the gauge kinetic function, but let us remark that both are related by the $\mathcal{N}=2$ structure of the theory. To do so, we compute the one-loop wave-function renormalization of the gauge fields and show that upon integrating out the tower of BPS states up to the species scale \eqref{eq:D0tower} we recover precisely eq. \eqref{eq:gaugekineticmatrix}.
		
For concreteness, we focus in this section on the case of a one-dimensional K\"ahler sector, namely $h^{1,1}=1$, with K\"ahler modulus $z=b+it$, and volume $ \mathcal{V}(X_3)=\mathcal{K}/6=t^3$. In this setting we are left with just two $U(1)$ gauge fields, the one corresponding to the graviphoton, and the one belonging to the unique vector multiplet. One could think of this e.g. in terms of the familiar toroidal orbifold $T^6/\mathbb{Z}_2\times \mathbb{Z}_2'$ (ignoring the blow-up modes) restricted to the isotropic case $t^1=t^2=t^3=t$, so as to have an explicit example in mind \cite{Camara:2005dc}. Still, in the more general scenario with $h^{1,1}>1$, one can similarly probe the large volume regime by splitting the K\"ahler moduli into the volume scalar, $\mathcal{V}(X_3)^{1/3}$, and the rescaled coordinates $M^a=t^a/\mathcal{V}^{1/3}$, as outlined after eq. \eqref{eq:massesD0D2}. Therefore, it is easy to see that the physics of the $\mathcal{V} \to \infty$ limit (keeping the $M^a$'s fixed) is captured by our current set-up. Other asymptotic limits, corresponding from the M-theory point of view to different boundaries on the space parameterized by the $M^a$'s, can also be explored. We leave a general analysis of this for future work, but we elaborate on some other interesting limits in section \ref{ss:generalizations}.
		
In our case, the gauge kinetic matrix around the large volume point from eq. \eqref{eq:gaugekineticmatrix} reads
\begin{equation} \label{eq:matrixonemodulus}
	\text{Im}\, \mathcal{N}\, = \frac{\mathcal{K}}{6} \, \left(
		\begin{array}{cc}
			1+3(b/t)^2 & -3b/t^2  \\
			-3b/t^2 & 3/t^2  \\
		\end{array}
	\right) \, ,
\end{equation}
where we have substituted the particular K\"ahler metric \eqref{eq:toroidalkahlermetric} for the isotropic case as explained above. Let us remark two important features of these gauge kinetic functions. First,  due to the off-diagonal terms, it implements the phenomenon of gauge kinetic-mixing\cite{Holdom:1985ag,delAguila:1988jz}. Second, in the large volume limit (i.e. $t \to \infty$) every entry blows up, and both vectors become weakly coupled.\footnote{Strictly speaking, when talking about weakly/strongly coupled gauge fields, one refers to the physical fields, that is those for which the kinetic functions adopt the canonical form and thus are diagonal. In our example it can be checked that the corresponding eigenvalues of the physical gauge couplings are also vanishing in the limit.} Also notice that the polynomial dependence on the divergent K\"ahler modulus is different between those terms involving the axion $b$ and those in which it is absent. In the following we explain how these facts arise naturally when integrating out the infinite tower of D0-D2 particles as well as the D0-brane bound states discussed above.
		 
\begin{figure}[tb]
			\begin{center}
				\includegraphics[scale=1.2]{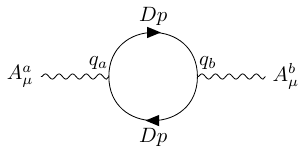}
				\caption{One-loop diagram with the BPS states circulating and giving rise to an effective low-energy kinetic mixing between the different $U(1)$ vector bosons.}
				\label{kineticmixing}
			\end{center}
		\end{figure}

We make use of the  renormalization group flow down to low energies, which induces kinetic mixing when we allow for heavy multi-charged particles to run in the loop, as displayed in fig. \ref{kineticmixing}. The gauge kinetic function $f_{AB}^{\text{UV}}$ at the UV scale, $\Lambda_{\text{UV}}$, is related to the low-energy gauge couplings (at an energy well below the masses of said charged particles) as follows
		\beq \label{eq:oneloopmixing}
		f_{AB}^{\text{UV}}\, =\, f_{AB} - \sum_i \frac{\beta_i}{8\pi^2 \kappa_4^2}\ q_{A}^{(i)} q_{B}^{(i)}\ \log \left(\frac{\Lambda_{\text{UV}}}{m_i}\right)\, ,
		\eeq
where $m_i$ and $q_{A}^{(i)}$ are the mass and the (dimensionless) charge of the particles in the tower whilst $\beta_i$ denotes the corresponding beta function coefficient associated to the (super)particle at the $i$-th step. Notice  that the $f_{AB}$ defined here includes a factor of $(\kappa_4)^{-2}$ that was explicitly extracted from the definition of $\text{Im}\, \mathcal{N}_{AB}$ in the action \eqref{eq:IIAaction4d}. Thus, the factor of $\kappa_4^2$ in the denominator appears because in our conventions the gauge fields are dimensionless. Additionally, the value of $\beta_i$ depends on the character and degeneracy of the particles running in the loop. For the BPS states corresponding to e.g. the D0-D2 tower, in the simplest one-modulus scenario, $\beta_i= 1$ since the tower is made of charged $\mathcal{N}=2$ hypermultiplets (see Appendix \ref{ap:Dpbranecontent} for details).

Before proceeding with the computation, let us make an important remark about the gauge charges of the states in the tower, since their precise form and value depends on the basis for the vector bosons that we use. In particular, the expression presented in eq. \eqref{eq:IIAaction4d} is in some sense adapted for the computation that is to follow, since in such basis the field strengths are (locally) exact and therefore well-suited for the renormalization of the propagator. Moreover, the charges of the states that run in the loops are quantized (they essentially count, modulo some signs and $\mathcal{O}(1)$ factors, the number of D$p$-particles comprising each bound-state hypermultiplet), and hence moduli independent \cite{Ceresole:1995ca}. In contrast, if one wants to talk about vector combinations which are `supermultiplet eigenstates' (the graviphoton and the orthogonal bosons in the vector multiplets) as well as their associated gauge charges (which would be now moduli  dependent),\footnote{The moduli dependent shifts in the charge vectors can be understood as e.g. induced D0-brane charges from the D2-particles arising due to a non-trivial $B_2$-flux along the 2-cycle that it wraps. This can be seen also from the Chern-Simons action of the D2-particle reduced along the non-trivial 2-cycle $ {S_{\text{CS}} = \mu_2 \int_{\mathcal{W}_{3}} P \left[ \sum_q C_q \wedge e^{B_2+2\pi \alpha' F} \right] \, }$, see eq. \eqref{eq:DBICSaction}.} one should first perform a `rotation' to go from one basis to the other. However, notice that in the latter basis the field strengths are not (locally) exact anymore, so that the one-loop renormalization that we aim to perform here would be rather cumbersome. For this reason we stick to the basis in which the 2-form field strengths are exact, at the cost of having more involved kinetic matrices a priori (c.f. eq. \eqref{eq:matrixonemodulus}).
		
\subsubsection*{The case $b=0$}
		
Let us now start with the case in which the axion VEV is set to zero, i.e. $b=0$. With this, the gauge kinetic functions \eqref{eq:matrixonemodulus} become diagonal, so that there is no kinetic mixing anymore, and the dependence with respect to the saxion $t$ is different for both gauge bosons. Let us now explain how such kinetic terms appear via Emergence, by integrating out the tower of D2-D0 BPS bound states with unit D2 charge and $n$ D0 charge.\footnote{In the notation of section \ref{sss:emergenceU(1)}, the graviphoton coupling to the D0 tower corresponds to $r=p=1$ and that of the D2-D0 to $r=1,\, p \to \infty$. For the other $U(1)$, the D2-D0 tower corresponds to $r=p \to \infty$.} We begin with the diagonal terms. First, from \eqref{eq:oneloopmixing} we obtain that the kinetic term for $A^0$ reads as follows 
\beq \label{eq:f00}
		f_{00}\, =\, f_{00}^{\text{UV}} + \frac{\beta}{16 \pi^2} \sum_{-n_{\text{max}}}^{n_{\text{max}}}  \left( \pi \Mpf^2 n^2 \right)\ \log \left(\frac{t^3 \,  \Lambda_{\text{QG}}^2}{\pi  \, (n^2+t^2) \, \Mpf^2}\right)\, ,
\eeq
where we have used eq. \eqref{eq:centralcharge} for the masses of the BPS states and the (quantized) charges under $A_0$ are given by $q_0^{(n)}\, =\, n  \sqrt{\pi} \Mpf$ (see Appendix \ref{ss:gaugecharges4d} for details about the correct dimensions and normalization of the charges in 4d Planck units). Also, as discussed around eq. \eqref{eq:D0tower}, $n_{\text{max}}$ is such that the mass of the heaviest BPS state of the tower is at the cut-off scale, namely $m_{n_{\text{max}}}^2 \sim \Lambda_{\text{QG}}^2 \sim 2 \pi \Mpf^2/ t$, which implies $n_{\text{max}} \sim t$. Therefore, by approximating the above sum with an integral (which is justified in the asymptotic limit) one recovers the following behaviour for the gauge kinetic function
\beq 
		f_{00}\, \simeq\, f_{00}^{\text{UV}} +  \beta \,  \Mpf^2 \, \left(\dfrac{3 \pi - 8}{144 \pi } \right)
		t^3 \sim  \,  t^3 \, \Mpf^2 \, ,
\eeq
where in the last step we have assumed (following the emergence prescription) that the UV-contribution is at most as large as the one-loop piece. Hence, we are able to generate the right asymptotic behavior in the first diagonal entry of eq. \eqref{eq:matrixonemodulus}. Similarly, including the contribution from the tower of D0-particles (which couples only to $A^0$), for which the mass in the denominator should be substituted as $(n^2+t^2) \, \to n^2$, the same asymptotic dependence $\sim t^3$ is generated.
		
Analogously, we can compute the one-loop contribution to the kinetic term of the $A^1$ gauge boson coming from the D0-D2 particles (note that the tower of D0 particles alone do not couple to $A^1$). In this case, the mass of the states in the tower is the same as before but the charge is now $q_1^{(n)}\, =\,  \sqrt{\pi} \Mpf$ for all the states (i.e. independent of $n$, as opposed to the previous case). The relevant contribution thus takes the form
\beq \label{eq:f11} 
		\begin{aligned} f_{11}\, &=\, f_{11}^{\text{UV}} + \frac{\beta}{16 \pi^2} \sum_{-n_{\text{max}}}^{n_{\text{max}}}  \left( \pi \Mpf^2 \right)\ \log \left(\frac{t^3 \,  \Lambda_{\text{QG}}^2}{\pi  \, (n^2+t^2) \, \Mpf^2}\right)\, , \\ &\simeq\, f_{11}^{\text{UV}} + \beta \Mpf^2 \left( \frac{4-\pi}{16 \pi }\right) t \sim   t \, \Mpf^2 \, ,
		\end{aligned}
\eeq
where in the last step we have assumed the UV piece to be again subleading. Interestingly, the saxion dependence of the second diagonal component of the gauge kinetic matrix   \eqref{eq:matrixonemodulus} presents again the same asymptotic behaviour as the one obtained via Emergence.
		
Let us now turn to the off-diagonal contributions. This is arguably the most interesting piece of the discussion, since it gives a couple of nice insights. First, recall the gauge kinetic matrix \eqref{eq:matrixonemodulus} is diagonal for  $b=0$. However, in principle, since the states that comprise the D0-D2 tower are charged with respect to both $U(1)$'s, kinetic mixing may be  generated via loops. Thus, if the emergence mechanism is behind the generation of the dynamics of the IR fields, it should produce a vanishing one-loop contribution when the full tower is included. In this case, such contribution indeed yields
\beq \label{eq:f01}
		f_{01}\, =\, \cancel{f_{01}^{\text{UV}}} + \frac{\beta \, \Mpf^2}{4\pi} \sum_{-n_{\text{max}}}^{n_{\text{max}}}    n\ \log \left(\frac{t^3 \,  \Lambda_{\text{QG}}^2}{\pi  \, (n^2+t^2) \, \Mpf^2}\right) \sim 0\,  ,
\eeq
where the \emph{exact} cancellation of the loop contributions is due to the fact that we sum from $-n_{\text{max}}$ to $n_{\text{max}}$, and hence the summation/integration is odd. Let us remark the importance of summing over all the states in the tower, as the individual contribution coming from each field is not vanishing, but the cancellation nicely arises when the full tower is taken into consideration. 
		
As a final comment, let us remark that this is also consistent with the expectations from our knowledge about the 5d M-theory limit that we are probing. In particular, since the 5d $\mathcal{N}=2$ theory obtained by CY$_3$ compactification of M-theory contains no axions (they arise in 4d as Wilson lines of the 5d vectors along the M-theory circle), and given that there is no kinetic mixing between the 5d graviphoton and the would-be KK photon (whose dynamics is encapsulated in the five-dimensional Einstein-Hilbert term, see eq. \eqref{eq:Mthyaction5d}), it is natural to impose  $f_{01}^{\text{UV}}=0$  in the present case.  
		
\subsubsection*{The case $b\neq0$}
		
Let us now turn on the axion VEV, $b \neq 0$, and discuss how the preceding results concerning \eqref{eq:matrixonemodulus} are modified. Notice that once we have non-vanishing axions in the vacuum the gauge kinetic functions include off-diagonal terms, and therefore there is some amount of kinetic mixing between the two gauge fields. This, together with the fact that also the $A^0$ gauge coupling is shifted by an axion-dependent amount, are the two main points of the present analysis, and as we will see they can be nicely accounted for in the framework of Emergence. The main difference with respect to the vanishing axion case is the $b$-induced shift in the mass of the states comprising the D0-D2 tower. More concretely, using eq. \eqref{eq:centralcharge}, we can see that this can be taken into account by the replacement
\begin{equation}
\label{eq:shifbneq0}
			m_n^2\, =\, \dfrac{\pi \Mpf^2}{t^3}\left[ n^2 +t^2 \right] \, \longrightarrow \dfrac{\pi \Mpf^2}{t^3}\left[ (n+b)^2 +t^2 \right] ,
\end{equation}
which basically means $n \to n+b$ in the denominators inside the logs of eqs. \eqref{eq:f00}-\eqref{eq:f01}. On the contrary, the charges with respect to the fields $A^0, A^1$ are left unchanged. At this point, one could be tempted to say that given the compactness of the axion, and the fact that in the asymptotic limit we are interested in the region where the saxion reaches the infinite-distance boundary of the moduli space (i.e. $t \to \infty$), such modification should not lead to an appreciable change of the results with respect to the $b=0$ situation. However, this intuition turns out to be actually incorrect, given that our one-loop computation was extremely sensitive to the cancellations taking place due to the symmetry upon exchanging $n$ to $-n$. Recall, this was in fact the reason why the off-diagonal terms of the gauge kinetic functions cancelled before. Now, on the other hand, the contribution to the mass that goes as $(n+b)^2$ ceases to be symmetric with respect to the D0 charge, and this is at the core of the generation of the axion-dependent terms in the gauge kinetic matrix.
		
Taking all this into account, the structure of the whole tower is very similar to the one before except from a key difference regarding which states lie below $\Lambda_{\text{QG}}$. In particular, if we now compute the values of $n_{\text{max}}$ and $n_{\text{min}}$ so that $m_{n_{\text{max}}}^2 \simeq \Lambda_{\text{QG}}^2 \simeq m_{n_{\text{min}}}^2$, we obtain the shifted quantities $n \in [-t-b, t-b]$, such that indeed $n_{\text{min}} \neq -n_{\text{max}}$.\footnote{Notice that CPT does not prevent this asymmetry in the D0-brane charge from happening, given that the would-be hypermultiplets inside the D0-D2 towers are already CPT invariant, such that for a given D0-brane charge $n$, the opposite one appears along with the anti-D2-brane wrapping the same 2-cycle.} Taking this into account, as well as the shift discussed around eq. \eqref{eq:shifbneq0}, we can proceed as in the previous section to obtain
\begin{align}
			f_{11}&\, \simeq\, f_{11}^{\text{UV}} + \beta \Mpf^2 \left( \frac{4-\pi}{16 \pi }\right) t\, \sim\,   t \, \Mpf^2 \, ,\\ 
			f_{01}&\, \simeq\, f_{11}^{\text{UV}} - \beta \Mpf^2 \left( \frac{4-\pi}{16 \pi }\right) b t\, \sim \, - b  t\, \Mpf^2 \, ,\\
			\begin{split}
				f_{00}&\, \simeq\, f_{00}^{\text{UV}} + \frac{\beta \Mpf^2}{144 \pi } \left[  \left( 3\pi -8  \right) \, t^3 + \left( 4 -\pi \right)\,  b^2 t \right]\, \sim\, (\, C \, t^3 +  \,  b^2 t)\, \Mpf^2\, .
			\end{split}
\end{align}
For $f_{11}$ we obtain the same result as in the case with $b=0$, as expected. For the mixed term, $f_{01}$, we get instead a non-vanishing result which is essentially $(-b)$-times the analogous contribution for $f_{11}$, and reproduces \eqref{eq:matrixonemodulus}. Notice that for this case it is even more clear that we should set the UV piece to zero, since as we discussed at the end of the $b=0$ section we do not expect such kinetic mixing to be present in the 5d parent theory. Finally, for the $f_{00}$ component it is interesting to see how the right structure of the gauge kinetic matrix is recovered also in this more general case, with the subleading but nevertheless diverging contributions depending on the axions appearing from the non-trivial dependence of the central charge, eq. \eqref{eq:centralcharge}, on the moduli fields. Notice that, apart from the usual order one prefactors, in this case we are also unable to fix the relative coefficient, $C$,  between the two terms in $f_{00}$. On the one hand, this is to be expected because we should also include the tower of D$0$-branes alone (which couple only to $A^0$), since they can be seen to contribute precisely to the term proportional to $t^3$, but not to the $b^2 t$ term, hence modifying $C$. Moreover, since all our computations (and even the precise definition of the species scale) are defined up to $\mathcal{O}(1)$ factors, one cannot expect to predict accurately the value for $C$. We however expect to get the correct dependence on  the fields, which is precisely what we are able to recover for \emph{every} component of the gauge kinetic matrix \eqref{eq:matrixonemodulus}.
		
\subsubsection*{Relationship with the Gopakumar-Vafa computation }
\label{sss:GV}	
It is natural to ask at this point how our results fit with the work\cite{Gopakumar:1998ii, Gopakumar:1998jq}. The main point of these two papers was that in fact one could compute exactly some particular (higher derivative) F-terms appearing in the type IIA 4d $\mathcal{N}=2$ effective action, which are encoded by certain $\mathcal{N}=2$ topological string partition functions (at different genus $g$) \cite{Bershadsky:1993cx,Antoniadis:1993ze}, by making use of type IIA/M-theory duality at large $\mathcal{V}$. Thus, their strategy consisted in integrating out at one-loop KK modes as well as M2-branes wrapped on certain 2-cycles when going from 5d to 4d, which from the type IIA perspective were seen as D0-branes and wrapped D2-branes (as well as bound states thereof), such that a simple Schwinger-type computation reproduced all said F-terms in 4d \emph{exactly}. The main idea we want to convey here is that our results are not in clash with \cite{Gopakumar:1998ii, Gopakumar:1998jq}, but instead they seem to nicely complement their discussion. 
		
\emph{The Schwinger-type computation}
		
Let us start by recalling what Gopakumar-Vafa (GV) compute. They focus on certain \emph{protected} F-terms in the 4d (euclidean) effective action, which only depend on the K\"ahler moduli, and which can be written in superspace as follows \cite{Gopakumar:1998ii}
\beq \label{eq:superspaceGV}
		\int d \theta^4\ \mathcal{W}^{2g}\, ,
\eeq
where $\mathcal{W}_{\mu \nu}= F^+_{\mu \nu} - R^+_{\mu \nu \lambda \rho} \theta \sigma^{\lambda \rho} \theta + \ldots$  is the (weight one) Weyl superfield and $F^+,\ R^+$ denote the self-dual pieces of the 4d graviphoton field strength and curvature tensor, respectively \cite{Antoniadis:1995zn}. Notice that, strictly speaking, such F-terms are really present for $g \geq 1$, since otherwise eq. \eqref{eq:superspaceGV} is identically zero. Thus, they all seem to correspond to \emph{higher-derivative} terms in the 4d lagrangian.
		
Once the series above is expanded in terms of their spacetime field components, they give rise to terms in the lagrangian of the form $\mathcal{F}_g R^2_+ F^{2g-2}_+$, where the index contractions are determined from \eqref{eq:superspaceGV}. The coefficients $\mathcal{F}_g$ are precisely the quantities calculated by the topological closed string theory at each genus $g$\cite{Gopakumar:1998ii, Gopakumar:1998jq}, which read
\beq \label{eq:TPF}
		\mathcal{F}(\lambda, z^a)= \sum_g \lambda^{2g-2} \mathcal{F}_g (z^a)\, ,
\eeq
with $\lambda$ denoting the topological string coupling and $z^a$ the K\"ahler moduli.
		
\begin{figure}[tb]
			\begin{center}
				\includegraphics[scale=1.5]{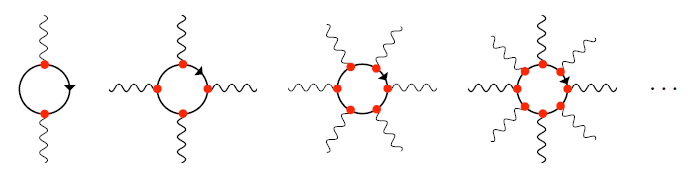}
				\caption{1PI diagrams arising from integrating out the D-particles at one loop in GV. The wiggly lines denote the external graviphotons radiated by the virtual particles. One should also include in each diagram four external gravitons as arising from the common factor $R^2_+$. Notice that the series starts at $g=2$.}
				\label{fig:EH}
			\end{center}
\end{figure}

One of the key insights from GV was that the $\mathcal{F}_g$ can be computed by integrating out at one loop, and in the large volume phase, towers of D0/D2-particles within some constant self-dual graviphoton field\footnote{These D-particles are charged under the 4d graviphoton, with physical charge given by their central charge, which in string units reads $e= m_{n_{2p}}=2 \pi |\mathcal{Z}_{\text{IIA}}|/g_s$ (see eq. \eqref{eq:centralcharge}).} (see fig. \ref{fig:EH}). Therefore, upon using the Schwinger prescription one arrives at the following one-loop contribution for e.g. an $\mathcal{N}=2$ hypermultiplet \cite{Gopakumar:1998jq}
\beq \label{eq:Schwinger}
		\sum_g \left( F_+ g_s \right)^{2g-2} \mathcal{F}^{\text{hyper}}_g= \frac{1}{4} \int_{\epsilon}^{\infty} \frac{d\tau}{\tau} \frac{1}{\text{sinh}^2 \left( \frac{\tau F_+}{2}\right)} e^{-\tau \frac{2 \pi |\mathcal{Z}_{\text{IIA}}|}{g_s}}\, ,
\eeq
where the mass of the BPS particles is measured in string units and a UV regulator $\epsilon \to 0^+$ has been introduced. Notice that this is nothing but a supergravity generalization of the Euler-Heisenberg effective action arising after integrating out the electron in QED \cite{Heisenberg:1936nmg}.

\emph{Gopakumar-Vafa and the Species Scale}
		
A crucial property of the one-loop determinant in \eqref{eq:Schwinger} is that it can be seen to converge for the terms with $g>1$ above, corresponding to the one-loop diagrams in fig. \ref{fig:EH} with two or more graviphoton insertions (which contribute to higher derivative terms in the 4d supergravity effective action), whilst the would-be terms associated to $g=0,1$, are indeed UV divergent. Note that this is to be expected since from the Euler-Heisenberg perspective those two terms would correspond,  respectively, to the contribution to the Casimir energy and wave-function renormalization of the 4d photon in QED (see e.g. \cite{Schwartz:2014sze}), and thus need to be regularized in some way. Moreover, the regulator $\epsilon$ introduced in eq. \eqref{eq:Schwinger} can be roughly related to a UV momentum cut-off as $\epsilon= m_{n_{2p}} \, \Lambda_{\text{UV}}^{-2}$ . The approach taken by \cite{Gopakumar:1998ii, Gopakumar:1998jq} in this regard was to regularize such terms by analytic continuation methods from the ones obtained for $g>1$, such that the total contribution from a given hypermultiplet to the topological partition function at genus $g$ would be (in string units)\footnote{Notice that for the $\mathcal{F}^{\text{hyper}}_0$ term, which is the relevant one for the emergence computations, we should also include the logarithmic factor depending on $|\mathcal{Z}_{\text{IIA}}|$ that would  mimic the one in eq. \eqref{eq:f00}. Moreover, for $\mathcal{F}^{\text{hyper}}_1$ there should also be an extra (dominant) logarithmic dependence on the central charge, which is indeed crucial to match the predictions of the (finite distance) conifold singularity in the gravitational $R^2$ term \cite{Vafa:1995ta}.}
\beq\label{eq:Fghyper} 
		\mathcal{F}^{\text{hyper}}_g = - \chi_g\,  \mathcal{Z}_{\text{IIA}}^{2-2g}\, ,
\eeq
where $\chi_g=2-2g$. Therefore, focusing on the genus-0 case in eq. \eqref{eq:Fghyper} together with the off-shell BPS content of the virtual particles running in the loop one could compute the contribution to the $\mathcal{N}=2$ \emph{prepotential} due to the infinite tower of D0-branes, D2-particles and bound states thereof . In particular, the term associated to the D0-brane hypermultiplets reads\cite{Gopakumar:1998ii}
\beq\label{eq:F0D0} 
		\mathcal{F}^{\text{D0, hyper}}_0 = -\chi_0
  \sum_{n \in \mathbb{Z} \setminus \lbrace0\rbrace} \left(2 \pi n \right)^{2}\, ,
\eeq
where we have not included the logarithmic factor (see eq. \eqref{eq:f00}) since it will not modify the final results once the summation is performed. Notice that in order to give a finite answer above (as well as for the D0-D2 tower) upon summing over the D0-brane charge $n$, GV employed a particular regularization which at the end of the day served to reproduced certain $\alpha'$-perturbative and non-perturbative terms in the type IIA prepotential \emph{exactly}, providing also for an interpretation of the coefficients in the expansion of the topological partition functions in terms of counting of BPS states.
		
Our aim here is to see how these results change when we impose the species scale cut-off in the Schwinger one-loop determinant, as well as restricting the sum over D0-brane charges up to the maximal one that falls below $\LQG$ (see discussion around eq. \eqref{eq:D0tower}). First, upon identifying $\epsilon= m_{n_{2p}}\,  \LQG^{-2}$ in \eqref{eq:Schwinger} and taking the asymptotic limit where essentially $\LQG \gg m_n$ for almost all $n<N_{\text{tot}}$, one can easily check that the F-terms of the form $\mathcal{F}_g R^2_+ F^{2g-2}_+$ for $g>1$ are again reproduced at leading order.
		
On the other hand, the imposition of a maximum charge $N_{\text{tot}}$ falling below $\LQG$ also allows one to perform the sumation in  \eqref{eq:F0D0}, but instead of giving a moduli-independent contribution, it turns out to scale as the tree-level part of the prepotential (which is not captured a priori by the topological string theory). This happens analogously for the contribution associated to the D0-D2 bound states, providing for an explanation of why from the emergence perspective they seem to give quantum corrections to the scalar metrics and gauge kinetic functions which scale like the tree-level piece (c.f. section \ref{ss:4doneloop}).

\subsection{Other possible infinite distance limits in 4d}
\label{ss:generalizations}		
Recall that in the analysis of section \ref{ss:4doneloop} we restricted ourselves to one-dimensional K\"ahler moduli spaces, in which the relevant infinite distance singularity was identified with the Large Volume Point. Moreover, the asymptotically massless tower of states giving rise to the singularity effectively implemented a circle decompactification to 5d M-theory on the same Calabi--Yau three-fold. However, as already commented around eq. \eqref{eq:matrixonemodulus}, the discussion there seems to apply equally well to higher dimensional moduli spaces, as long as we fix the rescaled K\"ahler parameters, $M^a = t^a/\mathcal{V}^{1/3}$, to be finite (and non-vanishing) whilst the volume modulus, $\mathcal{V}$, is taken to infinity.
		
One could also ask what happens now if we additionally move close to infinite distance boundaries in the $M^a$ directions as well. Indeed, what one expects is to approach another kind of singular limit, such that the QG resolution may change, with a different massless tower dominating the infinite distance regime, thus implementing some other `gravitational phase transition'. Notice, that given that we essentially explore the large volume phase of type IIA string theory on the CY three-fold, any sort of infinite distance singularity that we may encounter there should be intuitively present already in the vector multiplet moduli space of M-theory on the same three-fold \cite{Witten:1996qb}.
		
Before commenting on this, let us say that we will concentrate on two other possible infinite distance boundaries that appear in the large volume patch of the type IIA K\"ahler sector, as studied in \cite{Lee:2019wij, Corvilain:2018lgw}. These correspond to having a universal fibration structure, where the fibre shrinking (in rescaled coordinates) is given by an elliptic curve or a K3 surface \cite{Lee:2019wij}. These limits, which originally seemed to be in disagreement with the Emergence Proposal \cite{Grimm:2018ohb}, have been recently successfully re-analyzed from the prism of emergence in \cite{Marchesano:2022axe}. In the following, we revisit them, elaborating on several points to present complementary insights from our approach, but we encourage the interested reader to check \cite{Marchesano:2022axe} for the technical details.
		
\subsubsection{The F-theory limit}
\label{sss:IIA-Ftheorylimit}		
Let us start with the $T^2$-limits of \cite{Lee:2019wij}, in which apart from having $\mathcal{V} \to \infty$,\footnote{Actually, one can see that such $T^2$-limit belongs to ${\cal M}_{\rm VM}$ without having to take also $\mathcal{V} \to \infty$ (meaning that it is not obstructed by quantum corrections \cite{Lee:2019wij}). However, as noted in \cite{Corvilain:2018lgw, Aspinwall:2002nw, Alim:2012ss, Klemm:2012sx, Huang:2015sta}, such related infinite distance singularity is actually `T-dual' to the large volume one, such that we land again in 5d M-theory on the same (elliptic) CY$_3$, with the towers of D0 and D2-branes of section \ref{ss:preliminary} exchanged.} the CY$_3$ exhibits asymptotically some sort of elliptic fibration
\begin{equation}\label{eq:ellfibration}
			\begin{aligned}
				\pi: \qquad \mathbb{E} \rightarrow &\;X_{3} \\
				&\;\; \downarrow \qquad , \\ &\;B_{2}
			\end{aligned}
\end{equation}
where $\mathbb{E}$ denotes the elliptic fibre whose associated rescaled K\"ahler modulus, which we call $M^0$ here, vanishes asymptotically. Notice that this kind of limits correspond to the type III singularities in Mixed Hodge Structure (MHS) language, as discussed in \cite{Grimm:2018ohb, Corvilain:2018lgw}.
		
One can see \cite{Marchesano:2022axe} that in said infinite distance limits, contrary to what happens in our example above (see fig. \ref{fig:D2D0towers}), both the D0- and D2-particles present the same asymptotic mass scale (up to $\mathcal{O}(1)$ factors) in 4d Planck units. Moreover, the D2-branes wrapping the elliptic fibre provide now for an \emph{infinite} number of distinct BPS states, which geometrically is reflected in the fact that the Gopakumar-Vafa invariants associated to the $T^2$ fibre are (generically) non-zero and constant\cite{Klemm:2012sx, Klemm:1996hh} for each $n\in \mathbb{Z} \setminus \lbrace 0 \rbrace$
\beq \label{eq:GVinvariantsT2limit}
		\text{GV}_n = \chi(X_3) = 2 \left ( h^{1,1} (X_3) -  h^{2,1} (X_3) \right)\, ,
\eeq
where $n$ is nothing but the wrapping number of the D2-brane along the elliptic 2-cycle, $\mathbb{E}$. Additionally, both kind of towers can mix, forming bound states with D0 and D2 (quantized) charges, which can in principle run independently as long as the total mass remains below the species scale, $\LQG$. Note that this scenario corresponds to the case of multiplicative species (with two towers giving a total of $p=2$) discussed in section \ref{ss:MultipleTowers}, such that upon using our formulae \eqref{eq:massmixedsprectra}-\eqref{eq:NtotLQGeff}, one indeed reproduces the results in \cite{Marchesano:2022axe} for what they call the $w=2$ limit, as we review here quickly. Denoting the divergent modulus by $t$, the mass of the D0-branes, as well as that of D2-branes wrapping the toroidal fibre that is shrinking behave as $m_{\text{D}0}\sim m_{\text{D}2}\sim \Mpf/t$. Using the general formula \eqref{eq:NtotLQGeff} for $d=4$ and $p=2$ we obtain the following behaviour for the species scale and the total number of species
\begin{equation}
    \LQG\, \sim\,  \dfrac{\Mpf}{t^{\frac{1}{2}}}\, , \qquad \Ntot\, \simeq \, N_{\mathrm{D}0}\,  N_{\mathrm{D}2} \sim t\, .
\end{equation}
Performing now a similar calculation to the one in \eqref{eq:f00} (for the case of vanishing axions), in which we compute the contribution of the full tower of D2-D0 particles to the kinetic term of the graviphoton, we get
\beq \label{eq:f00F-theorylimit}
		f_{00}\, \sim\, \sum_{n_{\text{D}2}}^{N_{\mathrm{D}2}} \sum_{n_{\text{D}0}}^{N_{\mathrm{D}0}}  \left(  \Mpf^2\,  n_{\text{D}0}^2 \right)\, \log \left(\dfrac{t}{n_{\text{D}0}^2 + n_{\text{D}2}^2}\right)\, \sim \,  \Mpf^2 \, t^2\,  ,
\eeq
where once again we approximated the summations by integrals and used $N_{\mathrm{D}0} \, \sim \, N_{\mathrm{D}2} \sim t^{1/2}$. Similarly, for the kinetic term of the 1-form that couples to the D2-particle that becomes light we obtain the same leading parametric behaviour, namely
\beq \label{eq:f11F-theorylimit}
		f_{11}\, \sim\, \sum_{n_{\text{D}2}}^{N_{\mathrm{D}2}} \sum_{n_{\text{D}0}}^{N_{\mathrm{D}0}}  \left(  \Mpf^2\,  n_{\text{D}2}^2 \right)\, \log \left(\dfrac{t}{n_{\text{D}0}^2 + n_{\text{D}2}^2}\right)\, \sim \,  \Mpf^2 \, t^2\, ,
\eeq
whereas for the mixed terms (in the vanishing axion case) and the other $U(1)$'s (under which the light D2-D0-particles are not charged) we obtain vanishing contributions from the quantum corrections. Note that eqs. \eqref{eq:f00F-theorylimit} and \eqref{eq:f11F-theorylimit} indeed reproduce the right field dependence on the divergent modulus, $t$\cite{Grimm:2018ohb,Marchesano:2022axe}.

To finish discussing this limit, let us just add a few relevant comments about the QG resolution of the singularity. As discussed in \cite{Marchesano:2022axe} (and already remarked by \cite{Corvilain:2018lgw}), one expects a phase transition to 6d F-theory on the elliptic three-fold upon approaching this type III limit. The main intuition comes from type IIA/M-theory duality first, and then from M-/F-theory duality \cite{Vafa:1996xn} on the second place. Thus, recall that upon taking the $\mathcal{V} \to \infty$ limit and on top of that exploring singularities in the $M^a$ hyperspace, what we are effectively doing is to sample the 5d M-theory vector moduli space \cite{Witten:1996qb}. Therefore, since a $T^2$ (classical) limit there leads to a circle decompactification to F-theory \cite{Lee:2019wij}, the natural conclusion here would be to identify this type III singularity to a nested limit from 4d type IIA to 6d F-theory.
		
One non-trivial check that can be performed so as to provide evidence for the previous conclusion is to employ the super-quantum mechanical machinery (see Appendix \ref{ap:Dpbranecontent}) to deduce the field-theoretic content associated to the D2-branes wrapping the elliptic fibre, similar to what we did for the previous example in section \ref{ss:preliminary}. Therefore, what one actually finds is that the moduli space of the elliptic fibre $\mathbb{E}$ together with its flat connections is again an elliptically-fibered K\"ahler three-fold, such that the D2-particles contain for each $n_{\text{D}2}\in  \mathbb{Z} \setminus \lbrace 0 \rbrace$ precisely one massive spin-2 multiplet, $h^{1,1} (B_2)=h^{1,1} (X_3)-1$ massive vector multiplets and $h^{2,1} (X_3)+1$ massive hypermultiplets.
		
\subsubsection{The Emergent Heterotic String limit}
\label{sss:IIA-heteroticlimit}		
Finally, we discuss now the so-called $K3$-limits of \cite{Lee:2019wij}, which only exist at large volume (due to large $\alpha'$-corrections) and are characterized by the fact that the three-fold presents an asymptotic fibration structure of the form
\begin{equation}\label{eq:K3fibration}
			\begin{aligned}
				\pi: \qquad K3 \rightarrow &\;X_{3} \\
				&\;\; \downarrow \qquad , \\ &\;\mathbb{P}^1
			\end{aligned}
\end{equation}
where again it is the fibre the one shrinking the fastest in rescaled coordinates, upon taking the singular limit. Note that this set-up corresponds to the MHS type II singularities discussed in \cite{ Corvilain:2018lgw}.
		
The crucial difference here with respect to the previous limits is that the leading tower of states supposedly comes from a \emph{critical} emergent heterotic string\cite{Lee:2019wij}. This string would arise as an NS5-brane wrapping the $K3$ fibre in the type IIA side, which presents in its world-sheet effective theory precisely the spectrum associated to a would-be heterotic string compactification on $K3 \times T^2$ (or some free quotient thereof)\cite{Harvey:1995rn}. Therefore, if we \emph{assume} that this type II limit corresponds to an emergent heterotic string limit, which is supported by well-known dualities between 4d $\mathcal{N}=2$ type IIA/heterotic theories\cite{Ferrara:1995yx,Kachru:1995wm}, we can run our computations for critical string limits in section \ref{sss:emergenceU(1)} (more precisely around eq. \eqref{eq:oneloopstringtowergauge}) in order to reproduce the behaviour of the gauge kinetic matrix \eqref{eq:gaugekineticmatrix} in the present case. In particular, using the general formula for the emergent gauge coupling \eqref{eq:1-formemergencestringytowersgeneral} with the mass scale for the stringy tower in the case at hand, which reads $M_s\sim \Mpf/t^{1/2}$, we obtain
\beq
	\delta \left(\frac {1}{g^2}\right)\, \sim\, \frac {\Mpf^4}{M_s^2} \, \sim \, \Mpf^2\, t \, .
\eeq 
As expected, this gives the right asymptotic dependence (for vanishing axion VEVs), as explained in detail in \cite{Marchesano:2022axe} (in what they call the $w=1$ limit). In fact, upon using the more refined formulae from Appendix \ref{ap:Loops1-form}, we arrive at the same leading result.
		
\subsection{The type IIB side of the story}
\label{ss:4dtypeIIB}		
Up to now, we have discussed emergence in the vector multiplet moduli space by explicitly studying type IIA on $X_3$, where the limit under study was the large volume point and the towers of light states are comprised by D0 and D2-branes (as well as bound states thereof). In this section we elaborate on this from the type IIB perspective, by considering its compactification on the mirror $Y_3$. In fact, this is the set-up that was originally considered for a systematic study of the Swampland Distance Conjecture in \cite{Grimm:2018ohb, Grimm:2018cpv}. There, different infinite distance limits within the vector multiplet moduli space, corresponding to the space of complex structures of $Y_3$, were carefully studied (profiting from the fact that they receive no $\alpha'$ or $g_s$-corrections, and thus are known to encode non-trivial quantum corrections, see e.g. \cite{Strominger:1995cz, Vafa:1995ta}). In particular, it was found there that in the large complex structure (LCS) limit the tower of BPS-states fullfilling the SDC was comprised of D3-branes wrapping minimal (supersymmetric) 3-cycles (given close to the LCS point by special lagrangian 3-cycles, $\Sigma \subset Y_3$). The mass of such states is controlled by the $\mathcal{N}=2$ central charge $Z_{\text{IIB}}=e^{K_{\text{cs}}/2} \mathcal{Z}_{\text{IIB}}$, hence depending on the complex structure moduli, $z^i$, as follows
\beq \label{eq:centralchargeIIB}
		\frac{m_{\Sigma}}{\Mpf} = \sqrt{8 \pi} e^{K_{\text{cs}}/2} |\mathcal{Z}_{\text{IIB}} ({\Sigma})|\, ,
\eeq
where $\Sigma$ denotes the corresponding cycle wrapped by the D3-brane and ${\mathcal{Z}_{\text{IIB}} ({\Sigma}) = \int_{\Sigma} \Omega_3 (z^i)\,} $, is the period of the unique holomorphic ($3,0$)-form, $\Omega_3$, along that 3-cycle. In terms of an integral symplectic basis $(\gamma^K, \gamma_L)$ of $H_3(Y_3, \mathbb{Z})$, with $K, L=1, \ldots, h^{2,1}(Y_3)+1$, and polarized such that $\gamma^K \cap \gamma_L = \delta^K_L$, the central charge $\mathcal{Z}_{\text{IIB}} ({\Sigma})$ can be written as follows
\beq
		\mathcal{Z}_{\text{IIB}} ({\Sigma}) = n_K Z^K+m^L \mathcal{F}_L\, ,
\eeq
with $[\Sigma]= n_K [\gamma^K] + m^L [\gamma_L]$ ($n_K, m^L \in \mathbb{Z}$) and where we have introduced the holomorphic periods ${Z^K = \int_{\gamma^K} \Omega_3\,,} \quad {\mathcal{F}_L = \int_{\gamma_L}\Omega_3\, .}$
		
The original proposal in  \cite{Grimm:2018ohb} was that by precisely integrating out the appropriate tower of bound states of D3-branes getting asymptotically light (as we approach the LCS point) up to the species scale, one could reproduce the leading order behaviour of both the moduli space metric as well as the diagonal entries of the gauge kinetic matrix. However, as far as we are concerned, a microscopic understanding of the resolution of such infinite-distance singularity in QG was not proposed, neither a physical interpretation or rationale for the different towers becoming light was provided. The aim of this section is nothing but to try to fill this gap by relating the large complex structure singularity in type IIB with the already discussed large volume point in IIA through Mirror Symmetry (see e.g. \cite{Hosono:1994av, Hori:2003ic} for an introduction to the topic).
		
Our main claim regarding this is that the asymptotic physics of type IIB string theory compactified on the three-fold $Y_3$ around the LCS point is governed by the 5d $\mathcal{N}=2$ theory on the mirror $X_3$. What this is telling us is that indeed the resolution of both infinite distance singularities in QG requires from a five-dimensional M-theory description. In this regard, it is interesting that both type II compactifications lead to the same asymptotic physics in terms of a 5d gravitational theory, even though a priori only type IIA is directly connected to M-theory via circle compactification. Therefore, in some sense, type IIB inherits from type IIA (via Mirror Symmetry) such circle decompactification once we compactify on the three-fold $Y_3$. Note that this kind of phenomenon is commonplace in CY$_3$ compactifications, one additional example being the self-S-duality in 4d that type IIA inherits from type IIB (again through Mirror Symmetry), even though it is absent in the parent 10d theory \cite{Alvarez-Garcia:2021pxo, Baume:2019sry}. Besides this, the interpretation in terms of a five-dimensional M-theory also predicts a lighter (stable) tower in the type IIB set-up, in addition to the ones mainly considered in \cite{Grimm:2018ohb} (which were made from bound states of $n$ D3-branes wrapping the fastest shrinking cycle, $\gamma^0$, as well as an extra D3-brane wrapping once some subleading $\gamma^i$-cycle), that should correspond to the KK modes of the massless fields of the 5d M-theory. Here we argue that this additional tower of BPS states arising from bound states of just $n$ D3-branes wrapping  $\gamma^0$-cycle must also be considered.
		
\subsubsection*{The LCS limit}	
As shown by Strominger, Yau and Zaslow (SYZ) in \cite{Strominger:1996it}, Mirror Symmetry can be understood near the LCS regime in type IIB on $Y_3$ as performing (adiabatically) three T-dualities along the reference special lagrangian three-torus, $T^3$, by which the three-fold, $Y_3$, is fibered. Such special 3-cycle, which we call the SYZ-cycle, is nothing but the $\gamma^0$ that we referred to before. The base for the fibration would then be the (symplectic) dual cycle $\gamma_0$, which presents $S^3$-topology (see e.g. \cite{Alvarez-Garcia:2021pxo}). Moreover, one can see that these two 3-cycles are precisely the fastest shrinking and growing ones in the LCS regime, respectively.
		
Upon performing T-duality along the $T^3$ fibres, D3-branes which wrap such $\gamma^0$-cycle will turn into D0-branes in the type IIA picture, whilst the remaining D3-branes wrapping a linear combination of $\gamma^i$-cycles become the  D2-particles described in section \ref{ss:preliminary}. Specifically, this means that the field-theoretic content inside each kind of BPS D3-particle in type IIB on $Y_3$ should match the corresponding one in type IIA on $X_3$. Hence, there must be two distinguished (infinite) towers of BPS modes becoming light upon approaching the singularity: one consisting on bound states of D3-branes wrapping some $\gamma^i$-cycle only once as well as the SYZ-cycle $n$ times (corresponding to the mirror dual of the D0-D2 tower), and a second tower arising solely from bound states of D3-branes along the $T^3$-fibre (corresponding to the D0 tower). Moreover, this allows us to be slightly more precise about the microscopic structure of the towers. Thus, recall that Mirror Symmetry requires in particular for $Y_3$ to have a (supersymmetric) 3-cycle $\Sigma$ with the same moduli space of the D0-branes in type IIA, i.e. $\mathcal{M}_{\text{D}3}(\Sigma)=\mathcal{M}_{\text{D}0}$. Such moduli space is precisely the dual three-fold $X_3$, since at large volume the classical geometry should approximate the exact quantum moduli space well enough \cite{Strominger:1996it}. Therefore, according to the SYZ-conjecture, the moduli space of the aforementioned 3-cycle in $Y_3$ is just $\mathcal{M}_{\text{D}3}(\gamma^0)=X_3$. Hence, upon quantizing the non-linear sigma model of the D3-brane wrapping the $\gamma^0$-cycle one finds precisely the same states as we did for the D0-brane in the type IIA dual (see Appendix \ref{ss:4dDO} for details), thus signalling towards a decompactification to 5d M-theory on $X_3$.
		
Finally, let us make one last point regarding the behaviour of the relevant parameter of the Swampland Distance Conjecture in this well-known set-up, which fits nicely with the idea of having the additional light tower in this limit. According to the conjecture, for every infinite-distance point one expects the appearance of an infinite tower of states whose mass scale, $m_{\text{tow}}$, becomes exponentially small with respect to the moduli space distance ($\Delta$), i.e.
\begin{align}\label{eq:4dSDC}
			m_{\text{tow}} \sim \ \text{exp} (-\lambda \Delta)\, ,
\end{align}
where $\lambda$ is some a priori undetermined $\mathcal{O}(1)$ parameter in Planck units. There has been some recent interest in trying to constrain the possible values for $\lambda$ that can appear by essentially looking at string theory examples. In particular, the sharper lower bound for the \emph{lightest} tower in a $d$-dimensional theory,
\begin{align}\label{eq:boundlambda}
			\lambda_{\text{max}} \geq \frac{1}{\sqrt{d-2}}\, ,
\end{align}
has been recently proposed \cite{Etheredge:2022opl}. Specifically, the bound was shown to be saturated in several string theory examples with maximal/minimal supersymmetry in different number of spacetime dimensions.
		
In the original study of the LCS point in type IIB CY$_3$ compactifications \cite{Grimm:2018ohb}, the only tower whose existence could be rigorously proven (using arguments relying on the monodromy structure around that point, as well as a careful study of walls of marginal stability) indeed violates such strict lower bound, since it yields $\lambda=1/\sqrt{6}$ in 4d. However, a natural conclusion from our discussion above is that the bound in eq. \eqref{eq:boundlambda} is also satisfied in this scenario, precisely by the tower of D3-branes wrapping the SYZ-fibre (which is dual to the KK tower of the five-dimensional M-theory massless modes along the decompactifying circle) since it provides for a $\lambda=\sqrt{3/2} \geq \lambda_{\text{max}}=1/\sqrt{2}$.
		
\subsection{Emergence in the Hypermultiplet sector}
\label{ss:hypermultiplet4d}
		
Let us now turn our attention to the other (independent) sector of the moduli space of 4d $\mathcal{N}=2$ EFTs coming from CY$_3$ compactifications. In particular, our main goal here is to highlight some interesting points about the consistency of the Emergence Proposal with a ubiquitous phenomenon in these theories, namely that of instanton corrections (see also \cite{Hamada:2021yxy} for some nice complementary discussions about this point).
		
Due to $\mathcal{N}=2$ supersymmetry, the moduli space of supergravity theories factorises\footnote{\label{fn:modspaceprod}To get such a product structure one might need to consider a multiple cover of ${\cal M}_{\rm VM}$ or ${\cal M}_{\rm HM}$ \cite{Seiberg:1996ns}.} (even quantum-mechanically) at the two-derivative level as 
\begin{equation}\label{eq:N=2modulispace}
			{\cal M} = {\cal M}_{\rm VM} \times {\cal M}_{\rm HM}\, ,
\end{equation}
where the first component, the vector multiplet moduli space, is a projective special K\"ahler manifold that determines both the metric for the corresponding scalars as well as the gauge kinetic functions, which have been the main focus of the previous subsections. On the other hand, ${\cal M}_{\rm HM}$ represents the hypermultiplet moduli sector, which is a quaternionic K\"ahler manifold parameterized, in the particular case of type IIA on $X_3$, by the collective coordinates $q^p$ in eq. \eqref{eq:IIAaction4d}. In the type IIA set-up one finds precisely $\left( h^{2,1}(X_3)+1\right)$ hypermultiplets, with the universal one including the 4d dilaton, $\varphi_4$, and the extra $h^{2,1}$ ones comprising the complex structure moduli, $z^i$ (see e.g. \cite{Grimm:2005fa} for a precise definition of the 4d fields).
		
One of the features that makes the hypermultiplet sector qualitatively different from its vector counterpart is that it receives a plethora of perturbative and non-perturbative $g_s$-corrections (in the type IIA case). This is the so because, as we just mentioned, it includes $\varphi_4$ and $z^i$, which control the action of euclidean D2-branes (i.e. 4d instantons) wrapping minimal 3-cycles in the internal geometry. Therefore, the effective 4d action describing this sector can be heavily quantum-corrected depending on where we sit in moduli space, such that it is typically not enough to focus just on the tree-level part arising from direct CY$_3$ compactification from 10d type IIA. In particular, as nicely described in \cite{Marchesano:2019ifh,Baume:2019sry}, it may happen that a would-be infinite distance point which is present at tree-level in the moduli space geometry gets obstructed or excised in the \emph{exact} $\mathcal{N}=2$ quantum moduli space, precisely due to the infinitely many instanton corrections that become relevant.
		
\subsubsection*{The D4-brane Limit}
\begin{figure}[t]
		\begin{center}
			\subfigure[]{
				\includegraphics[height=4.5cm]{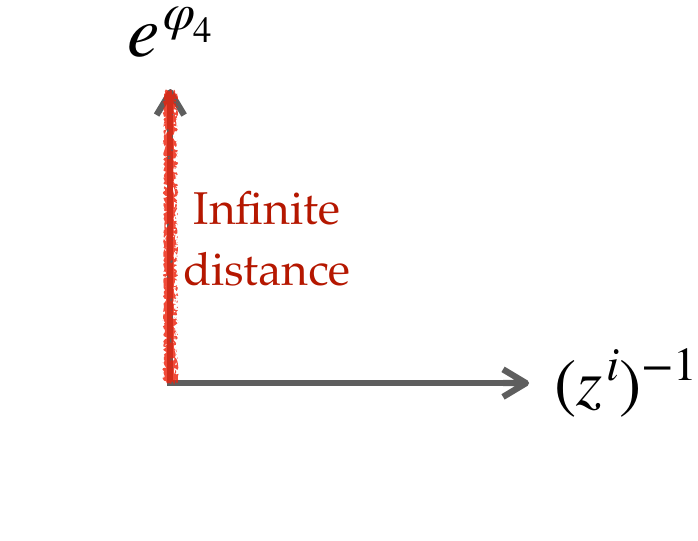} 
				\label{fig:Instantoncorrections1}
			}\qquad \qquad\qquad
			\subfigure[]{
				\includegraphics[height=4.5cm]{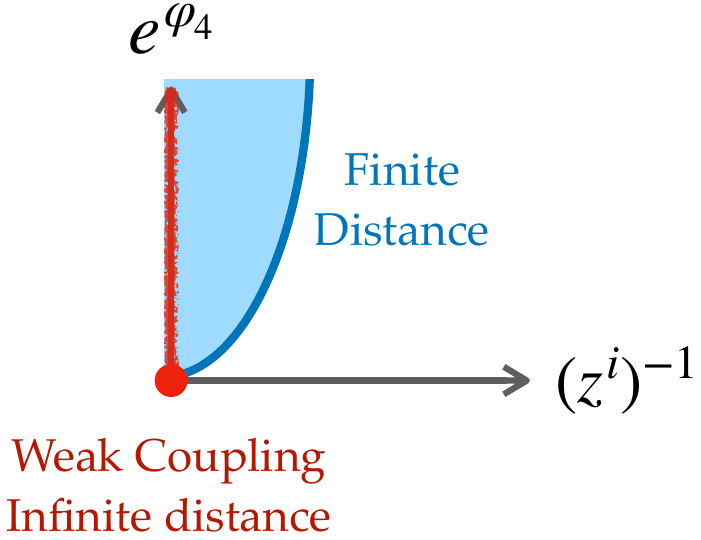}
				\label{fig:Instantoncorrections2}
			}
			\caption{(a) Classical infinite distance limit as $z^i \, \to \,i \,\infty$. (b) Including instanton corrections that become relevant along the limit actually obstruct the would-be classical singularity and only leave the weak coupling point as an infinite distance boundary.}			
		\label{fig:Instantoncorrections}
		\end{center}
\end{figure} 
Let us briefly review one instance in which such obstruction takes place, and try to understand it from the prism of emergence. We will concentrate on the infinite distance singularity in the mirror symmetric analogue of the D1-limit of \cite{Baume:2019sry}, which was closely investigated in \cite{Alvarez-Garcia:2021pxo} (we refer the interested reader to these references for the details of the examples).
		
Consider a trajectory moving only along the type IIA hypermultiplet moduli space. Probably the first such trajectory that comes to mind is the one that reaches the LCS point of the type IIB vector multiplet moduli space, embedded via the c-map within the type IIA hypermultiplet side, which means that we also keep the 4d dilaton fixed. As discussed in section \ref{ss:4dtypeIIB} (see also \cite{Grimm:2018ohb, Grimm:2018cpv}), in the type IIB side the LCS singularity may be understood from Emergence by integrating out D3-branes wrapping some combination of supersymmetric 3-cycles. However, in the type IIA frame we do not find any such D-brane particle states whose mass spectrum is controlled by the complex structure moduli, and hence it seems difficult to find an infinite tower which could give rise by quantum corrections to the singular behaviour of the tree-level 4d effective action. 
		
At this point, one could be tempted to conclude that the Emergence Proposal does not seem to hold in the present case. This is where the analysis of \cite{Baume:2019sry,Alvarez-Garcia:2021pxo} comes to our rescue. It can be seen that, even though we do not have the D3-particles here, there are indeed solitonic strings arising from wrapped D4-branes on non-trivial 3-cycles, together with euclidean D2-instantons, whose tension and actions, respectively, are controlled precisely by the complex structure moduli.\footnote{As an aside, both strings and instantons happen to be the c-duals of the same D3-particles in the type IIB side, hence explaining why their tension and action present the same moduli dependence.} The upshot is that along the limit at hand, the action of infinitely many euclidean D2-instantons decreases asymptotically, such that the tree-level geometry of the moduli space is heavily corrected and, in particular, the LCS point at fixed 4d dilaton is excised from the exact quantum moduli space, as schematically displayed in fig. \ref{fig:Instantoncorrections}. And this is precisely what we expected from the emergence perspective, since we could not find any tower that could generate the tree-level singularity. Thus, we can reconcile the absence of a tower with the fact that the infinite distance singularity is actually not present in the exact  moduli space.
		
Interestingly enough, even though the LCS singularity at fixed 4d dilaton does not belong to the quantum moduli space, one can see that it is actually traded for another (very different) infinite distance singularity, since the instanton corrections force the dilaton coordinate itself to run towards weak coupling if we insist into approaching the LCS point. The upshot, as explained in \cite{Alvarez-Garcia:2021pxo}, is that an emergent S-dual type IIA string arising from a D4-brane wrapping the SYZ-fibre (whose volume indeed decreases along the limit) becomes asymptotically tensionless, and in particular it does so faster than any other critical string, including the original fundamental one we started with. Therefore, at the end of the day one reaches a new kind of emergent string infinite distance singularity\cite{Lee:2019wij}, which can be reproduced by integrating out the oscillator modes associated to the D4-string (along with its KK-replicas \cite{Baume:2019sry}), in agreement with our expectations from Emergence.

\subsubsection*{Instanton corrections $\longleftrightarrow$ Emergence loop corrections}
		
In the previous discussion we have used the Emergence Proposal to predict, in some sense, the absence of an infinite distance singularity on the grounds of a lack of an infinite tower of particles which could give rise to it after being integrated out. To do so, we built upon several results from previous works to interpret the would-be infinite distance singularity to be precluded by the presence of an infinite tower of instanton corrections. Our aim in the following is to turn this logic around and use such instanton corrections (which are quantum in nature) to provide evidence for the picture advocated in the Emergence Proposal.
		
The strategy is to employ the c-map \cite{Cecotti:1988qn,Ferrara:1989ik}  to relate the D2-instanton corrections in e.g. the type IIA hypermultiplet sector to one-loop contributions of the infinite tower of D3-particles in its type IIB vector multiplet counterpart. For concreteness, here we skip the technical details of such construction, and focus on summarizing the logic behind it. Concretely, the c-map can be understood as a `dressed' T-duality which relates the vector multiplet moduli space of type IIA on a CY$_3$ with the hypermultiplet space of type IIB on the \emph{same} three-fold, and the vice versa. (This requires from a further $S^1$-compactification of both theories, and then performing the usual T-duality transformation along said 1-cycle.)
		
What is interesting about the c-map in this context is that it matches quantum contributions to the partition function of the theory as coming from D-instanton sums in one side of the duality, to one-loop corrections associated to integrating out a tower of D-particles on the other. In particular, one can essentially match the instanton number in one theory with the winding number of the particle worldline along the extra $S^1$ on the dual theory, being the latter moreover related upon Poisson resummation with the KK modes along the circle.\footnote{See e.g. \cite{Green:1997as} for a 10d analogue, where one-loop D0-brane (higher derivative) corrections in type IIA map under T-duality to D(-1)-instanton sums in type IIB string theory.} This explains why starting from the type IIA hypermultiplet sector, where we find that quantum corrections associated to a tower of euclidean D2-instantons smooth out the moduli space around a given point in ${\cal M}_{\rm HM}^{\rm IIA}$, one should  analogously obtain an infinite number of D3-particles that resolve the singularity in ${\cal M}_{\rm VM}^{\rm IIB}$, as required by the Emergence Proposal.
		
To close this section, let us mention that this is nothing but a generalization to infinite distances of a well-understood phenomenon in the finite distance case. Namely, as shown in \cite{Ooguri:1996me}, by using the c-map one can understand the resolution of the conifold singularity in type II string theory either by invoking the appearance of massless BH-states at that point in the vector multiplet moduli space (as in \cite{Strominger:1995cz}), or by a smoothing-out procedure due to a \emph{single} D-instanton in the hypermultiplet moduli space of the c-dual theory.

\section{Emergence in higher dimensional String Theory vacua}
\label{s:Emergenced>4}		
In this section we continue testing the emergence idea in several F-/M-/string theory constructions living in six, seven and ten (non-compact) spacetime dimensions, respectively. As we will see, each of these examples enjoys different singularity structures which may be resolved by different kinds of infinite towers of states becoming asymptotically light. We will again rely heavily on the machinery and formulae developed in section \ref{s:EmergenceQG}, such that we prefer to focus here on the physics discussions more than in the computations themselves, referring oftentimes to the material presented in that section for the technical details. 
		
\subsection{Emergence in 6d $\mathcal{N}=(1,0)$}
\label{ss:emergence6d}
		
In the present discussion we test our general analysis performed in section \ref{sss:emergenceU(1)} about the wave-function renormalization induced by a string tower on the $U(1)$ gauge field with respect to which some of its excitation modes are charged. There is indeed one concrete set-up in which it was claimed that the tower of oscillator modes of an asymptotically weakly-coupled \emph{critical} heterotic string was responsible for the emergence of an abelian $U(1)$ gauge coupling. It corresponds to finite volume limits in F-theory compactifications on an elliptically fibered CY$_3$ \cite{Vafa:1996xn,Morrison:1996na,Morrison:1996pp} in which also some gauge coupling tends to zero, as studied in \cite{Lee:2018urn}. There, it was shown that any such limit is indeed equi-dimensional \cite{Lee:2019wij}, such that it is accompanied by a critical heterotic string arising from a D3-brane wrapping a vanishing holomorphic 2-cycle. Moreover, (part of) the tower of excitation states associated to this fundamental dual string is the one satisfying both the SDC and the sub-Lattice WGC\cite{Heidenreich:2016aqi}. Let us summarize here the main results of \cite{Lee:2018urn} concerning the emergence phenomenon in this specific scenario. First of all, recall that the 6d Planck scale is indeed determined by the geometrical volume of the compactification space in F-theory, in particular by the volume\footnote{In this section we measure all dimension-full quantities in type IIB 10d Planck units.} of the complex K\"ahler surface $B_2$ which serves as the base for the elliptically-fibered CY$_3$
\beq
\label{eq:Planckmass6d}
	M_{\text{pl, 6}}^4 = 4 \pi \text{vol}_J (B_2)\, \Mpt^4 \ ,
\eeq
with $J$ denoting the appropriate K\"ahler form of $B_2$ in the 10d Einstein frame. On the other hand, the $U(1)$ in which we are interested arises from 7-branes wrapping complex curves, $\mathcal{C}$, on $B_2$ (i.e. the discriminant locus $\Delta$ of the elliptic fibration), whose volume sets the six-dimensional gauge coupling as follows
\beq
\label{eq:gaugecoupling6d}
	\frac{1}{g_{\text{IR}}^2} = 2 \pi \text{vol}_J (\mathcal{C})\, \Mpt^4 \ .
\eeq
It was shown in \cite{Lee:2018urn, Lee:2018spm} that weak coupling limits associated to such $U(1)$ gauge fields in the open string sector, with gravity remaining dynamical (which arise when we blow up the curve wrapped by the 7-brane while keeping the overall volume of the complex surface $B_2$ fixed and finite), require that $B_2$ contains some other rational curve (of vanishing self-intersection), $\mathcal{C}_0$, which intersects $\mathcal{C}$ and whose volume goes to zero as $\text{vol}_J (\mathcal{C}) \to \infty$. Additionally, a D3-brane wrapping such rational curve gives rise to a solitonic string in the six non-compact dimensions, whose tension is controlled by 
\beq
\label{eq:stringtension6d}
	T_{\mathcal{C}_0} = 2 \pi \text{vol}_J (\mathcal{C}_0)\ \Mpt^2 \sim 2 \pi \frac{\text{vol}_J (B_2)}{\text{vol}_J (\mathcal{C})}\ \Mpt^2 \ ,
\eeq
and which is shown to correspond to a weakly coupled dual heterotic string (compactified on some $K3$ surface), whose excitation states may be charged under the 7-brane gauge theory due to the fact that $\mathcal{C}_0$ and $\mathcal{C}$ intersect.
		
The purpose of this section is to relate the functional form of the gauge coupling in eq. \eqref{eq:gaugecoupling6d} to the appearance of the tower of asymptotically massless charged stringy states in the spirit of the Emergence Proposal. To do this we need two ingredients: first the 6d species scale for this set-up, together with its relation to the characteristic data of the tensionless heterotic string, and second, the one-loop renormalization of the inverse gauge coupling as well as the spectrum of stringy states charged under the $U(1)$ gauge field.
		
To determine the first part of the relevant information, let us use our results of section \ref{s:speciesscale}. In particular, eqs. \eqref{eq:speciescale}-\eqref{eq:Ns} when evaluated for the case of $d=6$ lead to an expression analogous to eq. \eqref{eq:maxstringlevel}
\beq
		\left ( \frac{M_{\text{pl, 6}}}{M_{\text{het}}}\right )^4 =\, N_{\text{het}}^2 \sum_{n=1}^{N_{\text{het}}} \text{exp}(\sqrt{n}) \, \sim 2\ N_{\text{het}}^{5/2} e^{\sqrt{N_{\text{het}}}}\, ,
\label{eq:maxstringlevel6d}
\eeq
where the quantities with the subscript `het' are associated to the emergent critical heterotic string. In particular, since it is obtained from the wrapped D3-brane above, we have $M_{\text{het}}^2\, =\, T_{\mathcal{C}_0}/2 \pi$. Recall also that for such a stringy tower $\LQG^2 \, \simeq \, N_{\text{het}} \, M_{\text{het}}^2$.
		
Additionally, the one-loop gauge coupling `running' at leading order is (see discussion around eq. \eqref{eq:oneloopstringtowergauge})
\beq
		\label{eq:runninggaugecoupling6d}
		\frac{1}{g_{\text{IR}}^2} = \frac{1}{g_{\text{UV}}^2} + \beta \Mpt^2 \sum_{k=1}^{k_{\text{max}}} q_k^2\, \Lambda_{\text{QG}}^2\, ,
\eeq
where $\beta$ is a positive numerical prefactor that depends on the specific type of state considered (e.g. whether it is a fermion or a boson, its tensor structure, etc.), and the sum runs over every stringy mode (labelled by the collective index $k$) charged under the $U(1)$ with quantized charge\footnote{Notice from eq. \eqref{eq:runninggaugecoupling6d} that we have extracted a factor of $\Mpt$ from the gauge charges associated to the oscillator modes of the wrapped D3-brane, so as to have the $q_k$ dimensionless. One can see that this is indeed the correct normalization from e.g. studying anomaly inflow in the worldsheet of the resulting EFT string\cite{Callan:1984sa,Heidenreich:2021yda}.} $q_k$ that appears below the species scale. In order to see how the above sum reproduces the expected divergence in eq. \eqref{eq:gaugecoupling6d} when we send $\text{vol}_J (\mathcal{C}) \to \infty$, we need a way to parameterize which $U(1)$ charges appear for each oscillator level $n$, which controls the mass of such states by the usual formula $m_n^2 = 8\pi (n-1) \,  T_{\mathcal{C}_0} $, as well as their degeneracy. Hence, we borrow some useful results from  \cite{Lee:2018urn}. First, one can estimate the highest charge  appearing at each excitation level $n$ by $q_{\text{max}}(n)= \sqrt{n}$, as frequently happens for heterotic compactifications.\footnote{\label{fn:modularity}This is essentially fixed by the modular properties of the heterotic string partition function on the worldsheet torus, see \cite{Lee:2018urn} and references therein.} Second, we notice that in concrete examples, such as e.g. F-theory on an elliptic CY$_3$ with base given by a Hirzebruch surface, $B_2=\mathbb{F}_1$, at a given $n$ for the heterotic string arising from the wrapped D3-brane, each value of the charges with $|q| < q_{\text{max}}(n)$ is populated by string states \cite{Lee:2018urn}. Taking this kind of behaviour as representative, we need to parameterize the degeneracy $d_{k,n}$ of charged states with charge $k \in \mathbb{Z}$ within each oscillator level $n$. In principle, one would need to extract the precise degeneracy from quantities such as the partition function of the tensionless string, but here we will take a simple (possibly oversimplified) parameterization of the form
\beq
\label{eq:dmn}
	d_{k,n} \sim f(k)\ \text{exp} (\sqrt{n}) \ ,
\eeq
with $f(k)$ any polynomial function with the only restriction that summing over all charges up to $q_{\text{max}}(n)$ within a given step in the tower reproduces the (asymptotic) level density $d_n \sim \text{exp} (\sqrt{n})$ (e.g. $f(k) \sim k^p/n^{(p+1)/2}$). With these ingredients at hand, we can then perform the summation implicit in eq. \eqref{eq:runninggaugecoupling6d}
\beq
\begin{aligned}
			\frac{1}{g_{\text{IR}}^2} & \sim \,  \Mpt^2 \,  \Lambda_{\text{QG}}^2 \sum_{n=1}^{N_{\text{het}}} \sum_{k=0}^{\sqrt{n}} k^2 d_{k,n}\, \sim\, \Mpt^2 \,  T_{\mathcal{C}_0} \,  N_{\text{het}} \sum_{n=1}^{N_{\text{het}}} n\ \text{exp} (\sqrt{n})\\
			&\sim\, \Mpt^2 \, T_{\mathcal{C}_0} \, N_{\text{het}}^{5/2} \text{exp} (\sqrt{N_{\text{het}}})\, .
		\end{aligned}
\eeq
Now, upon substituting eqs. \eqref{eq:maxstringlevel6d}-\eqref{eq:stringtension6d}, together with $M_{\text{het}}^2\, =\, T_{\mathcal{C}_0}/2\pi$, and using also the definition of the 6d Planck mass, eq. \eqref{eq:Planckmass6d}, we recover the same field dependence for the $U(1)$ gauge coupling as in eq. \eqref{eq:gaugecoupling6d}. The main difference with respect to the calculation performed in ref. \cite{Lee:2018urn} is that in principle we sum over all the string modes charged under the $U(1)$, with charges ranging from zero to the maximum value allowed for each oscillator level $n$, with the \emph{assumption} that the degeneracy for such gauge charges can be parameterized by the simple relation displayed in eq. \eqref{eq:dmn}. Instead, in the analysis of \cite{Lee:2018urn} it is assumed that the leading contribution to the wave-function renormalization of the 6d gauge field comes from the states of highest charged within each oscillator level $n$, whilst the rest do not enter neither in the one-loop correction nor in the computation of the species scale. 
		
\subsection{Emergence in 7d $\mathcal{N}=2$}
\label{ss:emergence7dN=2}
		
As our second example in this section we investigate the corner of the M-theory moduli space that corresponds to the F-theory limit, usually employed to actually define F-theory (see e.g. \cite{Weigand:2010wm,Weigand:2018rez,Cvetic:2018bni}). In particular, we focus on 7d compactifications of M-theory,\footnote{In \cite{Corvilain:2018lgw}, an analogous F-theory limit was investigated within the K\"ahler moduli space of M-theory compactified on elliptic CY three-folds. Our analysis in this section can be easily accommodated to include also that set-up.} where upon probing a certain infinite distance singularity we end up on an 8d supergravity theory as arising from F-theory. Before entering into the details of the analysis, let us quickly review some generalities about the F-/M-theory duality which will be useful later on. The reader who is familiar with this topic can safely skip this discussion. 
		
\subsubsection*{A Lightning review of F-/M-theory duality}
		
To establish the correspondence between F-theory and M-theory (or its 11d low-energy supergravity limit) we start by recalling that indeed M-theory compactified on a circle $S^1_a$ reduces to type IIA string theory in the limit where the radius $R_a$ of the circle vanishes, namely when $R_a\rightarrow 0$ \cite{Vafa:1996xn}. If we further compactify on an additional circle $S^1_b$, one effectively arrives at a $T^2$ compactification of M-theory, which after taking $R_a\rightarrow 0$, becomes type IIA on $S^1_b$. This is entirely equivalent to type IIB on the T-dual circle $\overline{S}^1_b$, with radius $\overline R_b =\alpha'/R_b$. Hence, if we additionally approach the small $R_b$ regime, we essentially decompactify $\overline R_b$ in type IIB dual frame. Therefore, one can conclude that M-theory on $T^2$ reduces to type IIB string theory in the limit of vanishing torus-volume, $\mathcal{V}_{T^2}$, with the complex structure of such torus, denoted $\tau$, left untouched. Additionally, one can see that the KK tower associated to the decompactifying circle in the type IIB frame corresponds, from the M-theory perspective, to M2-branes wrapping the whole torus-fibre $n \in \mathbb{Z}$ times, whose mass
\begin{align}\label{eq:massM2branes}
			m_{\text{M2}}\,  (n T^2) = \left | \int_{n T^2} J \right |= |n\, \mathcal{V}_{T^2}|= \left |\frac{n}{\overline R_b} \right |\, ,    
\end{align}
goes then to zero in the limit $\overline R_b \to \infty$.
		
Consider now more generally M-theory compactified on an elliptically-fibered $n$-fold $X_{n}$, over some $(n-1)$-fold base $C_{n-1}$, which we denote as follows
\begin{equation}\label{eq:fibration}
			\begin{aligned}
				\pi: \qquad T^2 \rightarrow &\;X_{n} \\
				&\;\; \downarrow \qquad . \\ &\;C_{n-1}
			\end{aligned}
\end{equation}
One can easily see that in order to preserve some unbroken supersymmetry in the low energy EFT, the $n$-fold needs to have at least vanishing first Chern class, i.e. it has to be Calabi--Yau. This results in a supersymmetric EFT in $\mathbb{R}^{1,10-2n}$. Performing now adiabatically the duality discussed above\cite{Vafa:1996xn}, namely by taking the limit of zero fibre volume from the M-theory perspective, one obtains type IIB string theory compactified on $C_{n-1}$, possibly with some non-perturbative defects (i.e. 7-branes) wrapping certain cycles of the internal geometry (and given by the singular loci of the torus fibration).
		
\subsubsection*{M-theory on $K3$}
		
Let us thus consider M-theory compactified on a $K3$ surface down to seven (non-compact) spacetime dimensions. We will closely follow the analysis performed in ref. \cite{Lee:2019xtm}, where the F-theory limit in such set-up was carefully studied.
		
The low energy EFT presents $\mathcal{N}=2$ supersymmetry,\footnote{\label{fn:symplecticmajorana}Recall that the spinor irrep of $\mathfrak{so}(1,6)$ naturally admits a quaternionic representation, $\mathbb{H}$, but we choose to count here the number of \emph{symplectic-Majorana} spinors, thus the notation $\mathcal{N}=2$, even though we have minimal supersymmetry in 7d.} namely it preserves half of the original 32 supercharges of 11d supergravity. For simplicity (and in order to directly translate the results from \cite{Lee:2019xtm}) we focus on attractive $K3$'s \cite{Moore:1998pn}, i.e. those CY$_2$ in which the rank of the Picard group, $\text{Pic}(K3)= H^{1,1}(K3) \cap H_2(K3,\mathbb{Z})$ is maximal, namely $\text{rk(Pic(K3))=20}$. This translates into the fact that all integral curve classes in $H_2(K3,\mathbb{Z})$ admit holomorphic representatives, or in more physical terms, they could all give rise in principle to distinguished states from wrapped M2-branes whose masses are controlled by the periods of the K\"ahler form $J$ along said cycles.
		
The main advantage of focusing on such attractive $K3$ surfaces is that their complex structure is completely fixed, such that every quantity that will appear in the following will depend solely on its K\"ahler degenerations. With this restriction in mind, let us review some useful mathematical properties of $K3$ two-folds (see e.g. \cite{Aspinwall:1996mn}). Their Hodge numbers read
\beq\label{eq:hodgeK3}
		h^{2,0}(K3) = h^{0,2}(K3) = 1\,,\qquad h^{1,1}(K3) = 20\, ,
\eeq
to which we can associate a basis of harmonic 2-forms, i.e. $\lbrace \omega_A \rbrace= \lbrace \Omega_2, \overline \Omega_2, \omega_a \rbrace$, where $\Omega_2 \in H^{2,0}(K3)$ is the unique holomorphic (2,0)-form and $\omega_a \in H^{1,1}(K3)$, $a=1, \ldots, h^{1,1}(K3)$.
		
Upon reducing the 11d 3-form field $\hat C_3$ on the basis for harmonic 2-forms in $K3$ one obtains $h^2(K3)=22$ 1-form gauge fields in total, three of which belong to the gravity multiplet. We will focus on the components along $H^{1,1}(K3)$, since those are the ones which depend on the K\"ahler sector once we restrict ourselves to the attractive case \cite{Lee:2019xtm}, i.e.
\beq\label{eq:7dreductionC3}
		\hat C_3=A^a \wedge \omega_a\, .
\eeq
Furthermore, reducing the 11d supergravity action on the two-fold produces the following kinetic term for these 1-forms after performing the  Weyl rescaling $g_{\mu \nu} \to \mathcal{V}_{K3}^{-2/5} g_{\mu \nu}$ in order to go to the 7d Einstein frame (we measure all dimension-full quantities in terms of the 7d Planck length, $\ell_{7}$)
\begin{align}\label{eq:gaugekineticmatrix7d}
			S^{\text{7d}}_{\text{M-th}} \supset -\frac{2 \pi}{\ell_{7}^5} \int \frac{1}{2} g_{a b}\ dA^a\wedge \star dA^b\, .    
\end{align}
The gauge kinetic matrix $g_{a b}$ takes the form
\begin{align}\label{eq:matrix1forms7dMtheory}
			g_{a b}= \frac{1}{\mathcal{V}_{K3}^{3/5}}\int_{K3} \omega_a \wedge \star \omega_b = \frac{1}{\mathcal{V}_{K3}^{8/5}} j_a j_b - \frac{1}{\mathcal{V}_{K3}^{3/5}}\Omega_{a b}\, ,    
\end{align}
with $J=j^a \omega_a$ the K\"ahler form of the complex surface, $\Omega_{a b}=\omega_a \cdot \omega_b  \equiv \int_{K3} \omega_a \wedge \omega_b$ and $j_a= \Omega_{a b} j^b\, $. This also means that in such a frame the 7d Planck scale simply reads $M_{\text{pl}, 7}^5= \frac{4 \pi}{\ell^5_{7}}$. Notice that the moduli space of 11d supergravity on the $K3$ surface is classically exact \cite{Witten:1995ex, Cadavid:1995bk} and it is given by the group coset $O(3,19) / O(3) \times O(19)$, in contrast to say type IIA on the same two-fold, which probes the quantum $H^2(K3)$-cohomology due to the extra modes arising from the $B_2$-field.
		
The crucial point now is that in order to have an infinite-distance, weak coupling point at fixed internal volume (so as to keep gravity dynamical), one of the entries in $g_{a b}$ has to blow-up. Indeed, as demonstrated in \cite{Lee:2019xtm}, this requires that the K\"ahler form must behave asymptotically as follows
\begin{align}
			J= t J_0 + \sum_i \frac{a_i}{2t} J_i\, , \qquad t \to \infty\, ,    
\end{align}
where $J_0, J_i$ are generators of the K\"ahler cone, $a_i$ are some constant numerical factors and $t$ is the K\"ahler modulus that scales to infinity in the limit. Additionally, the finite volume requirement imposes the following restrictions on the generators $J_0, J_i$ \cite{Lee:2019xtm}
\begin{align}
			J_0 \cdot J_0 = 0\, , \qquad \sum_i \frac{a_i}{2} J_i \cdot J_0= \mathcal{V}_{K3} + \mathcal{O} (1/t^2)\, .    
\end{align}
Therefore, taking as a basis for $H^{1,1}(K3)$ precisely these 2-forms, $\lbrace \omega_a \rbrace = \lbrace J_0, J_i \rbrace$ one can see that the vector that becomes weakly coupled when we take $t \to \infty$ is precisely $A^0$, whose kinetic term behaves as
\begin{align}\label{eq:gaugecoupling7d}
			g_{0 0} = \frac{t^2}{\mathcal{V}_{K3}^{8/5}} - \frac{1}{\mathcal{V}_{K3}^{3/5}}C_0 \cdot C_0\, ,    
\end{align}
where the second term is independent of $t$ and both $g_{i j}$ and $g_{0 i}$ do not blow up and hence become negligible (when compared to $g_{0 0}$) upon entering the asymptotic regime. In eq. \eqref{eq:gaugecoupling7d} above we have introduced a curve class $C_0$ which belongs the the set $\lbrace C_0, C_i \rbrace$ dual to $\lbrace J_0, J_i \rbrace = \lbrace C^0, C^i \rbrace$, namely it satisfies
\begin{align}
			C_0 \cdot C^0 = 1\, , \qquad C_0 \cdot C^i = 0\, , \qquad C_i \cdot C^0 = 0\, , \qquad C_i \cdot C^j = \delta_i^j\, .    
\end{align}
Moreover, it can be shown that the divisor class defined by $C^0=J_0$ contains necessarily a genus-one holomorphic curve whose volume tends to zero in the limit at a rate $\text{vol}(C^0) = \mathcal{V}_{K3}/t + \mathcal{O} (1/t^3)$. This means, in particular, that our $K3$ surface admits an elliptic fibration (over a $\mathbb{P}^1$-base in this case) of the form depicted in eq. \eqref{eq:fibration}.
		
Additionally, one obtains a tower of asymptotically light BPS-like particles arising from wrapped M2-branes on the shrinking curve, $C^0$. Moreover, thanks to the $T^2$-topology of said curve, one can see that there are indeed bound states of $n$ M2-particles for each $n \in \mathbb{Z}$, such that what we obtain is essentially a tower of light particles with constant degeneracy at each mass level in the spectrum.\footnote{A more rigorous way to motivate the existence of the tower of states is to use the fact that the Gopakumar-Vafa invariants \cite{Gopakumar:1998ii, Gopakumar:1998jq} for a 2-cycle wrapped $n$ times on a genus-one curve on $K3$ are constant and non-vanishing for each $n \in \mathbb{Z}$, namely $\text{GV}_n = \chi(K3) = 24$ \cite{Katz:1999xq}.} These constitute precisely the KK-like tower that effectively implements the circle decompactification in the F-theory limit. Their mass and charge can be computed from the dimensional reduction of the Nambu-Gotto plus Chern-Simons (effective) action for the M2-brane in M-theory (see e.g. \cite{Weigand:2018rez})
\begin{equation}\label{eq:DBICSactionM2}
			\begin{aligned}
				S_{\text{M2}} = -\frac{2 \pi}{\ell_{11}^3} \int_{\text{M2}} \sqrt{-\hat g} + \frac{2 \pi}{\ell_{11}^3} \int_{\text{M2}} \hat C_3\, ,
			\end{aligned}
\end{equation}
such that after compactifying 11d supergravity on $K3$, reducing the M2-brane effective action on the genus-one curve and performing the 7d Weyl rescaling of the metric as discussed after eq. \eqref{eq:7dreductionC3}, one finds for the characteristic scale of the tower the following quantity
\begin{align}\label{eq:M2particlemassafterWeylrescaling}
			m_{\text{M2}} = \frac{2 \pi}{\ell_{7}} \text{vol} (C^0)\, \mathcal{V}_{K3}^{-1/5} \sim \frac{2 \pi}{\ell_{7}} \frac{\mathcal{V}_{K3}^{4/5}}{t} \qquad \text{as}\ \ t \to \infty\, .    
\end{align}
Hence, we can now use our general formulae for the `one-loop renormalization' of the inverse gauge coupling (squared) in the presence of a tower of light states in $d$-dimensions, namely eq. \eqref{eq:gaugeemergenceddimensions}, and apply it to the seven-dimensional case at hand, with $p=r=1$ (we have just one KK tower associated to the extra decompactifying circle in F-theory), which yields
\beq\label{eq:oneloopgaugecoupling7d}
		\frac{1}{g_{\text{IR}}^2} = g_{0 0} \sim M_{\text{pl}, 7}^5 \left( \frac{2 \pi}{\ell_{7}\, m_{\text{M2}}} \right)^2 \sim M_{\text{pl}, 7}^5\, \frac{t^2}{\mathcal{V}_{K3}^{8/5}}\, .
\eeq
Above we have used the fact that the M2-particle charges are given by $q_n\, = \,  \frac{2 \pi n}{\ell_{7}}$, 
as well as the definition of the species scale \eqref{species}. Thus, the weak coupling singularity in eq. \eqref{eq:gaugecoupling7d} can in principle be obtained via the emergence mechanism.
		
\subsubsection*{Classical vs Quantum in Gravity}	
The example we just discussed is very suggestive of the presumably `fundamental' character of the Emergence Proposal. Typically, when studying some particular example in which Emergence can be quantitatively tested, one arrives at the following question: to which extent the moduli space singularity can be associated to a purely \emph{quantum} phenomenon? 
		
Let us be more specific about this concern. Imagine we started from a $(d+1)$-dimensional EFT weakly coupled to Einstein gravity, in which the dynamics of the spacetime geometry are encoded in the usual Einstein-Hilbert lagrangian
\beq
		\mathcal{L}^{d+1}_{\text{EH}}\, =\,  \dfrac{1}{2\kappa_{d+1}^2}\, \sqrt{-\hat g}\, \hat R\, .
\eeq
Now, perform a simple $S^1$-compactification, upon which our original light fields (if any) are replicated such that they arrange themselves in the familiar KK prescription. In particular, upon dimensionally reducing the Einstein-Hilbert term above (and performing the Weyl rescaling $g_{\mu \nu} \, \to \, \rho^{-\frac{2}{d-2}}\, g_{\mu \nu}$ in order to go to the $d$-dimensional Einstein frame), one obtains the following lagrangian (see e.g. \cite{vanBeest:2021lhn})
\beq
		\mathcal{L}^{d}\, =\, \dfrac{1}{2\kappa_{d}^2}\,  \sqrt{- g} \left\{ R -\rho^{-2 \, \left(\frac{d-1}{d-2}\right)}\ F^2 - \left( \frac{d-1}{d-2} \right)\frac{(\partial \rho)^2}{\rho^2}\right\}\, ,
\eeq
which encodes the dynamics of gravity, the KK-photon $A_{\mu}$ (with field strength $F_{\mu \nu}$) as well as the radius field $\rho$. If we now focus on the lower dimensional theory, we can explore (at least) two kinds of infinite distance singularities, associated to the large/small radius regime. For concreteness, we choose to explore the large radius point, i.e. $\rho \to \infty$, in which our theory naturally decompactifies, and hence the KK tower encoding the replicas of the $d$-dimensional fields become asymptotically light in Planck units. However, by simple dimensional reduction (meaning without caring a priori about would-be quantum corrections associated to the massive towers) one can clearly see that both the radius field metric singularity as well as the weak coupling behaviour of the KK photon are already apparent at `tree-level'. In other words, both singular behaviours seem to be already present at the classical level.
		
On the other hand, when trying to compute the first quantum corrections induced by the towers becoming light below the species scale according to the emergence prescription, what we find (as we reviewed in section \ref{s:EmergenceQG}) is that they produce the \emph{same} kind of singular behaviour (i.e. the same moduli dependence) as the one found already at tree-level. Therefore, one cannot be sure about whether the singularity itself arises at the classical or quantum level. In other words, the IR behavior of the scalar metric (weak gauge couplings) obtained by KK reduction at tree-level is `as if' it had been obtained from an EFT with cut-off given by the species scale, in which the metric (gauge coupling) was not necessarily divergent (vanishing), by integrating out the tower of light KK particles that are also present in that theory. The whole point of the stronger versions of the Emergence Proposal is that such divergent (vanishing) behaviour could arise completely as a quantum-mechanical effect, at least in some duality frame.
		
This is precisely why the apparently simple example that we just studied can be very illuminating. As we saw before, the precise singularity in the K\"ahler sector of the M-theory compactification that we analyzed, which is also a weak coupling limit for some particular $U(1)$ gauge field, is supposedly associated to infrared divergent loop contributions from (infinitely many) charged states arising from wrapped M2-branes. Such non-perturbative states, which when sufficiently far away from the singularity are well described by a semi-classical black hole geometry (i.e. with an horizon larger than the Planck length/species scale), indeed become asymptotically massless (in 7d Planck units) when approaching the point in moduli space where the wrapped 2-cycle shrinks to zero size. Moreover, as we saw by e.g. looking at the `one-loop renormalization' of the $U(1)$ gauge coupling in eq. \eqref{eq:oneloopgaugecoupling7d}, the divergences in the effective lagrangian \eqref{eq:gaugekineticmatrix7d} had precisely the right field dependence to be produced by integrating out the tower of would-be BH's up to the species scale. Therefore, in a sense, this seems like a natural generalization to the infinite-distance case of the celebrated conifold singularity arising in the complex structure sector of type IIB string theory compactified on a CY$_3$ \cite{Strominger:1995cz}. There it was also the case that an apparently classical singularity arising just from Calabi--Yau data was indeed `resolved' by inherently quantum-mechanical effects associated to BH's becoming massless at the conifold point.
		
However, one should remember that the exact same singularity could be understood as a decompactification from 7d to 8d in F-theory on the elliptic $K3$ surface (times the $S^1$). In such duality frame, our argumentation from the beginning of the present discussion applies, where the tower responsible from the infinite distance singularity as seen from the seven-dimensional theory is nothing but the KK replica of the F-theory fields along the circle. From this perspective, one could have been tempted to claim that the singularity was classical,\footnote{Let us stress that in F-/M-theory duality it is known that in order to precisely match both effective actions at the two-derivative level, it is sometimes crucial to take into account quantum corrections associated to the KK modes. This happens e.g. in 6d F-theory compactifications on elliptic CY$_3$, where upon reducing on the circle some 5d Chern-Simons terms (which appear to be classical from the M-theory perspective) are fully generated by the KK tower at the quantum level \cite{Bonetti:2011mw,Bonetti:2012fn,Bonetti:2013ela}.} since it already appeared at tree-level just from dimensional reduction.
		
The idea we want to convey though, is that it may not make sense to actually distinguish between classical and quantum gravity, given that the very concept of what is \emph{classical} and what is \emph{quantum-mechanical} is not duality invariant (see e.g. \cite{Vafa:1995fj}). This allows us to speculate with the idea that string theory, or gravity itself, may need from quantum mechanics for consistency from the beginning, such that it may be intrinsically quantum in nature and even do not have a classical limit in the usual sense \cite{Strominger:1995qi}.

\subsection{Emergence in 10d $\mathcal{N}=(1,1)$}
\label{ss:Emergence10dST}
		
Ten-dimensional string theories present the simplest possible moduli spaces, comprised e.g. in type IIA by a single modulus VEV, namely that of the dilaton, $\phi_0 \equiv \braket{\phi} \in \mathbb{R}$. Therefore, it seems natural to try to understand the infinite distance limits in such a simple moduli space from the point of view of Emergence. In the present section we focus on type II strings, explicitly performing the computations in type IIA, but we expect the same kind of logic to be applicable to heterotic and type I strings as well.

Before going into the details, let us recall the bosonic part of the relevant 10d actions in the Einstein frame, namely (see e.g. \cite{Polchinski:1998rq,Polchinski:1998rr})
\begin{equation}\label{eq:IIA10d}
			\begin{aligned}
				S_\text{IIA, E}^{\text{10d}} =\frac{1}{2\kappa_{10}^2} \int \text{d}&^{10}x\sqrt{-g} \left(R-\frac{1}{2}(\partial \phi)^2\right)-\frac{1}{4\kappa_{10}^2}\int e^{-\phi} H_3\wedge \star H_3 \\
				&-\frac{1}{4\kappa_{10}^2}\int \left[e^{\frac{3}{2}\phi}F_2 \wedge \star F_2 + e^{\frac{1}{2}\phi} \tilde F_4 \wedge \star \tilde F_4 + B_2\wedge F_4 \wedge F_4\right]\, , 
			\end{aligned}
\end{equation}
for type IIA, and
\begin{equation}\label{eq:IIB10d}
			\begin{aligned}
				S_\text{IIB, E}^{\text{10d}} = & \frac{1}{2\kappa_{10}^2} \int \text{d}^{10}x\sqrt{-g} \left(R-\frac{1}{2}(\partial \phi)^2\right) -\frac{1}{4\kappa_{10}^2}\int e^{-\phi} H_3\wedge \star H_3 \\
				&-\frac{1}{4\kappa_{10}^2}\int \left[e^{2 \phi}F_1 \wedge \star F_1 + e^{\phi} \tilde F_3 \wedge \star \tilde F_3 + \frac{1}{2} \tilde F_5 \wedge \star \tilde F_5  +C_4\wedge H_3 \wedge F_3\right]\,,
			\end{aligned}
\end{equation}
for type IIB. Here, $H_3=dB_2$, $\tilde{F}_4=d C_3-C_1 \wedge H_3$, $\tilde{F}_3=dC_2-C_0 H_3$ and $\tilde{F}_5= dC_4-\frac{1}{2} C_2 \wedge F_3 + \frac{1}{2} B_2 \wedge H_3$. Notice that the gravitational strength, dictated by the (dimension-full) quantity $2\kappa_{10}^2= 2 \Mpt^{-8} = (2 \pi)^{-1} M_s^{-8}\ e^{2 \phi_0}$, is controlled by the dilaton VEV $\phi_0$, which determines the Planck-to-string scale ratio. It is also important to realize that the exponential couplings involving the dilaton field $\phi$ in the above actions capture, once we perform a Taylor series expansion around its VEV, not only the kinetic term for the RR $p$-forms, but also an infinite number of polynomial couplings between such field and the field strengths squared. 
		
Within the moduli space parameterized by the dilaton VEV, there are two infinite distance points,\footnote{Actually, this is only true for type IIA string theory since type IIB enjoys a larger $SL(2, \mathbb{Z})$ duality group, which in particular relates such pair of infinite distance points to infinitely many others (described in terms of the axio-dilaton $\tau=C_0+ie^{-\phi}$) where the (dual) emergent weakly coupled strings are given by different $(p, q)$ 1-branes, see e.g. \cite{Weigand:2018rez}.} $\phi_0 \rightarrow \pm \infty$. In the type IIA side, when going into the direction $\phi_0 \rightarrow \infty$ (strong coupling) the kinetic terms for the RR $p$-forms, $C_1$ and $C_3$, grow whereas that of the NSNS 2-form, $B_2$, decreases (and analogously for type IIB). One thus would expect that in the strong coupling limit, emergence should give rise to the kinetic terms for the RR $p$-forms. On the other hand, at weak coupling, namely $\phi_0\to -\infty$, we expect to be able to reproduce instead the kinetic term for the Kalb-Ramond 2-form via Emergence. 
		
From now on we will focus on 10d type IIA string theory, and discuss both the weak and strong coupling singularities within the dilaton moduli space. Before doing so, let us make some digression so as to elaborate a bit about the peculiarities associated to theories with maximal supersymmetry as seen from the prism of the Emergence Proposal.
		
\subsubsection*{Maximal supergravity and uniqueness of the action}
Our aim will be to provide evidence for how the emergence mechanism comes at play in the aforementioned set-ups. However, before entering into the details of the calculations, it is worth commenting on some subtlety associated to maximal supergravity in any $d \geq 4$. Indeed, one may argue that with such large amount of supersymmetry, the two-derivative action is completely fixed, such that no significant quantum corrections may appear which could destroy its strongly constrained structure. Therefore, two possibilities come quickly two our mind: either there are strong cancellations in the loop integrals which naively correct these terms of the theory, or rather there can exist `quantum corrections' of the kind we are interested from the Emergence Proposal, which at the end of the day, result in a common wave-function renormalization of the whole two-derivative lagrangian.
		
Regarding the first possibility, as we already commented at the beginning of section \ref{s:EmergenceQG}, such cancellations on e.g. 10d $\mathcal{N}=(1,1)$ supergravity are not expected to occur on general physical grounds. The reason being that the quantum loop corrections we focus on in here arise from \emph{charged} BPS particles, whose field-theoretic content should contribute positively, regardless of its spin or statistics, both for the graviton and graviphoton kinetic terms. The simplest way to see this is via toroidal compactification down to 4d, which leads to an $\mathcal{N}=8$ supergravity theory\cite{Cremmer:1979up}. In that scenario, the same kind of cancellations associated to BPS particles that happen already in 10d should also be at work. However, if that is the case it would mean e.g. that for certain $U(1)$ gauge fields, namely the graviphotons, some kind of `anti-screening' should happen, contrary to our experience from QED, where both scalar and spin-$\frac{1}{2}$ fields contribute with the \emph{same} sign to the wave-function renormalization of the $U(1)$ vector. This contradicts physical intuition, since it is well-known that such `anti-screening' in the vacuum polarization diagrams only happens for non-abelian gauge fields. Similar comments apply to the graviton wave-function renormalization, to which all point-particles contribute positively at one loop, including massive, charged spin-1 fields \cite{Anber:2011ut,Donoghue:1994dn, Han:2004wt}. As for the second option, it seems plausible that the quantum loops due to charged particles do affect the kinetic terms of the supergravity lagrangian in a way consistent with maximal supersymmetry, namely  that the wave-function renormalization be shared by all fields belonging to the same supermultiplet\cite{Bilal:2001nv}. Our computations in this section are indeed compatible with this latter picture.
		
To finish, let us mention an intriguing prediction from the Emergence Proposal in set-ups enjoying maximal supersymmetry. One can argue that the moduli spaces associated to such theories should contain no finite-distance singularities of any kind, but only \emph{infinite distance} ones instead. The reason being that with that much supersymmetry, even short BPS supermultiplets contain spin-2 particles, such that having one of them in the theory implicitly requires an infinite number of `replicas' due to gauge invariance\cite{Duff:1989ea}. Therefore, if a singularity appears in the moduli space where one of such multiplets becomes massless (in Planck units), an infinite number of them must become light as well, thus generating an infinite distance singularity of KK type.\footnote{For singularities associated to higher-spin towers becoming massless similar considerations should apply, therefore signalling some emergent string arising upon reaching the asymptotic boundary in moduli space.} Similar ideas have been proposed and studied \cite{Bianchi:2009wj, Green:2007zzb}, pointing towards our expectations from the Emergence Proposal to be on the right track.

\subsubsection{Weak Coupling Limit}
\label{sss:IIAweakcoupling}
The weak coupling point (i.e. $g_s \to 0$) corresponds to a trivial emergent string limit in which the critical string becoming asymptotically tensionless (in Planck units) is just the fundamental type IIA string. Thus, an increasing number of oscillators become light upon approaching the singularity, falling below the species scale, as we computed in section \ref{s:speciesscale} (see also Appendix \ref{ss:typeII10d} for a more refined derivation).
		
From the 10d low energy action \eqref{eq:IIA10d} we see that when $g_s$ goes to 0 (i.e. $\phi_0 \to -\infty$) the distance in moduli space becomes singular,\footnote{This fact becomes more apparent upon using non-canonical coordinates, i.e. $g_s \in \mathbb{R}_{\geq0}$, to parameterize the dilaton moduli space, $\mathcal{M}_{\text{dilaton}}$. The metric of $\mathcal{M}_{\text{dilaton}}$ in this coordinates becomes $G_{g_s g_s}= 1/(g_s)^2$.} and the $B_2$-field presents a weak-coupling behaviour as well. From emergence, both kinetic terms in this limit should thus be generated by quantum corrections induced by the light tower of oscillator modes. We only discuss here the generation of the dilaton metric along said infinite-distance limit, since the couplings that these fields present with respect to the massless scalar field are universal and arise from the moduli-dependence of their masses. The explicit analysis for $B_2$ is more involved, since we do not know a priori the precise interaction rules, but we can use supersymmetry to argue that if the right kinetic term is generated for the dilaton, the same should happen for the $B_2$-field.

The tower of string excitations in 10d type IIA is given by the following masses and (asymptotic) degeneracies
\beq
\label{eq:10dweakcouplingspectrum}
		m_n^2\, =\, 8 \pi (n-1) \, T_s \, , \qquad  d(n)\, \sim \,  n^{-11/2}\, e^{4\pi \sqrt{2} \sqrt{n}}\, ,
\eeq
where $T_s=2\pi M_s^2$ and $M_s= M_s(\phi)$ in 10d Planck units. Following the general analysis in section \ref{sss:emergencemodulimetric} for the special case of $d=10$ and with the spectrum given above,\footnote{To connect directly with section \ref{sss:emergencemodulimetric}, we take the asymptotic degeneracy of states to be simply $d(n)\sim e^{\sqrt{n}}$, but we emphasize that the leading field dependence obtained with the full expression in eq. \eqref{eq:10dweakcouplingspectrum} is the same.} the total contribution for the metric due to e.g. fermionic loops goes roughly as (c.f. discussion around eq. \eqref{eq:scalarloopfermionssummary})
\beq
		\delta g_{\phi \phi} \, \sim \, \sum_{n=1}^{N_s}\mu_n^2\,  d(n)\,  \LQG^{6}\, \simeq \, (\partial_\phi M_s)^2\,  \sum_{n=1}^{N_s}  n\, d(n)\, \LQG^{6}\, .
\eeq
Upon substituting now $\LQG ^2 \simeq N_s\, M_s^2$ and approximating the sum by an integral we obtain
\beq
		\delta g_{\phi \phi} \,
		\sim \, (\partial_\phi M_s)^2\,  M_s^{6}\, N_s^{9/2}\, e^{\sqrt{N_s}}\, .
\eeq
Using the relation \eqref{eq:Ns}, which in 10d takes the form $N_s^{\frac{9}{2}}e^{\sqrt{N_s}}\, \simeq \, \left(\Mpt/ M_s\right)^{8}\,$, we finally get the following one-loop metric
\beq\label{eq:IIAweakcouplingmetric}
		\delta g_{\phi \phi}\, \sim \, \left( \dfrac {\partial_\phi M_s}{M_s^2} \right)^2\, \Mpt^{8}\, ,
\eeq
which indeed matches the behaviour in eq. \eqref{eq:IIA10d}. Repeating the same analysis for the bosonic loops yields the same field dependence.
		
\subsubsection{Strong Coupling Limit}
\label{sss:IIAstrongcoupling}
		
Let us discuss now the strong coupling limit, namely $g_s\to \infty$, in which the non-perturbative D0-branes become lighter than any other object in the theory and dominate the dynamics \cite{Witten:1995ex}. We show how the process of integrating out the tower of D0-branes up to the species scale reproduces the divergent behaviour both of the dilaton metric  and the gauge couplings for the the RR $p$-forms, $C_1$ and $C_3$. To perform the computation we need to `resolve' the D0-brane content in terms of massive ten-dimensional fields as well as to find the interactions between such states and the massless spectrum of 10d type IIA.
		
\subsubsection*{The D0-brane field content}
		
The essential point we want to remark here is that the field content associated to any step $n\in \mathbb{Z} \setminus \lbrace0\rbrace$ of the D0-brane tower corresponds to a spin-2 field, a Majorana spin-$\frac{3}{2}$ gravitino, and a rank-3 antisymmetric tensor field, all of them with a mass given by that of the corresponding bound state of $n$ D0-branes, namely
\begin{align}\label{eq:10dD0mass}
			m_n= n\, m_{\text{D}0} = 2\pi n\, \frac{M_s}{g_s}\, . 
\end{align}
Notice that, even though D0-branes  give rise to semi-classical black hole solutions (i.e. with a horizon)  at weak coupling ($g_s \ll 1$) \cite{Horowitz:1991cd}, upon approaching the strong coupling singularity they become asymptotically massless. Therefore, the fact that they have been presumably integrated out in string theory so as to arrive at the low-energy EFT \eqref{eq:IIA10d}, renders the theory inconsistent in the strong coupling limit by introducing the singular behaviour observed in the kinetic terms.
		
A minimal way to motivate the counting of dof outlined above is to realize that indeed D0-branes are BPS particles of maximal 10d $\mathcal{N}=(1,1)$ supergravity, such that their description in field theory language should correspond to a \emph{short} multiplet. This fixes completely its field content, since the number of local degrees of freedom comprising the D0 then has to be equal to that of a massless supermultiplet, and there is only one of these in maximal (non-chiral) 10d supergravity, the gravity multiplet itself. Now, in terms of representations of $SO(8)$ (or rather the appropriate spinorial double cover), as befits a massless multiplet, the low-lying field content of type IIA string theory is precisely \cite{Green:2012oqa} 
\beq\label{eq:dofcountingIIA}
		(\textbf{8}_v \oplus \textbf{8}_s) \otimes (\textbf{8}_v \oplus \textbf{8}_s)= (\textbf{1} \oplus \textbf{28} \oplus \textbf{35} \oplus \textbf{8}_v \oplus \textbf{56}_t)_{\text{bos}} \oplus (\textbf{8}_s \oplus \textbf{8}_c \oplus \textbf{56}_c \oplus \textbf{56}_s)_{\text{ferm}}\ ,
\eeq
where the bosonic part contains the dilaton, the metric tensor and the $B_2$-field in the NS sector and the 1-form and 3-form in the RR sector, whereas the fermionic piece is comprised by two gravitini and two dilatini of opposite chirality. Therefore, in terms of representations of the massive little group $SO(9)$, these states arrange themselves via the Stückelberg mechanism so as to form a symmetric traceless representation of $SO(9)$ ($\textbf{1}\, \oplus\, \textbf{8}_v\, \oplus\, \textbf{35} \to \textbf{44}$), a massive 3-form potential ($\textbf{28}\, \oplus\, \textbf{56}_t \to \textbf{84}_t$), and as a massive Majorana spin-$\frac{3}{2}$ field ($\textbf{8}_s\, \oplus\, \textbf{8}_c\, \oplus\, \textbf{56}_c\, \oplus\, \textbf{56}_s \to \textbf{128}$) \cite{Green:1999by}.
		
Another way to argue for the multiplet content associated to the D0 tower is also standard, and it is based on the duality between M-theory and type IIA at strong coupling\cite{Witten:1995ex}. At low energies, M-theory reduces to the unique $\mathcal{N}=1$ supergravity theory in eleven dimensions, with field content given by the metric, an anti-symmetric 3-form gauge field, and the single gravitino of spin-$\frac{3}{2}$. Now, upon Kaluza-Klein compactification on a circle, one can check\cite{Witten:1995ex} that by truncating to the zero-mode sector we recover exactly type IIA supergravity, i.e. eq. \eqref{eq:IIA10d}. However, if one insists on keeping track of the KK replica of the 11d gravity multiplet, one realizes that the D0-branes, which supposedly correspond to such KK tower, are indeed made of the massive replica of the 11d massless fields, namely the graviton, 3-form and gravitino, hence matching with our expectations above.
		
There is yet another, very explicit way to check the conjectured D0-brane field-theoretic content, by quantizing the super-quantum mechanical theory describing the effective dynamics in the presence of such non-perturbative states. By doing so, one reaches similar conclusions, as detailed in Appendix \ref{ss:10dDO}. 
		
\subsubsection*{The interaction vertices}
	
Having established the field content of the tower of states becoming light in the asymptotic limit, we now need to compute their interaction vertices with the massless modes in \eqref{eq:IIA10d}. To do so, we rely on M-theory/type IIA duality and keep track of the couplings between the zero and non-zero Fourier modes upon compactifying 11d $\mathcal{N}=1$ supergravity on a circle. 
		
We focus on the explicit computation of the couplings between the KK replica of the 11d 3-form gauge field  and the massless type IIA RR $p$-forms, since this is enough to illustrate how the emergence mechanism could be at work in the present set-up. The analysis of the couplings associated to the massive spin-$\frac{3}{2}$ and spin-2 fields can be obtained along similar lines\cite{Huq:1983im}, although we do not expect any significant change in the conclusions. 
		
In order to find such couplings, we focus on the bosonic part of the 11d M-theory action\footnote{\label{fn:hat}We use hats to denote fields/operators that live/act on eleven dimensions, while those without them live/act on ten dimensions.}
\begin{align}\label{eq:Mthyactionbos}
			S^{\text{11d}}_{\text{M-th}} = \frac{1}{2\kappa_{11}^2} \int \hat{R} \ \hat{\star} 1-\frac{1}{2}  d\hat{C}_3\wedge \hat{\star}  d\hat{C}_3 -\frac{1}{6} \hat C_3 \wedge d\hat{C}_3 \wedge d\hat{C}_3 \, ,    
\end{align}
and take the usual ansatz for the metric upon compactification on $S^1$
\begin{align}\label{eq:KKansatz11dmetric}
			ds^2_{11} = ds^2_{10}+ e^{2\varphi} (dz-C_1)^2\, ,   
\end{align}
where $z \in [0, 2 \pi \ell_{11})$ parameterizes the circular direction, $C_1$ denotes the KK photon and $\varphi$ is the radion, which is related to the 10d dilaton as $\varphi=2\phi/3$\cite{Witten:1995ex}. Analogously, in type IIA supergravity the KK photon becomes the RR 1-form, hence the notation. By introducing this into the action \eqref{eq:Mthyactionbos} and performing the Weyl rescaling of the 10d metric $g_{\mu \nu}\to e^{-\varphi/4} g_{\mu \nu}$ to go to the 10d Einstein frame, we recover the action for the massless sector given in \eqref{eq:IIA10d}. Similarly, expanding the field strength $\hat{G}_4= d\hat{C}_3$ into its Fourier modes we can obtain the relevant action for the massive $C_3^{(n)}$, including their interaction terms with respect to the RR $p$-forms (see Appendix \ref{ap:MtheoryKKcompactS1} for a detailed discussion on the $S^1$ compactification to the 10d string frame, with the more familiar identifications from IIA/M-theory duality). After making a field redefinition of $C_3^{(n)}$ so as to obtain conventional kinetic terms for the states in the tower, the action reads
\begin{equation}
\label{eq:finalbosonicaction}
	\begin{aligned}
		S^{\text{10d}}_{C_3^{(n)}} \, =\, & -\frac{1}{4\kappa^2_{10}} \int  \sum_{n \neq 0} \left [ \mathcal{D} C^{(n)}_3 \wedge \star \mathcal{D} C^{(-n)}_3 + e^{-\frac{3}{2}\phi_0} \left(\frac{2\pi n}{\ell_{10}}\right)^2\, C^{(n)}_3 \wedge \star C^{(-n)}_3 \right] \\
		& - \frac{1}{4\kappa^2_{10}} \int \sum_{n \neq 0} e^{-\frac{1}{2}\phi_0}\, \frac{2\pi in}{\ell_{10}} \left [ C_3 \wedge C^{(n)}_3 \wedge dC^{(-n)}_3 - C_3\wedge dC^{(n)}_3 \wedge C^{(-n)}_3 \right] + \ldots\, ,
	\end{aligned}
\end{equation}
where $\mathcal{D} C^{(n)}_3= dC^{(n)}_3 + \frac{in}{\ell_{10}}\, C_1 \wedge C^{(n)}_3$ refers to the covariant derivative with respect to the $U(1)$ graviphoton, $C_1$, under which the whole tower is charged, as expected on general grounds. Moreover, $\ell_{10}$ denotes the 10d Planck length, whilst the ellipsis indicates that there may be additional interaction terms involving the dilaton field, massless $p$-forms and massive KK replica.

The first line of eq. \eqref{eq:finalbosonicaction} above descends from the reduction of the Einstein and 3-form pieces of the 11d action, and therefore includes the couplings to the graviphoton, whereas the second line comes from the Chern-Simons term and gives rise to the interactions between the massless and massive 3-forms (see Appendix \ref{ap:MtheoryKKcompactS1} for details). This means, in particular, that the (trilinear) interaction vertices between the massive 3-forms and both the RR $C_1$ and $C_3$ fields in eq. \eqref{eq:finalbosonicaction} above are controlled respectively by
\begin{equation}\label{eq:C3nverticesRR}
			V_{C_1}^{(n)}\simeq \frac{n}{\ell_{10}^{9}}\, , \qquad V_{C_3}^{(n)}\simeq \frac{n\, e^{-\frac{\phi_0}{2}}} {\ell_{10}^{9}}\, . 
\end{equation}
Also, from the quadratic term of the $C_3^{(n)}$ it can be seen that its mass indeed matches that of the state given by $n$ D0-branes, which in 10d Planck units is precisely $m_n^2=n^2\, m^2_{\text{D}0}\,\propto e^{-\frac{3}{2}\phi_0}\, \Mpt^2$.
		
\subsubsection*{Emergence of the RR kinetic terms}
		
Let us now show how to reproduce the parametric behaviour of the kinetic terms associated to the RR 1-form and 3-form fields of type IIA string theory (c.f. \eqref{eq:IIA10d}) in the strong coupling limit via quantum corrections arising from the D0-brane tower. We focus on the one-loop contribution induced by the massive 3-forms, whereas in principle one should also take into account that of the fermions and massive spin-2 KK replica. It is reasonable to expect, though, the latter corrections to contribute with with the same asymptotic (moduli) dependence.

First of all, upon using the definition of the species scale, eq. \eqref{species}, as well as the mass formula of the tower of D0-branes, we get the following expressions for the $\phi_0$-scaling of the species scale and the number of species 
\beq \label{eq:QGscaleandN} 
		\Lambda_{\text{QG}} \sim m_{\text{D}0}^{\frac{1}{9}}\ \Mpt^{\frac{8}{9}} \sim  e^{-\frac{1}{12}\phi_0} \Mpt\ , \qquad N \sim  m_{\text{D}0}^{-\frac{8}{9}}\ \Mpt^{\frac{8}{9}}  \sim  e^{\frac{2}{3}\phi_0}\, .
\eeq
Using the expression \eqref{eq:1-formloopscalarssummary} for the contribution of a loop of scalars as a proxy, substituting the charges by the vertices given in \eqref{eq:C3nverticesRR}, and summing over all the tower, we obtain the following result for the metric of the RR 1-form
\beq \label{eq:sumovertoweroneform}
		\dfrac{1}{g^2_{C_1}}\, \sim\, \dfrac{1}{\ell_{10}^2} \sum_n n^2 \LQG^6\,  \sim\, \dfrac{1}{\ell_{10}^2} N^3 \LQG^6\, \sim\, 
		\dfrac{e^{\frac{3}{2}\phi_0}}{\kappa_{10}^2}\, ,
\eeq
and
\beq \label{eq:sumovertowerthreeform}
		\dfrac{1}{g^2_{C_3}}\, \sim\, \dfrac{1}{\ell_{10}^2} \sum_n e^{-\phi_0}\, n^2 \LQG^6\,  \sim\, \dfrac{1}{\ell_{10}^2} e^{-\phi_0}\,  N^3 \LQG^6\, \sim\, 
		\dfrac{e^{\frac{\phi_0}{2}}}{\kappa_{10}^2}\, ,
\eeq
for the RR 3-form, which agree with the classical dependence as seen from \eqref{eq:IIA10d}. Notice that a factor of $(2\kappa_{10}^2) \propto \ell_{10}^{8}$ in eqs. \eqref{eq:sumovertoweroneform} and \eqref{eq:sumovertowerthreeform} above cancels exactly between the propagators of the $C_3^{(n)}$ fields and the vertices \eqref{eq:C3nverticesRR}. Analogous results can be obtained by using the couplings of the tower of massive gravitini instead.

\subsubsection*{Emergence of the dilaton metric}
		
To close up this subsection, we reproduce by a simple one-loop computation again the metric of the 10d dilaton field. Following the general procedure outlined in section \ref{sss:emergencemodulimetric}, we compute the wave-function renormalization induced by the coupling of the D0-branes to the scalar $\phi$ through its modulus-dependent mass, and reproduce its kinetic term in the 10d action  \eqref{eq:IIA10d}. As it is clear from the bosonic action in eq. \eqref{eq:finalbosonicaction} (as well as from our knowledge from string theory. c.f. eq. \eqref{eq:10dD0mass}), the \emph{off-shell} mass of the D0-particles (in Planck units) depends on the dilaton modulus, such that after expanding around its VEV, one finds familiar couplings between e.g. the massive 3-forms and gravitini to the scalar field. Hence, considering the individual contributions from bosons, \eqref{eq:scalarloopscalarssummary}, and fermions, \eqref{eq:scalarloopfermionssummary} (with $\lambda_n\, =\, 2 n\, m_{\text{D}0}(\partial_\phi m_{\text{D}0})$ and $\mu_n \, = \, \partial_\phi m_{\text{D}0}$), and summing over the full tower, we get
\beq
		\delta g_{\phi \phi}^{C_3^{(n)}}\, \sim\,  \sum_n \lambda_n^2 \Lambda_{\text{QG}}^{4}\, =\, \Lambda_{\text{QG}}^{4} (\partial_\phi m_{\text{D0}})^2 m_{\text{D0}}^2 \sum_n n^4\, \sim\, M_{\text{pl, 10}}^8 \left( \frac{\partial_\phi m_{\text{D0}}}{m_{\text{D0}}} \right)^2\, ,
\eeq
and 
\beq
		\delta g_{\phi \phi}^{\psi_{\mu}^{(n)}}\, \sim\,  \sum_n \mu_n^2 \Lambda_{\text{QG}}^{6}\, =\, \Lambda_{\text{QG}}^{6} (\partial_\phi m_{\text{D0}})^2 \sum_n n^2\, \sim\, M_{\text{pl, 10}}^8 \left( \frac{\partial_\phi m_{\text{D0}}}{m_{\text{D0}}} \right)^2\, ,
\eeq
respectively. Again, as in eqs. \eqref{eq:sumovertoweroneform} and \eqref{eq:sumovertowerthreeform} above, there is an exact cancellation of factors of $(2\kappa_{10}^2)$ between the vertices and propagators due to the non-canonical normalization of the massive fields, c.f. eq. \eqref{eq:finalbosonicaction}. Therefore, we obtain the expected result for the (field-independent) kinetic metric for the dilaton in \eqref{eq:IIA10d}, written here in a (moduli space) covariant manner. 

\section{Emergence of scalar potentials}
\label{s:Scalarpotential}
		
In this section we study scalar potentials in the context of Emergence. To do so, we rely on the dual formulation of such potentials in terms of $(d-1)$-forms in $d$ spacetime dimensions. In its simplest incarnation, this formulation can be understood in the presence of a cosmological constant, since this may be dualized to a background VEV for a $d$-dimensional field strength, namely $\frac{1}{g}F^{\mu_1 \ldots \mu_d}= f_0 \, \epsilon^{\mu_1 \ldots \mu_d}$, with $\Lambda_{\mathrm{c.c.}}=g^2 f_0^2/2$. Hence, in the dual formulation the cosmological-constant-like term would arise from the kinetic term associated to the non-dynamical $(d-1)$-form.  Since $(d-1)$-forms couple naturally to codimension-1 branes, it can be seen that differentiated vacua with different values of the cosmological constant are thus separated by such membranes, with their corresponding (quantized) charge under the $(d-1)$-form, $q$, giving precisely the shift in the value of the cosmological constant, i.e. $2 \Lambda_{\mathrm{c.c.}}= g^2 f_0^2 \to g^2(  f_0 + q)^2\, $.\footnote{Note that this relation between the cosmological constant and the background field strength of a $(d-1)$-form is the familiar picture used for the neutralization of the cosmological constant in the Bousso-Polchinski mechanism \cite{BP} (building on previous ideas by Brown and Teitelboim \cite{BT, BT2}).} 
		
This dual formulation in terms of $(d-1)$-forms can be generalized from the case of a cosmological constant to a more intricate field-dependent scalar potential by essentially allowing the gauge coupling, $g$, to depend on the dynamical scalar fields. Moreover, it is possible to even capture negative contributions to the potential in the kinetic terms of the dual $(d-1)$-forms, as the ones arising from e.g. O$p$-planes in string theory. In particular, it was shown in \cite{Herraez:2018vae} (see also \cite{Bielleman:2015ina,Carta:2016ynn} for previous analysis on the topic, or \cite{Farakos:2017jme, Bandos:2018gjp, Lanza:2019xxg} for lower dimensional supergravity arguments) that the full type IIA flux potential on a CY$_3$ (orientifold) compactification  in the Large Volume Limit can alternatively be expressed in terms of 3-forms, whose field strength VEVs correspond to the quantized fluxes of the theory. Additionally, this has also been argued to be the case close to any infinite distance boundary in the moduli space of F-theory on CY$_4$\cite{Grimm:2019ixq}, by making use of asymptotic Hodge Theory techniques. In this more general scenario, the potential in the presence of $q_A$ (quantized) fluxes takes the generic form
\begin{equation}
			\label{eq:potential4forms}
			V=\dfrac{1}{2}Z^{AB}q_A q_B\, ,
\end{equation}
which is nothing but a generalization of the $g^2 f_0^2$ above since $Z^{AB}$ denotes precisely the inverse of the kinetic matrix for the $(d-1)$-form field strengths. Thus, in the very same way as other $p$-form gauge kinetic terms are expected to be generated via Emergence by integrating out the relevant towers of charged states becoming light in the weak coupling limit, the same idea may be applied for $(d-1)$-forms, hence generating $Z_{AB}$, which encodes all (singular) field dependence in the scalar potential.
		
In what follows we will study in detail the case of the emergence of a flux potential for a type IIA set-up in four non-compact dimensions. We will show how, by integrating out towers of massive gravitini (and superpartners), the required emergence mechanism takes place. Finally, in subsection \ref{ss:generalfluxpot}, we will consider the general case of $d$-dimensional flux potentials and show how Emergence is actually consistent with both the AdS/dS Distance Conjecture and the (asymptotic) dS Conjecture.

\subsection{The IIA flux potential in 4d}
\label{ss:fluxpotential4d}
		
\begin{figure}[tb]
			\begin{center}
				\includegraphics[scale=0.42]{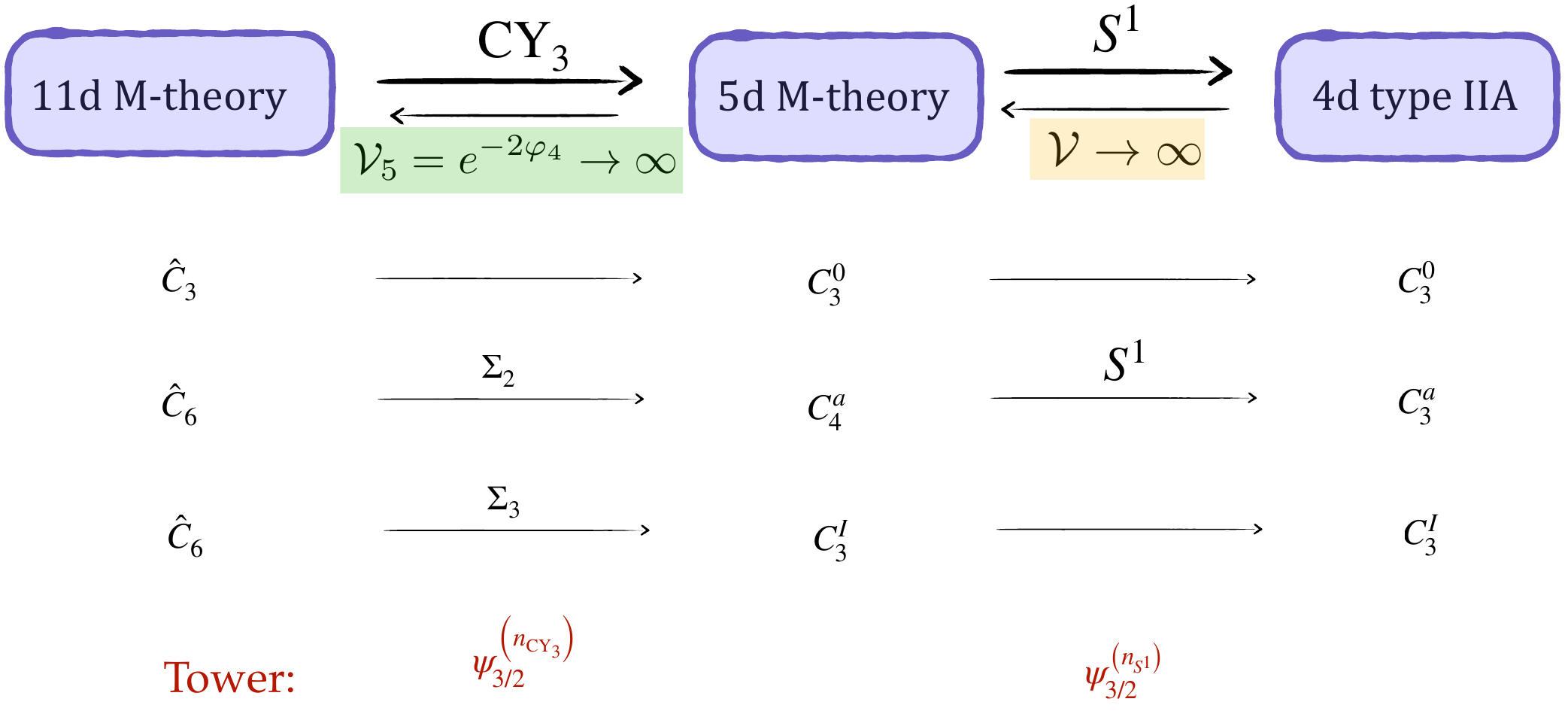}
				\caption{Summary of the two-step decompactification that we perform in order to obtain the emergent kinetic terms of the $p$-forms that contribute to the 4d scalar potential.}
				\label{fig:emergentpotentials}
			\end{center}
		\end{figure}

Let us now focus on the flux potential that arises upon compactification of type IIA on a CY$_3$, which will serve as a proof of concept that scalar potentials may be generated via Emergence. We consider this potential in the presence of of 6-form and 4-form RR fluxes, as well as 3-form NSNS fluxes, denoted collectively as $q_A=\left( e_0, e_a, h_I\right)$,  whose dual 4-form field strengths are denoted by $F_4^A=\left( F_4^0, F_4^a, H_4^I \right)=\left( dC_3^0, dC_3^a, d\mathcal{C}_3^I \right)$ and obtained, respectively, upon expanding the IIA ten-dimensional field strenghts $\hat{F}_4$, $\hat{F}_6$,  $\hat{H}_7$ on a basis of harmonic 0-forms, 2-forms and 3-forms, respectively. For simplicity, we have restricted ourselves to the case of vanishing VEVs for the axions and leave a more general analysis of the axion-dependent pieces of the potential for future work. The kinetic terms for the aforementioned 3-forms then read \cite{Herraez:2018vae}
\begin{equation}
			S^{4d}_{\text{IIA}} \supset  - \frac{1}{2\kappa_4^2} \int Z_{AB} F_4^A \wedge \star F_4\,, \qquad Z_{AB}= \dfrac{e^{-K}}{8} \left(
			\begin{array}{c  c c  }
				1 & &  \\
				& 4g_{ab}&  \\
				& & c_{IJ}\\ 
			\end{array}
			\right)\, ,
\label{eq:S4form4d}
\end{equation}
where $e^K=e^{4\varphi_4}/8\mathcal{V}$, while $g_{ab}$ and $c_{IJ}$ are respectively the metrics of the K\"ahler and complex structure sectors. Notice that we are not including the 4-forms that are dual to the RR 2-form and 0-form fluxes, since in the limits we are going to consider their kinetic terms do not diverge (i.e. they do not correspond to weak coupling points for said 3-form potentials). For concreteness, we study the case with one modulus in each sector, namely $\mathcal{V}$ and $\varphi_4$ (this corresponds to  $h^{1,1}=1$ and $h^{2,1}=0$). Still, let us remark that in the more general scenario the current analysis still applies, after redefining the moduli so as to extract the dependence on the overall volume and 4d dilaton, with the assumption that they both become large asymptotically. In the present case, the kinetic terms in eq. \eqref{eq:S4form4d} reduce to
\begin{equation}
			S^{4d}_{\text{IIA}} \supset  - \frac{1}{4\kappa_4^2} \int e^{-4\varphi_4} \mathcal{V} \left( F_4^0 \wedge \star F_4^0\, +\, \dfrac{1}{\mathcal{V}^{2/3}}\,  F_4^a \wedge \star F_4^a\, +\,  e^{2\varphi_4} \, H_4^I \wedge \star H_4^I\right) \, ,
\label{eq:S4form4dsimple}
\end{equation}
where the indices $a$ and $I$ are not being summed over, but just used to keep track of the different 4-forms in the game. The limits $\mathcal{V}\to \infty$  and $e^{-\varphi_4} \to \infty$ correspond to weak coupling points for all the 3-forms, so according to the emergence conjecture we expect to generate them by integrating out towers of `charged' particles. In particular, we know that the large volume limit corresponds to a decompactification to 5d M-theory on the same CY$_3$, such that the relevant tower should be indeed the subsector of KK modes along the circle that couples to the 3-forms, which in IIA language should be described by D0-brane bound states, as described in section \ref{s:emergence4dN=2}. Moreover, in addition to the volume dependence of the 3-form kinetic terms, they also seem to depend on $\varphi_4$, which was not the case in their 1-form counterparts due to the product structure of the $\mathcal{N}=2$ moduli space (see eq. \eqref{eq:N=2modulispace}). The limit $e^{-\varphi_4}\to \infty$ can be understood from the 5d M-theory perspective to correspond to some further decompactification to eleven dimensions, since $\mathcal{V}_5=e^{-2\varphi_4}$. Hence, we would expect this field-dependent part of the kinetic terms to be generated by the KK tower associated to the CY$_3$ in the decompactification limit of M-theory. Therefore, we expect to generate the full field dependence of the kinetic terms in \eqref{eq:S4form4dsimple} by studying this two step decompactification of M-theory on the CY$_3$ times $S^1$, in which we first decompactify the circle and then the full CY$_3$. Our strategy is thus to start from 11d M-theory and identify the relevant couplings of the KK towers to the fields that will eventually give rise to the 3-forms in 4d, as summarized in fig. \ref{fig:emergentpotentials}. In particular, we will use the KK towers of gravitini (first on the CY$_3$ and then on the $S^1$) as a proxy, since as we discuss below they couple to all the relevant $p$-forms. Let us also mention that this process implies a choice of path in moduli space, in which we send $\mathcal{V}\to \infty $ first and $e^{-\varphi_4} \to \infty$ afterwards. In principle though, the generation of the kinetic terms in our gravitaitonal EFT should be valid for any (asymptotically geodesic) path approaching the same  singularity, at which the relevant weak coupling emerges. In the present case, we have chosen this particular path because it makes the computation easier, as we can sequentially follow the different towers of gravitini along with their contributions to the kientic matrix $Z_{A B}$ in \eqref{eq:S4form4d}.  

\subsection{Emergence in 5d M-theory}
\label{ss:emergencepotential5d}
		
In this section we discuss the first step summarized in fig. \ref{fig:emergentpotentials}, namely the compactification of 11d M-theory on a CY$_3$, focusing on the kinetic terms that are relevant for the 4d potential and how their dependence on $\mathcal{V}_5=e^{-2\varphi_4}$ can be generated via Emergence by keeping track of their couplings with the Kaluza-Klein tower of gravitini. In particular, we focus on the emergence of the kinetic terms for the two five-dimensional 3-forms that appear upon reduction of the M-theory $\hat{C}_3$ and its magnetic dual $\hat{C}_6$ along a point and a 3-cycle, respectively, as well as on the 4-form that arises upon reducing  $\hat{C}_6$ on a 2-cycle.\footnote{One could analogously start with the eleven-dimensional $\hat{C}_3$ only and perform the dualization in the 5d theory, where the scalars $\xi^K$ are dual to 3-forms and the fluxes to 4-forms.} Note that only the latter contributes to the M-theory scalar potential, but the three of them contribute to the 4d type IIA scalar potential since they give rise to 3-forms upon further reduction on $S^1$.
		
The bosonic part of the 5d effective action of M-theory compactified on a Calabi--Yau three-fold is displayed in eq. \eqref{eq:Mthyaction5d}, and its field content is discussed in general in Appendix \ref{ap:5dMtheory}. Recall that in 5d M-theory, the volume modulus belongs to the hypermultiplet sector, and in the Einstein frame it has a kinetic term of the form
\begin{equation}
			S^{\text{5d}}_{\text{M-th}} \supset - \dfrac{1}{4 \kappa_5^2} \int \dfrac{1}{\mathcal{V}_5^2} \, d\mathcal{V}_5 \wedge \star d \mathcal{V}_5 \, .
\end{equation}
As explained at the end of Appendix \ref{ap:5dMtheory}, the presence of internal $\hat{G}_4$-flux can be understood in a dual formulation in terms of 4-form gauge potentials in 5d. From the 11d M-theory perspective, this arises upon reducing the `democratic' (pseudo-)action in the presence of both $\hat{G}_4$ and its magnetic dual $\hat{G}_7$ (complemented with the additional on-shell constraint $\hat G_7= \star \hat G_4\, $). Moreover, from this reduction we can also follow the 4-form field strengths that will eventually contribute to the scalar potential in 4d, so we expand the 11d field strengths as
\begin{equation}
			\label{eq:11dp-formsexpansion}
			\hat{G}_4= F_4^0+\ldots \, ,  \qquad \hat{G}_7= F_5^a \wedge \omega_a + H_4^I \wedge \alpha_I+\ldots\, ,
\end{equation}
where $\omega_a$ and $\alpha_I$ are a basis of harmonic 2 and 3-forms, respectively. As discussed above, we restrict from now on to the case in which the only modulus in the hypermultiplet sector is $\mathcal{V}_5$ (i.e. $h^{2,1}=0$), for which the relevant piece of the action reads
\begin{equation}
			S^{\text{5d}}_{\text{M-th}} \supset  - \frac{1}{4\kappa_5^2} \int  \mathcal{V}_5^2 \left( F_4^0 \wedge \star F_4^0\, +\,  F_5^a \wedge \star F_5^a\, +\,  \dfrac{1}{ \mathcal{V}_5} \, H_4^I \wedge \star H_4^I\right) \, .
			\label{eq:Spform5dsimple}
\end{equation}
Let us clarify that this corresponds, in the presence of more moduli, to decompactifying the volume without probing any of the other boundaries of the corresponding moduli space, which would capture the finer structure in the kinetic terms, which we leave for future work. In particular, the volume prefactor for the metric of $F_4^0$ and $F_5^a$ would not change, since its metric depends on the vector multiplet moduli, which factorizes and does not include the volume modulus.

\subsubsection*{Computing 3-form and 4-form emergence from fermionic couplings}
		
In the following we explain how the moduli-dependence in the kinetic terms of both the 3-form and 4-forms in eq. \eqref{eq:Spform5dsimple} can be reproduced by a one-loop computation of a tower of (massive) particles coupled to such gauge fields. Incidentally, notice that the part corresponding to the 4-forms already serves as a first example of a potential that can be generated via Emergence. To do so, the crucial question is which kind of particles couple to those $p$-forms. Focusing on the gravitino dependent part of the 11d M-theory action, we can see that it couples to the 4-form field strength, $\hat{G}_4$ (and its corresponding 7-form magnetic dual, $\hat{G}_7$) in the following way (see e.g. \cite{Ortin:2015hya} for more details)
\beq
\label{eq:MthyactiongravitinoG4coupling}
		\begin{aligned}
			S_{\text{M-th}}^{11d} \supset  \,  \frac{1}{2\kappa_{11}^2} \int \text{d}^{11}x\ \hat{e} & \left( -i \Bar{\psi}_{M} \, \Gamma^{M N P} \, \mathcal{D}_{N} \psi_{P} + \frac{1}{192} \Bar{\psi}_{N}\,  \Gamma^{[N}\,  \Gamma_{M_1 \ldots M_4}\,  \Gamma^{P]}\,  \psi_{P} \, \hat{G}_4^{M_1\ldots M_4} \right.  \\
			& \left. \quad + \frac{1}{192} \Bar{\psi}_{N}\, \Gamma^{[N}\, \Gamma_{M_1\ldots M_7 }\, \Gamma^{P]}\, \psi_{P} \, \hat{G}_7^{M_1\ldots M_7} \right) \, ,
		\end{aligned}
\eeq
where $\hat e_{M}^{\mathcal{A}}$ denotes the vielbein with $\mathcal{A}=0, \ldots, 10$ being the tangent space indices, $\hat e$ its determinant, and $\mathcal{D}_{N}$ is the covariant derivative.\footnote{More specifically, the covariant derivative acting on the gravitino would be $\mathcal{D}_{N} \psi_{P}= \partial_N\psi_{P}-\hat{\Gamma}^Q_{N P}\psi_{Q}+ \frac{1}{4} \hat{\omega}^{\mathcal{A} \mathcal{B}}_N \Gamma_{\mathcal{A} \mathcal{B}} \psi_{P}$, where $\hat{\Gamma}^Q_{N P}$ is the affine connection.}
		
As already anticipated at the beginning of this section,  upon compactification of this piece of the action on a CY$_3$ we expect to obtain couplings between the massive KK towers of gravitini and the different $p$-form field strengths that arise in 5d. Of course, the type and number of KK excitations along a particular CY$_3$ (beyond the fact that there is at least one tower of massive spin-2 modes, i.e. the graviton replica) is not known in full generality, but on general grounds we expect that such massive fields arrange themselves in terms of 5d $\mathcal{N}=2$ \emph{long} multiplets, which have twice as many on-shell dofs as the analogous massless multiplets we were discussing in the previous section. This stems from the fact that the Calabi--Yau does not posses any non-trivial harmonic 1-forms, which implies that the KK towers are not charged and thus cannot be BPS.
		
Luckily for us, the details of such KK gravitino towers are not necessary  to extract the volume dependence in the interaction between them and the massless $p$-form field strengths  we are interested in, namely those in \eqref{eq:11dp-formsexpansion}. As explained in Appendix \ref{ap:dimreductiongravitino}, we obtain the following volume dependence for the relevant interaction vertices (c.f. eq. \eqref{eq:5dp-formpsipsicouplings})\footnote{These can also be obtained from the general eq. \eqref{eq:gravitinovertices2} with $D=11$, $k=6$, and the right combinations of $P=4,\, 7$ and $p=4,\, 5$ that correspond to the expansion \eqref{eq:11dp-formsexpansion}.}
\begin{equation}\label{eq:5dgravitinop-formvertices}
			V^{(F_4^0)}_{\text{5d},\,n}\simeq \mathcal{V}_5\, , \qquad \qquad V^{(F_5^a)}_{\text{5d},\,n}\simeq \mathcal{V}_5 \, , \qquad \qquad V^{(H_4^I)}_{\text{5d},\,n}\simeq \mathcal{V}_5^{1/2} \, . 
\end{equation}
Note that they are $n$-independent, since the interaction they come from in 11d does not include any derivative of the massive modes. Furthermore, the relevant one-loop diagram includes two extra powers of the external momenta with respect to the computations in Appendix \ref{ap:Loops1-form}, since the gravitino bilinear couples directly to the field strength, instead of the gauge potential itself. This implies that the piece that corrects the propagator gets an extra factor of the cut-off squared ($\Lambda^2$) with respect to eq. \eqref{eq:1-formloopfermionssummary}. For the 3-form field strengths, including the contribution from the full gravitino tower up to the species scale yields
\beq \label{eq:threeform5d}
		\frac {\partial \Pi^{\epsilon \eta \kappa}_{\mu \nu \rho}(p)}{\partial p^2} \bigg\rvert_{p=0} \sim - \sum_{n=1}^N \delta^{[\epsilon \eta \kappa]}_{[\mu \nu \rho]}  \, V_{\text{5d},\,n}^2  \, \Lambda_{\text{QG}}^{3} \sim - \delta^{[\epsilon \eta \kappa]}_{[\mu \nu \rho]}\, V_{\text{5d},\,n}^2\, N\,  \Lambda_{\text{QG}}^{3} \sim - \delta^{[\epsilon \eta \kappa]}_{[\mu \nu \rho]}\, V_{\text{5d},\,n}^2\,  M_{\text{pl,}\, 5}^3\, ,
\eeq
where we have only used the fact that the vertices $V_n$ do not depend on $n$ and the definition of the 5d species scale, namely $M_{\text{pl,}\, 5}^3=N \LQG^3$. The same applies for the 4-form field strength upon adjusting the tensorial structure $\delta^{[\epsilon \eta \kappa]}_{[\mu \nu \rho]}\to \delta^{[\epsilon \eta \kappa\delta]}_{[\mu \nu \rho \sigma]}$. Notice that for the previous computation to be carried out in general it was crucial that it did not depend explicitly on the masses of the KK gravitini, as is the case for the leading correction to the propagator under consideration, which due to the two extra power of the external momenta discussed above, only depends on $\LQG^3$ in 5d. Finally, substituting the vertices \eqref{eq:5dgravitinop-formvertices} into \eqref{eq:threeform5d}, and extracting the tensorial structure, we obtain the following kinetic terms for the different $p$-forms via Emergence
\begin{equation}
			\label{eq:emergent5dkineticterms}
			Z^{\text{5d}}_{00}\sim \dfrac{\mathcal{V}^2_5}{\kappa_5^2} \, , \qquad  \qquad  Z^{\text{5d}}_{aa}\sim  \dfrac{\mathcal{V}_5^2}{\kappa_5^2} \, , \qquad \qquad  Z^{\text{5d}}_{II}\sim  \dfrac{\mathcal{V}_5}{\kappa_5^2} \, , 
\end{equation}
which indeed reproduces the  5d kinetic terms in \eqref{eq:Spform5dsimple}. Also it reduces to the right $\varphi_4$-dependence in 4d, as shown in \eqref{eq:S4form4dsimple}, after the identification $\mathcal{V}_5=e^{-2\varphi_4}$. Hence, we have not only reproduced via the emergence procedure the $\mathcal{V}_5$-dependence of the 5d M-theory flux potential, encapsulated in \eqref{eq:kineticterms4forms}, but also the correct dependence for the kinetic terms of the 3-forms that do not directly contribute to the potential in 5d but do so in 4d. As a last remark, let us stress that this supports the idea that all $p$-forms should be treated democratically from the point of view of Emergence. We have just illustrated with this example that dynamical $p$-form fields, such as 3-forms in five dimensions, can turn into the non-dynamical ones responsible for the scalar potential in 4d, but all their kinetic terms may be obtained in a similar way from the point of view of Emergence.		
		
\subsection{Emergence of the type IIA flux potential in 4d}
\label{ss:4dpotential}
		
We now turn on to the computation of the $\mathcal{V}$-dependent piece in \eqref{eq:S4form4dsimple} from integrating out the tower of gravitini along the circle, which corresponds to the second step summarized in fig. \ref{fig:emergentpotentials}. From the 4d point of view, this corresponds to the $\mathcal{V} \to \infty$ limit when $\varphi_4$ is kept constant (i.e. we only move along the 4d vector multiplet moduli space). On the other hand, from the 11d perspective, recalling that $\mathcal{V}_5=e^{-2\varphi_4}$, this limit corresponds to keeping the volume of the CY$_3$ in 11d Planck units fixed whereas the circle direction is varied. This being the case, the dependence on $\varphi_4$ is not modified by this new tower of `light' states and thus it is the same than the one which can be directly read off from eqs. \eqref{eq:emergent5dkineticterms} by simply rewriting them in terms of $\varphi_4$ now.
		
The relevant 4-forms as well as their kinetic terms in \eqref{eq:S4form4dsimple} are easily obtained in this set-up from dimensional reduction of the 5-form and 4-form field strengths in \eqref{eq:11dp-formsexpansion} along the M-theory circle and a point, respectively.\footnote{We are slightly abusing the notation here by denoting the five-dimensional field strengths and their four-dimensional counterparts with the same symbols, but the distinction should be straightforward from the context.} Similarly, we can also reduce the five-dimensional interactions between these 5-form and 4-form field strengths and the 5d massless gravitini (which comes from the 5d zero modes ($n=0$) in the interaction terms \eqref{eq:5dp-formpsipsicouplings}) to obtain the couplings between the corresponding four-dimensional 4-form fields strengths and the KK tower of gravitini along the circle. This is performed in detail in Appendix \ref{ap:dimreductiongravitino}, and the relevant vertices including the $n$-th KK mode of the gravitino tower along the circle can be obtained by substituting $D=5$, $d=4$, $P=4,5$ and $p=4$ in eq. \eqref{eq:gravitinovertices2}, and are given by
\begin{equation}\label{eq:4dgravitinop-formvertices}
			V^{(F_4^0)}_{\text{4d},\,n}\simeq  R_5^{3/2}\, , \qquad \qquad V^{(F_4^a)}_{\text{4d},\,n}\simeq R_5^{1/2} \, , \qquad \qquad V^{(H_4^I)}_{\text{4d},\,n}\simeq R_5^{3/2} \, ,
\end{equation}
where $R_5$ is the radius of the $S^1$ in 5d Planck units. Adding up the contribution of the whole tower up to the species scale, including the extra factors of the external momenta as in the 5d M-theory computation, we obtain the following contribution to the propagator of the 3-forms
\beq \label{eq:threeform4d}
		\frac {\partial \Pi^{\epsilon \eta \kappa}_{\mu \nu \rho}(p)}{\partial p^2} \bigg\rvert_{p=0} \sim - \sum_{n=1}^N \delta^{[\epsilon \eta \kappa]}_{[\mu \nu \rho]}  \, V_{\text{4d},\,n}^2  \, \Lambda_{\text{QG}}^{2} \sim - \delta^{[\epsilon \eta \kappa]}_{[\mu \nu \rho]}\, V_{\text{4d},\,n}^2\, N\,  \Lambda_{\text{QG}}^{2} \sim - \delta^{[\epsilon \eta \kappa]}_{[\mu \nu \rho]}\, V_{\text{4d},\,n}^2\,  \Mpf^2\, ,
\eeq
where once again we have crucially used the fact that the interactions are independent of $n$ as well as the relation between the Planck mass and the species scale in four dimensions, $\Mpf^2=N\LQG^2$. Inserting now the dependence on the volume for each of the 4-forms, given in eq. \eqref{eq:4dgravitinop-formvertices}, we end up with the following kinetic terms for the four dimensional 3-forms
\begin{equation}\label{eq:ZAB4d}
			Z^{\text{4d}}_{00}\sim \dfrac{R_5^3}{\kappa_4^2} \, , \qquad  \qquad  Z^{\text{4d}}_{aa}\sim  \dfrac{R_5}{\kappa_4^2}\, , \qquad \qquad  Z^{\text{4d}}_{II}\sim   \dfrac{R_5^3}{\kappa_4^2}\, .
\end{equation}
Including now the field dependent contributions from both \eqref{eq:emergent5dkineticterms} and \eqref{eq:ZAB4d}, and rewriting them in terms of the appropriate 4d variables using $R_5^3=\mathcal{V}$ and $\mathcal{V}_5=e^{-2\varphi_4}$, we obtain the following result for the total one-loop kinetic terms of the four-dimensional 3-forms
\begin{equation}\label{eq:ZABtotal}
			Z_{00}\sim \dfrac{1}{\kappa_4^2}e^{-4\varphi_4}\, \mathcal{V} \, , \qquad    Z_{aa}\sim  \dfrac{1}{\kappa_4^2}e^{-4\varphi_4}\, \mathcal{V}^{1/3} \, , \qquad   Z_{II}\sim   \dfrac{1}{\kappa_4^2}e^{-2\varphi_4}\, \mathcal{V}\, .
\end{equation}
This indeed gives the right dependence on the 4d dilaton and volume of the Calabi--Yau, as can be seen by directly comparing with  eq. \eqref{eq:S4form4dsimple}.

\subsection{Emergent potentials and Swampland Conjectures}
\label{ss:generalfluxpot}
We will now take a wider scope and consider the emergence of flux potentials in $d$ spacetime dimensions as coming from $(d-1)$-forms. We will show how Emergence may provide for a microscopic understanding of the AdS/dS Distance Conjecture\cite{Lust:2019zwm} and the (asymptotic) de Sitter Swampland conjecture \cite{Obied:2018sgi,Ooguri:2018wrx,Garg:2018zdg}, at least in the context of flux potentials. From the discussion above it transpires that a flux potential will thus appear if one-loop corrections are able to generate a metric $Z_{AB}$ for the relevant $(d-1)$-forms in the system. As already mentioned, the loop computations for $p$-forms coupled to a tower of states is expected to be analogous to the ones we did for 1-form gauge fields, up to $\mathcal{O}(1)$ factors. Therefore, diagrams like those in fig. \ref{fig:1-formpropagator} will give rise to an emergent contribution for a single $(d-1)$-form with a \emph{diagonal} metric (c.f. section \ref{sss:emergenceU(1)})
\beq
		Z_{AA}\, \sim\, \sum_{n=1}^N n_A^{2/r}\, \LQG^{d-4}\, \sim\, \left( \frac {\Mpd^{d-2}}{\Mt^2}\right)^{\alpha_{d,r}/\alpha_{d,p}}\, ,
\eeq
in complete analogy with the metric of a 1-form coupled to a tower of states with characteristic mass scale $\Mt$. We will have in mind the coupling of the $(d-1)$-forms to towers of massive $(d-1)$-forms, analogous to those in eq. \eqref{eq:finalbosonicaction}. Recall that $\alpha_{d,r}$ and $\alpha_{d,p}$ depend on the `internal structure' of the tower as well as the charges of their particles, and are given by eq. \eqref{eq:dosalphas}. Now, if we turn on some fluxes associated to this $(d-1)$-forms, a scalar potential is generated, which takes the form
\beq
		V (\phi)\, \sim\, Z^{AA}(q_A)^2\, \sim\, (q_A)^2 \left(\frac {\Mt^2}{\Mpd^{d-2}}\right)^{\alpha_{d,r}/\alpha_{d,p}}\, ,
\eeq 
where $\phi$ denotes collectively the moduli of the theory. Equivalently, one can write
\beq
		\Mt\, \simeq\, V(\phi)^{\frac {\alpha_{d,p}} {2\alpha_{d,r}} }\, \Mpd^{1-d(\alpha_{d,p}/2\alpha_{d,r})}\, .
		\label{eq:adsconjecture}
\eeq
This expression tell us that, as the potential approaches zero, a tower of states with characteristic mass scale $\Mt$ becomes asymptotically massless. This is indeed consistent with the purported AdS/dS Distance Conjecture proposed in \cite{Lust:2019zwm}. The latter states that a tower of particles should become massless (in Planck units) as the vacuum energy of a theory $V_0$ goes to zero like that in eq. \eqref{eq:adsconjecture}
\beq
		\Mt\, \lesssim\, V_0^\delta \Mpd^{1-d\delta}\, ,
\eeq
with $\delta$ a positive constant with $\delta \leq 1/2$ in the case of AdS vacua. Thus emergence in the context of fluxes is consistent with this conjecture with $\delta = {\frac {\alpha_{d,p}} {2\alpha_{d,r}}}$. Interestingly, in ref. \cite{Castellano:2021mmx} it was found an expression similar to  that in eq. \eqref{eq:adsconjecture} but for $\alpha_{d,r}=1/2$, based on completely different arguments related to the covariant entropy bound\cite{Bousso:1999xy}.
		
We can have several $(d-1)$-forms $C_\beta$, $\beta=1, \ldots, N$, that may couple to different towers with characteristic scale $\Mti$, which may appear in the given asymptotic limit. In this case the full potential will be a sum
\beq
		V_{\text{tot}}\, \simeq\, \sum_\beta\sum_i  c_{\beta, i}\, \left(\frac {\Mti^2}{\Mpd^{d-2}}\right)^{\alpha_{d,r_i}/\alpha_{d,p_i}}\, ,
\label{eq:potencialgeneral}
\eeq
with $c_{\beta,i}$ coefficients proportional to flux (squared) parameters. 
		
The example discussed above of type IIA CY$_3$ vacua with a single K\"ahler modulus and  e.g. RR-fluxes may be described in these terms. In that example, with tower spectroscopy discussed in detail in section \ref{s:emergence4dN=2}, there are two relevant infinite sets of states in the large volume limit, the tower of (bound states) of D0's and the tower of bound states of D2-D0 particles. In our notation the first has $p=1$ (i.e. $m_n=n\, m_{\text{D0}}$) and the second $p \to \infty$ ($m_n \simeq m_{\text{D0-D2}}$ for all $n\in \mathbb{Z}$). The 3-form $C_3^0$ couples to both towers with $r=1$ (meaning here that $q_n=n$) whereas the 3-forms $C_3^a$ couple to the D0-D2 tower with $r \to \infty$. Then the potential has an asymptotic dependence on the single K\"ahler modulus $t$ like
\beq
		V_{\text{flux}}\, \sim\,  c_{0,00}\, (m_{\text{D0}})^2\, +\,  c_{0,02}\, (m_{\text{D0-D2}})^6\, +\, c_{a,02}\, (m_{\text{D0-D2}})^2\, ,
\eeq
where the subindices `00' and `02' refer to the D0 and D0-D2 towers respectively. Here $m_{\text{D0}}$ and $m_{\text{D0-D2}}$ are the characteristic scale of the two towers respectively. Note that the first two terms correspond to the contribution of the $C_3^0$ form and behave like $\sim 1/t^3$ and the last comes from $C_3^a$ and behaves like $\sim 1/t$. These are in agreement with the discussion in the previous section.
		
With the above expression eq. \eqref{eq:potencialgeneral} for an asymptotic flux potential we can test other swampland conjectures involving scalar potentials. In particular, the (asymptotic) dS Swampland Conjecture \cite{Obied:2018sgi,Ooguri:2018wrx,Garg:2018zdg} states that  a scalar potential weakly coupled to Einstein gravity in dS space verifies
\beq
		\Mpd\, \frac {|\nabla V|}{V}\, \geq\, c\, ,
\eeq
where the gradient is taken with respect to any asymptotic modulus and $c$ is a constant of order one. If we consider the tower leading to the largest contribution to the potential to be characterized by a scale $\Mt$, the potential is given by eq. \eqref{eq:adsconjecture} and then one finds
\beq
		\Mpd\, \frac {|\nabla V|}{V}\,  \simeq\, 2\frac {\alpha_{d,r}}{\alpha_{d,p}} \frac {|\nabla \Mt|}{\Mt}\, \simeq\, 2\frac {\alpha_{d,r}}{\alpha_{d,p}}\, \lambda\, ,
\eeq
where $\lambda$ is the coefficient of the exponent of the Swampland Distance Conjecture. Thus, one can conclude that emergence is actually in agreement with the dS swampland conjecture with 
\beq
		c\, \lesssim\,  2\frac {\alpha_{d,r}}{\alpha_{d,p}}\, \lambda \, .
\eeq
Thus the values for the bound are in the range $2\lambda/(d-1) \leq c\ \leq 2\lambda (d-1)$. In particular for $r=p$ one has $c=2\lambda$, which is the conjectured result in reference \cite{Andriot:2020lea}. On the other hand, for $r=1$ one recovers the generalized result $c = \lambda/\alpha_{d,p}$ conjectured in \cite{Castellano:2021mmx}. 
		
In summary, in the context of Emergence, kinetic terms for $(d-1)$-forms may also be generated, creating a scalar potential if the corresponding (dual) fluxes are turned on in the vacuum. Such emergent potentials are moreover consistent with both the AdS/dS Distance Conjecture and the dS (asymptotic) Swampland Conjecture. In addition, specific values for the relevant $\mathcal{O}(1)$ factors in those are obtained. We find remarkable that the Emergence Proposal is not only able to provide for a microscopic rationale for the SDC and the (magnetic) WGC, but also of these other two important Swampland conjectures. 	
		
\section{Conclusions and outlook}
\label{s:Conclusions}
		
In the present paper we have taken the first steps towards a systematic analysis of the idea of Emergence in Quantum Gravity, as discussed at the beginning of section \ref{s:EmergenceQG}. We started by analyzing the notion of {\it species scale}, which plays an important role within the emergence mechanism. We computed it both for the case of (multiple) Kaluza-Klein as well as for string towers. In the latter case we found that keeping track of the $\log N$ terms (with $N$ being the number of species) is actually important in order to make sense of the state counting between $\LQG$ and $M_s$. We also found that in string theory, the asymptotic number of states below the cut-off $\LQG$ in the weak coupling limit is directly related to the dilaton VEV (in any spacetime dimension). This is in agreement with the results in \cite{Dvali:2009ks,Dvali:2010vm}, which suggest that string theory may be viewed as a theory of roughly $1/g_s^{2}$ species weakly coupled to gravity in the small $g_s$-limit.
 
One-loop self-energy computations play an important role in the practical calculations of Emergence and we address them in section \ref{s:EmergenceQG}. We performed such computations in $d$-dimensions for moduli fields, gauge bosons and (chiral) fermions in the external legs. The crucial ingredient to get emergent kinetic terms is to use the species scale as the physical cut-off in these loop computations, as well as to sum over all states belonging to the relevant tower(s) and falling below $\LQG$ (for each specific asymptotic limit under consideration). Remarkably, even though each individual contribution to the kinetic terms is a quantum loop effect, the final resummed metric turns out to be classical (i.e. independent of $\hbar$), due to this rather special choice of UV cut-off. We obtain that indeed one can reproduce the correct leading field dependence for the known non-linear sigma model metrics of string theory examples for moduli scalars, 1-forms and spin-$\frac{1}{2}$ fields. In particular, we reproduce the asymptotic field dependence in the Swampland Distance Conjecture as well as the (magnetic) Weak Gravity Conjecture for a $U(1)$ vector field, in the presence of either KK or string towers. We also found that fermion fields can indeed get large contributions to their wave-function renormalization, a fact that may be important for the application of the Emergence Proposal to phenomenology, since most of the SM particles are actually described by spin-$\frac{1}{2}$ fields. Additionally, we argued that in the context of Emergence, small Yukawa couplings in 4d may imply the existence of light towers of states, and we elaborate more on these ideas in \cite{Castellano:2023qhp}.

The Emergence Proposal should be easily falsifiable since it gives a very definite prediction, namely whenever there is a singularity at infinite distance in the metric of any consistent theory of Quantum Gravity, there should be a tower of states which, upon quantum integration, should reproduce the original singular behaviour. This bears certain resemblance with the  well-know (finite distance) conifold singularity in the vector multiplet moduli space of 4d $\mathcal{N}=2$ theories \cite{Strominger:1995cz}, which in this case can be seen to arise from integrating out one BPS state becoming massless at that point. However, a crucial difference to keep in mind is that in the case of Emergence that we study here the number of states becoming (asymptotically) massless indeed grows in an unbounded fashion, pointing towards a more drastic \emph{gravitational} phase transition.  
 
Due to its predictive power, it seems important to check the emergence idea against as many examples of string theory vacua as possible. This was one of the main aims of this paper. We have tested the Emergence Proposal in a number of selected examples that display different features both in terms of the number of dimensions where the EFT lives, the amount of supersymmetry and the structure of the relevant towers appearing along the relevant limits. In particular, we have revisited the large volume limit of 4d $\mathcal{N}=2$ theories arising from type IIA Calabi--Yau three-fold compactifications, in which the towers include both D0 and D2-particle systems (or their type IIB D3-particle mirror duals), which corresponds to decompactification to 5d M-theory. We have performed a detailed analysis of the structure of the towers and tested the emergence mechanism in this set-up, being able to reproduce also the axion-dependent parts within the gauge kinetic functions. We have also revisited other limits in this vector multiplet moduli space, emphasizing the string-theoretic interpretation of such limits and also reproducing the required field dependence. Furthermore, we have investigated the case of infinite distance singularities obstructed by non-perturbative instanton corrections in the hypermultiplet moduli space of such compactifications, finding a nice compatibility between the emergence picture and the presence of such (infinite) instanton corrections. We have also studied higher dimensional examples, including the case of a 6d  F-theory compactification in which the kinetic terms for the gauge couplings arise (from the emergence point of view) from a solitonic tensionless heterotic string, and a 7d example from M-theory on (attractive) $K3$'s. We considered as well the case of 10d string theories, focusing on the emergence mechanism in  $d=10$ type IIA, which is particularly rich. In all these cases, despite the different structures, Emergence seems to give results which reproduce the right field dependence providing for the leading singular behaviour.

One of the open questions in the last few years was whether flux potentials can be generated by Emergence. This is equivalent to the question of whether non-propagating $(d-1)$-forms (in $d$ spacetime dimensions) can get emergent kinetic terms. We found that indeed this may be the case by analizing in detail the generation of the flux potential  in type IIA CY$_3$ compactifications for such vacua in the relevant limits. In practice, this is most easily studied by decompactifying the theory first up to five dimensions in M-theory, and then growing the whole Calabi--Yau large upon going to 11d supergravity. We found that the required kinetic terms for the 3-form fields in 4d may arise from quantum loops involving KK towers of massive gravitini and their bosonic superpartners. 
 
All the examples we have checked in string theory so far seem to be naively consistent with Emergence, at least at the level of properly generating the leading (divergent) field dependence. We have found so far no counterexamples to the emergence idea and we have moreover checked that infinite distance singularities may arise non-trivially in QG as an intrinsic `IR phenomenon', as originally proposed in \cite{Vafa:1995ta, Ooguri:2006in}. 

Many important questions remain open before claiming full success and generality of the Emergence Proposal, so let us finally comment on some of them. First, it would be very interesting to study how Emergence applies for general CY$_3$ type IIA compactifications with multiple K\"ahler moduli in full generality, as well as in the hypermultiplet sector. Also, studying Emergence in other 10d string theories seems particularly relevant, especially in the two heterotic strings (which we have not discussed in this paper in detail), as well as other F-theory and M-theory vacua. In this regard, understanding in full detail the absence of \emph{exact} cancellations within the light particle self-energy whenever our theories enjoy many supersymmetries is also a concrete computation that may play a crucial role here, and can in principle be addressed with the current tools at our disposal. In any event, as discussed in detail both in sections \ref{s:EmergenceQG} and \ref{ss:Emergence10dST}, one does not expect such cancellations to the kinetic terms to occur based on general physical arguments. More generally, understanding how this idea could be realized in AdS spacetimes, upon using the AdS/CFT correspondence\cite{Maldacena:1997re}, also seems particularly interesting. Moreover, assuming Emergence to pass every test one may think of, the question arises of what is the fundamental origin of this property as well as its role in the context of the Swampland Program. We have seen in this paper that up to four Swampland conjectures are in one way or another related to this proposal, namely the Swampland Distance Conjecture, the (magnetic) Weak Gravity Conjecture, the AdS/dS Distance Conjecture and the  (asymptotic) dS Conjecture. This points towards some sort of unifying picture in which Emergence and the concept of the species scale could be at the core of some aspects of the Swampland Program. On the other hand, if the strong formulation of Emergence is true, it would be important to try to get some insights about the fundamental {\it topological} theory which could underlie this rather counter-intuitive phenomenon.
		
		
\vspace{1.5cm}
\centerline{\bf Acknowledgments}
\vspace{0cm}
		
We would like to thank G. Bossard, J. Calder\'on-Infante, M. Delgado, J. Donoghe, N. Gendler, M. Graña, D. van de Heisteeg, N. Kovensky, W. Lerche, F. Marchesano, J. McNamara, L. Melotti, M. Montero, T. Ortín, E. Palti, J. Parra-Mart\'inez, A. Uranga,  C. Vafa, I. Valenzuela and M. Wiesner for useful discussions and correspondence. A.C. would like to thank the Institut de Physique Th\'eorique CEA/Saclay, as well as the Simons Center for Geometry and Physics, Stony Brook University for hospitality and support during the last stages of this work. This work is supported through the grants CEX2020-001007-S, PGC2018-095976-B-C21 and PID2019-108892RB-I00, funded by MCIN/AEI/10.13039/501100011033 and by ERDF A way of making Europe. The work of A.C. is supported by the Spanish FPI grant No. PRE2019-089790 and by the Spanish Science and Innovation Ministry through a grant for postgraduate students in the Residencia de Estudiantes del CSIC. The work of A.H. is supported by the ERC Consolidator Grant 772408-Stringlandscape. 
		

\newpage
\appendix
		
\section{Species scales for Type II, Heterotic and Bosonic strings}
\label{ap:speciesST}

In this appendix we want to refine some of the computations performed in section \ref{s:speciesscale} regarding the calculation of the species scale associated to an emergent weakly coupled string limit. The main point here is to show that the approximations taken by neglecting some powers and numerical prefactors are indeed justified (see discussion around eq. \eqref{eq:leveldensity}). In particular, we perform the derivation of the species scale for type II and heterotic superstrings in 10d, as well as for the bosonic string in 26d, by taking a more accurate expression for the asymptotic level density of states. We also discuss the fact that our results seem to be essentially unaltered when including the leading logarithmic corrections to the definition of the species scale in \eqref{species}, as discussed around eq. \eqref{eq:speciesW-1}.
		
\subsection{Type II String Theory in 10d}
\label{ss:typeII10d}
		
Let us start with the refined version of the analysis of section \ref{s:speciesscale} for the case of the superstring of either type IIA or type IIB. To do so, we just need the analytic expression for the level density of the closed superstring, i.e. $d_n^{\text{cl}}$. It is easy to see that the level matching condition imposes the closed-string states at the $n$-th massive level to be given by the tensor product of the would-be open-string states at level $n$ with themselves. Thus, the asymptotic density for closed-string states reads\cite{Green:2012oqa}
\beq
		d_n^{\text{cl}}=(d_n^{\text{op}})^2\, \sim\, n^{-11/2}\, \text{exp}(4\pi \sqrt{2n})\, .
\label{eq:leveldensityclosed}
\eeq
Therefore, we can follow the same steps as in section \ref{s:speciesscale} in order to compute both $\Ns$ and $\LQG$ by taking now into account the refined level density $d_n^{\text{cl}}$. We particularize here to the case of a 10d flat background, since it is there where the exact quantization of the superstring leads to the specific spectrum that we all know and love. Using \eqref{eq:leveldensityclosed} above one can thus compute the total number of string states up to a maximum excitation level $\Ns$
\beq
		\sum_{n=1}^{\Ns} d_n^{\text{cl}}\, \sim\, 8388608\, \sqrt{2} \pi^9\, \Gamma (-9, -4 \sqrt{2} \pi \sqrt{\Ns})\, \sim\, \frac{\text{exp}(4\pi \sqrt{2\Ns})}{2 \sqrt{2} \pi\, \Ns ^5} \left( 1+ \mathcal{O}(\Ns^{-1/2}) \right)\, ,
\label{eq:refinedsumofstates}
\eeq
where $\Gamma(s,z)$ is the (upper) incomplete gamma function, which is defined as follows
\beq
		\Gamma(s,z)=\int_z^{\infty} dt\ t^{s-1}\ e^{-t }\, .
\eeq
Next, upon using the relations \eqref{eq:speciescale} and \eqref{eq:Ns}, one arrives at the analogue of eq. \eqref{eq:maxstringlevel}
\beq
		\left ( \frac{\Mpt}{\Ms}\right )^8 \simeq\, \Ns^4\, \sum_{n=1}^{\Ns} d_n^{\text{cl}}\, \sim\, \frac{\text{exp} ({4\pi \sqrt{2\Ns}})}{2\pi \sqrt{2} \Ns}\, \sim\, \frac{1}{g_s^2}\, ,
\label{eq:maxstringlevel10d}
\eeq
which can be solved for $\Ns$ again explicitly in terms of the Lambert $W$ function (see footnote \ref{fn:Lambert})
\beq\label{eqap:NsmaxtypeII}
		\Ns\, \sim\, \frac{1}{8\pi^2}\ W_{-1} \left [ -\frac{2^{3/4} \sqrt{\pi}}{(\Mpt/\Ms)^4} \right ]^2\, .
\eeq
Let us make a couple of interesting remarks about the new solution found for the maximum excitation level of the string. First, and in contrast to what happens when we ignore from the beginning the polynomial dependence on $n$ in the level density $d_n^{\text{cl}}$, one actually obtains two distinct (real) solutions for eq. \eqref{eq:maxstringlevel10d}. In one of them $N_s$ asymptotes to 0 when we take the limit $\Ms/\Mpt \to 0$, and is given in terms of the principal branch of Lambert $W$ function. This is of course meaningless since for the approximations taken above to be valid and in order to obtain eq. \eqref{eq:maxstringlevel10d} itself, it was necessary to assume that $\Ns \to \infty$ from the start. The second solution, which is the one displayed in eq. \eqref{eqap:NsmaxtypeII}, provides us with a $\Ns$ that diverges asymptotically, and therefore justifies \emph{a posteriori} the approximations taken so far. Moreover, using the relevant expansion for the $W$ function (see eq. \eqref{eq:Wexpansion}) we can give a precise estimation for the asymptotic functional dependence of $\Ns$, which turns out to be still logarithmic
\beq \label{eq:asymptoticNs}
		\sqrt{\Ns}\, \sim\, \frac{1}{2 \sqrt{2} \pi} \left ( 4\ \log(\Mpt/\Ms)- \log\left( \log(\Mpt/\Ms) \right )\right) + \ldots\, .
\eeq
Thus, the conclusions of section \ref{s:speciesscale} regarding the behaviour of both the QG cut-off associated to a fundamental string and the number of oscillators getting light are essentially unaltered, with the logarithmic dependence minimally modified, since the dominant source in the level density $d_n^{\text{cl}}$ is indeed the exponential in $\sqrt{n}$. 
		
\subsection{Heterotic String Theory in 10d}
\label{ss:heterotic10d}
		
Let us proceed now with other non-trivial example in 10d, the case of the heterotic superstring. Again, to perform a similar analysis as for the type II case we just need the asymptotic level density for the spectrum of the string. Thus, we can simply use the formulae available in the literature \cite{Green:2012oqa} or repeat an alternative derivation along the lines of \cite{Heidenreich:2017sim}. We will take the latter approach here for illustrative purposes. Recall that the 10d heterotic string spectrum is determined by the following mass formula
\beq \label{eq:heteroticmass}
		\frac{\alpha'}{4} m^2 = N_L +\frac{1}{2}\mathbf{Q}^2-1=N_R\, ,
\eeq
where $\alpha'$ is the Regge slope parameter and $N_{L, R} \in \mathbb{Z}_{\geq 0}$ denote the oscillator numbers for the left- and right-movers, respectively. These come, respectively, with the degeneracy of the closed bosonic string and open superstring. Therefore, the total number of states at a given mass level $\alpha' m^2/4 = n$ would be
\beq \label{eq:heteroticdensity}
		d^{\text{het}}_n\, =\, \sum_{\mathbf{Q} \in \Gamma}^{\mathbf{Q}^2 \leq 2(n+1)}d_L(n+1-\mathbf{Q}^2/2)\, d_R(n)\, ,
\eeq
with $\Gamma$ being the corresponding even self-dual `internal' lattice of the heterotic string under consideration. In order to estimate the asymptotic level density, we can make use of those associated to the bosonic and open superstring, namely
\beq \label{eq:densitiesheterotic} 
		d_L(n)\, \sim\, n^{-27/4}\, \text{exp}(4 \pi \sqrt{n})\, , \qquad d_R(n)\,\sim\, n^{-11/4}\, \text{exp}(2\pi \sqrt{2n})\, .
\eeq
Therefore, upon approximating the summation above by an integral, we obtain for the heterotic string the following total asymptotic level density of states
\beq 
		d^{\text{het}}_n\, \sim\, d_R(n) \int_{\mathbf{Q}^2 \leq 2n} d^{16}Q\, \frac{e^{4 \pi \sqrt{n+1-\mathbf{Q}^2/2}}}{(n+1-\mathbf{Q}^2/2)^{27/4}}\, \sim\,  d_R(n) \int d^{16}Q\, \frac{e^{4 \pi \sqrt{n} -\frac{\pi}{\sqrt{n}}\mathbf{Q}^2}}{n^{27/4}}\, \sim\, \frac{e^{2 \pi (2+\sqrt{2}) \sqrt{n}}}{n^{11/2}}\, ,
\eeq
where in the second step we have used the fact that the integrand is essentially dominated by those states with $\mathbf{Q}^2 \lesssim \mathcal{O}(\sqrt{n})$. Notice that the level density is very similar to that of the type II superstring (c.f. eq. \eqref{eq:leveldensityclosed}), the only difference being a numerical coefficient in the exponential dependence on $\sqrt{n}$. Therefore, we expect to obtain essentially similar results for the heterotic string when computing the species scale at asymptotic weak coupling, i.e. when $g_h \to 0$. 
		
For instance, the total number of string excitations up to some (large) level $N_{\text{het}}$ would be 
\beq\label{eq:refinedsumofstatesheterotic}
		\sum_{n=1}^{N_{\text{het}}} d_n^{\text{het}}\, \sim\, 1024\, (2+\sqrt{2})^9 \pi^9\, \Gamma (-9, -2(2+ \sqrt{2}) \pi \sqrt{N_{\text{het}}})\, \sim\, \frac{\text{exp}(2 \pi (2+\sqrt{2}) \sqrt{N_{\text{het}}})}{(2+ \sqrt{2}) \pi\, N_{\text{het}} ^5}\left( 1+ \mathcal{O}(N_{\text{het}}^{-1/2}) \right)\, ,
\eeq
where we have approximated one more time the summation by an integral.
	
Next, upon using the relations \eqref{eq:speciescale} and \eqref{eq:Ns}, one arrives at
\beq
		\left ( \frac{\Mpt}{\Ms}\right )^8 \simeq\, N_{\text{het}}^4\, \sum_{n=1}^{N_{\text{het}}} d_n^{\text{het}}\, \sim\, \frac{\text{exp} (2 \pi (2+\sqrt{2}) \sqrt{N_{\text{het}}})}{(2+ \sqrt{2}) \pi\, N_{\text{het}}}\, \sim\, \frac{1}{g_h^2}\, ,
\label{eq:maxstringlevelhet}
\eeq
which in terms of Lambert $W$ function leads to
\beq
		N_{\text{het}}\, \sim\, \frac{1}{2(3+2 \sqrt{2})\pi^2}\, W_{-1} \left [ -\frac{\sqrt{(2+\sqrt{2}) \pi}}{(\Mpt/\Ms)^4} \right ]^2\, .
		\eeq
Again, we can give a precise estimation of the functional dependence of $N_{\text{het}}$ by means of the asymptotic expansion of Lambert $W$ function
\beq \label{eq:asymptoticNshet}
		\sqrt{N_{\text{het}}}\, \sim\, \frac{1}{\sqrt{2(3+2 \sqrt{2})} \pi} \left ( 4\ \log(\Mpt/\Ms)- \log \left(\log(\Mpt/\Ms) \right) \right) + \ldots\, .
\eeq

\subsection{Bosonic String Theory}
\label{ss:bosonic26d}
		
Finally, let us also study for completeness the case of the bosonic string theory in 26d. Recall that we have already used the asymptotic level density of states for the open bosonic string when performing the analysis in the heterotic case above (see eq. \eqref{eq:densitiesheterotic}). Taking into account the level matching condition of the closed string, one finds for the level density the following functional form\cite{Green:2012oqa}
\beq
		d_n^{\text{cl}}=(d_n^{\text{op}})^2\, \sim\, n^{-27/2}\ \text{exp}(8\pi \sqrt{n})\, .
\label{eq:leveldensityclosedbos}
\eeq
Hence, starting from this expression one can compute the total number of string states up to a maximum excitation level denoted again by $\Ns$
\beq
		\sum_{n=1}^{\Ns} d_n^{cl}\, \sim\, 75557863725914323419136\, \pi^{25}\, \Gamma (-25, -8 \pi \sqrt{\Ns})\, \sim\, \frac{\text{exp}(8\pi \sqrt{\Ns})}{4 \pi\, \Ns ^{13}}\left( 1+ \mathcal{O}(\Ns^{-1/2}) \right)\, ,
\label{eq:refinedsumofstatesbos}
\eeq
Next, upon using the relations \eqref{eq:speciescale} and \eqref{eq:Ns}, one arrives at
\beq
		\left ( \frac{M_{\text{pl, 26}}}{\Ms}\right )^{24} \simeq\, \Ns^{12}\ \sum_{n=1}^{\Ns} d_n^{\text{cl}}\, \sim\, \frac{\text{exp} ({8\pi \sqrt{\Ns}})}{4\pi \Ns}\, \sim\, \frac{1}{g_s^2}\, ,
\label{eq:maxstringlevel26d}
\eeq
which can be solved for $\Ns$ in terms of Lambert $W$ function
\beq
		\Ns\, \sim\, \frac{1}{16\pi^2}\ W_{-1} \left [ -\frac{2 \sqrt{\pi}}{(M_{\text{pl, 26}}/\Ms)^{12}} \right ]^2\, .
\eeq
This allows us once again to give a precise estimation of the functional dependence of $\Ns$ by means of the asymptotic expansion of Lambert $W$ function
\beq \label{eq:asymptoticNsbos}
		\sqrt{\Ns}\, \sim\, \frac{1}{4 \pi} \left ( 12\ \log(M_{\text{pl, 26}}/\Ms)- \log \left(\log(M_{\text{pl, 26}}/\Ms) \right) \right) + \ldots\, .
\eeq

\subsection{Logarithmic corrections to the species bound}
	
To close this appendix, we briefly elaborate in the following on the leading logarithmic corrections to the species bound and whether they could potentially modify some of the results regarding the computation of the species scale associated to critical strings. Recall that such corrections, motivated either from the perturbative corrections to the exact graviton propagator or the non-perturbative BH decay (see section \ref{s:speciesscale}) essentially amount to some kind of $\log N$ correction in the defining expression for the species bound, eq. \eqref{species}. Hence, by taking into account these modifications one finds that e.g. for type II superstring, the maximum excitation level $\Ns$ below the species scale (c.f. eq. \eqref{eq:maxstringlevel10d}) has the expression 
\beq
		\left ( \frac{\Mpt}{\Ms}\right )^8 \simeq\, \frac{\Ns^4}{8}\, \sum_{n=1}^{\Ns} d_n^{\text{cl}} \left [- W_{-1} \left ( \frac{-8}{\sum_{n=1}^{\Ns} d_n^{\text{cl}}}\right) \right]\, \sim\, \frac{\text{exp} ({4\pi \sqrt{2\Ns}})}{4 \sqrt{\Ns}}\, ,
\label{eq:newmaxstringlevel10d}
\eeq
where the last equality holds asymptotically and to leading order in $\Ns$. This implies that the functional dependence of $\Ns$ remains essentially unchanged, and therefore our results do not seem to be a mirage of the approximated expression eq. \eqref{species} frequently used for the definition of the species scale when applied to weak-coupling string limits.
		
\section{Loop calculations}
\label{ap:Loops}
		
In this appendix we perform the detailed computations of the one-loop diagrams presented in section \ref{s:EmergenceQG}, giving the general results as well as the leading asymptotic expansions in all the relevant regimes. In particular, we will focus on the diagrams that contribute to the wave-function renormalization of a scalar, a 1-form gauge field and a Weyl spinor coming from loops of massive scalars and fermions. To do so, we compute the amputated one-loop Feynman diagram corresponding to each of these processes, shown in figs. \ref{fig:scalarpropagator}-\ref{fig:kineticfermionsbas}. As an important remark, even though we discuss our set-up for the different relevant cases in Lorentzian spacetimes, when performing any such loop calculation we will analytically continue our integrals so as to work with Euclidean signature instead, which simplifies our analysis considerably.
		
\subsection{Self-energy of a modulus}
\label{ap:Loopsscalar}
		
Let us begin by considering a real modulus, $\phi$, coupled to massive (real) scalars, $\sigma^{(n)}$, or Dirac fermions, $\psi^{(n)}$, through their mass terms as follows
\begin{align}
			S_{\mathrm{kin,} \phi}\, &=\, -\dfrac{1}{2} \int   d\phi \wedge \star d\phi \, , \label{eq:Skinphi} \\
			S_{\sigma^{(n)}} \, &=\,  - \dfrac{1}{2} \int   \left( d\sigma^{(n)} \wedge \star d\sigma^{(n)}\,  +\,  m_n(\phi)^2 \sigma^{(n)} \sigma^{(n)} \right) \star 1\, , \label{eq:Ssigman}\\
			S_{\psi^{(n)}} \, &= \,  \int  \left( i \overline{\psi^{(n)}}\,  \slashed{\partial} \, \psi^{(n)} \, -\,  m_n(\phi) \, \overline{\psi^{(n)}}\psi^{(n)} \right) \star 1\, . \label{eq:Spsin}
\end{align}
We keep in mind that the label $n \in \mathbb{N}$ will eventually denote the step in the tower in which either the scalars $\sigma^{(n)}$ or the fermions $\psi^{(n)}$ are organized, with their masses $m_n(\phi)$ increasing accordingly, but for now the computation is meant to be quite general. In the context of Emergence, we are interested in the computation of the wave-function renormalization of the scalar field $\phi$ in $d$ spacetime dimensions due to scalar and fermionic loops. The idea is thus to extract the momentum-dependent part of the exact propagator of our massless modulus $\phi$ at $\mathcal{O}(\hbar)$ in the Wilsonian effective action after integrating out the heavy fields, which takes the form
\beq\label{eq:exactpropscalar}
		D(p^2)=\frac{1}{p^2-\Pi(p^2)}\, ,
\eeq
after deforming the contour of integration and analytically extending the results to Euclidean signature, i.e. $\Bar{g}_{\mu \nu}= \delta_{\mu \nu}$ (see e.g. \cite{Nair:2005iw}). Here, $\Pi(p^2)$ corresponds to the (amputated) one-loop Feynman diagram displayed in fig \ref{fig:scalarpropagator}.
		
\subsubsection*{Scalar loop}
		
Let us begin by considering the contribution due to a loop of scalars $\sigma^{(n)}$, which is shown in fig. \ref{fig:scalarloopscalar} and  reads (taking into account the overall $1/2$ symmetry factor of the diagram)
\beq
		\Pi_n(p^2) \ = \frac{\lambda_n^2}{2} \int \frac {\text{d}^dq}{(2\pi)^d} \frac {1}{(q^2+m_n^2)} \frac {1}{\left((q-p)^2+m_n^2\right)}\, ,
\label{eq:selfenergyscalar(ap)}
\eeq
with the coupling $\lambda_n=2m_n(\partial_\phi m_n)$ coming from the trilinear vertex arising after expanding the mass term in eq. \eqref{eq:Ssigman} around the modulus VEV at linear order. Since we are interested in the correction to the propagator, we need to extract the term proportional to $p^2$, so that we take a derivative with respect to $p^2$ and evaluate the result at $p=0$ to obtain\footnote{Notice that naively one would also obtain a term proportional to $1/|p|$ after taking the derivative with respect to $p^2$, but this would correct the linear term in the momentum expansion, which can be seen to be absent when the detailed computation is performed (as required by Lorentz invariance).} 
\beq
		\frac {\partial \Pi_n(p^2)}{\partial p^2} \bigg\rvert_{p=0}  = - \frac{\lambda_n^2}{2} \int \frac {\text{d}^dq}{(2\pi)^d} \frac {1}{(q^2+m_n^2)^3}\, .
\label{eq:sigmaa_ap}
\eeq
From this expression we expect the integral to be divergent for $d\geq 6$ and convergent otherwise. However, since we will always keep in mind the idea of introducing the UV cut-off associated to QG, namely the species scale, we perform the momentum integral up to a maximum scale $\Lambda$, which yields the following general expression
\beq
		\begin{split}
			\frac{\partial \Pi_n(p^2)}{\partial p^2} \bigg\rvert_{p=0}  \, =  \, - \lambda_n^2  \ \frac{ \pi^{d/2}  }{8\,  (2 \pi)^d \, \Gamma(d/2) } \ \frac{\Lambda^d}{m_n^6} \ & \left[ - \frac{(d-6)m_n^4+(d-4)m_n^2\Lambda^2}{(\Lambda^2+m_n^2)^2} + \right.  \\
			& \quad \left.  + \left(d+ \frac{8}{d}-6\right) \ _2{\cal F}_1\left( 1,\frac{d}{2};\frac{d+2}{2}; -\frac{\Lambda^2}{m_n^2}\right) \right]  \, ,
		\end{split}
\label{eq:scalarloopscalarexact}
\eeq
with $_2{\cal F}_1(a,b;c;d)$ the ordinary (or Gaussian) hypergeometric function. Given the kind of towers that we are dealing with (c.f. section \ref{s:speciesscale}), the two relevant asymptotic limits for this expression are $\Lambda \gg m_n$ (for most states of KK-like towers) and $\Lambda \simeq m_n$ (for most states of stringy towers). In order to study each of these limits in turn, we will also distinguish between $d>6$, $d=6$ and $d<6$, given that the divergence of the corresponding expressions in the large $\Lambda$ limit is different for these three cases.
		
Let us begin by considering the limit, $\Lambda \gg m_n$, which dominates the contributions coming from KK-like towers. In this case, the integral diverges polynomially  with $\Lambda$ for $d>6$ as
\beq
		\frac {\partial \Pi_n^{(d>6)}(p^2)}{\partial p^2} \bigg\rvert_{p=0}\, = \, - \lambda_n^2 \ \frac{\pi^{d/2}}{(2 \pi)^d\ \Gamma\left( d/2 \right) \ (d-6)}   \ \Lambda^{d-6} \, + \, \mathcal{O}\left(\Lambda^{d-8}\, m_n^2\right) + \mathrm{const.} \ ,
\label{eq:scalarloopscalarsd>6}
\eeq
such that the leading term goes like $\Lambda^{d-6}$. For $d<6$ one can expand eq. \eqref{eq:scalarloopscalarexact} to obtain
\beq
		\frac {\partial \Pi_n^{(d<6)}(p^2)}{\partial p^2} \bigg\rvert_{p=0}\, = \, - \lambda_n^2 \ \frac{\pi^{\frac{d+2}{2}}}{16 \ (2 \pi)^d  \ \Gamma\left( d/2 \right)} \frac{(d-2)(d-4)}{\sin\left( d \pi/2\right)} \ \frac{1}{m_n^{6-d}} + \, \mathcal{O}\left(\frac{1}{\Lambda^{6-d}}\right)  \ .
\label{eq:scalarloopscalarsd<6}
\eeq
		\begin{table}[t]\begin{center}
				\renewcommand{\arraystretch}{2.00}
				\begin{tabular}{|c||c|c|c|c|c|}
					\hline
					$d$ & 2 & 3 & 4 & 5 & 6 \\
					\hline 
					$\dfrac {\partial \Pi_n}{\partial p^2} \bigg\rvert_{p=0}$ &
					$-\dfrac{1}{16 \pi}\dfrac{\lambda_n^2}{m_n^4}$ & 
					$ -\dfrac{1}{64 \pi  }\dfrac{\lambda_n^2}{m_n^3}$ &
					$ -\dfrac{1}{64 \pi^2}\dfrac{\lambda_n^2}{m_n^2}$ & 
					$-\dfrac{9}{128 \pi^2}\dfrac{\lambda_n^2}{m_n}$ &
					$ -\dfrac{\lambda_n^2 }{256 \pi ^3} \log \left(\dfrac{\Lambda ^2}{m_n^2}\right) $ \\
					\hline 
					\hline
					$d$ &  7 & 8 & 9 & 10 & 11\\
					\hline 
					$\dfrac {\partial \Pi_n}{\partial p^2} \bigg\rvert_{p=0}$ &
					$  -\dfrac{\lambda_n^2  \, \Lambda }{240 \pi^4}  $ &
					$-\dfrac{\lambda_n^2  \, \Lambda^2  }{3072 \pi^4} $ &
					$ -\dfrac{\lambda_n^2  \, \Lambda^3  }{10080 \pi^5}$ &
					$ -\dfrac{\lambda_n^2  \, \Lambda^4   }{98304 \pi^5}$ &
					$  -\dfrac{\lambda_n^2  \, \Lambda^5 }{302400 \pi^6} $  \\
					\hline
				\end{tabular}
				\caption{Leading contribution to the wave-function renormalization of a modulus field due to a loop of massive scalars, as given by eq. \eqref{eq:scalarloopscalarexact}, in the limit $\Lambda\gg m_n$ for different number of spacetime dimensions $2 \leq d \leq 11$.}
				\label{tab:scalarloopscalarLambda>>m}\end{center}
		\end{table}  
Note that the leading term here is actually the one denoted as `const.' in eq. \eqref{eq:scalarloopscalarsd>6}, which was irrelevant there but gives instead the leading correction for $d<6$. Additionally, notice that the piece $\frac{(d-2)(d-4)}{\sin\left( d \pi/2\right)}$ must be defined as a limit for some integers $d$, and it takes a value of $4/\pi$ for $d=2,\, 4$, the value $3$ for $d=1,\, 5$, and a value of $1$ for $d=3$. Let us also remark that for $d<6$ the loop integral is convergent such that, at the QFT level, no UV cut-off (or UV regulator whatsoever) is actually necessary. Finally, for the marginal case, $d=6$, we get the expected leading logarithmic correction
\beq
		\frac {\partial \Pi_n^{(d=6)}(p^2)}{\partial p^2} \bigg\rvert_{p=0} \, =\, -  \frac{\lambda_n^2 }{256 \pi^3 }\log \left( \frac{\Lambda^2}{m_n^2} \right) \ + \ \mathcal{O}\left( \Lambda^0 \right)  \, .
\label{eq:scalarloopscalard=6}
\eeq
A summary of the relevant leading term for different numbers of dimensions in this limit can be found in table \ref{tab:scalarloopscalarLambda>>m}.
		
Consider now the the alternative limiting case, $\Lambda \simeq m_n$, which gives an upper bound for the states whose contribution to the loop must be included. Notice that this is the dominant contribution for towers of stringy modes (up to logarithmic corrections). In this case, we can expand eq. \eqref{eq:scalarloopscalarexact} for any $d$ and the expression reads
\beq
		\begin{split}
			\frac {\partial \Pi_n(p^2)}{\partial p^2} \bigg\rvert_{p=0} \, = \, -  \frac{\lambda_n^2 \  \pi^{d/2}}{32 \ (2 \pi)^d\ \Gamma(d/2) }   &\left\{ 10-2d+(d-2)(d-4) \left[\psi\left( \frac{d+2}{4}\right)-  \psi\left( \frac{d}{4} \right)  \right] \right\} \Lambda^{d-6}   \\
			& \quad + \mathcal{O}(\Lambda-m_n)   \, ,
		\end{split}
\label{eq:scalarloopscalarLambda=m}
\eeq
where $\psi(z)$ represents the digamma function.\footnote{The digamma function, $\psi(z)$, is defined as the logarithmic derivative of the familiar gamma function $\Gamma(z)$ with respect to its argument, namely
\beq
			\notag \psi(z) = \frac{d}{dz} \log \left( \Gamma(z)\right) = \frac{\Gamma'(z)}{\Gamma(z)}\, .
\eeq
} Notice that, since in this limit $\Lambda \simeq m_n$ we recover the same leading asymptotic dependence with $\Lambda$ and $m_n$ than the one in eqs. \eqref{eq:scalarloopscalarsd>6}-\eqref{eq:scalarloopscalard=6}. The precise form of the leading term for different number of spacetime dimensions is summarized in table \ref{tab:scalarloopscalarLambda=m}.
		\begin{table}[t]\begin{center}
				\renewcommand{\arraystretch}{2.00}
				\begin{tabular}{|c||c|c|c|c|c|}
					\hline
					$d$ & 2 & 3 & 4 & 5 & 6 \\
					\hline 
					$\dfrac {\partial \Pi_n}{\partial p^2} \bigg\rvert_{p=0}$ &
					$-\dfrac{3}{64 \pi}\dfrac{\lambda_n^2}{m_n^4}$ & 
					$ -\dfrac{1}{128 \pi}\dfrac{\lambda_n^2}{m_n^3}$ &
					$ -\dfrac{1}{256 \pi^2}\dfrac{\lambda_n^2}{m_n^2}$ & 
					$-\dfrac{(3\pi-8)}{768 \pi^3}\dfrac{\lambda_n^2}{m_n}$ &
					$ -\dfrac{8\log(2)-5}{2048 \pi ^3} \lambda_n^2  $ \\
					\hline 
					\hline
					$d$ &  7 & 8 & 9 & 10 & 11\\
					\hline 
					$\dfrac {\partial \Pi_n}{\partial p^2} \bigg\rvert_{p=0}$ &
					
					$\begin{aligned}[t] -\tfrac{(16-5\pi) }{2560 \pi^4} \times  & \\  \lambda_n^2  \, \Lambda & \end{aligned}$ &
					$\begin{aligned}[t]-\tfrac{(17-24\log(2)) }{24576 \pi^4} \times & \\ \lambda_n^2  \, \Lambda^2 & \end{aligned}$  &
					$ \begin{aligned}[t]-\tfrac{(105\pi-328)}{322560 \pi^5} \times & \\ \lambda_n^2  \, \Lambda^3 & \end{aligned}$ &
					$ \begin{aligned}[t]-\tfrac{(16\log(2)-11)  }{131072 \pi^5} \times & \\ \lambda_n^2  \, \Lambda^4 & \end{aligned}$ &
					$ \begin{aligned}[t] -\tfrac{(992-315\pi) }{9676800 \pi^6} \times & \\ \lambda_n^2  \, \Lambda^5 & \end{aligned}$  \\
					\hline
				\end{tabular}
				\caption{Leading contribution to the wave-function renormalization of a modulus field due to a loop of massive scalars, given by eq. \eqref{eq:scalarloopscalarexact}, in the limit $\Lambda \simeq m_n$ for different number of spacetime dimensions $2 \leq d \leq 11$.}
				\label{tab:scalarloopscalarLambda=m}\end{center}
		\end{table}  

Let us remark that the leading asymptotic dependence with the relevant energy scale (i.e. with the UV cut-off or the mass of the particle running in the loop) is the same for the two limiting cases, $\Lambda \gg m_n$ and $\Lambda\simeq m_n$, whilst only the numerical $\mathcal{O}(1)$ prefactors differ between the two expressions. Thus, since these two limits bound the contribution of a particle to the loop, we can safely use the given asymptotic dependence on $\Lambda$ or $m_n$ in order to calculate the field dependent contribution of the towers to the relevant kinetic terms at the level of emergence, since only order one factors may change slightly (which we are not sensitive to in any event).

\subsubsection*{Fermionic loop}
		
We now consider the contribution to the scalar metric from a loop of fermions, with a coupling induced by the mass term of the fermion as specified in the action \eqref{eq:Spsin}. The calculation is similar to the scalar loop above, and the corresponding Feynman diagram, displayed in fig. \ref{fig:scalarloopfermion}, gives the following contribution
\begin{equation}\label{eq:selfenergyfermion(ap)}
			\begin{aligned}
				\Pi_n(p^2) \ &= -\mu_n^2  \int \frac {\text{d}^dq}{(2\pi)^d}\ \text{tr} \left (\frac {1}{i \slashed{q}+m_n}\ \frac {1}{i (\slashed{q}-\slashed{p})+m_n} \right) = \\
				&= -\mu_n^2  \int \frac {\text{d}^dq}{(2\pi)^d} \text{tr} \left ( \frac{(-i \slashed{q} + m_n)(-i(\slashed{q}-\slashed{p})+m_n)}{(q^2+m_n^2)((q-p)^2+m_n^2)} \right)\ . 
			\end{aligned}
\end{equation}
Here, the relevant coupling constant coming from the trilinear coupling that arises after expanding the mass term in the action is $\mu_n = \partial_\phi m_n(\phi)$, and notice that there is an extra minus sign with respect to eq. \eqref{eq:selfenergyscalar(ap)} due to the fact that the particle is of fermionic nature. By recalling that the dimensionality of the Dirac matrices in $d$ spacetime dimensions is $\fdim$ (where $\lfloor x \rfloor$ denotes the largest integer less than or equal to $x$), and using the following identities about their traces,
\begin{equation}
			\label{eq:scalarloopfermionstraces}
			\text{tr} \left(\gamma^{\mu} \right)=\ 0\,  , \qquad
			\text{tr} \left(\gamma^{\mu} \gamma^{\nu}\right)=\ \fdim \delta^{\mu \nu}\ ,
\end{equation}
we can explicitly perform the trace in eq. \eqref{eq:selfenergyfermion(ap)}, which leads to
\begin{equation}
			\text{tr}\left\{ (-i \slashed{q} + m_n)(-i(\slashed{q}-\slashed{p})+m_n) \right\} \, =\, -\fdim (q^2 - p\cdot q -m_n^2)\, .
\end{equation}
Thus, by extracting the part that is linear in $p^2$ we arrive to
\begin{equation}\label{eq:fermionloopddim}
			\frac{\partial \Pi_n(p^2)}{\partial p^2} \bigg\rvert_{p=0} \, = \,   -\mu_n^2\, \fdim  \int \frac {\text{d}^dq}{(2\pi)^d} \frac{1}{(q^2+m_n^2)^2} \ + \ 2 m_n^2 \, \mu_n^2\ \fdim \int \frac {\text{d}^dq}{(2\pi)^d} \frac{1}{(q^2+m_n^2)^3} \, ,
\end{equation}
where we have used the fact that some terms quadratic in $q$ cancel identically between themselves and that the terms linear in $q$ vanish after integration along the angular directions. Notice that the second piece is exactly the same as the contribution from $\fdim$ real scalars as considered above (recall that $\lambda_n= 2 m_n (\partial_\phi m_n)=2 m_n \mu_n$), but with \emph{opposite} sign. Thus, we can use all the results from our previous computations in order to to evaluate the its exact contribution, and in the case in which the number of fermionic degrees of freedom equals the bosonic ones (as e.g. in supersymmetric set-ups) there is an exact cancellation between these two pieces. The first term in eq. \eqref{eq:fermionloopddim}, however, has a different (although similar) structure, and it is expected to be divergent for $d\geq 4$. Its precise form after imposing a UV cut-off ($\Lambda$) for the momentum integral is therefore
\begin{equation}
			\frac{\partial \Pi_n(p^2)}{\partial p^2} \bigg\rvert_{p=0}  \, =  \, - \mu_n^2  \ \frac{ \fdim \pi^{d/2}  }{  (2 \pi)^d \, \Gamma(d/2) } \ \frac{\Lambda^d}{m_n^4} \  \left[  \frac{m_n^2}{\Lambda^2+m_n^2}  + \left(\frac{2}{d}-1\right) \ _2{\cal F}_1\left( 1,\frac{d}{2};\frac{d+2}{2}; -\frac{\Lambda^2}{m_n^2}\right) \right]  \, .
\label{eq:scalarloopfermionexact}
\end{equation}
		\begin{table}[t]\begin{center}
				\renewcommand{\arraystretch}{2.00}
				\begin{tabular}{|c||c|c|c|c|c|}
					\hline
					$d$ & 2 & 3 & 4 & 5 & 6 \\
					\hline 
					$\dfrac {\partial \Pi_n}{\partial p^2} \bigg\rvert_{p=0}$ &
					$-\dfrac{1}{2 \pi}\dfrac{\mu_n^2}{m_n^2}$ & 
					$ -\dfrac{1}{4 \pi  }\dfrac{\mu_n^2}{m_n}$ &
					$ -\dfrac{\mu_n^2 }{4 \pi^2 } \log \left(\dfrac{\Lambda ^2}{m_n^2}\right)$ & 
					$ -\dfrac{\mu_n^2  \, \Lambda }{3 \pi^2}$ &
					$ -\dfrac{\mu_n^2  \, \Lambda^2 }{16 \pi^3} $ \\
					\hline 
					\hline
					$d$ &  7 & 8 & 9 & 10 & 11\\
					\hline 
					$\dfrac {\partial \Pi_n}{\partial p^2} \bigg\rvert_{p=0}$ &
					$  -\dfrac{\mu_n^2  \, \Lambda^3 }{45 \pi^4}  $ &
					$-\dfrac{\mu_n^2  \, \Lambda^4  }{192 \pi^4} $ &
					$ -\dfrac{\mu_n^2  \, \Lambda^5  }{525 \pi^5}$ &
					$ -\dfrac{\mu_n^2  \, \Lambda^6   }{2304 \pi^5}$ &
					$  -\dfrac{\mu_n^2  \, \Lambda^7 }{6615 \pi^6} $  \\
					\hline
				\end{tabular}
				\caption{Leading contribution to the wave-function renormalization of a modulus field due to a loop of massive fermions, as given by eq. \eqref{eq:scalarloopfermionexact}, in the limit $\Lambda\gg m_n$ for different number of spacetime dimensions $2 \leq d \leq 11$.}
				\label{tab:scalarloopfermionLambda>>m}\end{center}
		\end{table}  
Now, in the limit $\Lambda \gg m_n$, which as we said is particularly relevant for most states in a KK-like tower, the leading contribution from the fermionic loop to the propagator for $d>4$ takes the form
\begin{equation}
			\frac {\partial \Pi_n^{(d>4)}(p^2)}{\partial p^2} \bigg\rvert_{p=0}\, = \, -\mu_n^2\,  \frac{2^{\lfloor \frac{d+2}{2} \rfloor} \pi^{d/2}}{(2 \pi)^d\ \Gamma\left( d/2 \right) \ (d-4)}   \ \Lambda^{d-4} \, + \, \mathcal{O}\left(\Lambda^{d-6}\, m_n^2\right) + \mathrm{const.} \ ,
			\label{eq:scalarloopfermionsd>4}
\end{equation}
which is very similar to the scalar contribution \eqref{eq:scalarloopscalarsd>6} but with a different power for the cut-off. Similarly, for $d<4$ the dominant piece (which corresponds to the `const.' piece in the previous expansion) reads\footnote{Notice that in the context of the Swampland Program one typically studies EFTs in $d\geq 4$, but we also include here the results in lower dimensions for completeness.}
\begin{equation}
			\frac {\partial \Pi_n^{(d<4)}(p^2)}{\partial p^2} \bigg\rvert_{p=0}\, = \, - \mu_n^2 \ \frac{2^{\lfloor\frac{d-2}{2} \rfloor} \ \pi^{\frac{d+2}{2}} }{(2 \pi)^d  \ \Gamma\left( d/2 \right)} \ \frac{(2-d)}{\sin\left( d \pi/2\right)} \ \frac{1}{m_n^{4-d}} + \, \mathcal{O}\left(\frac{1}{\Lambda^{4-d}}\right)  \ ,
\label{eq:scalarloopfermionsd<4}
\end{equation}
where once again for $d=2$ the quotient $\frac{(2-d)}{\sin\left( d \pi/2\right)}$ is defined as a limit and takes a value of $2/ \pi$. For the marginal case, we recover the expected logarithmic divergence
\beq
		\frac {\partial \Pi_n^{(d=4)}(p^2)}{\partial p^2} \bigg\rvert_{p=0} \, =\, -  \frac{\mu_n^2 }{4 \pi^2 }\log \left( \frac{\Lambda^2}{m_n^2} \right) \ + \ \mathcal{O}\left( \Lambda^0 \right)  \, .
\label{eq:scalarloopfermionsd=6}
\eeq
The precise leading contributions for the relevant values of $d$ are summarized in table \ref{tab:scalarloopfermionLambda>>m}.
		
Taking now the other relevant limit, namely $\Lambda \simeq m_n$, we can similarly expand eq. \eqref{eq:scalarloopfermionexact} to obtain the following expression
\begin{equation}
			\frac {\partial \Pi_n(p^2)}{\partial p^2} \bigg\rvert_{p=0} \, = \, - \mu_n^2 \ \frac{2^{\lfloor\frac{d}{2}-2 \rfloor}  \pi^{d/2}}{ (2 \pi)^d\ \Gamma(d/2) }   \left\{ 2+(d-2)\left[\psi\left( \frac{d}{4}\right)-  \psi\left( \frac{d+2}{4} \right)  \right] \right\} \Lambda^{d-4} + \, \mathcal{O}(\Lambda-m_n)   \, .
\label{eq:scalarloopfermionLambda=m}
\end{equation}
As in the scalar case, since we have $\Lambda\simeq m_n$ the asymptotic dependence with the relevant scale is the same as the one in the $\Lambda \gg m_n$ limit, and only the numerical prefactors change. The relevant leading terms for $2\leq d \leq 11$ are outlined in table \ref{tab:scalarloopfermionLambda=m}.
		\begin{table}[t]\begin{center}
				\renewcommand{\arraystretch}{2.00}
				\begin{tabular}{|c||c|c|c|c|c|}
					\hline
					$d$ & 2 & 3 & 4 & 5 & 6 \\
					\hline 
					$\dfrac {\partial \Pi_n}{\partial p^2} \bigg\rvert_{p=0}$ &
					$-\dfrac{1}{4 \pi}\dfrac{\mu_n^2}{m_n^2}$ & 
					$ -\dfrac{(\pi -2)}{8 \pi^2  }\dfrac{\mu_n^2}{m_n}$ &
					$ -\frac{(2\log(2)-1)}{8 \pi^2 } \mu_n^2 $ & 
					$ \begin{aligned}[t]-\tfrac{(10-3\pi)}{24 \pi^3}\times &\\ \mu_n^2  \, \Lambda & \end{aligned}$ &
					$ \begin{aligned}[t]-\tfrac{(3-4\log(2))}{32 \pi^3}\times &\\ \mu_n^2  \, \Lambda^2 & \end{aligned}$ \\
					\hline 
					\hline
					$d$ &  7 & 8 & 9 & 10 & 11\\
					\hline 
					$\dfrac {\partial \Pi_n}{\partial p^2} \bigg\rvert_{p=0}$ &
					$ \begin{aligned}[t] -\tfrac{ (15\pi-46)}{360 \pi^4}\times &\\  \mu_n^2  \, \Lambda^3 & \end{aligned}$ &
					$\begin{aligned}[t]-\tfrac{(3\log(2)-2)}{96 \pi^4}\times &\\ \mu_n^2  \, \Lambda^4  & \end{aligned}$ &
					$ \begin{aligned}[t]-\tfrac{(334-105\pi) }{12600 \pi^5}\times &\\ \mu_n^2  \, \Lambda^5 & \end{aligned}$ &
					$ \begin{aligned}[t]-\tfrac{ (17-24\log(2)) }{4608 \pi^5}\times &\\ \mu_n^2  \, \Lambda^6 & \end{aligned}$ &
					$ \begin{aligned}[t] -\tfrac{(315\pi-982) }{264600 \pi^6}\times &\\  \mu_n^2  \, \Lambda^7 & \end{aligned}$  \\
					\hline
				\end{tabular}
				\caption{Leading contribution to the wave-function renormalization of a modulus field due to a loop of massive fermions, as given by eq. \eqref{eq:scalarloopfermionexact}, in the limit $\Lambda\simeq m_n$ for different number of spacetime dimensions $2 \leq d \leq 11$.}
				\label{tab:scalarloopfermionLambda=m}\end{center}
		\end{table}  

\subsection{Self-energy of a gauge 1-form}
\label{ap:Loops1-form}
		
We consider now a 1-form, $A_\mu$, with field strength $F_{\mu \nu }\, =\,2\  \partial_{[\mu} A_{\nu]}$, coupled to massive (complex) scalars, $\chi^{(n)}$, or fermions, $\psi^{(n)}$, through the following action
\begin{align}
			S_{\mathrm{kin,} A_1}\, &= \, -\dfrac{1}{4\, g^2} \int d^d x \sqrt{-g} \   F_{\mu\nu} F^{\mu \nu}\, , \label{eq:SkinA1} \\
			S_{\sigma^{(n)}} \, &=\,  - \dfrac{1}{2} \int d^d x \sqrt{-g} \ \left(  D_\mu \chi^{(n)} \overline{D^\mu \chi^{(n)}}\,  +\,  m_n^2 \ \chi^{(n)} \overline{\chi^{(n)}} \right) \, , \label{eq:SchinA1}\\
			S_{\psi^{(n)}} \, &= \,  \int d^d x \sqrt{-g} \   \left( i \overline{\psi^{(n)}}\,  \slashed{D} \, \psi^{(n)} \, -\,  m_n \ \overline{\psi^{(n)}}\psi^{(n)} \right) \, . \label{eq:SpsinA1}
\end{align}
Here, the overline denotes complex conjugation for the scalars as well as Dirac conjugation for the fermions, whilst $D_\mu$ represents the appropriate covariant derivative of the fields minimally coupled to $A_1$, defined as 
\begin{equation}
			D_\mu \chi^{(n)} \, = \, \left(\partial_\mu -i q_n A_\mu \right) \chi^{(n)} \, , \qquad D_\mu \psi^{(n)} \, = \, \left(\partial_\mu -i q_n A_\mu \right) \psi^{(n)}\, .
\end{equation}
Once again, we are interested in the corrections to the propagator of $A_1$ in $d$ spacetime dimensions induced by quantum loops from integrating out heavy scalar and fermion fields. As a remark, we will not elaborate on the subtleties associated to gauge invariant regularization, which are made manifest specially when imposing a UV cut-off, $\Lambda$.\footnote{Notice that a similar problem arises upon imposing a UV cut-off at or below the Planck scale in a gravitational theory, since doing so is a priori inconsistent with diffeomorphism invariance. This becomes evident e.g. when we compactify such theory on a circle, where the gauge invariances of the spin-2 KK massive modes mix between each other, such that simply retaining a finite number of them explicitly breaks the gauge symmetries\cite{Dolan:1983aa,Duff:1989ea}.} Let us just mention that gauge invariance in the presence of a hard cut-off can be ensured rigorously (see e.g. \cite{Costello,Strassler:1992zr}), but we will take a pragmatic approach here by focusing only on the dependence of the required amplitudes with $\Lambda$, instead of watching carefully that the correct tensorial structure is maintained even at the quantum level (which is of course related to the preservation of gauge invariance). To do so, we use the Lorenz gauge (i.e. $\partial_\mu A^\mu=0$), since it can also be easily generalized to arbitrary $p$-form gauge fields. The propagator then takes the form (on a flat background with Euclidean metric $\Bar{g}_{\mu \nu}= \delta_{\mu \nu}$)\footnote{\label{fn:FeynmantHooftgauge}Strictly speaking, in order to fix the tensorial structure of the propagator as in eq. \eqref{eq:A1propagatorapp}, one has to impose additionally the Feynman-`t Hooft gauge, which is an instance of the more general $R_{\xi}$-gauges, with $\xi$ fixed to be equal to 1.} 
\begin{equation}
\label{eq:A1propagatorapp}
			D^{\mu \nu} (p^2) \, = \, \left( \dfrac{p^2}{g^2} \delta ^{\mu \nu} - \Pi^{\mu \nu}(p^2) \right)^{-1}\, ,
\end{equation}
where $\Pi^{\mu \nu}$ is zero at tree level, and gives the amputated Feynman diagram from the loops shown in fig. \ref{fig:1-formpropagator}. By using again our gauge choice, we can extract the tensorial dependence as follows 
%
\begin{equation}
			\label{eq:A1loopamplitudeapp}
			\Pi^{\mu \nu} (p^2) \, = \, \Pi(p^2) \delta ^{\mu \nu} \, .
\end{equation}
We are thus interested in extracting the piece proportional to $p^2$ within $\Pi(p^2)$, as arising from the aforementioned loop corrections.
		
\subsubsection*{Scalar loop}
		
We begin by considering the coupling of the 1-form to a complex scalar, $\chi^{(n)}$, with mass $m_n$ and charge $q_n$, as given by the action \eqref{eq:SchinA1}. The relevant one-loop Feynamn diagram is shown in fig. \ref{fig:1-formloopscalar}, and it reads
\beq
		\Pi^{\mu \nu}_n(p) \, =\,   g^2 \, q_n^2 \int \frac {\text{d}^dq}{(2\pi)^d} \frac {(2q-p)^{\mu} (2q-p)^{\nu}}{(q^2+m_n^2)\left( (q-p)^2+m_n^2\right)} \, .
\label{eq:A1scalar}
\eeq
From all the terms in the numerator, we only need to keep track of the ones $\propto q^\mu q^\nu$. The reason being that the ones proportional to $p^\mu p^\nu$ amount essentially to a change of gauge, which as we argued is not important for our purposes here, whilst the ones linear in $q^\mu$ instead turn out to either cancel identically or produce also linear terms in $q^\mu$ after taking the derivative with respect to $p^2$ and setting $p$ to zero, which then also cancel after the angular integration (due to Lorentz invariance). Moreover, we can explicitly use the the angular integration (or Lorentz invariance) to replace
\begin{equation}
\label{eq:qmuqnuaverage}
			q^\mu q^\nu \ \longrightarrow \ \dfrac{q^2}{d} \, \delta^{\mu \nu}\, ,
\end{equation}
under the integral over all $q$. Notice that this gives rise at the end of the day to the tensor structure announced in \eqref{eq:A1loopamplitudeapp}. Thus, the precise form of the relevant piece of the amputated Feynman diagram yields
\begin{equation}\label{eq:1-formscalarloopprop}
			\frac{\partial \Pi^{\mu \nu}_n(p^2)}{\partial p^2} \bigg\rvert_{p=0} \, = \, -g^2\,  q_n^2\,   \frac{4}{d} \, \delta^{\mu\nu} \, \int \dfrac{d^d q}{(2\pi)^d} \dfrac{q^2}{(q^2+m_n^2)^3} \, .
\end{equation}
As happened with the modulus case, we expect this integral to diverge depending on the dimension of the spacetime our theory lives in. In particular, it seems to diverge for $d\geq4$, but we will introduce a cut-off for any $d$ since at the end of the day we are interested in integrating up to a physical UV cut-off beyond which our EFT weakly coupled to Einstein gravity stops being valid. The exact expression gives therefore
\beq
		\begin{split}
			\frac{\partial \Pi_n(p^2)}{\partial p^2} \bigg\rvert_{p=0}   =  \, - g^2 \, q_n^2  \ \frac{ \pi^{d/2}  }{d\,  (2 \pi)^d \, \Gamma(d/2) } \ \frac{\Lambda^{d+2}}{m_n^6} \ & \left[ - \frac{(d-4)m_n^4+(d-2)m_n^2\Lambda^2}{(\Lambda^2+m_n^2)^2} + \right.  \\
			& \quad \left.  + \dfrac{d(d-2)}{d+2} \ _2{\cal F}_1\left( 1,\frac{d+2}{2};\frac{d+4}{2}; -\frac{\Lambda^2}{m_n^2}\right) \right]  \, ,
		\end{split}
\label{eq:1-formloopscalarexact}
\eeq
where $\Pi_n(p^2)$ captures the part of the diagram after extracting the tensorial piece (c.f. eq. \eqref{eq:A1loopamplitudeapp}).
		
In analogy with the massless scalar case, the two relevant asymptotic limits that we take for this expression are $\Lambda \gg m_n$ (for most states of KK-like towers) and $\Lambda \simeq m_n$ (for most states of stringy towers). In the limit, $\Lambda \gg m_n$, the integral diverges polynomially with $\Lambda$ for $d>4$ as 
\begin{equation}
			\frac {\partial \Pi_n^{(d>4)}(p^2)}{\partial p^2} \bigg\rvert_{p=0} = \, -g^2 \, q_n^2\  \frac{8\ \pi^{d/2}}{(2 \pi)^d\ \Gamma\left( d/2 \right) \ d\, (d-4)}    \ \Lambda^{d-4} \, + \, \mathcal{O}\left(\Lambda^{d-6}\, m_n^2\right) + \mathrm{const.} \ ,
\label{eq:1formloopfermionsd>4}
\end{equation}
whereas in lower dimensions it is convergent and the leading contribution is given by
\begin{equation}
			\frac {\partial \Pi_n^{(d<4)}(p^2)}{\partial p^2} \bigg\rvert_{p=0} = \, - g^2 \, q_n^2 \ \frac{\pi^{\frac{d+2}{2}} }{2 \, (2 \pi)^d  \ \Gamma\left( d/2 \right)} \ \frac{(2-d)}{\sin\left( d \pi/2\right)} \ \frac{1}{m_n^{4-d}} + \, \mathcal{O}\left(\frac{1}{\Lambda^{4-d}}\right)  \ ,
\label{eq:1formloopfermionsd<4}
\end{equation}
with the quotient $(2-d)/\sin\left( d \pi/2\right)$ defined as a limit with value $2/ \pi$ for $d=2$. Finally, for the marginal case, we get the expected logarithmic behaviour familiar from (scalar) QED
\beq
		\frac {\partial \Pi_n^{(d=4)}(p^2)}{\partial p^2} \bigg\rvert_{p=0} \, =\, -  \frac{g^2 \, q_n^2 }{16 \pi^2 }\log \left( \frac{\Lambda^2}{m_n^2} \right) \ + \ \mathcal{O}\left( \Lambda^0 \right)  \, .
\label{eq:1formloopscalard=4}
\eeq
These precise leading contributions for different values of $d$ are summarized in table \ref{tab:1-formloopscalarLambda>>m}.
		\begin{table}[t]\begin{center}
				\renewcommand{\arraystretch}{2.00}
				\begin{tabular}{|c||c|c|c|c|c|}
					\hline
					$d$ & 2 & 3 & 4 & 5 & 6 \\
					\hline 
					$\dfrac {\partial \Pi_n}{\partial p^2} \bigg\rvert_{p=0}$ &
					$-\dfrac{1}{4 \pi}\dfrac{g^2 \, q_n^2}{m_n^2}$ & 
					$ -\dfrac{1}{8 \pi  }\dfrac{g^2 \, q_n^2}{m_n}$ &
					$ -\dfrac{g^2 \, q_n^2 }{16 \pi^2 } \log \left(\dfrac{\Lambda ^2}{m_n^2}\right)$ & 
					$ -\dfrac{g^2 \, q_n^2  \, \Lambda }{15 \pi^2}$ &
					$ -\dfrac{g^2 \, q_n^2  \, \Lambda^2 }{192 \pi^3} $ \\
					\hline 
					\hline
					$d$ &  7 & 8 & 9 & 10 & 11\\
					\hline 
					$\dfrac {\partial \Pi_n}{\partial p^2} \bigg\rvert_{p=0}$ &
					$  -\dfrac{g^2 \, q_n^2  \, \Lambda^3 }{630 \pi^4}  $ &
					$-\dfrac{g^2 \, q_n^2  \, \Lambda^4  }{6144 \pi^4} $ &
					$ -\dfrac{g^2 \, q_n^2  \, \Lambda^5  }{18900 \pi^5}$ &
					$ -\dfrac{g^2 \, q_n^2  \, \Lambda^6   }{184320 \pi^5}$ &
					$  -\dfrac{g^2 \, q_n^2  \, \Lambda^7 }{582120 \pi^6} $  \\
					\hline
				\end{tabular}
				\caption{Leading contribution to the wave-function renormalization of a gauge 1-form due to a loop of massive charged complex scalars, as given by eq. \eqref{eq:1-formloopscalarexact}, in the limit $\Lambda\gg m_n$, for different number of spacetime dimensions $2 \leq d \leq 11$. }
				\label{tab:1-formloopscalarLambda>>m}\end{center}
		\end{table}  
In the other relevant limit, namely when $\Lambda \simeq m_n$, the expansion of eq. \eqref{eq:1-formloopscalarexact} produces instead 
\begin{equation}
			\begin{split}
				\frac {\partial \Pi_n(p^2)}{\partial p^2} \bigg\rvert_{p=0}  =  - g^2 q_n^2 \ \frac{ \pi^{d/2}}{ 4 d\, (2 \pi)^d\ \Gamma(d/2) }  & \left\{ 2(d-3) +d(d-2)\left[\psi\left( \frac{d+2}{4}\right)-  \psi\left( \frac{d+4}{4} \right)  \right] \right\} \Lambda^{d-4} \\
				& \quad +  \mathcal{O}(\Lambda-m_n)   \, .
			\end{split}
\label{eq:1-formloopscalarLambda=m}
\end{equation}
The precise values for this expression in different number of spacetime dimensions are summarized in table \ref{tab:1-formloopscalarLambda=m}.
		\begin{table}[t]\begin{center}
				\renewcommand{\arraystretch}{2.00}
				\begin{tabular}{|c||c|c|c|c|c|}
					\hline
					$d$ & 2 & 3 & 4 & 5 & 6 \\
					\hline 
					$\dfrac {\partial \Pi_n}{\partial p^2} \bigg\rvert_{p=0}$ &
					$-\dfrac{1}{16 \pi}\dfrac{g^2 \, q_n^2}{m_n^2}$ & 
					$ -\dfrac{(3\pi -8)}{48 \pi^2  }\dfrac{g^2 \, q_n^2}{m_n}$ &
					$ \begin{aligned}[t]-\tfrac{(8\log(2)-5)}{128 \pi^2 } \times &\\g^2 \, q_n^2 &\end{aligned}$  & 
					$ \begin{aligned}[t]-\tfrac{(16-5\pi)}{160 \pi^3}\times &\\ g^2 \, q_n^2  \, \Lambda & \end{aligned}$ &
					$ \begin{aligned}[t]-\tfrac{( 17-24\log(2))}{1536 \pi^3}\times &\\ g^2 \, q_n^2  \, \Lambda^2 & \end{aligned}$ \\
					\hline 
					\hline
					$d$ &  7 & 8 & 9 & 10 & 11\\
					\hline 
					$\dfrac {\partial \Pi_n}{\partial p^2} \bigg\rvert_{p=0}$ &
					$ \begin{aligned}[t] -\tfrac{ (105\pi-328)}{20160 \pi^4}\times &\\  g^2 \, q_n^2  \, \Lambda^3 & \end{aligned}$ &
					$\begin{aligned}[t]-\tfrac{(16\log(2)-11)}{8192 \pi^4}\times &\\ g^2 \, q_n^2  \, \Lambda^4  & \end{aligned}$ &
					$ \begin{aligned}[t]-\tfrac{(992-315\pi ) }{604800 \pi^5}\times &\\ g^2 \, q_n^2  \, \Lambda^5 & \end{aligned}$ &
					$ \begin{aligned}[t]-\tfrac{ (167-240\log(2)) }{1474560 \pi^5}\times &\\ g^2 \, q_n^2  \, \Lambda^6 & \end{aligned}$ &
					$ \begin{aligned}[t] -\tfrac{(385\pi-1208) }{10348800 \pi^6}\times &\\  g^2 \, q_n^2  \, \Lambda^7 & \end{aligned}$  \\
					\hline
				\end{tabular}
				\caption{Leading contribution to the wave-function renormalization of a gauge 1-form due to a loop of massive charged complex scalars, given by eq. \eqref{eq:1-formloopscalarexact}, in the limit $\Lambda \simeq m_n$, for different number of spacetime dimensions $2 \leq d \leq 11$.}
				\label{tab:1-formloopscalarLambda=m}\end{center}
		\end{table}  

\subsubsection*{Fermionic loop}
		
Let us consider now the effect of the coupling of the 1-form to a spin-$\frac{1}{2}$ fermion, $\psi^{(n)}$, with mass $m_n$ and charge $q_n$ (c.f. action \eqref{eq:SpsinA1}. The corresponding one-loop Feynamn diagram is displayed in fig. \ref{fig:1-formloopfermion}, and it takes the form
\begin{equation}\label{eq:A1selfenergyfermion}
			\begin{aligned}
				\Pi^{\mu\nu}_n(p^2) \ &= - (ig)^2 \, q_n^2  \int \frac {\text{d}^dq}{(2\pi)^d}\ \text{tr} \left (\frac {1}{i \slashed{q}+m_n}\ \gamma^\mu \ \frac {1}{i (\slashed{q}-\slashed{p})+m_n} \ \gamma^\nu  \right) = \\
				&= g^2\ q_n^2  \int \frac {\text{d}^dq}{(2\pi)^d} \text{tr} \left ( \frac{(-i \slashed{q} + m_n)\, \gamma^\mu \, (-i(\slashed{q}-\slashed{p})+m_n)\, \gamma^\nu }{(q^2+m_n^2)((q-p)^2+m_n^2)} \right)\ .  
			\end{aligned}
\end{equation}
In order to perform the traces of the numerator we make use of eq. \eqref{eq:scalarloopfermionstraces}, as well as
\begin{equation}
			\text{tr} \left(\gamma^{\mu} \gamma^{\nu} \gamma^{\rho} \gamma^{\sigma}\right)= \fdim \left( \delta^{\mu \nu}\delta^{\rho \sigma}-\delta^{\mu \rho}\delta^{\nu \sigma}+\delta^{\mu \sigma}\delta^{\rho \nu} \right)\ ,
\label{eq:traceidentityfour}
\end{equation}
to obtain
\begin{equation}
\label{eq:traces1-formloop}
			\begin{split}
				\text{tr}& \left ( (-i \slashed{q} + m_n)\, \gamma^\mu \, (-i(\slashed{q}-\slashed{p})+m_n)\, \gamma^\nu \right)\, = \\
				& \qquad \qquad \qquad  =\, \fdim \, \left( \delta^{\mu \nu} \, (q^2 - q\cdot p + m_n^2)- 2\,  q^\mu q^\nu+p^\mu q^\nu+p^\nu q^\mu  \right) \, . 
			\end{split}
\end{equation}
By differentiating with respect to $p^2$ so as to select the piece that contributes to the propagator, and after taking into account the fact that  linear and cubic  terms in $q^\mu$ yield zero after performing the angular integration over $q$, together with the identical cancellations of the term proportional to $(q\cdot p)$ and eq. \eqref{eq:qmuqnuaverage}, we get
\begin{equation}\label{eq:1formfermionloopcompleteexpression}
			\frac{\partial \Pi^{\mu \nu}_n(p^2)}{\partial p^2} \bigg\rvert_{p=0}  = \, -\fdim\, g^2\,  q_n^2 \, \delta^{\mu\nu}  \int \dfrac{d^d q}{(2\pi)^d} \dfrac{1}{(q^2+m_n^2)^2}\, + \,  \fdim\,  g^2\,  q_n^2\,   \frac{2}{d} \, \delta^{\mu\nu}  \int \dfrac{d^d q}{(2\pi)^d} \dfrac{q^2}{(q^2+m_n^2)^3}\, .
\end{equation}
Similarly to the modulus case, the second term has the same form as the scalar contribution but with an opposite sign. In fact, by taking into account that we are now considering a complex scalar, with two real degrees of freedom, it can be seen that in the presence of an equal number of fermionic and bosonic degrees of freedom with identical mass and charge, the cancellation between the scalar contribution and this second term from the fermions would be exact. Therefore, the precise expression for this term (along with its asymptotic expansions) can be easily obtained from eqs. \eqref{eq:1-formloopscalarexact}-\eqref{eq:1-formloopscalarLambda=m} by simply multiplying by a factor of $-\fdim/2$.
		
Let us now focus on the first term. Notice that, after extracting the tensorial structure, it gives exactly the same contribution as \eqref{eq:fermionloopddim} upon substituting $\mu_n^2 \,  \to \,  g^2 \, q_n^2$. Hence, we can use, \emph{mutatis mutandis}, the corresponding formulae from the modulus section, that we summarize here for completeness. By introducing a UV cut-off, $\Lambda$, and performing the integral the first term in eq. \eqref{eq:1formfermionloopcompleteexpression} reads
\begin{equation}
			\frac{\partial \Pi_n(p^2)}{\partial p^2} \bigg\rvert_{p=0}  \, =  \, - g^2\, q_n^2  \ \frac{ \fdim \pi^{d/2}  }{  (2 \pi)^d \, \Gamma(d/2) } \ \frac{\Lambda^d}{m_n^4} \  \left[  \frac{m_n^2}{\Lambda^2+m_n^2}  + \left(\frac{2}{d}-1\right) \ _2{\cal F}_1\left( 1,\frac{d}{2};\frac{d+2}{2}; -\frac{\Lambda^2}{m_n^2}\right) \right]  \, .
\label{eq:1-formloopfermionexact}
\end{equation}
		\begin{table}[t]\begin{center}
				\renewcommand{\arraystretch}{2.00}
				\begin{tabular}{|c||c|c|c|c|c|}
					\hline
					$d$ & 2 & 3 & 4 & 5 & 6 \\
					\hline 
					$\dfrac {\partial \Pi_n}{\partial p^2} \bigg\rvert_{p=0}$ &
					$-\dfrac{1}{2 \pi}\dfrac{g^2\, q_n^2}{m_n^2}$ & 
					$ -\dfrac{1}{4 \pi  }\dfrac{g^2\, q_n^2}{m_n}$ &
					$ -\dfrac{g^2\, q_n^2 }{4 \pi } \log \left(\dfrac{\Lambda ^2}{m_n^2}\right)$ & 
					$ -\dfrac{g^2\, q_n^2  \, \Lambda }{3 \pi^2}$ &
					$ -\dfrac{g^2\, q_n^2  \, \Lambda^2 }{16 \pi^3} $ \\
					\hline 
					\hline
					$d$ &  7 & 8 & 9 & 10 & 11\\
					\hline 
					$\dfrac {\partial \Pi_n}{\partial p^2} \bigg\rvert_{p=0}$ &
					$  -\dfrac{g^2\, q_n^2  \, \Lambda^3 }{45 \pi^4}  $ &
					$-\dfrac{g^2\, q_n^2  \, \Lambda^4  }{192 \pi^4} $ &
					$ -\dfrac{g^2\, q_n^2  \, \Lambda^5  }{525 \pi^5}$ &
					$ -\dfrac{g^2\, q_n^2  \, \Lambda^6   }{2304 \pi^5}$ &
					$  -\dfrac{g^2\, q_n^2  \, \Lambda^7 }{6615 \pi^6} $  \\
					\hline
				\end{tabular}
				\caption{Leading contribution to the wave-function renormalization of a gauge 1-form due to a loop of massive charged fermions, as given by eq. \eqref{eq:1-formloopfermionexact}, in the limit $\Lambda\gg m_n$ for different number of spacetime dimensions $2 \leq d \leq 11$.}
				\label{tab:1-formloopfermionLambda>>m}\end{center}
		\end{table}  
In the limit $\Lambda \gg m_n$, the leading piece from the fermionic loop to the propagator when $d>4$ is
\begin{equation}
			\frac {\partial \Pi_n^{(d>4)}(p^2)}{\partial p^2} \bigg\rvert_{p=0}\, = \, -g^2\, q_n^2\,  \frac{2^{\lfloor \frac{d+2}{2} \rfloor} \pi^{d/2}}{(2 \pi)^d\ \Gamma\left( d/2 \right) \ (d-4)}   \ \Lambda^{d-4} \, + \, \mathcal{O}\left(\Lambda^{d-6}\, m_n^2\right) + \mathrm{const.}\, .
\label{eq:1-formloopfermionsd>4}
\end{equation}
Similarly, for $d<4$, the dominant contribution takes the form
\begin{equation}
			\frac {\partial \Pi_n^{(d<4)}(p^2)}{\partial p^2} \bigg\rvert_{p=0}\, = \, - g^2\, q_n^2 \ \frac{2^{\lfloor\frac{d-2}{2} \rfloor} \ \pi^{\frac{d+2}{2}} }{(2 \pi)^d  \ \Gamma\left( d/2 \right)} \ \frac{(2-d)}{\sin\left( d \pi/2\right)} \ \frac{1}{m_n^{4-d}} + \, \mathcal{O}\left(\frac{1}{\Lambda^{4-d}}\right)\, ,
\label{eq:1-formloopfermionsd<4}
\end{equation}
with the quotient $\frac{(2-d)}{\sin\left( d \pi/2\right)}$  defined as a limit with value $2/ \pi$ for $d=2$. For the marginal case,  the expected logarithmic divergence is obtained
\beq
		\frac {\partial \Pi_n^{(d=4)}(p^2)}{\partial p^2} \bigg\rvert_{p=0} \, =\, -  \frac{g^2\, q_n^2 }{4 \pi^2 }\log \left( \frac{\Lambda^2}{m_n^2} \right) \ + \ \mathcal{O}\left( \Lambda^0 \right)\, .
\label{eq:1-formloopfermionsd=}
\eeq
These results are summarized in table \ref{tab:1-formloopfermionLambda>>m}.
		
Taking instead the limit $\Lambda \simeq m_n$ in eq. \eqref{eq:1-formloopfermionexact}, one arrives at
\begin{equation}
			\frac {\partial \Pi_n(p^2)}{\partial p^2} \bigg\rvert_{p=0} \, = \, - g^2\, q_n^2 \ \frac{2^{\lfloor\frac{d}{2}-2 \rfloor}  \pi^{d/2}}{ (2 \pi)^d\ \Gamma(d/2) }   \left\{ 2+(d-2)\left[\psi\left( \frac{d}{4}\right)-  \psi\left( \frac{d+2}{4} \right)  \right] \right\} \Lambda^{d-4} + \, \mathcal{O}(\Lambda-m_n)\, .
\label{eq:1-formloopfermionLambda=m}
\end{equation}
As happened with the scalar modulus before, since we have $\Lambda\simeq m_n$ the asymptotic dependence with the relevant scale coincides with the $\Lambda \gg m_n$ limit, and only the numerical prefactors change. The relevant leading terms for $2\leq d \leq 11$ are shown in table \ref{tab:1-formloopfermionLambda=m}.
		\begin{table}[t]\begin{center}
				\renewcommand{\arraystretch}{2.00}
				\begin{tabular}{|c||c|c|c|c|c|}
					\hline
					$d$ & 2 & 3 & 4 & 5 & 6 \\
					\hline 
					$\dfrac {\partial \Pi_n}{\partial p^2} \bigg\rvert_{p=0}$ &
					$-\dfrac{1}{4 \pi}\dfrac{g^2\, q_n^2}{m_n^2}$ & 
					$ -\dfrac{(\pi -2)}{8 \pi^2  }\dfrac{g^2\, q_n^2}{m_n}$ &
					$ -\frac{(2\log(2)-1)}{8 \pi^2 } g^2\, q_n^2 $ & 
					$ \begin{aligned}[t]-\tfrac{(10-3\pi)}{24 \pi^3}\times &\\ g^2\, q_n^2  \, \Lambda & \end{aligned}$ &
					$ \begin{aligned}[t]-\tfrac{(3- 4\log(2))}{32 \pi^3}\times &\\ g^2\, q_n^2  \, \Lambda^2 & \end{aligned}$ \\
					\hline 
					\hline
					$d$ &  7 & 8 & 9 & 10 & 11\\
					\hline 
					$\dfrac {\partial \Pi_n}{\partial p^2} \bigg\rvert_{p=0}$ &
					$ \begin{aligned}[t] -\tfrac{ (15\pi- 46)}{360 \pi^4}\times &\\  g^2\, q_n^2  \, \Lambda^3 & \end{aligned}$ &
					$\begin{aligned}[t]-\tfrac{(3\log(2)-2)}{96 \pi^4}\times &\\ g^2\, q_n^2  \, \Lambda^4  & \end{aligned}$ &
					$ \begin{aligned}[t]-\tfrac{(334-105\pi) }{12600 \pi^5}\times &\\ g^2\, q_n^2  \, \Lambda^5 & \end{aligned}$ &
					$ \begin{aligned}[t]-\tfrac{ (17-24\log(2)) }{4608 \pi^5}\times &\\ g^2\, q_n^2  \, \Lambda^6 & \end{aligned}$ &
					$ \begin{aligned}[t] -\tfrac{(315\pi -982) }{264600 \pi^6}\times &\\  g^2\, q_n^2  \, \Lambda^7 & \end{aligned}$  \\
					\hline
				\end{tabular}
				\caption{Leading contribution to the wave-function renormalization of a gauge 1-form due to a loop of massive charged fermions, as given by eq. \eqref{eq:1-formloopfermionexact}, in the limit $\Lambda\simeq m_n$ for different number of spacetime dimensions $2 \leq d \leq 11$.}
				\label{tab:1-formloopfermionLambda=m}\end{center}
		\end{table} 

\subsection{Self-energy of a Weyl fermion}
\label{ap:LoopsWeylfermion}
	
To close up this appendix, we will consider a \emph{chiral} (i.e. we restrict to even-dimensional spacetimes) spin-$\frac{1}{2}$ field, $\chi$, coupled to massive (complex) scalars, $\phi^{(n)}$ \emph{and} Dirac fermions, $\Psi^{(n)}$, through the following Yukawa-like interactions
\begin{equation}\label{eq:fermion&bosonYukawas(ap)}
			Y_n\ \overline{\phi^{(n)}} \left(\psi^{(n)}\chi \right)\, ,  
\end{equation}
where $Y_n$ denotes the coupling constant and $n \in \mathbb{Z} \setminus \lbrace0\rbrace$ labels the massive fields. We also use $\psi^{(n)}$ to denote the Weyl fermion of the same chirality as $\chi$, which pairs up with its charge conjugate (say the one labeled by $-n$) so as to form the aforementioned (massive) Dirac spin-$\frac{1}{2}$ field, i.e. $\Psi^{(n)}=\left(\psi^{(n)}, \overline{\psi^{(-n)}} \right)^{\text{T}}$. 
		
In the following, and for simplicity, we will use Dirac fermions all along so as to perform the relevant loop integrals. Therefore, in order to take into account that the massless field $\chi$ is chiral, we define a new Dirac fermion, $\mathcal{X}$, which reduces to $\chi$ upon using the familiar chirality projector $P_{-}=\frac{1}{2}(1-\gamma^{d+1})$, i.e. $\chi = P_{-}\ \mathcal{X}$. $\gamma^{d+1}$ is nothing but the proper generalization of the four-dimensional $\gamma^5$ to $d$ dimensions.
\beq
		\gamma^{d+1}=-i^{\frac{d}{2}-1} \prod_{\mu=0}^{d} \gamma^{\mu}\, .
\eeq
With all this in mind, it is easy to see that the interaction \eqref{eq:fermion&bosonYukawas(ap)} above can be written in terms of $\Psi^{(n)}, \mathcal{X}$ as follows
\begin{equation}\label{eq:interactionsfermion2(ap)}
			Y_n\ \phi^{(n)} \left( \overline{\Psi^{(n)}} P_{-} \mathcal{X} \right) + \text{h.c.}\, .  
\end{equation}
The idea then is to extract again the momentum-dependent part of the exact propagator of our massless fermion $\chi$ at $\mathcal{O}(\hbar)$ in the effective action, which after analytically extending to Euclidean signature reads formally
\beq\label{eq:Euclexactpropagator(ap)}
		S(\slashed{p})= \frac{1}{i\slashed{p}}\, P_{-} + \frac{1}{i\slashed{p}}\, P_{-}\, \left(i\Sigma(\slashed{p})\right)\, \frac{1}{i\slashed{p}}\, P_{-} + \ldots\, ,
\eeq
where the fermion self-energy $i\Sigma(\slashed{p})$ corresponds in this case to the (amputated) one-loop Feynman graph displayed in fig. \ref{fig:kineticfermionsbas}. (Notice that this is nothing but the fermionic analogue of $\Pi(p^2)$ in eq. \eqref{eq:exactpropscalar}.) 
		
\subsubsection*{Loop computation}
		
We will concentrate on the first diagram\footnote{The analysis involving massive vectors as in fig. \ref{fig:kineticfermionsbas}(b) should give us analogous results.} in fig. \ref{fig:kineticfermionsbas} involving Dirac fermions $\Psi^{(n)}$ of mass $m_n^{{\text{f}}}$ as well as complex bosonic scalars $\phi^{(n)}$ with mass given by $m_n^{{\text{b}}}$, which reads
\beq
		i\Sigma_n(\slashed{p}) \ = |Y_n|^2 \int \frac{\text{d}^dq}{(2\pi)^d} \frac{P_{-} \left( -i\slashed{q} + m_n^{\text{f}}\right) P_{+}}{q^2+(m_n^{\text{f}})^2} \frac {1}{(q-p)^2+(m_n^{{\text{b}}})^2}\, ,
\label{eq:selfenergychi(ap)}
\eeq
where the projection operators $P_{\pm}$ arise from the Feynman rules associated to the interaction \eqref{eq:interactionsfermion2(ap)}. There are several interesting things to notice before moving on with the loop computation. First, and due to the anti-commutation properties between $\gamma^{d+1}$ and the $\gamma^{\mu}$ (namely $\lbrace \gamma^{\mu}, \gamma^{d+1} \rbrace=0$), the operators $P_{\pm}$ project out the term proportional to $m_n^{{\text{f}}}$ in the numerator of eq. \eqref{eq:selfenergychi(ap)} above whilst keeping the one $\propto \slashed{q}$. This ultimately translates into the fact that the self-energy provides no net contribution at $\mathcal{O}(\hbar)$ for the mass of the chiral field $\chi$.\footnote{This is actually ensured to be true at all orders in perturbation theory due to the chirality of the fermionic field $\chi(x)$.} We also notice that the self-energy includes the projector $P_{+}$, as it should since it is associated to the \emph{chiral} massless fermion, $\chi$.
		
Thus, in order to extract the wave-function renormalization one needs to focus on the piece in the self-energy linear in $p$. Therefore, one can mimic the discussion in the preceding sections by taking a derivative with respect to $p^{\mu}$, and then evaluating the resulting expression at $p=0$. Doing so one finds
\beq
		\begin{aligned}\label{eq:Euclwavefunctionfermion(ap)}
			\frac{\partial \Sigma_n(\slashed{p})}{\partial p^{\mu}} \bigg\rvert_{p=0} = \frac{-2 |Y_n|^2 \delta_{\mu \nu} \gamma^{\nu}\ P_{+}}{d} \int \frac{\text{d}^dq}{(2\pi)^d} \frac{q^2}{\left[ q^2 + (m_n^{{\text{f}}})^2 \right] \left[ q^2 + (m_n^{{\text{b}}})^2 \right]^2}\, .
		\end{aligned}
\eeq
Notice that this has the correct sign so as to renormalize the wave-function of the massless fermion appropriately in eq. \eqref{eq:Euclexactpropagator(ap)}.

Now, in order to study the kind of corrections induced by the above diagram, we will first specialize to the easier case in which both towers present identical masses, namely when $m_n^{{\text{b}}} = m_n^{{\text{f}}} = m_n$. One is thus lead to perform the following integral in momentum space (after introducing a UV cut-off, $\Lambda$), which we already encountered in section \ref{ap:Loops1-form} before (c.f. eq. \eqref{eq:1-formscalarloopprop})
\beq\label{eq:momentumintegralfermionsamemasses}
		\begin{aligned}
			\frac{\partial \Sigma_n(\slashed{p})}{\partial p^{\mu}} \bigg\rvert_{p=0} = \frac{-2 |Y_n|^2 \delta_{\mu \nu} \gamma^{\nu}\ P_{+}}{d} \int_{|q| \leq \Lambda} \frac{\text{d}^dq}{(2\pi)^d} \frac{q^2}{\left(q^2 + m_n^2 \right)^3} \, .
		\end{aligned}
\eeq
Of course, this is not a coincidence, since one place in which the kind of diagrams that we studied in this section naturally appears is in supersymmetric gauge theories, see discussion in section \ref{sss:emergencefermion} in the main text. The behaviour of such integral depends, among various things, on the ratio $\Lambda/m_n$ as well as the spacetime dimension, $d$. For, concreteness, let us show in here the explicit results for the case in which $\Lambda/m_n \gg 1$. For $d>4$, the integral diverges polynomially as 
\begin{equation}
			\frac{\partial \Sigma_n^{(d>4)}(\slashed{p})}{\partial p^{\mu}} \bigg\rvert_{p=0} = \, -|Y_n|^2 \delta_{\mu \nu} \gamma^{\nu}\ P_{+}\  \frac{4\ \pi^{d/2}}{(2 \pi)^d\ \Gamma\left( d/2 \right) \ d\, (d-4)}    \ \Lambda^{d-4} \, + \, \mathcal{O}\left(\Lambda^{d-6}\right) + \mathrm{const.}\, ,
\label{eq:fermionloopd>4}
\end{equation}
whereas in lower dimensions it is convergent and the leading contribution is given by
\begin{equation}
			\frac{\partial \Sigma_n^{(d<4)}(\slashed{p})}{\partial p^{\mu}} \bigg\rvert_{p=0} = \, - |Y_n|^2 \delta_{\mu \nu} \gamma^{\nu}\ P_{+} \ \frac{\pi^{\frac{d+2}{2}} }{4 \, (2 \pi)^d  \ \Gamma\left( d/2 \right)} \ \frac{(2-d)}{\sin\left( d \pi/2\right)} \ \frac{1}{m_n^{4-d}} + \, \mathcal{O}\left(\frac{1}{\Lambda^{4-d}}\right)\, ,
\label{eq:fermionloopd<4}
\end{equation}
with the quotient $(2-d)/\sin\left( d \pi/2\right)$ defined as a limit with value $2/ \pi$ for $d=2$. Finally, for the marginal case, we get the usual logarithmic behaviour
\beq
		\frac{\partial \Sigma_n^{(d=4)}(\slashed{p})}{\partial p^{\mu}} \bigg\rvert_{p=0} \, =\, -  \frac{|Y_n|^2 \delta_{\mu \nu} \gamma^{\nu}\ P_{+}}{32 \pi^2 }\, \log \left( \frac{\Lambda^2}{m_n^2} \right) \ + \ \mathcal{O}\left( \Lambda^0 \right)\, .
\label{eq:fermionloopd=4}
\eeq

Let us come back to the more general expression, i.e. eq. \eqref{eq:Euclwavefunctionfermion(ap)}, in which we take the states running in the loop to have different masses. Performing the momentum integral we arrive at the analogue of eq. \eqref{eq:momentumintegralfermionsamemasses} 
\begin{equation}
			\begin{aligned}
				\frac{\partial \Sigma_n(\slashed{p})}{\partial p^{\mu}} \bigg\rvert_{p=0} &= \, -|Y_n|^2 \delta_{\mu \nu} \gamma^{\nu}\ P_{+}\  \frac{2\ \pi^{d/2}}{(2 \pi)^d\ \Gamma\left( d/2 \right)\ d} \ \frac{\Lambda^{d+2}}{\left[ (m_n^{{\text{f}}})^3 -m_n^{{\text{f}}} (m_n^{{\text{b}}})^2 \right]^2} \left[  \frac{(m_n^{{\text{f}}})^2\left[(m_n^{{\text{f}}})^2 - (m_n^{{\text{b}}})^2 \right]}{(m_n^{{\text{b}}})^2\ (\Lambda^2+(m_n^{{\text{b}}})^2)} + \right.  \\
				& \quad \left.  + \dfrac{2}{d+2} \ _2{\cal F}_1\left( 1,\frac{d+2}{2};\frac{d+4}{2}; -\frac{\Lambda^2}{(m_n^{{\text{f}}})^2}\right) + \right.  \\
				& \quad \left.  + \dfrac{(m_n^{{\text{f}}})^2\left[(m_n^{{\text{b}}})^2(d-2) - (m_n^{{\text{f}}})^2 d\right]}{(m_n^{{\text{b}}})^4\ (d+2)} \ _2{\cal F}_1\left( 1,\frac{d+2}{2};\frac{d+4}{2}; -\frac{\Lambda^2}{(m_n^{{\text{b}}})^2}\right) \right] \ ,
			\end{aligned}
\label{eq:momentumintegralfermiondiffmasses(ap)}
\end{equation}
which of course reduces to the previous expressions whenever the masses are taken to be equal.

\section{The D$p$-particle content. A Super-Quantum Mechanical Analysis}
\label{ap:Dpbranecontent}
		
In this appendix we provide for some additional support for the conjectured field-theoretic content associated to the towers of D0- as well as D2-branes (and bound states thereof) becoming light upon approaching the strong coupling/large volume singularity of type IIA string theory in ten and four spacetime dimensions, respectively. The basic tool that we employ is a super-quantum mechanical description associated to the effective dynamics of such non-perturbative states. The material here is complementary to some discussions appearing in the main text, specifically in sections \ref{s:emergence4dN=2} and \ref{ss:Emergence10dST}.

\subsection{The 10d D0-brane content}
\label{ss:10dDO}

The analysis of section \ref{sss:IIAstrongcoupling} regarding the strong coupling singularity in 10d type IIA string theory along the lines of the Emergence Proposal rests on the field-theoretic content associated the D0-brane tower. Therefore, as argued in that section, the important point to stress is that any step of the D0-brane tower should contain a massive spin-2 field and 3-form potential in the bosonic sector, as well as a single massive Majorana Rarita-Scwhinger field in the fermionic counterpart.
	
Of course, the main motivation for this fact comes from the conjectured duality between M-theory and type IIA string theory at strong coupling\cite{Witten:1995ex}. Thus, as explained below eq. \eqref{eq:dofcountingIIA}, upon Kaluza-Klein compactification of M-theory on a circle, one can see that by truncating to the zero-mode sector we recover exactly type IIA supergravity (see also Appendix \ref{ap:MtheoryKKcompactS1}), whilst the KK replica of the 11d massless fields exhibit a number of similarities with bound states of D0-branes, e.g. their mass and charge.
	
In what follows, our aim will be to provide for some additional support in favour of this important statement by making use of a simple \emph{super-quantum mechanical} analysis. For this discussion we will closely follow the steps outlined in section 2 of \cite{Sethi:1997pa}. Thus, the main idea is to try to quantize the theory describing the effective low energy dynamics in the presence of a \emph{single} D0-brane (namely the supersymmetric version of the Dirac-Born-Infeld and Chern-Simons action, see eq. \eqref{eq:DBICSaction} below). Notice that, while there are many subtleties associated to such quantization process, in the following we will mostly neglect these complications, and turn instead to a quantum-mechanical analysis of the D0-brane effective action in the so-called static gauge. In this simplified scenario, as it is well-known, the effective QM theory is essentially the dimensional reduction of 10d $\mathcal{N}=1$ Super Yang-Mills (SYM) theory to the one-dimensional worldline of the D0-brane \cite{Witten:1995im}.
	
Therefore, let us take as our starting point the super-QM action obtained by dimensional reduction of 10d $SU(n)$ gauge theory with 16 supercharges. The bosonic degrees of freedom can be seen to be the `transverse' coordinates of the D0-particle, $x^i_a$, for $i=1, \ldots,9$, which take values in the adjoint representation of the gauge algebra,\footnote{Recall that the transverse coordinates appear from the reduction of the gauge field $A_{\mu}$ in the 10d SYM theory\cite{Polchinski:1998rq,Polchinski:1998rr}.} such that $a$ is the gauge index. These scalars along with their conjugate canonical momenta, $p_{j, a}$, satisfy the following commutation relations
\beq
	[x^i_a,p_{j, b}]=i \delta_{ab} \delta^i_j \ .
	\label{eq:bosonicalgebra}
\eeq
For completeness, let us say that the generators of the adjoint representation $t^a$ have been normalized so that $\text{Tr}\,(t^a t^b)=n \delta^{ab}$. Hence, the Hamiltonian for the system adopts the form\cite{Banks:1996vh}
\beq
	H= \frac{1}{2n} \sum_{i=1}^9 \text{Tr}\, (p^i p_i) + V(x^i)+H_F\ ,
	\label{eq:1dHamiltonian}
\eeq
where $V(x^i)$ is a (gauge invariant) potential which depends polynomially on $x^i$ and $H_F$ is some Hamiltonian associated to the fermionic degrees of freedom, which is typically a bilinear in the fermions and depends linearly on the bosonic coordinates. (Below we will specify the form of such potentials for the D0-brane system, see eq. \eqref{eq:D0Hamiltonian}.) Moreover, as happens with e.g. the familiar Maxwell theory, the $A_0$ equation of motion introduces simply a constraint on the system, which in this case is valued in the adjoint of $\mathfrak{su}(n)$, such that it obeys the algebra $[C_a, C_b]=i f^c_{ab} C_c$ and of course commutes both with the Hamiltonian \eqref{eq:1dHamiltonian} and the supercharges, ensuring that if we start from a state verifying the constraints in the initial `time slice', they will still be satisfied at all later times. Thus, the physical Hilbert space for the quantized system is a subspace of the original one, $\mathcal{H}_{\text{phys}} \subset \mathcal{H}$, whose states are annihilated by the constraints, i.e. $\ket{\Psi} \in \mathcal{H}_{\text{phys}}$ iff $C_a \ket{\Psi}=0$.
	
Now, for the specific case of a (system of) D0-brane(s), we should have $\mathcal{N}=16$ supersymmetry in 1d, given that it is a BPS state of the bulk theory and thus preserves half of the original supersymmetries. Hence, in addition to the bosonic degrees of freedom described around eq. \eqref{eq:bosonicalgebra}, the SQM theory also contains 16 \emph{real} Grassman-valued fields which arise from the reduction of the unique Majorana-Weyl spinor in 9d (again transforming in the adjoint of the gauge group). Therefore, such fermionic dof's should transform in the $\textbf{16}$ representation of the $Spin(9)$ flavour or internal group\footnote{Notice that such internal group simply descends from the 10d Lorentz group of the parent SYM theory. The symmetry is generated (in the abelian case) by $J^{ij}=i\psi_{\alpha} (S^{ij})^{\alpha}_{\beta} \psi^{\beta}$ plus a bosonic piece, where $S^{ij}=\frac{1}{4} [\gamma^i,\gamma^j]$ are the spinor generators of the group of spatial rotations in 10d.} (in contrast to the bosonic ones, that transform in the $\textbf{9}_v$ of $SO(9)$, due to spin-statistics in 10d), i.e. our fermions are described by real Grassmann-valued fields, $\psi_{a}^{\alpha}$, where $\alpha=1, \ldots, 16$, with associated $\gamma$-matrices, $\gamma^i_{\alpha \beta}$, satisfying the 9d Clifford algebra
\beq
	\lbrace \gamma^i,\gamma^j\rbrace=2 \delta^{ij} \ .
	\label{eq:cliffordalgebra}
\eeq
Hence, specializing to this QM system, the Hamiltonian \eqref{eq:1dHamiltonian} becomes\cite{Sethi:1997pa}
\beq
	H_{\text{D}0}= \frac{1}{2n} \sum_{i=1}^9 \text{Tr}\, (p^i p_i) - \frac{1}{4n}\sum_{i,j} \text{Tr} \left ( [x^i, x^j]^2 \right ) -\frac{1}{2n} \sum_i \text{Tr}\,(\psi \gamma^i[x_i, \psi])\ ,
	\label{eq:D0Hamiltonian}
\eeq
where upon quantizing the fermions $\psi_a^{\alpha}$, the states in the Hilbert space are forced to group into representations of the $Spin(16)$ algebra
\beq
	\lbrace \psi_a^{\alpha},\psi_b^{\beta}\rbrace= \delta_{ab} \delta^{\alpha \beta} \ ,
	\label{eq:16dcliffordalgebra}
\eeq
on which the fermions act (modulo a factor of $\sqrt{2}$) precisely as Dirac $\gamma$-matrices.
	
For completeness, let us also state here the form of the constraints $C_a$
\beq
	C_a= f_a^{bc} \sum_i (x^i_b p_{i,c})-\frac{i}{2} f_a^{bc}\sum_{\alpha} \psi^{\alpha}_b \psi_{\alpha,c}\ ,
	\label{eq:constraints}
\eeq
as well as that of the supercharges
\beq
	Q_{\alpha}= \frac{1}{n} \sum_i \gamma^i_{\alpha \beta} \text{Tr}\,(\psi^{\beta} p_i)- \frac{i}{4n} \sum_{i,j} \text{Tr} \left([\gamma^i, \gamma^j] \psi [x_i, x_j] \right)_{\alpha}\ ,
	\label{eq:supercharges}
\eeq
which satisfy the susy algebra
\beq
	\lbrace Q_{\alpha},Q_{\beta}\rbrace=2 \delta^{\alpha \beta} H_{\text{D}0} + \sum_{i,a} 2 \gamma^i_{\alpha \beta} x_{i,a} C_a  \ .
	\label{eq:susyalgebra}
\eeq
Notice that the supersymmetry algebra closes on the Hamiltonian iff the constraints are set to zero, meaning that supersymmetry is realized only when acting on the physical Hilbert space $\mathcal{H}_{\text{phys}}$ \cite{Sethi:1997pa}.
	
Interestingly enough though, when we focus on the simplest possible case, namely by studying the SQM associated to a single D0-brane propagating in 10d, all the complications in finding a supersymmetric ground state disappear, essentially because the gauge group becomes abelian and there is no constraint on the Hilbert space (i.e. $C_a=0$ in eq. \eqref{eq:constraints}) neither a potential for the bosons nor the fermions in eq. \eqref{eq:D0Hamiltonian}. Thus, the ground state is simply the zero-momentum (super-)particle $\ket{\Psi}_{\text{gs}}= \ket{p^i=0} \otimes \mathcal{X}$, where $\mathcal{X}$ is a spinor of $Spin(16)$. Therefore, such `vacuum' is clearly degenerate, and it contains precisely $2^{8}= 256$ states, matching with our expectations in e.g. eq. \eqref{eq:dofcountingIIA}. Now, notice that as we argued before (see discussion around \eqref{eq:cliffordalgebra}), the flavour group of the SQM we have just described is clearly $Spin(9)$. However, one should recall that there is a nice correlation between fermion number in the 1d QM (which is determined by the chirality of the corresponding spinor $\mathcal{X}$) and flavour representation, which essentially follows from spin-statistics in 10d. Thus, states with odd fermion number in the SQM Hilbert space transform under representations of $Spin(9)$, while those of even fermion number arrange into representations of $SO(9)$, instead. Hence, one can see that at the end of the day, the $\textbf{256}$ of $Spin(16)$ becomes
\beq
	\textbf{256}= (\textbf{44} \oplus \textbf{84}_t)_{\text{bos}} \oplus (\textbf{128})_{\text{ferm}}\ ,
	\label{eq:dofcountingD0}
\eeq
in terms of representations of the little group $SO(9)$.

\subsection{The 4d D0-brane content}
\label{ss:4dDO}
	
A crucial point in our analysis in section \ref{s:emergence4dN=2} has to do with the field-theoretic content associated to a tower of D0-branes, which we \emph{claim} that should exist (and be stable) at least close to the large volume point of type IIA string theory on a CY$_3$ (which we denote $X_3$ in the following). Those states were seen to contribute significantly to the one-loop renormalization of the $U(1)^{h^{1,1}+1}$ gauge sector in the vector multiplet moduli space. Hence, it is important for us to be able to give some theoretical support in favour of this conjecture as well as to provide for some more details about their specific four-dimensional spectrum.
	
As already explained in the main text, the basic statement is that within the D0-brane tower lives a complicated spectrum consisting of a massive spin-2 multiplet, ($h^{1,1}-1$) massive vector multiplets as well as ($h^{2,1}+1$) massive hypermultiplets. Intuitively, one can understand this latter fact again by calling on type IIA/M-theory duality, since going to the large volume regime in type IIA (while staying put in the hypermultiplet moduli space, $\mathcal{M}_{\text{HM}}$) corresponds to an effective $S^1$-decompactification to 5d M-theory on the \emph{same} three-fold (see Appendix \ref{ap:5dMtheory} for details). Therefore, as it is customary in KK compactifications, we expect to find the massive replicas of all the light fields appearing in the higher-dimensional theory, which in this case consist of a gravity multiplet, ($h^{1,1}-1$) vector multiplets as well as ($h^{2,1}+1$) hypermultiplets \cite{Cadavid:1995bk}, where $h^{p,q}= \text{dim}\ H^{p,q} (X_3)$ denote the Hodge numbers associated to the three-fold $X_3$ (see \eqref{Hodgediamond} below).
	
Notice, however, that contrary to what happens in the reduction from 11d to 10d as discussed in section \ref{ss:Emergence10dST}, where we have the maximal amount of supersymmetry, the reduced number of supercharges here does not uniquely determine the supermultiplet content of BPS states. Therefore, we need other means to investigate more closely the field-theoretic components hidden within the D0-brane. To do so, we will resort again to a simple SQM analysis. Consider then IIA string theory compactified on a large Calabi--Yau space $X_3$, and let us concentrate on those BPS states arising from the dimensional reduction of the 10d D0-brane. Notice that the moduli space, $\mathcal{M}_{\text{D0}}$, of such non-perturbative state is a copy of $X_3$ itself. The low-energy effective dynamics of the D0-brane can be described by a super-quantum mechanical theory (a supersymmetric non-linear sigma model with target space $\mathbb{R}^3 \times X_3$ in the non-relativistic limit) with four supercharges \cite{Strominger:1996it} (see also \cite{Witten:1996qb, Kachru:2018nck} and references therein). Therefore, if we focus on the collective coordinates of the brane corresponding to the moduli space $\mathcal{M}_{\text{D0}}$, the associated SQM Hilbert space can be seen to be isomorphic to the space of square-integrable (complexified) differential forms in $X_3$. They arrange into a bosonic and a fermionic piece, $\mathcal{H}=\mathcal{H}_B \oplus \mathcal{H}_F$, such that
\beq\label{eq:SQMHilbert}
	\mathcal{H}_B= \mathbb{C} \otimes \left(\, \bigoplus_{p\ \text{even}} \Omega^p(X_3)\, \right)\, , \qquad  \mathcal{H}_F= \mathbb{C} \otimes \left(\, \bigoplus_{p\ \text{odd}} \Omega^p(X_3)\, \right)\, ,
\eeq
whilst the supersymmetric ground states are in one-to-one correspondence with the cohomology of $X_3$.
	
In order to identify the supersymmetric ground states along with their $SO(1,3)$ quantum numbers, one should first recall that the D0-brane breaks half of the 8 supercharges of the bulk $\mathcal{N}=2$ theory in an $SO(3)$ invariant fashion,\footnote{Actually, we will label our quantum numbers in terms of its universal cover extension $SU(2)$, which is more appropriate to include also fermionic representations.} which is the 4d massive little group. Hence, in terms of $SO(3)$, the supercharges group themselves into two independent spinors, each of them forming\footnote{We use the notation $\textbf{j}$ here with $j \in \mathbb{Z}_{\geq0}/2$ to denote the corresponding irreducible representation of $SU(2)$ with highest weight equal to $j$.} a $\textbf{1/2}$ irrep of $SU(2)$. Without loss of generality, we take any of such irreps as being the supercharges broken by the D0-brane state. These supercharges give rise to four fermion zero modes (one can think of them as Goldstinos), which are indeed related by the unbroken supersymmetries to 4d spatial translations (the associated Goldstone bosons). Now, when acting on a supersymmetric ground state of the 0-brane with these zero modes one obtains the following quantum numbers \cite{Kachru:2018nck}
\beq\label{eq:qnumbers}
	(2 \times \textbf{0}) \oplus \textbf{1/2}\, ,
\eeq
i.e. half an hypermultiplet of 4d $\mathcal{N}=2$ supersymmetry. This is not the end of the story though, since the supersymmetric ground states of the SQM theory, which were indeed given by the `de Rahm cohomology' of the moduli space $X_3$, arrange themselves into further multiplets of $SU(2)$ according to the Hodge-Lefschetz decomposition\cite{Figueroa-OFarrill:1997djj} of said cohomology (recall that $X_3$ is K\"ahler). More precisely, the four unbroken supercharges can be seen to act on the SQM Hilbert space (see eq. \eqref{eq:SQMHilbert} above) as the Dolbeaut operators and their adjoints: $\partial, \Bar{\partial}, \partial^{\dagger}, \Bar{\partial}^{\dagger}$. On the other hand, the `Hodge-Lefschetz' $SU(2)$ group acts on the $X_3$ cohomology as follows\cite{Griffiths:433962}
\beq
	L_0\, \omega =\frac{p+q-3}{2}\, \omega, \qquad \omega \in H^{p,q} (X_3)\, ,
\eeq
whilst $L_{\pm}$ act by taking the exterior product or either contracting with the K\"ahler form $J$, respectively.\footnote{One can see quickly just by separating into columns the Hodge diamond \eqref{Hodgediamond} which irreps of $SU(2)$ do indeed appear. $L_{\pm}$ would then act as raising and lowering operators within each column.} Hence one can decompose the Hodge diamond for the CY three-fold $X_3$ 
\begin{align}\label{Hodgediamond}
		\begin{matrix}&&&h^{0,0} & & &\\ &&h^{1,0} & &h^{0,1} && \\ &h^{2,0}&&h^{1,1}&&h^{0,2}&\\ h^{3,0}&&h^{2,1} &&h^{1,2}&&h^{0,3}\\ &h^{3,1}&&h^{2,2}&&h^{1,3}&\\ &&h^{3,2}&&h^{2,3}&& \\ &&&h^{3,3} &&&\end{matrix}\; \stackrel{ X_3}{=}\;\begin{matrix}&&&1& & &\\ &&0 & &0&& \\ &0&&h^{1,1}&&0&\\ 1&&h^{2,1} &&h^{2,1}&&1\\ &0&&h^{1,1}&&0&\\ &&0&&0&& \\ &&&1 &&&\end{matrix}\,
\end{align}
in terms of representations of $SU(2)$ as
\begin{equation}
		\begin{aligned}
			& (1, J, J^2, J^3) \to  \textbf{3/2}\, ,\\
			& (h^{1,1}-1) \ \text{primitive 2-forms in} \ H^2(X_3) \to (h^{1,1}-1) \times \textbf{1/2}\, ,\\
			& (2h^{2,1}+2) \ \text{3-forms in} \ H^3(X_3) \to (2h^{2,1}+2) \times \textbf{0}\, .
		\end{aligned}
\end{equation}
Now, tensoring with the quantum numbers obtained before in eq. \eqref{eq:qnumbers}, we get the following multiplets contained within a single D0-brane
\begin{equation}\label{eq:D0multiplets}
		\begin{aligned}
			& \textbf{3/2} \otimes \left[ (2 \times \textbf{0}) \oplus \textbf{1/2} \right]=\left(2 \times \textbf{3/2} \right) \oplus \textbf{1} \oplus \textbf{2}\, ,\\
			& (h^{1,1}-1) \times \textbf{1/2}\otimes \left[ (2 \times \textbf{0}) \oplus \textbf{1/2} \right]= (h^{1,1}-1) \times \left[ (2 \times \textbf{1/2}) \oplus \textbf{0} \oplus \textbf{1} \right]\, ,\\
			& (2h^{2,1}+2) \times \textbf{0}\otimes \left[ (2 \times \textbf{0}) \oplus \textbf{1/2} \right]= (2h^{2,1}+2) \times \left[ (2 \times \textbf{0}) \oplus \textbf{1/2} \right]\, ,
		\end{aligned}
\end{equation}
together with additional states arising from the anti-D0 brane. Interestingly, what we have obtained is nothing but the (complex) massive KK replica of the 5d $\mathcal{N}=2$ supergravity multiplets belonging to M-theory compactified on the three-fold $X_3$, arranged in terms of 4d massive representations of course. They consist of a single massive spin-2 multiplet, $(h^{1,1}-1)$ massive vector multiplets as well as $(h^{2,1}+1)$ massive hypermultiplets. This means that inside the D0-brane (along with its anti-D0 partner) we essentially have all the complex KK-modes of the five-dimensional massless fields, as one would expect from M-theory/type IIA duality.
	
Notice that the analysis of the D0-brane content in 10d discussed in section \ref{ss:10dDO} above can also be recast along the same lines followed in here by observing that the moduli space in that case is simply a point. Hence, the cohomology is trivial and the quantum numbers of the ten-dimensional D0-brane states can be obtained by just acting on a supersymmetric ground state with the 16 fermion zero modes associated to the broken supercharges. These arrange into a $\textbf{16}$ irrep of $Spin(9)$, thus giving rise to the 256 states corresponding to the BPS supermultiplet in 10d $\mathcal{N}=(1,1)$ supergravity, c.f. eq. \eqref{eq:dofcountingD0}. 
	
\subsection{The 4d D2-D0 bound state content}
\label{ss:4dD0D2}
	
Recall that apart from the proposed tower of D0-branes, the large volume point of type IIA compactified on the Calabi--Yau requires from an additional tower made from bound states of a single D2-brane and $n$ D0's, with $n \in \mathbb{Z}$ being essentially arbitrary as long as the total mass remains below the species scale, (see e.g. eq. \eqref{eq:D0tower}). Now, given that we have already argued that the D0-brane content alone should provide for the KK massive replica of the 5d supergravity fields, it is natural to ask the analogous question for the content of the D2-D0 bound states. This is what we turn to next.

For the D2-branes the possibilities are richer,\footnote{We acknowledge A. Uranga for discussions regarding this issue.} depending essentially on the moduli space associated to the holomorphic 2-cycle wrapped by the brane together with its possible $U(1)$ flat connections and gauge bundles over the complex surface, the latter accounting for the number of D0-branes bounded to the D2. We will not provide here for an extensive account of all such possibilities, since this is presented elsewhere in the literature (see e.g. \cite{Witten:1996qb, Gopakumar:1998ii, Gopakumar:1998jq}) and is beyond the scope of our work, but rather we prefer to comment on some generalities that can be useful for understanding some of the arguments regarding the BPS content of the D2-particles in section \ref{s:emergence4dN=2}.

The basic idea is that for a D2-brane wrapping some genus $g$ surface, $\Sigma_g$, one needs to quantize the supersymmetric one-dimensional sigma-model on $\mathcal{M}_{\text{D2}}$, which consists generically of a $T^{2g}$-fibration over $\mathcal{M}_{\Sigma_g}$
\begin{equation}\label{eq:T2gfibration}
			\begin{aligned}
				\pi: \qquad T^{2g} \rightarrow &\;\mathcal{M}_{\text{D2}} \\
				&\;\; \downarrow \qquad , \\ &\;\mathcal{M}_{\Sigma_g}
			\end{aligned}
\end{equation}
where $\mathcal{M}_{\Sigma_g}$ denotes the moduli space of deformations of $\Sigma_g$ within the three-fold $X_3$\cite{Gopakumar:1998ii, Gopakumar:1998jq}. Thus, analogously as in the D0-brane case before, the $SO(3)$ quantum numbers associated to the cohomology of $\mathcal{M}_{\text{D2}}$ tensored with \eqref{eq:qnumbers} provide for the 4d spin multiplets contained within the D0-D2 bound states. Therefore, one is prompted to look at such moduli space for each specific case in turn.

For illustrative purposes, let us consider here a couple of simple examples. First, suppose that D2-brane wraps some $\mathbb{P}^1$ cycle inside $X_3$, whose moduli space is given just by a point. The cohomology of $\mathcal{M}_{\text{D2}}$ is then trivial and, according to the aforementioned Lefschetz decomposition, we are left with
\begin{equation}\label{eq:D2qnumbers}
		\textbf{0}\otimes \left[ (2 \times \textbf{0}) \oplus \textbf{1/2} \right]= (2 \times \textbf{0}) \oplus \textbf{1/2}\, ,
\end{equation}
i.e. with half a massive hypermultiplet. To complete the supermultiplet one needs to take into account also the CPT conjugate, given precisely by the anti-D2 brane wrapping the \emph{same} curve. This means that in this set-up we get an additional tower of charged hypermultiplets as well as their KK replica.

It is also interesting to consider the F-theory limit discussed in section \ref{ss:generalizations}, where the tower of D2-branes wrapping the elliptic fibre provides for the KK replica of the 6d massless fields along the circle connecting F-theory to 5d M-theory on $X_3$. In order to see this, one should recall that upon taking said limit in the 4d type IIA vector multiplet sector, the elliptic fibre $\mathbb{E}$ shrinks whilst the two-fold base $B_2$ blows up (both in rescaled coordinates). Therefore, one could argue that the moduli space of $\mathbb{E}$ inside $X_3$ is well approximated in this case by the base of the fibration (c.f. eq. \eqref{eq:ellfibration}), such that according to \eqref{eq:T2gfibration}, the cohomology of $\mathcal{M}_{\text{D2}}$ should coincide with that of $X_3$, thus accounting for the supermultiplet content proposed by the end of section \ref{sss:IIA-Ftheorylimit}, namely one massive spin-2 multiplet, $h^{1,1} (B_2)=h^{1,1} (X_3)-1$ massive vector multiplets and $h^{2,1} (X_3)+1$ massive hypermultiplets.
		
\section{M-theory compactifications}
\label{ap:Mthycompactifications}
		
In this appendix we include detail formulae and derivations relevant for the discussions involving M-theory compactifications down to ten and five spacetime dimensions. We start with the reduction of 11d M-theory to ten dimensions, which plays an important role for the emergence mechanism to work in 10d type IIA at strong coupling, as discussed in section \ref{ss:Emergence10dST} in the main text. We then turn in section \ref{ap:5dMtheory} to M-theory compactified on a CY$_3$, which is relevant for Emergence in 4d $\mathcal{N}=2$, as discussed in section \ref{s:emergence4dN=2} as well as for the emergent potentials of section \ref{s:Scalarpotential}. We conclude in section \ref{ap:dimreductiongravitino} by deriving some general formulae describing the volume dependence of the massive gravitino couplings to the appropriate $p$-form gauge fields appearing in the different low-energy M-theory effective actions in several dimensions, which are relevant for the emergence computations associated to the flux potential in 5d and 4d.
		
\subsection{M-theory on $S^1$}
\label{ap:MtheoryKKcompactS1}
		
In the present section we review with some more detail the compactification of 11d supergravity on a circle so as to obtain the 10d massless type IIA effective action along with the relevant kinetic terms and interactions associated to the Kaluza-Klein replica of the 11d gravity multiplet. 
The discussion here is complementary to the analysis performed in section \ref{ss:Emergence10dST}, although the computations will be done following a different approach than the one employed in the main text. The reason for this is twofold: first, upon doing so we will see that the identifications between 10d and 11d fields become simply the familiar ones (see e.g. \cite{Witten:1995ex}) and second, this alternative approach will serve also to show how our results are independent of the specific units or conventions adopted, as they should be.
		
\subsubsection*{Duality with 10d type IIA}
		
Let us start then by writing here the type IIA supergravity action in the 10d string frame
\begin{equation}
			\begin{aligned}
				S_\text{IIA, s}^{\text{10d}} = &\frac{2\pi}{\ell_s^8} \int \text{d}^{10}x\sqrt{-g}\ e^{-2 \phi} \left(R+4(\partial \phi)^2\right)-\frac{2\pi}{\ell_s^8}\int \frac{e^{-2\phi}}{2} H_3\wedge \star H_3 \\
				&-\frac{2\pi}{\ell_s^8}\int \frac{1}{2} \left[F_2 \wedge \star F_2 + \tilde F_4 \wedge \star \tilde F_4 + B_2\wedge F_4 \wedge F_4\right]\, , 
			\end{aligned}
\end{equation}
where $H_3=dB_2$, $\tilde{F}_4=d C_3-C_1 \wedge H_3$ and $F_2=dC_1$ in the previous expression. (Recall that $\ell_s=2\pi \sqrt{\alpha'}$ denotes the fundamental string length.) Next, upon performing a usual Weyl rescaling to the 10d metric
\begin{align}\label{eqap:10dWeylrescaling}
			g_{\mu \nu} \to g_{\mu \nu}\, e^{(\phi-\phi_0)/2}\ , 
\end{align}
one arrives at the following Einstein-framed action
\begin{equation}
			\begin{aligned}\label{eqap:IIA10daction}
				S_\text{IIA, E}^{\text{10d}} = &\frac{1}{2\kappa_{10}^2} \int \text{d}^{10}x\sqrt{-g} \left(R-\frac{1}{2}(\partial \phi)^2\right)-\frac{1}{4\kappa_{10}^2}\int e^{-(\phi-\phi_0)}\, H_3\wedge \star H_3 \\
				&-\frac{e^{2 \phi_0}}{4\kappa_{10}^2}\int \left[e^{\frac{3}{2}(\phi-\phi_0)}\, F_2 \wedge \star F_2 + e^{\frac{1}{2}(\phi-\phi_0)}\, \tilde F_4 \wedge \star \tilde F_4 + B_2\wedge F_4 \wedge F_4\right]\, , 
			\end{aligned}
\end{equation}
where one can now clearly see that the gravitational strength, encapsulated by the (dimension-full) factor $2\kappa_{10}^2= 2 \Mpt^{-8}= (2 \pi)^{7} \alpha'^4\ e^{2 \phi_0}$, is essentially controlled by the dilaton VEV $\phi_0 \equiv \braket{\phi}$, which thus determines the Planck-to-string scale ratio. Notice that this action is similar but not quite the same as the one displayed in eq. \eqref{eq:IIA10d}, the difference being the reference scale in which we choose to measure the ten-dimensional fields, namely string units here vs Planck units there.
		
In what follows we will start from 11d supergravity and compactify the theory on a circle, retaining part of the massive spectrum (as seen from the lower dimensional theory) in an explicit way within our action functional. This let us obtain some of the (minimal) couplings that such fields present with respect to the massless modes of the 10d theory, which subsequently are used to repeat again the emergence computations of section \ref{sss:IIAstrongcoupling}, using this time different units and conventions.
		
Hence, we take the 11d $\mathcal{N}=1$ supergravity action describing the low energy limit of M-theory, which reads
\begin{align}\label{eqap:Mthyaction}
			S^{\text{11d}}_{\text{M-th}} = \frac{1}{2\kappa_{11}^2} \int \hat{R} \ \hat{\star} 1-\frac{1}{2}  d\hat{C}_3\wedge \hat{\star}  d\hat{C}_3 -\frac{1}{6} \hat C_3 \wedge d\hat{C}_3 \wedge d\hat{C}_3 \, ,    
\end{align}
in the bosonic sector. Upon compactifying on a $S^1$, one should first expand the different 11d fields in a Fourier series, following the Kaluza-Klein prescription (see e.g. \cite{Duff:1986hr}). Similarly as we did in section \ref{ss:Emergence10dST}, we concentrate our efforts here in computing explicitly the couplings that the massive replica of the 3-form gauge field present with respect to the massless type IIA Ramond-Ramond $p$-forms, since this will be enough to illustrate later on how the emergence mechanism applies to this particular set-up. 
		
As it is customary in $S^1$ compactifications, we impose the familiar ansatz for the metric
\begin{align}
			ds^2_{11} = ds^2_{10}+ e^{2\varphi} (dz-C_1)^2\, ,    
\end{align}
where $z \in [0, 2 \pi R)$ parameterizes the circular direction, $\varphi$ is the `radion' field and $C_1=(C_1)_{\mu}\, dx^{\mu}$ denotes the KK photon.\footnote{The fields $g_{\mu \nu} (x)$, $\varphi(x)$ and $C_1(x)$ represent the zero (Fourier) modes corresponding to the appropriate components of the 11d metric along the $S^1$.} Notice that the notation has been chosen in order to make the matching with type IIA supergravity more explicit, where the KK photon becomes nothing but the RR 1-form. The dimensional reduction of the Einstein-Hilbert on a $S^1$ is straightforward but cumbersome, so we skip it and refer the interested reader to the existing literature\cite{Duff:1986hr}. Thus, we focus on the last two terms in eq. \eqref{eqap:Mthyaction} above.
		
We start with the first one, corresponding to the kinetic term of the massless 3-form potential. Hence, upon reducing such term by expanding the field strength $\hat G_4= d\hat C_3$ into its Fourier modes and integrating over the compact direction, we arrive at the following action in 10d (after fixing some `unitary' gauge for the massive modes\cite{Duff:1986hr})
\begin{equation}
			\label{eq:3formddimgravity}
			\begin{aligned}
				S^{\text{10d}}_{\text{kin,} C_3^{(n)}} = &-\frac{2 \pi R}{4\kappa_{11}^2} \int e^{\varphi} \left( dC_3-C_1 \wedge dB_2 \right) \wedge \star \left( dC_3-C_1 \wedge dB_2 \right) + e^{-\varphi}\, dB_2 \wedge \star dB_2\\
				& - \frac{2 \pi R}{4\kappa_{11}^2} \int \sum_{n \neq 0} \left [ e^{\varphi}\, \mathcal{D} C^{(n)}_3 \wedge \star \mathcal{D} C^{(-n)}_3 + e^{-\varphi}\, \frac{n^2}{R^2}\, C^{(n)}_3 \wedge \star C^{(-n)}_3 \right]\, ,
			\end{aligned}
\end{equation}
where $B_2$, $C_3$ are 2-form and 3-form fields, respectively, that arise essentially from the zero-mode components of $\hat C_3$ with/without one leg in the $z$-th direction, whilst $\mathcal{D} C^{(n)}_3= dC^{(n)}_3 + \frac{in}{R}\, C_1 \wedge C^{(n)}_3$ refers to the covariant derivative with respect to the $U(1)$ gauge field $C_1$, under which the whole tower is charged (as expected on general grounds). Notice that in the above expression we are keeping inside $\varphi(x)$ its asymptotic value, i.e. the constant VEV $\varphi_0 \equiv \braket{\varphi}$. Thus, the physical radius of the eleven-th dimension would be given in this case by $R_{\text{phys}}= R\, e^{\varphi_0}$.
		
On the other hand, the Chern-Simons term appearing in eq. \eqref{eqap:Mthyaction}, when reduced on the M-theory circle, can be seen to give the following \emph{topological} contribution to the 10d action
\begin{equation}
			\label{eq:CSterm}
			\begin{aligned}
				S^{\text{10d}}_{\text{top,} C_3^{(n)}} \supset -\frac{2 \pi R}{4\kappa_{11}^2} \int  B_2 \wedge dC_3 \wedge dC_3 + \sum_{n \neq 0} \frac{in}{R} \left ( C_3 \wedge C^{(n)}_3 \wedge dC^{(-n)}_3 - C_3\wedge dC^{(n)} \wedge C^{(-n)}_3 \right)\, ,
			\end{aligned}
\end{equation}
where the first summand leads to the Chern-Simons term in type IIA supergravity (see e.g. \cite{Polchinski:1998rq,Polchinski:1998rr}), while the second one encapsulates the (trilinear) interactions between each 3-form field in the KK tower and the massless RR 3-form potential. 
		
Now, in order to extract the precise Feynman rules associated to the kind of interactions between the $C_3^{(n)}$ and the massless fields $C_1$ and $C_3$, it is necessary to first match the zero mode action of the M-theory circle compactification with the two-derivate type IIA supergravity lagrangian, c.f. eq. \eqref{eqap:IIA10daction}. Upon taking into account the Weyl rescaling \eqref{eqap:10dWeylrescaling}, where the 10d dilaton $\phi$ is related to the radion field by $\varphi=2 \phi/3$ \cite{Witten:1995ex},\footnote{This is precisely where the discussion departs from the one in section \ref{sss:IIAstrongcoupling}. There, a similar Weyl rescaling is performed to the 10d metric but including not only the fluctuation $\tilde \phi=\phi-\phi_0$, but also its asymptotic value, see discussion around eq. \eqref{eq:KKansatz11dmetric}.} we arrive at the following 10d bosonic action
\begin{equation}
\label{eqap:finalbosonicaction}
	\begin{aligned}
			S^{\text{10d}}_{\text{bos}} =& \tilde{S}_\text{IIA, E}^{\text{10d}} -\frac{2 \pi R_{\text{phys}}}{4\kappa^2_{11}} \int  \sum_{n \neq 0} \left [ e^{\frac{\phi-\phi_0}{2}}\, \mathcal{D} C^{(n)}_3 \wedge \star \mathcal{D} C^{(-n)}_3 + e^{-(\phi-\phi_0)}\, \frac{n^2}{R^2_{\text{phys}}}\, C^{(n)}_3 \wedge \star C^{(-n)}_3 \right] \\
			 &-  \frac{2 \pi R_{\text{phys}}}{4\kappa^2_{11}} \int \sum_{n \neq 0} \frac{in}{R_{\text{phys}}} \left [ C_3 \wedge C^{(n)}_3 \wedge dC^{(-n)}_3 - C_3\wedge dC^{(n)}_3 \wedge C^{(-n)}_3 \right] + \ldots \, ,
	\end{aligned}
\end{equation}
where $\tilde{S}_\text{IIA, E}^{10}$ can be seen to almost agree with eq. \eqref{eqap:IIA10daction} above. Let us write its expression explicitly here
\begin{equation}\label{eq:almostIIA10d}
		\begin{aligned}
				\tilde{S}_\text{IIA, E}^{\text{10d}} = &\frac{2 \pi R_{\text{phys}}}{2\kappa^2_{11}} \int \text{d}^{10}x\sqrt{-g} \left(R-\frac{1}{2}(\partial \phi)^2\right)-\frac{2 \pi R_{\text{phys}}\, e^{-\frac{4}{3}\phi_0}}{4\kappa^2_{11}}\int e^{-(\phi-\phi_0)}\, dB_2 \wedge \star dB_2 \\
				&-\frac{2 \pi R_{\text{phys}}}{4\kappa^2_{11}}\int \left[e^{\frac{4}{3}\phi_0} e^{\frac{3}{2}(\phi-\phi_0)}\,dC_1 \wedge \star dC_1 + e^{\frac{1}{2}(\phi-\phi_0)}\, \tilde F_4 \wedge \star \tilde F_4 + e^{-\frac{2}{3}\phi_0}\, B_2\wedge dC_3 \wedge dC_3\right]\, . 
		\end{aligned}
\end{equation}
In order to match completely the above action with the `conventional' one as displayed in eq. \eqref{eqap:IIA10daction}, it is thus necessary to make the following redefinitions of massless fields (see e.g. sect. 16 of \cite{Ortin:2015hya})
\begin{align}
			\label{eq:CBredefs}
			C_1 \to C_1\, e^{\frac{1}{3}\phi_0}\ , \qquad C_3 \to C_3\, e^{\phi_0}\ , \qquad B_2 \to B_2\, e^{\frac{2}{3}\phi_0} , 
\end{align}
as well as some identifications of the relevant scales and constants appearing in our set-up\cite{Witten:1995ex, Ortin:2015hya}
\begin{align}\label{eq:vevmatching}
			\frac{2 \pi R_{\text{phys}}}{2\kappa^2_{11}}=\frac{1}{2\kappa^2_{10}}= \frac{2 \pi}{g_s^2 \ell_s^8}\, , 
\end{align}
where $(2\kappa^2_{11})^{-1}= \frac{2 \pi}{g_s^3 \ell_s^9}$ and $2 \pi R_{\text{phys}}= \ell_s g_s$. (Recall that $R_{\text{phys}}= R\, e^{2 \phi_0/3}$, with $g_s=e^{\phi_0}$.)
		
\subsubsection*{Revisiting emergence in 10d}
		
From eq. \eqref{eqap:finalbosonicaction} above one can clearly see that the interaction vertices between the massive 3-forms and both the RR $C_3$ and $C_1$ fields are controlled essentially by the same quantity, namely $e^{\phi_0}/(2\kappa^2_{10} R_{\text{phys}})$. What we want to do now is to repeat the emergence computations of section \ref{sss:IIAstrongcoupling} with these new interaction rules. Notice from \eqref{eqap:IIA10daction} that the kinetic terms of both $p$-form gauge fields are controlled by the same quantity
\begin{align}\label{eqap:coupling1formIIA}
		\frac{e^{2 \phi_0}}{2\kappa^2_{10}}=\frac{g_s^2 \Mpt^8}{2}\, , 
\end{align}
which should be what one obtains after summing over quantum corrections induced by the tower of D0-branes up to the species scale. We will consider the corrections induced by the KK replica of the 3-form potential, bearing in mind that according to our general discussion in section \ref{ss:Emergence10dST}, the massive spin-2 and spin-$\frac{3}{2}$ should contribute similarly.
		
Thus, for the RR 1-form one obtains the following one-loop correction to the self-energy in the usual Lorenz gauge
\beq \label{eqap:oneformfermion10d}
	\frac {\partial \Pi_{n,\mu \nu}^{(10d)}(p)}{\partial p^2} \bigg\rvert_{p=0}\, \sim\, - (2\kappa^2_{10})^2\, q_n^2\, g^2\, \delta_{\mu \nu} \Lambda_{\text{QG}}^{6}\ ,
\eeq
whilst the expression for the 3-form is essentially the same up to the tensorial structure, which becomes the one appropriate for a rank-3 antisymmetric gauge field. Notice that we have neglected contributions of $\mathcal{O}(1)$ above and we have moreover included a factor of $(2\kappa^2_{10})^2$, which appears through the propagators of the massive 3-form fields, $C^{(n)}_3$ (see eq. \eqref{eqap:finalbosonicaction}). From the discussion above one can properly identify the physical $U(1)$ charges as
\beq
	q_n\, g= e^{\phi_0}\, \frac{n\, m_{\text{D}0}}{2\kappa^2_{10}}\, ,
\eeq
where we have used the relations displayed around eq. \eqref{eq:vevmatching} as well as the fact that $ m_{\text{D}0}=1/R_{\text{phys}}$. All in all, after summing over the entire tower of D0-branes, we obtain the following one-loop wave-function renormalization for the $C_1$-field
\beq \label{eqap:sumovertoweroneform}
	\sum_n \frac {\partial \Pi_{n,\mu \nu}^{(10d)}(p)}{\partial p^2} \bigg\rvert_{p=0} \sim - e^{2 \phi_0}\, m_{\text{D}0}^2\, \Lambda_{\text{QG}}^{6}\, \delta_{\mu \nu} \sum_n n^2 \sim - \delta_{\mu \nu}\ e^{2 \phi_0}\, m_{\text{D}0}^2\, \Lambda_{\text{QG}}^{6}\ N^3\, ,
\eeq
where $N$ denotes the total number of steps of the D0-brane tower whose mass is at or below the QG scale, $\Lambda_{\text{QG}}$. Now, upon using the definition of the species scale, eq. \eqref{species}, as well as the mass formula of our BPS tower, one can show that both $\Lambda_{\text{QG}}$ and $N$ can be written (up to order one factors) in terms of the mass of a single D0-brane as follows
\beq \label{eqap:QGscaleandN} 
	\Lambda_{\text{QG}}\, \sim\, m_{\text{D}0}^{\frac{1}{9}}\, \Mpt^{\frac{8}{9}}\, , \qquad N\, \sim\,  m_{\text{D}0}^{-\frac{8}{9}}\, \Mpt^{\frac{8}{9}}\, .
\eeq
Substituting these back into eq. \eqref{eqap:sumovertoweroneform} we find
\beq
	\sum_n \frac {\partial \Pi_{n,\mu \nu}^{(10d)}(p)}{\partial p^2} \bigg\rvert_{p=0} \sim - \delta_{\mu \nu}\, e^{2 \phi_0}\, \Mpt^8\, ,
\eeq
which indeed agrees with what we wanted to reproduce, c.f. eq. \eqref{eqap:coupling1formIIA}.

Concerning the wave-function renormalization of the 10d dilaton field, the calculation is exactly the same as the one performed in section \ref{sss:IIAstrongcoupling}, the only difference being the units in which we measure the mass of the D0-branes, which does not affect the final result after summing over all the contributions from the tower. 
		
\subsection{M-theory on a $\mathrm{CY}_3$}
\label{ap:5dMtheory}
		
We turn now to M-theory compactifications with eight (unbroken) supercharges down to five dimensions. The content of this subsection is complementary to the discussions in sections \ref{s:emergence4dN=2} and \ref{s:Scalarpotential} from the main text, although some extra material is presented exclusively here so as to not deviate the attention of the reader from the main ideas developed there.

\subsubsection*{The 5d $\mathcal{N}=2$ EFT}
Let us start by reviewing the 5d $\mathcal{N}=2$ supergravity effective action that arises from compactifying M-theory on a Calabi--Yau threefold, $X_3$. Its bosonic part before introducing the background $\hat{G}_4$-fluxes reads as follows\footnote{In our conventions $(2\kappa^2_5)^{-1}= 2 \pi/ \ell_{5}^3$, where $\ell_{5}$ denotes the five-dimensional Planck length.} \cite{Cadavid:1995bk}
\begin{equation}\label{eq:Mthyaction5d}
		\begin{aligned}
			S^{\text{5d}}_{\text{M-th}}= \dfrac{1}{2\kappa^2_5} \int &  R \star 1 -  G_{a b} \left( dM^a\wedge \star dM^b  + F^a \wedge \star F^b \right) \\ 
			&- \frac{1}{6} \mathcal{K}_{a b c} A^a \wedge F^b \wedge F^c- 2 h_{pq}d  q^p \wedge \star d  q^q \, , 
		\end{aligned}
\end{equation}
where $M^a=t^a/\mathcal{V}_5^{1/3}$, $a=1, \ldots, h^{1,1}$, are real scalars within the vector multiplets (which arise from the K\"ahler deformations, $t^a$, of the compact three-fold), $F^a=dA^a$ denote the corresponding field strengths of the $U(1)$ gauge bosons plus the graviphoton, and the scalars in the various hypermultiplets are represented by $q^p$. Recall that the scalars in the vector multiplet moduli space are subject to the constraint
\beq
		\mathcal{F}=\frac{1}{3!} \mathcal{K}_{a b c} M^a M^b M^c \stackrel{!}{=} 1\, ,
\eeq
where $\mathcal{K}_{a b c}$ are the triple intersection numbers of the Calabi--Yau three-fold. These scalars parameterize a manifold with \emph{very special geometry}, such that it is endowed with a metric (which also appears in the kinetic terms of the $h^{1,1}$ gauge fields) given explicitly by
\beq\label{eq:5dVMmetric}
		G_{a b}=\frac{1}{2 \mathcal{V}_5^{1/3}} \int_{X_3} \omega_a \wedge \star \omega_b=- \frac{1}{2} \left( \partial_{M^a} \partial_{M^a} \mathcal{F}\right)\bigg\rvert_{\mathcal{F}=1} \, ,
\eeq
thus depending on the inner product of the harmonic $(1,1)$-forms in $X_3$, denoted $\omega_a\, $. Taking the constraints into account, the 5d theory has $(h^{1,1}-1)$ vector multiplets, $(h^{2,1}+1)$ hypermultiplets and as usual, one gravity multiplet.
		
Let us now be more specific about the hypermultiplet sector. In principle one obtains the following classical metric after reducing on the three-fold (and performing the Weyl rescaling $g_{\mu \nu} \to \mathcal{V}_5^{-2/3}g_{\mu \nu}$ to go to the 5d Einstein frame)
\begin{align}\label{eq:classicalhypermetric}
			\nonumber h_{pq} dq^p \wedge \star dq^q =\, &\frac{1}{4\mathcal{V}_5^2}d\mathcal{V}_5\wedge \star d\mathcal{V}_5 + G_{l \bar{k}} dz^l \wedge \star d\bar z^k + \frac{1}{4} \mathcal{V}_5^2 dC_3\wedge \star dC_3 \\
			\nonumber&+ \frac{1}{4}\left(\xi^K d \tilde{\xi}_K- \tilde{\xi}_K d \xi^K\right) \wedge dC_3 \\
			&-\frac{1}{4\mathcal{V}_5} (\text{Im}\ \mathcal{M}^{-1})^{KL} \left(d \tilde{\xi}_K-\mathcal{M}_{KN}d \xi^N  \right)\wedge \star \left(d \tilde{\xi}_L-\mathcal{M}_{LM}d \xi^M  \right)\, ,
\end{align}
where $\mathcal{V}_5$ denotes the volume of the three-fold in 11d Planck units, the $z^l$, $l=1, \ldots, h^{2,1}$ refer to the complex structure moduli and the rest of the fields arise from the piece in the reduction of the original 3-form of M-theory ($\hat{C}_3$) that is not part of the $U(1)$ gauge fields, namely
\beq \label{eq:3-formexpansion}
		\hat{C}_3-A^a \wedge \omega_a= \xi^K \alpha_K - \tilde{\xi}_L \beta^L + C_3\, .
\eeq
Here, we have introduced a set of real 3-forms $\alpha_K$, $K=0,1,\dots, h^{2,1}$ that together with their duals $\beta^K$ form a symplectic basis of $H^3(X_3, \mathbb{Z})$, such that $\int_{X_3} \alpha_K \wedge \beta^L =\delta_K^L\,$. Moreover, the (complex) matrix $\mathcal{M}^{KL}$ appearing in eq. \eqref{eq:classicalhypermetric} depends explicitly on the complex structure of the $X_3$, but its precise form is not needed for our purposes. Notice that in order to have an actual quaternionic-K\"ahler metric in the hypermultiplet sector above it is necessary to first dualize the 3-form $C_3$ into a scalar, $\sigma$, which belongs to the so-called universal hypermultiplet, as so does $\mathcal{V}_5$ \cite{Cadavid:1995bk}, giving a total of $(h^{2,1}+1)$ hypermultiplets.
		
\subsubsection*{Adding 4-form fluxes}

We can also consider the inclusion of some background flux for the 4-form field strength of 11d supergravity, $\hat{G}_4$,  along some internal cycles of $X_3$. As usual, this flux can be expanded in a basis of harmonic 4-forms,  $\tilde{\omega}^a$, as follows
\beq\label{eq:G4flux}
		\hat{G}^{\text{flux}}_4 = f_a \, \tilde{\omega}^a\, ,
\eeq
where the coefficients $f_a$ are quantized integers\cite{Grimm:2013fua}.\footnote{Strictly speaking, the correct quantization condition for the $\hat{G}_4$-flux along the $X_3$ is  ${\hat G_4 +\frac{1}{2} c_2(X_3) \in H^4(X_3, \mathbb{Z})\,} $, such that in general the flux is half-integer quantized  \cite{Witten:1996md}.}  Repeating the $X_3$ reduction with this extra ingredient, one recovers essentially the same effective action with two  important modifications. First, the term in the second line of the hypermultiplet metric \eqref{eq:classicalhypermetric} now reads
\beq
		h_{pq} dq^p \wedge \star dq^q \supset \frac{1}{4}\left(\xi^K d \tilde{\xi}_K- \tilde{\xi}_K d \xi^K+2A^a f_a\right) \wedge dC_3\, ,
\eeq
which produces some gaugings of our 5d EFT upon dualization of $C_3$. Second, a scalar potential for both $M^a$ and $\mathcal{V}_5$ is induced
\beq \label{eq:5dpotential}
		V^{\text{M}}_{\text{flux}}= \frac{1}{8 \kappa^2_5} \frac{G^{a b}}{\mathcal{V}_5^2} f_a f_b\, ,
\eeq
where $G^{a b}$ denotes the inverse matrix of the vector multiplet metric \eqref{eq:5dVMmetric}. Notice that this potential is quadratic in $f_a$, and written in this form it is very reminiscent of the one arising in type II string theory on $X_3$ when RR and NS fluxes are included (see e.g. \cite{Herraez:2018vae}). There is an important difference which renders the analysis in 5d easier to handle than its 4d counterpart, namely the fact that the scalars $M^a$ are real, and hence do not present axionic partners. This is the reason why the flux potential \eqref{eq:5dpotential} appears as a bilinear on the flux quanta directly, instead of on some axion-dependent polynomials (denoted by $\rho$'s in \cite{Herraez:2018vae}). Also notice that the structure of the allowed fluxes in M-theory on the $X_3$ is simpler than the one in the type IIA scenario, since there is neither Romans mass or any extra possible fluxes arising only after further circle reduction (e.g. the NS $H_3$-flux which depends on the complex structure sector or the ones associated to the RR 1-form $C_1$, which are sourced by KK-monopoles from the M-theory perspective).

\subsubsection*{4-form formulation of the scalar potential in 5d M-theory}	
Let us now try to reexpress the 5d flux potential in eq. \eqref{eq:5dpotential} above in terms of a set of (non-propagating) 4-form potentials $C^a_4$, whose `kinetic terms' may be obtained (or corrected) through an emergence-like computation. Following the same logic in \cite{Bielleman:2015ina, Herraez:2018vae} we expect these 4-form fields to be obtained upon reduction of the 6-form potential $\hat{C}_6$ in 11d supergravity along harmonic (1,1)-forms as $\hat{C}_6= C^a_4 \wedge \omega_a + \ldots\, $. Therefore, let us try to dualize the flux quanta $f_a$ by treating those as dynamical fields and adding a Lagrange-multiplier-like term in the action, which forces them to satisfy the Bianchi identity $df_a=0$
\beq\label{eq:lagrangemult} 
		\Delta S^{\text{5d}}_{\text{M-th}} = -\frac{1}{4\kappa^2_5} \int f_a \wedge dC^a_4\, .
\eeq
Thus, varying the action with respect to $C^a_4$ implies that the condition $df_a=0$ is to be satisfied.\footnote{Strictly speaking, one should add an extra boundary term to \eqref{eq:lagrangemult} of the form $\Delta S^{\text{5d}}_{\text{bound}} = \frac{1}{4\kappa^2_5} \int d \left( C^a_4 \wedge f_a \right)$ for the variational problem associated to $C^a_4$ to be well posed at infinity. In any event, this does not change the formulae in eqs. \eqref{eq:eomf_a}-\eqref{eq:kineticterms4forms}, but just results in the appearance of the correct boundary term in the dual action associated to the `kinetic terms' for the 4-forms, $C^a_4$.} If instead we solve for $f_a$ through its equations of motion we find
\beq \label{eq:eomf_a}
		\frac{1}{\mathcal{V}_5^2}G^{a b} f_b \star 1+dC^a_4 + 2A^a \wedge dC_3 =0 \quad \Longleftrightarrow \quad  f_a= \mathcal{V}_5^2 G_{a b} \star \left(dC^b_4 + 2 A^b \wedge dC_3 \right)\, .
\eeq
Upon substituting such relations in the action plus the Lagrange multiplier \eqref{eq:lagrangemult} we arrive at 
\beq\label{eq:kineticterms4forms}
		-\frac{ G^{a b}}{8\mathcal{V}_5^2} f_a f_b \star 1-\frac{1}{4} f_a \wedge dC^a_4 - \frac{1}{2} f_a A^a \wedge dC_3=-\frac{\mathcal{V}_5^2}{8} G_{a b} \big( dC^a_4+ 2 A^a \wedge dC_3 \big)  \wedge \star \big( dC^b_4 + 2 A^b \wedge dC_3 \big)\, .
\eeq
Interestingly enough, apart from the fact that such kinetic terms mix the gauge transformations of the 4-forms $C^a_4$ with those of the $U(1)$ vectors $A^a$, they have a factor $\mathcal{V}_5^2$ in front which was already present in the kinetic term of the 3-form, c.f. first line of eq. \eqref{eq:classicalhypermetric}. Moreover, the metric in the kinetic term for such 4-forms is precisely the same as the one in the vector multiplets, eq. \eqref{eq:5dVMmetric}, suggesting that they share the same weak-coupling points whose origin could be thus related to similar kinds of infinite towers of states. Notice that this is not entirely surprising since the origin of such 4-form gauge fields lies in the reduction of the dual $\hat{C}_6$ along the same harmonic 2-forms of the three-fold.
		
Hence, if we start from a `democratic' formulation of 11d supergravity where both $\hat{C}_3$ and its magnetic dual $\hat{C}_6$ appear and reduce the kinetic terms of the 6-form, we should find (after the proper Weyl rescaling) a term of the form \eqref{eq:kineticterms4forms}. For completeness, let us also point out that the 5d domain walls which in principle interpolate between the distinct `vacua' generated by different choices of flux numbers $f_a$, are precisely given by M5-branes wrapping the 2-cycles dual to the different harmonic $(1,1)$-forms.

\subsection{Dimensional reduction of the $P$-form-gravitino coupling}
\label{ap:dimreductiongravitino}
Motivated by the gravitino coupling with the field strength $\hat{G}_4$ in 11d M-theory (see eq. \eqref{eq:MthyactiongravitinoG4coupling}), in this appendix we consider the reduction of a coupling of the same form in $D$ dimensions after compactification on a $k=D-d$ dimensional manifold. To be as general as possible, we begin with a coupling between a $P$-form field strength and a gravitino bilinear in  $D$ dimensions and obtain the $d$-dimensional couplings between the reduced $p$-form field strength ($p\leq P$) and the corresponding gravitino bilinear KK replicas. The goal is to extract the dependence on the internal volume of the $d$-dimensional coupling between the $p$-form and the KK modes of the gravitino, as this is what we use in the main text to compute the  one-loop contribution to the kinetic term of the corresponding massless $(p-1)$-form. As we show, this volume dependence can be extracted very generally, and the specific details of the compactification would only introduce different order-one factors and dependence on other moduli, but not on the overall volume. Our results are thus applicable to both Calabi--Yau and circle compactifications, which are the ones we use in the main text, and we discuss some more specific properties of them at the end of the section, after presenting the general formulae. Let us start with the coupling
\begin{equation}
			\label{eq:P-formgravitinogeneral}
			S^D_{G_p \bar{\psi} \psi} = \frac{1}{16\kappa_{D}^2}\int \text{d}^{D}x \sqrt{-g_{D}}\dfrac{1}{(P)!}G_{M_1 \ldots M_{P}} \overline{\psi}_N \Gamma^{[N|}\Gamma^{M_1 \ldots M_{P}} \Gamma^{|Q]} \psi_Q\, , 
\end{equation}
where $\Gamma^M$ denote the $D$-dimensional gamma matrices and $\Gamma^{M_1 \ldots M_{n}}$ their anti-symmetrized combinations. Here we use $M=0, \ldots, D-1$, $\mu=0, \ldots, d-1$, $i=1, \ldots, k$. First of all, we can expand the $D$-dimensional gravitino field as
\begin{equation}	\label{eq:gravitinoexpansiongeneral}
			\psi_{\mu}\, =\,  \sum_{n,\alpha}  \psi^{(n)}_{\mu,\alpha } \otimes \xi^{(n)}_\alpha\, ,
\end{equation}
where $\xi^{(n)}_{\alpha }$ denote the internal spinors that are eigenstates of the Dirac operator ($i \slashed{\nabla}_{\text{CY}}$) along the internal space
\begin{equation}
			\label{eq:Diraceigenvalueproblemgeneral}
			-i \gamma^m \nabla_m \xi^{(n)}_{\alpha} =  \lambda_{(n)} \xi^{(n)}_{\alpha}\, .
\end{equation}
Similarly, we can obtain the zero modes of the lower-dimensional $p$-form field strengths upon expanding the $P$-form field strengths in the corresponding basis of internal harmonic $(P-p)$-forms, denoted generically by $\omega_a$, namely
\begin{equation}	\label{eq:P-formexpansiongeneral}
			G_P= \sum_a F_p^a \wedge \omega_a\, , \qquad a=1, \ldots, b^{P-p}\, .
\end{equation}
Introducing now eqs. \eqref{eq:gravitinoexpansiongeneral} and \eqref{eq:P-formexpansiongeneral} into the $P$-form-gravitino bilinear, \eqref{eq:P-formgravitinogeneral} we obtain the following lower dimensional interaction term (note that this includes the case $P=p$, in which case $\omega_a=1$)
\begin{equation}
			\label{eq:gravitinovertices1}
			\sqrt{g_{d}} \sum_{n,\alpha,\beta} (F_p)^a_{\mu_1 \ldots \mu_p}\ \left( \overline{\psi^{(n)}_{\nu,\alpha}} \gamma^{[\nu |}\gamma^{\mu_1 \ldots \mu_p} \gamma^{| \rho ]} \psi^{(n)}_{\rho,\beta} \right) \int d^ky \sqrt{g_{k}} (\omega_a)_{i_1\ldots i_{P-p}}\ \xi^{(n), \dagger}_\alpha \gamma^{i_1\ldots i_{P-p}} \xi^{(n)}_\beta   \, .
\end{equation}
The integral over the internal dimensions is a bilinear on the internal spinors $\xi^{(n)}_\alpha$ and the choice of each pair of $\alpha$ and $\beta$ depends on the number of internal gamma matrices, i.e. $(P-p)$,  through generalizations of the orthogonality condition $\xi^{\dagger}_\alpha \xi_{\beta}=\delta_{\alpha \beta}$ (for particular cases, such as  CY$_3$ compactifications, one can be more precise and obtain relations such as  \eqref{eq:xiproperties}). Since we are interested in the dependence of such interaction on the internal volume, $\mathcal{V}_k$, we do not need to perform the computation in more detail, since keeping track of the $\gamma^i$ structure is enough for this purpose, as we illustrate in the following.
		
To begin with, we can extract the dependence of the internal integral on the overall volume by rewriting all the quantities depending on the internal metric in terms of a related one of unit volume, i.e. a new metric $\tilde{g}_k$ satisfying the constraint 
\begin{equation}
			\int_{\mathcal{M}_k} d^ky \sqrt{\tilde{g}_k} = 1\, .
\end{equation}
It is easy to see that both metrics are hence related as $(g_k)_{ij}=(\tilde{g}_k)_{ij} \mathcal{V}_k^{2/k}$. This means, in particular, that we can analogously define \emph{unimodular} (internal) vielbeins $\tilde{e}^{i}_a=\mathcal{V}_k^{1/k} e^{i}_a$, such that the final dependence on the volume scalar of the internal integral appearing in eq. \eqref{eq:gravitinovertices1} becomes simply a prefactor of the form $\mathcal{V}_k^{1-\frac{P-p}{k}}$. In addition, we need to take into account the Weyl rescaling of the $d$-dimensional metric to go to the Einstein frame, namely $(g_d)_{\mu \nu} \to\, \mathcal{V}_k^{-\frac{2}{d-2}}\, (g_d)_{\mu \nu} $, together with the field redefinition of the massive modes of the gravitino in order for them to have canonical kinetic terms (so that we can directly apply the results from the diagrams computed in section \ref{s:EmergenceQG} and Appendix \ref{ap:Loops}), which yields $\psi_{\mu , \alpha}^{(n)}\to \, \mathcal{V}_k^{-\frac{1}{2d-4}} \, \psi_{\mu , \alpha}^{(n)}$. Therefore, putting together all these rescalings and redefinitions, the $d$-dimensional interactions between the tower of gravitini and the corresponding zero-mode $p$-form field strength read
\begin{equation}
			\label{eq:gravitinovertices2}
			S^{d}_{F_p \bar{\psi} \psi }= \mathcal{V}_k^\theta\sqrt{g_{d}} \sum_{n,\alpha,\beta} (F_p)^a_{\mu_1 \ldots \mu_p}\ \left( \overline{\psi^{(n)}_{\nu, \alpha}} \gamma^{[\nu |}\gamma^{\mu_1 \ldots \mu_p} \gamma^{| \rho ]} \psi^{(n)}_{\rho,\beta} \right) M^{\alpha \beta}  \, , \qquad \theta=\dfrac{p-d+1}{d-2}-\dfrac{P-p}{k}+1 \, ,
\end{equation}
where $M^{\alpha \beta}$ represents the volume-independent part that comes from the internal integration with respect to the internal unimodular volume element and can only depend on other internal moduli, which  we do not take to be divergent in the emergence computations in the main text for simplicity.
		
\subsubsection*{11d M-theory on a Calabi--Yau threefold}
	
For concreteness, we let us now specify to the case of a  11d M-theory on a Calabi--Yau threefold. First, we can decompose the $11$d $\Gamma$-matrices as \cite{Grana:2020hyu,Held:2010az}
\begin{equation}
			\label{eq:gammadecomposition}
			\Gamma^{\mu}=\gamma^{\mu} \otimes \gamma_{7}\, , \qquad \Gamma^{4+m}=\mathbb{I} \otimes \gamma^{m}\, ,
\end{equation}
where $\gamma^\mu$ and $\gamma^m$ denote the 5d and 6d $\gamma$-matrices, respectively and similarly with the chirality matrices. (In particular, one has for the chirality operator in the six-dimensional internal space the following $\gamma_{7}=i \gamma^{123456}$.) Moreover, recall that on a Calabi--Yau three-fold there is one \emph{globally} defined internal spinor, which turns out to be also covariantly constant. We denote it by $\xi_+$, together with its charge conjugate $\xi_+^C=\xi_-$, of opposite chirality (i.e. $\gamma_7\  \xi_{\pm}=\pm \xi_{\pm}$). The following bilinears of the internal spinors turn out to be particularly useful (see e.g. Appendix A of \cite{Grana:2020hyu})
\begin{equation}
			\label{eq:xiproperties}
			\begin{array}{rl}
				\xi_+^\dagger \xi_+\, =\, \xi_-^\dagger \xi_- \, =\, 1\\
				\xi_{\pm}^{ \dagger} \gamma^{m_{1} \ldots m_{n}} \xi_{\mp}=0  & \quad \text{for}\ n=0,2,4,6\, ;\\
				\xi_{\pm}^{ \dagger} \gamma^{m_{1} \ldots m_{n}} \xi_{\pm}=0 &  \quad \text{for}\ n =1,3,5\, ,\\
			\end{array}
\end{equation}
and one can also construct both the K\"ahler 2-form and the (unique) holomorphic (3,0)-form of the three-fold in the following way \cite{Koerber:2010bx}:
\begin{equation}
			\label{JOdefinitions}
			J_{i j}=i \xi_+^{\dagger} \gamma_{i j} \xi_+=-i \xi_-^{\dagger} \gamma_{i j} \xi_-\, , \qquad \Omega_{ijk}=\xi_-^{\dagger} \gamma_{ijk} \xi \, .
\end{equation}

These (covariantly) constant spinorial modes are part of a larger group of eigenstates of the Dirac operator ($i \slashed{\nabla}_{\text{CY}}$), so that eq. \eqref{eq:Diraceigenvalueproblemgeneral} now takes the simplified form
\begin{equation}
			\label{eq:Diraceigenvalueproblem}
			-i \gamma^m \nabla_m \xi^{(n)}_{\pm} = \pm \lambda_{(n)} \xi^{(n)}_{\pm}\, ,
\end{equation}
and the eigenspinors satisfy the chirality condition 
\begin{equation}
			\gamma_7\ \xi^{(n)}_{\pm}= \pm \xi^{(n)}_{\pm}\, .
\end{equation}
Of course, the series of eigenvalues starts with $\lambda_{(0)}=0$, whose eigenspinors are precisely the covariantly constant ones, i.e. $\xi^{(0)}_{\pm}=\xi_{\pm}$. Notice that given that $\gamma_7$ is purely imaginary, the eigenfunctions of the hermitian operator $(i \slashed{\nabla}_{\text{CY}})^2$ come in pairs of opposite chirality, even for the zero modes\cite{Green:2012pqa}. They can also be related by complex (charge) conjugation, as it was the case for the zero modes as well as satisfy similar orthogonality conditions as the corresponding zero modes, see eq. \eqref{eq:xiproperties}.
		
Thus, the expansion of the gravitino into into its KK modes, eq. \eqref{eq:gravitinoexpansiongeneral}, now reads
\begin{equation}
\label{eq:gravitinoexpansion+-}
	\begin{split}
				\psi_{\mu}\, =\,  \sum_{n=0}^\infty \left( \psi^{(n)}_{\mu} \otimes \xi^{(n)}_+ +  \psi^{(n),\ C}_{\mu} \otimes \xi^{(n)}_-\right) \, , 
	\end{split}
\end{equation}
where  $\psi^{(n)}_{\mu}, \psi^{(n),\ C}_{\mu}$ are an infinite number of five-dimensional gravitini together with their charge conjugates. These comprise in particular the two massless 5d $\mathcal{N}=2$ symplectic-Majorana gravitini (see e.g. \cite{Aspinwall:1996mn} and footnote \ref{fn:symplecticmajorana}), as well as the tower of massive quaternionic spinors. Notice that this is the most general expansion compatible with the Majorana character of the single 11d gravitino. Of course, we are restricting to the $\mu$ indices since these are the ones that yield the 5d gravitini, but there should also be analogously massless and massive dilatini following from the internal components of the 11d gravitino. 
		
Finally, from the 11d Dirac conjugate we can read the 5d and 6d Dirac conjugates, given by
\begin{equation}
			\overline{\psi_M}=\psi^{\dagger}_M\  \Gamma^0 \ \Longrightarrow \ \overline{\psi^{(n)}_\mu}=\psi^{(n),\ \dagger}_\mu\ \gamma^0 \, , \quad \overline{\xi^{(n)}_\pm}= \xi_\pm^{(n),\ \dagger} \gamma_7\, ,
\end{equation}
where we have used the decomposition reviewed in eq. \eqref{eq:gammadecomposition}.
		
Dimensionally reducing eq. \eqref{eq:P-formgravitinogeneral} in such setup, with $D=11$, $k=6$, and the values of $P$ and $p$ that correspond to the expansion \eqref{eq:11dp-formsexpansion} we obtain the following schematic terms\footnote{Note that we are using here $\mathcal{V}_6$ to denote the volume of the six-dimensional internal manifold in 11d Planck units, but this is denoted $\mathcal{V}_5$ in the main text since it is the one that enters the 5d EFT, in order to distinguish it from the one that appears in the 4d EFT, which is denoted $\mathcal{V}$ and corresponds to the volume of the internal manifold measured in 10d string units.}
\begin{equation}\label{eq:5dp-formpsipsicouplings}
	 \begin{split}
				S^{4d}_{F_4^0\bar{\psi}\psi}=& \, \mathcal{V}_6 \sum_{n,\alpha,\beta} (F_4^0)_{\mu_1 \ldots \mu_4}\ \left( \overline{\psi^{(n)}_{\nu, \alpha}} \gamma^{[\nu |}\gamma^{\mu_1 \ldots \mu_4} \gamma^{| \rho ]} \psi^{(n)}_{\rho,\beta} \right) M_{F_4^0}^{\alpha \beta} \, , \\
				S^{4d}_{F_5^a\bar{\psi}\psi}=& \, \mathcal{V}_6 \sum_{n,\alpha,\beta} (F_5^a)_{\mu_1 \ldots \mu_5}\ \left( \overline{\psi^{(n)}_{\nu, \alpha}} \gamma^{[\nu |}\gamma^{\mu_1 \ldots \mu_5} \gamma^{| \rho ]} \psi^{(n)}_{\rho,\beta} \right) M_{F_5^a}^{\alpha \beta} \, , \\
				S^{4d}_{H_4^I\bar{\psi}\psi}=& \, \mathcal{V}_6^{1/2} \sum_{n,\alpha,\beta} (H_4^I)_{\mu_1 \ldots \mu_4}\ \left( \overline{\psi^{(n)}_{\nu, \alpha}} \gamma^{[\nu |}\gamma^{\mu_1 \ldots \mu_4} \gamma^{| \rho ]} \psi^{(n)}_{\rho,\beta} \right) M_{H_4^I}^{\alpha \beta} \, , \\
		\end{split}
\end{equation}
(with $\alpha\, , \beta=\pm$) which give the volume dependence displayed in eqs. \eqref{eq:5dgravitinop-formvertices} and where we have not explicitly shown the expressions for $M^{\alpha \beta}$  since they are not relevant for our purposes, but they can be easily obtained. For $n=0$ these also serve as the starting point for the reduction from 5d to 4d to extract the dependence on the volume of the $S^1$, denoted $R_5$, which is needed in section \ref{ss:4dpotential}.

\section{Comments about gauge charges in supergravity}
\label{ap:gaugechargessugra}
The aim of this appendix is to provide for an intuitive and perhaps more familiar way of obtaining the electric-like couplings that the different fields comprised by the D0- and D2-particle towers in ten and four dimensions present with respect to the massless 1-forms within each theory. This also avoids the complications associated to keeping track of the details in the relevant circle compactifications involving the 11d and 5d M-theory duals. The main advantage of this method is that it provides us for the precise value of the gauge charges (along with their correct normalization) regardless of our conventions for the Planck scale, Weyl rescalings (with or without the VEVs included), etc. 	
		
\subsection{Gauge charges in 10d $\mathcal{N}=(1,1)$ supergravity}
\label{ss:gaugecharges10d}		
Let us start by recalling that in the presence of a (single) D$p$-brane, the effective action of the ten-dimensional theory gets two additional terms, the so-called Dirac-Born-Infeld (DBI) and Chern-Simons (CS) actions, which in the string frame read as follows (see e.g. \cite{Polchinski:1998rq,Polchinski:1998rr})
\begin{equation}\label{eq:DBICSaction}
		\begin{aligned}
				S_{\text{DBI+CS}} =& -\mu_p \int_{\mathcal{W}_{p+1}}\text{d}^{p+1} \xi\  e^{-\phi} \sqrt{- \text{det} \left( P(g+B_2) +2\pi \alpha' F \right)}\\
				&+ \mu_p \int_{\mathcal{W}_{p+1}} P \left [ \sum_q C_q \wedge e^{B_2+2\pi \alpha' F} \right ]\, ,
		\end{aligned}
\end{equation}
where $P (\cdot)$ denotes the pull-back to the ($p+1$)-dimensional worldvolume of the D$p$-brane, $\mathcal{W}_{p+1}$, which we choose to parameterize with the local coordinates $\xi^{\alpha}$, $\alpha=0, \ldots, p$. The quantity $\mu_p=2 \pi/\ell_s^{p+1}$ is nothing but the $p$-brane charge density (measured in the string frame), whilst its physical tension $\tau_p$ is related to the former as $\tau_p=\mu_p e^{-\phi_0}$. Particularizing now to the type IIA case, notice that for the D0-brane such quantities are simply given by 
\begin{align}
			\label{eq:D0tensionandcharge10d}
			\mu_0= \frac{2 \pi}{\ell_s}\ , \qquad \tau_0= \frac{2 \pi e^{-\phi_0}}{\ell_s}\ , 
\end{align}
where $\ell_s=2 \pi \sqrt{\alpha'}$ denotes the physical string length. In particular, since both in string theory and supergravity the massless fields are conventionally defined so as to be dimensionless, the gauge charges of the different particles arising in the theory are indeed dimension-full, as in \eqref{eq:D0tensionandcharge10d} above.
		
In a next step, we should perform a Weyl rescaling in order to go to the 10d Einstein frame. This allows us to directly obtain the relevant masses and charges (upon using appropriate units) that the D0-brane field-theoretic content presents, which should then enter in the relevant interaction vertices when performing the emergence computations in section \ref{ss:Emergence10dST} as well as in Appendix \ref{ap:MtheoryKKcompactS1}.

\subsection{Gauge charges in 4d $\mathcal{N}=2$ supergravity}
\label{ss:gaugecharges4d}		
Similar considerations apply to the 4d $\mathcal{N}=2$ set-up thoroughly discussed in section \ref{s:emergence4dN=2}. There we considered the four-dimensional theory that arises upon compactifying 10d type IIA string theory on a Calabi--Yau three-fold $X_3$, whose bosonic action reads as
\begin{equation}\label{eq:4dactionapp}
		\begin{aligned}
				S_{\text{IIA}}^{\text{4d}} = \frac{1}{2\kappa^2_4} \int & R \star 1 - \frac{1}{2} \text{Re}\, \mathcal{N}_{AB} F^A \wedge F^B - \frac{1}{2} \text{Im}\, \mathcal{N}_{AB} F^A \wedge \star F^B\\
				&-2G_{a\bar b} dz^a\wedge \star d\bar z^b - 2h_{pq}d  q^p \wedge \star d  q^q \, ,
		\end{aligned}
\end{equation}
where $(2\kappa^2_4)^{-1}= \Mpf^2/2=2 \pi/ \ell_4^2$ and the precise functional form for the different metrics and quantities in the vector multiplet sector was displayed in eqs. \eqref{eq:gaugetopologicalterm}-\eqref{eq:kahlersectormetric} in the main text. It is important to stress here that in order to get to the expression above it was necessary to perform after compactifying on the three-fold the following Weyl rescaling of the 4d metric\cite{Bodner:1990zm}
\begin{align}
			\label{eq:Weylrescaling4d}
			g_{\mu \nu} \to g_{\mu \nu}\, e^{2\varphi_4}\ , 
\end{align}
where $\varphi_4$ is the 4d dilaton, defined in eq. \eqref{eq:4dilaton}. Its VEV, namely $\varphi^0_4 \equiv \braket{\varphi_4}$, can be seen to control the quotient between the Planck and string scales in four dimensions, i.e. $\Mpf^2/M_s^2= 4 \pi e^{-2\varphi^0_4}$. Now, in order to perform the relevant one-loop computation of section \ref{ss:4doneloop} which we recall here for the comfort of the reader,
\beq \label{eq:onelooprunning4d}
		f_{AB}^{\text{UV}} = f_{AB} - \sum_i \frac{\beta_i}{8\pi^2}\ q_{A}^{(i)} q_{B}^{(i)}\ \log \left(\frac{\Lambda_{\text{UV}}}{m_i}\right)\, ,
\eeq
it is therefore necessary to know the mass and charge of each particle running in the loop. Since those states are nothing but D0- and D2-branes (wrapping non-contractible 2-cycles), both quantities can be read very easily from their associated effective dynamics, which is captured by the compactification down to 4d of the DBI and CS actions (see eq. \eqref{eq:DBICSaction} above). Thus, for the D0-brane one finds (in the string frame)
\begin{align}
			\label{eq:D0tensionandcharge4d}
			\mu_0= \frac{2 \pi}{\ell_s}\ , \qquad \tau_0= \frac{2 \pi }{\ell_s\, e^{\phi_0}}\ , 
\end{align}
whilst for the D2-particle one has instead
\begin{align}
			\label{eq:D2tensionandcharge}
			\mu_2^a= \frac{2 \pi}{\ell_s}\ , \qquad \tau_2^a= \frac{2 \pi |z^a|}{\ell_s\, e^{\phi_0}}\ . 
\end{align}
We use the notation $z^a = \int_ {\gamma^a_2} J_c$ to denote the vacuum expectation value of the (complexified) volume associated to the 2-cycle wrapped by the D2-brane, $\gamma_2^a$, as measured in units of $\ell_s$ and with respect to the (string frame) K\"ahler form, $J$. In this appendix we will proceed as in the main text and restrict ourselves to the one-modulus case, where the mass of the \emph{unique} D2-particle becomes the following 
\begin{align}
			\tau_2= \frac{2 \pi \mathcal{V}^{1/3}}{\ell_s\, e^{\phi_0}}\ . 
\end{align}
Notice that for simplicity we have set the VEV of the axion $b$ equal to zero. However, after performing the Weyl rescaling, eq. \eqref{eq:Weylrescaling4d} so as to go to the 4d Einstein frame, both the D0- and D2-brane masses catch via the DBI action some extra dependence on the 4d dilaton VEV
\begin{align}
\label{eq:newD0D2tension}
			\tau_0 \to \frac{\ell_s}{\ell_4}\tau_0\ e^{\varphi_4^0}= \frac{2 \pi}{\ell_4\, \mathcal{V}^{1/2}}\,=\, \frac{ \sqrt{\pi} \Mpf}{\mathcal{V}^{1/2}}\ , \qquad \tau_2\to \frac{\ell_s}{\ell_4}\tau_2\ e^{\varphi_4^0}= \frac{2 \pi }{\ell_4\, \mathcal{V}^{1/6}} \, = \, \frac{\sqrt{\pi} \Mpf }{ \mathcal{V}^{1/6}}\, .
\end{align}
whereas the gauge charges with respect to the $U(1)$ vectors become
\begin{align}
			\mu_0= \frac{2 \pi}{\ell_4}\ , \qquad \mu_2^a= \frac{2 \pi}{\ell_4}\, . 
\end{align}
With these, one could now insert the gauge charges of the BPS states contributing to the one-loop renormalization above, eq. \eqref{eq:onelooprunning4d}, and perform the summation over the entire tower up to the species scale, taking also into account the different masses appearing in the game.
		
\bibliography{refs-emergence}
\bibliographystyle{JHEP}

\end{document}